\pdfoutput=1
\documentclass[12pt,a4paper]{article}

\usepackage{ifthen} \newboolean{pdflatex}
\setboolean{pdflatex}{true} 

\newboolean{articletitles}
\setboolean{articletitles}{true} 

\newboolean{uprightparticles}
\setboolean{uprightparticles}{false}

\def\paperauthors{LHCb collaboration} \def\paperasciititle{Amplitude analysis of B2psi2SKpipi decays} \def\papertitle{Amplitude analysis of $\signal$ decays} \def\paperkeywords{{High Energy Physics}, {LHCb}} \def\papercopyright{\the\year\ CERN for the benefit of the LHCb collaboration} \def\paperlicence{CC BY 4.0 licence}
\def\paperlicenceurl{https://creativecommons.org/licenses/by/4.0/}

\newif\ifEnableSectionTOCLinks
\EnableSectionTOCLinksfalse

\usepackage[top=1in, bottom=1.25in, left=1in, right=1in]{geometry}

\usepackage{microtype}
\usepackage{lineno}  \usepackage{xspace} \usepackage{caption}

\usepackage{graphicx}  \usepackage{color}
\usepackage{colortbl}
\graphicspath{{./figs/}} 

\usepackage{amsmath} \usepackage{amssymb}
\usepackage{amsfonts}
\usepackage{upgreek} 

\newcommand*\patchAmsMathEnvironmentForLineno[1]{\expandafter\let\csname old#1\expandafter\endcsname\csname #1\endcsname
\expandafter\let\csname oldend#1\expandafter\endcsname\csname
end#1\endcsname
 \renewenvironment{#1}{\linenomath\csname old#1\endcsname}{\csname oldend#1\endcsname\endlinenomath}}
\newcommand*\patchBothAmsMathEnvironmentsForLineno[1]{\patchAmsMathEnvironmentForLineno{#1}\patchAmsMathEnvironmentForLineno{#1*}}
\AtBeginDocument{\patchBothAmsMathEnvironmentsForLineno{equation}\patchBothAmsMathEnvironmentsForLineno{align}\patchBothAmsMathEnvironmentsForLineno{flalign}\patchBothAmsMathEnvironmentsForLineno{alignat}\patchBothAmsMathEnvironmentsForLineno{gather}\patchBothAmsMathEnvironmentsForLineno{multline}\patchBothAmsMathEnvironmentsForLineno{eqnarray}}

\usepackage{hyperxmp}

\usepackage[pdftex,
            pdfauthor={\paperauthors},
            pdftitle={\paperasciititle},
            pdfkeywords={\paperkeywords},
            pdfcopyright={Copyright (C) \papercopyright},
            pdflicenseurl={\paperlicenceurl}]{hyperref}

\usepackage[colorinlistoftodos,textsize=scriptsize]{todonotes}

\usepackage[bottom,flushmargin,hang,multiple]{footmisc}

\usepackage[all]{hypcap} 

\usepackage{xspace} 
\usepackage{upgreek}

\def\lhcb   {\mbox{LHCb}\xspace}

\def\MagUp {\mbox{\em Mag\kern -0.05em Up}\xspace}

\ifthenelse{\boolean{uprightparticles}}{

 \def\Pmu         {\ensuremath{\upmu}\xspace}

 \def\Ppi         {\ensuremath{\uppi}\xspace}

 \def\Pphi        {\ensuremath{\upphi}\xspace}                 
                  
 \def\Pchi        {\ensuremath{\upchi}\xspace}                 
 \def\Ppsi        {\ensuremath{\uppsi}\xspace}                 
 \def\Pomega      {\ensuremath{\upomega}\xspace}                 

 \def\PDelta      {\ensuremath{\Delta}\xspace}                 
 \def\PXi         {\ensuremath{\Xi}\xspace}                 
 \def\PLambda     {\ensuremath{\Lambda}\xspace}                 
 \def\PSigma      {\ensuremath{\Sigma}\xspace}                 
 \def\POmega      {\ensuremath{\Omega}\xspace}                 
 \def\PUpsilon    {\ensuremath{\Upsilon}\xspace}
 \let\oldPi\Pi
 \def\PPi         {\ensuremath{\oldPi}\xspace}

 \def\PB      {\ensuremath{\mathrm{B}}\xspace}                 
                  
 \def\PD      {\ensuremath{\mathrm{D}}\xspace}

 \def\PJ      {\ensuremath{\mathrm{J}}\xspace}                 
 \def\PK      {\ensuremath{\mathrm{K}}\xspace}

 \def\Pb      {\ensuremath{\mathrm{b}}\xspace}                 
 \def\Pc      {\ensuremath{\mathrm{c}}\xspace}

 \def\Pi      {\ensuremath{\mathrm{i}}\xspace}

 \def\Ps      {\ensuremath{\mathrm{s}}\xspace}

 \def\thebaroffset{0.0em}
}
{

 \def\Pmu         {\ensuremath{\mu}\xspace}

 \def\Ppi         {\ensuremath{\pi}\xspace}

 \def\Pphi        {\ensuremath{\phi}\xspace}                 
                  
 \def\Pchi        {\ensuremath{\chi}\xspace}                 
 \def\Ppsi        {\ensuremath{\psi}\xspace}                 
 \def\Pomega      {\ensuremath{\omega}\xspace}                 
 \mathchardef\PDelta="7101
 \mathchardef\PXi="7104
 \mathchardef\PLambda="7103
 \mathchardef\PSigma="7106
 \mathchardef\POmega="710A
 \mathchardef\PUpsilon="7107
 \mathchardef\PPi="7105
                  
 \def\PB      {\ensuremath{B}\xspace}                 
                  
 \def\PD      {\ensuremath{D}\xspace}

 \def\PJ      {\ensuremath{J}\xspace}                 
 \def\PK      {\ensuremath{K}\xspace}

 \def\Pb      {\ensuremath{b}\xspace}                 
 \def\Pc      {\ensuremath{c}\xspace}

 \def\Pi      {\ensuremath{i}\xspace}

 \def\Ps      {\ensuremath{s}\xspace}

 \def\thebaroffset{0.18em}
}
\newcommand{\offsetoverline}[2][\thebaroffset]{\kern #1\overline{\kern -#1 #2}}

\makeatletter
\ifcase \@ptsize \relax \newcommand{\miniscule}{\@setfontsize\miniscule{4}{5}}\or \newcommand{\miniscule}{\@setfontsize\miniscule{5}{6}}\or \newcommand{\miniscule}{\@setfontsize\miniscule{5}{6}}\fi
\makeatother

\DeclareRobustCommand{\optbar}[1]{\shortstack{{\miniscule (\rule[.5ex]{1.25em}{.18mm})}
  \\ [-.7ex] $#1$}}

\def\mup        {{\ensuremath{\Pmu^+}}\xspace}

\def\mumu       {{\ensuremath{\Pmu^+\Pmu^-}}\xspace}

\def\Z      {{\ensuremath{\PZ}}\xspace}

\def\squark    {{\ensuremath{\Ps}}\xspace}

\def\cquark    {{\ensuremath{\Pc}}\xspace}

\def\bquark    {{\ensuremath{\Pb}}\xspace}

\def\pion   {{\ensuremath{\Ppi}}\xspace}

\def\pip    {{\ensuremath{\pion^+}}\xspace}
\def\pim    {{\ensuremath{\pion^-}}\xspace}
\def\pipm   {{\ensuremath{\pion^\pm}}\xspace}
\def\pimp   {{\ensuremath{\pion^\mp}}\xspace}

\def\kaon    {{\ensuremath{\PK}}\xspace}
\def\Kbar    {{\ensuremath{\offsetoverline{\PK}}}\xspace}
\def\Kb      {{\ensuremath{\Kbar}}\xspace}
\def\KorKbar {\kern \thebaroffset\optbar{\kern -\thebaroffset \PK}{}\xspace}

\def\Kp      {{\ensuremath{\kaon^+}}\xspace}
\def\Km      {{\ensuremath{\kaon^-}}\xspace}

\def\KS      {{\ensuremath{\kaon^0_{\mathrm{S}}}}\xspace}

\def\Kstarz  {{\ensuremath{\kaon^{*0}}}\xspace}
\def\Kstarzb {{\ensuremath{\Kbar{}^{*0}}}\xspace}

\newcommand{\phiz}{\ensuremath{\Pphi}\xspace}
\newcommand{\omegaz}{\ensuremath{\Pomega}\xspace}

\def\Dbar    {{\ensuremath{\offsetoverline{\PD}}}\xspace}
\def\D       {{\ensuremath{\PD}}\xspace}

\def\DorDbar {\kern \thebaroffset\optbar{\kern -\thebaroffset \PD}\xspace}
\def\Dz      {{\ensuremath{\D^0}}\xspace}
\def\Dzb     {{\ensuremath{\Dbar{}^0}}\xspace}
\def\Dp      {{\ensuremath{\D^+}}\xspace}
\def\Dm      {{\ensuremath{\D^-}}\xspace}

\def\DpDm    {\ensuremath{\Dp {\kern -0.16em \Dm}}\xspace}

\def\Dsmp    {{\ensuremath{\D^{\mp}_\squark}}\xspace}

\def\B       {{\ensuremath{\PB}}\xspace}
\def\Bbar    {{\ensuremath{\offsetoverline{\PB}}}\xspace}

\def\BorBbar {\kern \thebaroffset\optbar{\kern -\thebaroffset \PB}\xspace}
\def\Bz      {{\ensuremath{\B^0}}\xspace}
\def\Bzb     {{\ensuremath{\Bbar{}^0}}\xspace}
\def\Bd      {{\ensuremath{\B^0}}\xspace}

\def\BdorBdbar {\kern \thebaroffset\optbar{\kern -\thebaroffset \Bd}\xspace}
\def\Bu      {{\ensuremath{\B^+}}\xspace}

\def\Bp      {{\ensuremath{\Bu}}\xspace}

\def\Bs      {{\ensuremath{\B^0_\squark}}\xspace}
\def\Bsb     {{\ensuremath{\Bbar{}^0_\squark}}\xspace}
\def\BsorBsbar {\kern \thebaroffset\optbar{\kern -\thebaroffset \Bs}\xspace}

\def\jpsi     {{\ensuremath{{\PJ\mskip -3mu/\mskip -2mu\Ppsi}}}\xspace}
\def\psitwos  {{\ensuremath{\Ppsi{(2S)}}}\xspace}

\def\psires  {{\ensuremath{\Ppsi}}\xspace}

\def\chicone  {{\ensuremath{\Pchi_{\cquark 1}}}\xspace}

\def\Y#1S{\ensuremath{\PUpsilon{(#1S)}}\xspace}

\def\LorLbar     {\kern \thebaroffset\optbar{\kern -\thebaroffset \PLambda}\xspace}

\newcommand{\decay}[2]{\ensuremath{#1\!\to #2}\xspace} 

\def\to                 {\ensuremath{\rightarrow}\xspace}

\def\eps   {{\ensuremath{\varepsilon}}\xspace}

\def\CP                {{\ensuremath{C\!P}}\xspace}

\def\AT#1     {\ensuremath{A_{\mathrm{T}}^{#1}}\xspace}

\def\C#1      {\ensuremath{\mathcal{C}_{#1}}\xspace}                       \def\Cp#1     {\ensuremath{\mathcal{C}_{#1}^{'}}\xspace}                    \def\Ceff#1   {\ensuremath{\mathcal{C}_{#1}^{\mathrm{(eff)}}}\xspace}        \def\Cpeff#1  {\ensuremath{\mathcal{C}_{#1}^{'\mathrm{(eff)}}}\xspace}       \def\Ope#1    {\ensuremath{\mathcal{O}_{#1}}\xspace}                       \def\Opep#1   {\ensuremath{\mathcal{O}_{#1}^{'}}\xspace}

\newcommand{\nospaceunit}[1]{\ensuremath{\text{#1}}}       
\newcommand{\aunit}[1]{\ensuremath{\text{\,#1}}}

\newcommand{\tev}{\aunit{Te\kern -0.1em V}\xspace}
\newcommand{\gev}{\aunit{Ge\kern -0.1em V}\xspace}
\newcommand{\mev}{\aunit{Me\kern -0.1em V}\xspace}
\newcommand{\kev}{\aunit{ke\kern -0.1em V}\xspace}
\newcommand{\ev}{\aunit{e\kern -0.1em V}\xspace}
 
\newcommand{\mevc}{\ensuremath{\aunit{Me\kern -0.1em V\!/}c}\xspace}
\newcommand{\gevc}{\ensuremath{\aunit{Ge\kern -0.1em V\!/}c}\xspace}
\newcommand{\mevcc}{\ensuremath{\aunit{Me\kern -0.1em V\!/}c^2}\xspace}
\newcommand{\gevcc}{\ensuremath{\aunit{Ge\kern -0.1em V\!/}c^2}\xspace}

\def\mum  {\ensuremath{\,\upmu\nospaceunit{m}}\xspace}

\def\fb   {\ensuremath{\aunit{fb}}\xspace}
\def\invfb   {\ensuremath{\fb^{-1}}\xspace}

\def\ps   {\ensuremath{\aunit{ps}}\xspace}

\def\gsim{{~\raise.15em\hbox{$>$}\kern-.85em
          \lower.35em\hbox{$\sim$}~}\xspace}
\def\lsim{{~\raise.15em\hbox{$<$}\kern-.85em
          \lower.35em\hbox{$\sim$}~}\xspace}

\def\pt         {\ensuremath{p_{\mathrm{T}}}\xspace}

\def\ptot       {\ensuremath{p}\xspace}

\def\evtgen     {\mbox{\textsc{EvtGen}}\xspace}

\def\geant      {\mbox{\textsc{Geant4}}\xspace}

\def\photos     {\mbox{\textsc{Photos}}\xspace}

\def\pythia     {\mbox{\textsc{Pythia}}\xspace}

\def\tell1  {TELL1\xspace}
\def\ukl1   {UKL1\xspace}

\newcommand{\eg}{\mbox{\itshape e.g.}\xspace}
\newcommand{\ie}{\mbox{\itshape i.e.}\xspace}

\newcommand{\cf}{\mbox{\itshape cf.}\xspace}

\newcommand{\lhcborcid}[1]{\href{https://orcid.org/#1}{\hspace*{0.1em}\raisebox{-0.45ex}{\includegraphics[width=1em]{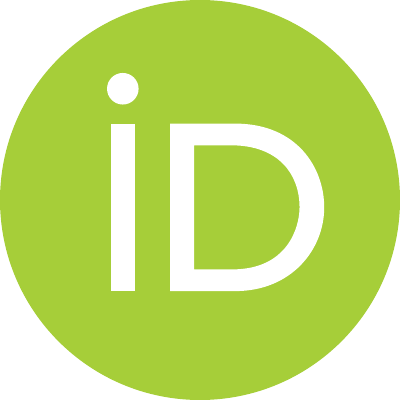}}}}

\hypersetup{
  colorlinks   = true, urlcolor     = blue, linkcolor    = blue, citecolor    = red   }

\ifEnableSectionTOCLinks
    \usepackage[explicit]{titlesec} 

\let\oldcontentsline\contentsline
    \renewcommand

\titleformat{\section}{\normalfont\Large\bf}{\hyperlink{tocsection.\thesection}{{\thesection} \parbox[t]{\dimexpr\textwidth-1pc}{#1}}}{1pc}{}

\titleformat{\subsection}{\normalfont\bf}{\hyperlink{tocsubsection.\thesubsection}{{\thesubsection} \parbox[t]{\dimexpr\textwidth-1pc}{#1}}}{1pc}{}

\titleformat{name=\section,numberless}[display]{}{}{0pt}{\normalfont\Huge\bfseries #1}
\fi

\usepackage{cite} \usepackage{mciteplus}
 
\newcommand{\phs}{\ensuremath{\Phi}}  \newcommand{\phsd}{\ensuremath{\phi}} \newcommand{\dphs}{\ensuremath{\mathrm{d}\phs}}  

\newcommand{\phsPoint}{\ensuremath{\mathbf{x}}}
\newcommand{\dphsPoint}{\ensuremath{\mathrm{d}^7x}}

\newcommand{\prt}[1]{\ensuremath{#1}}

\newcommand{\signal}{\prt{\Bp \to \psitwos \Kp\pip\pim}}
\newcommand{\norm}{\prt{\Bp \to \jpsi \Kp\pip\pim}}

\def\Xz   {{\ensuremath{X^0}}\xspace}

\def\XSone   {{\ensuremath{ T_{c\bar c0}^*(4475)^0 }}\xspace}
\def\XStwo   {{\ensuremath{ T_{c\bar c0}^*(4710)^0 }}\xspace}
\def\XAone   {{\ensuremath{ T_{c\bar c1}(4650)^0 }}\xspace}
\def\XVone   {{\ensuremath{ T_{c\bar c1}^*(4800)^0 }}\xspace}

\def\Xsz      {{\ensuremath{T_{c \bar c \bar s}^0}}\xspace}
\def\Xspp      {{\ensuremath{T_{c \bar c \bar s}^{++}}}\xspace}
\def\XsA      {{\ensuremath{T_{c \bar c \bar s 1}}}\xspace}
\def\XsV      {{\ensuremath{T^{*}_{c \bar c \bar s 1}}}\xspace}

\def\XsAone      {{\ensuremath{ \XsA(4600)^0 }}\xspace}
\def\XsAtwo      {{\ensuremath{ \XsA(4900)^0 }}\xspace}
\def\XsVone      {{\ensuremath{ \XsV(5200)^0 }}\xspace}

\def\Z   {{\ensuremath{T_{c\bar c}}}\xspace}
\def\Zp   {{\ensuremath{T_{c\bar c}^+}}\xspace}
\def\Zm   {{\ensuremath{T_{c\bar c}^-}}\xspace}
\def\Zpm   {{\ensuremath{T_{c\bar c}^\pm}}\xspace}

\def\ZA   {{\ensuremath{T_{c\bar c1}}}\xspace}
\def\ZV   {{\ensuremath{T^{*}_{c\bar c1}}}\xspace}

\def\ZAone   {{\ensuremath{ \ZA(4200) }}\xspace}
\def\ZAtwo   {{\ensuremath{ \ZA(4430) }}\xspace}

\def\ZVone   {{\ensuremath{ \ZV(4055) }}\xspace}

\def\ZsA   {{\ensuremath{T_{c\bar c \bar s 1}}}\xspace}

\def\Zsp   {{\ensuremath{T_{c\bar c \bar s}^+}}\xspace}

\def\ZsAone   {{\ensuremath{ \ZsA(4000)^+ }}\xspace}
\def\ZsAtwo   {{\ensuremath{ \ZsA(4220)^+ }}\xspace}

\makeatletter
\g@addto@macro\bfseries{\boldmath}
\makeatother

\usepackage{longtable} \usepackage{subcaption}
\usepackage{rotating}

\begin{document}

\renewcommand{\thefootnote}{\fnsymbol{footnote}}
\setcounter{footnote}{1}

\begin{titlepage}
\pagenumbering{roman}

\vspace*{-1.5cm}
\centerline{\large EUROPEAN ORGANIZATION FOR NUCLEAR RESEARCH (CERN)}
\vspace*{1.5cm}
\noindent
\begin{tabular*}{\linewidth}{lc@{\extracolsep{\fill}}r@{\extracolsep{0pt}}}
\ifthenelse{\boolean{pdflatex}}{\vspace*{-1.5cm}\mbox{\!\!\!\includegraphics[width=.14\textwidth]{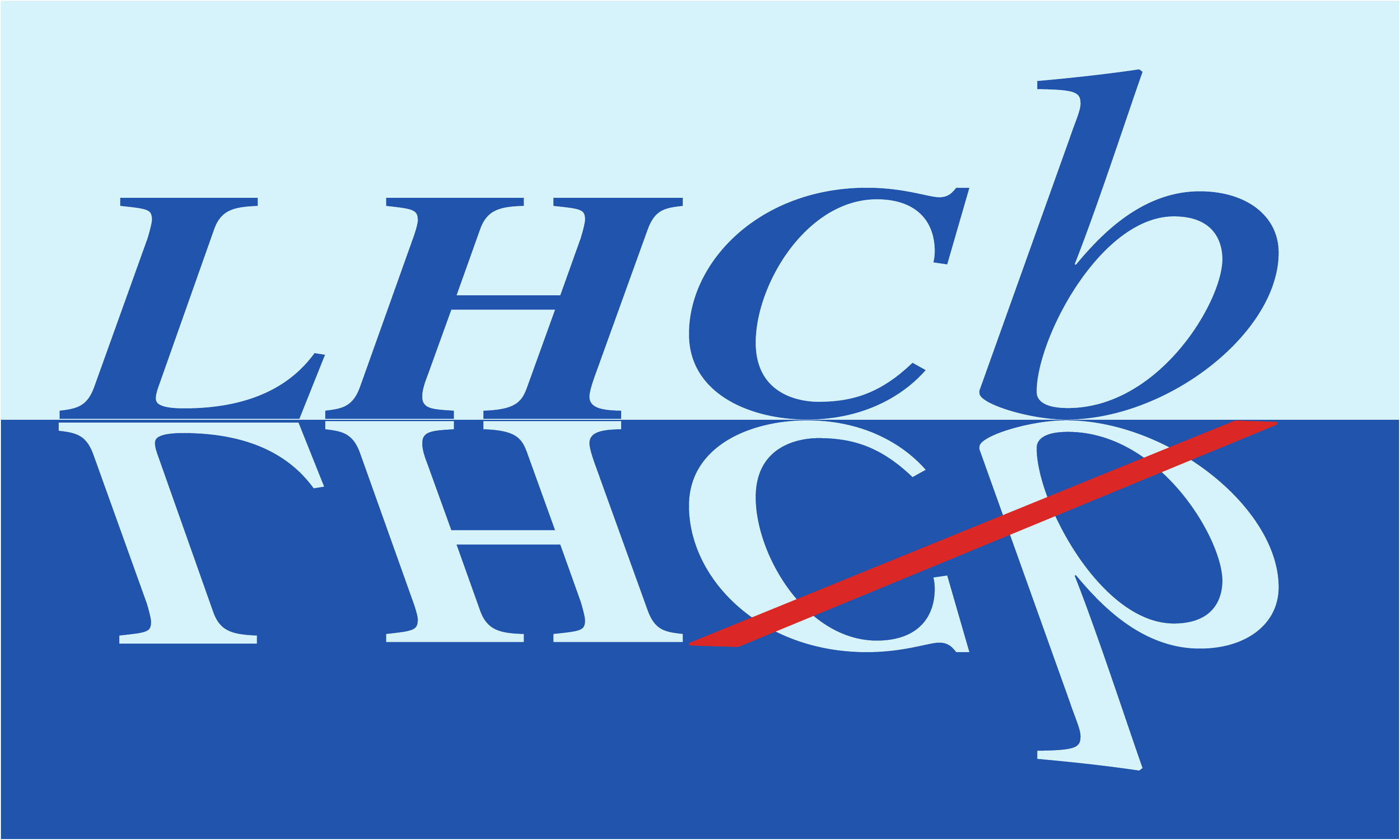}} & &}{\vspace*{-1.2cm}\mbox{\!\!\!\includegraphics[width=.12\textwidth]{figs/lhcb-logo.eps}} & &}\\
 & & CERN-EP-2024-177 \\  & & LHCb-PAPER-2024-014 \\  & & January 8, 2025 \\ & & \\
\end{tabular*}

\vspace*{4.0cm}

{\normalfont\bfseries\boldmath\huge
\begin{center}
\papertitle 
\end{center}
}

\vspace*{2.0cm}

\begin{center}
\paperauthors\footnote{Authors are listed at the end of this paper.}
\end{center}

\vspace{\fill}

\begin{abstract}
  \noindent
  The first full amplitude analysis of $B^+ \to \psi(2S) K^+ \pi^+ \pi^-$ decays
   is performed
  using proton-proton collision data corresponding to an integrated luminosity of $9\,\text{fb}^{-1}$ recorded with the LHCb detector.
   The rich $K^+ \pi^+ \pi^-$ spectrum is studied and
  the branching fractions of the resonant substructure associated with the prominent $K_1(1270)^+$ contribution are measured.  
  The data cannot be described by conventional strange and charmonium resonances only.  
  An amplitude model with 53 components is developed comprising 11 hidden-charm exotic hadrons.
  New production mechanisms for charged charmonium-like states are observed.
  Significant resonant activity with spin-parity $J^P = 1^+$ in the $\psi(2S) \pi^+$ system is confirmed and
  a multi-pole structure is demonstrated.
  The spectral decomposition of the $\psi(2S) \pi^+ \pi^-$ invariant-mass structure, dominated by $X^0 \to \psi(2S) \rho(770)^0$ decays, 
  broadly resembles the $J/\psi \phi$ spectrum observed in $B^+ \to J/\psi \phi K^+$ decays.
  Exotic $\psi(2S) K^+ \pi^-$ resonances are observed for the first time.
  
\end{abstract}

\vspace*{2.0cm}

\begin{center}
  Published in  JHEP 01 (2025) 054
\end{center}

\vspace{\fill}

{\footnotesize 
\centerline{\copyright~\papercopyright. \href{\paperlicenceurl}{\paperlicence}.}}
\vspace*{2mm}

\end{titlepage}

\newpage
\setcounter{page}{2}
\mbox{~}

\renewcommand{\thefootnote}{\arabic{footnote}}
\setcounter{footnote}{0}

\cleardoublepage

\pagestyle{plain} \setcounter{page}{1}
\pagenumbering{arabic}


\section{Introduction}
\label{sec:Introduction}

Multibody beauty to charmonium decays
have played a pivotal role in hadron spectroscopy.
Two decades ago, the Belle collaboration observed the narrow $\chi_{c1}(3872)$ state in \mbox{$\Bp \to \chi_{c1}(3872) \Kp \to [\jpsi \pip \pim] \Kp$} decays\footnote{Inclusion of charge-conjugate modes is implied throughout this paper except where explicitly stated.}~\cite{XBelle}.
This discovery challenged our understanding of the strong interaction 
and marked the beginning of spectroscopic studies of hadrons formed by more than three quarks, known as exotic hadrons.
Since then, the $\chi_{c1}(3872)$ meson's quantum numbers,
lineshape
and decay modes
have been extensively studied~\cite{Johnson:2024omq}.
However, its true nature remains unclear, with interpretations ranging from conventional charmonium ($c\bar c$) \cite{Xcharm} 
to exotic states such as $D^{*0} \Dzb$ molecules \cite{Xmolecule}, tetraquarks ($c \bar c q \bar q$) \cite{Xquark} or quark-gluon hybrids ($c \bar c g$) \cite{Xhybrid}. 
The discovery of the charged charmonium-like $\ZA(4430)^-$ state in \mbox{$B^{0} \to \ZA(4430)^- \Kp \to [\psi(2S) \pi^{-}] \Kp$} decays~\cite{Belle:2009lvn,LHCb-PAPER-2014-014,Belle:2013shl} 
represents another milestone for the study of exotic-hadron spectroscopy.
The charged nature of the $\ZA(4430)^-$ state eliminates the possibility of a conventional charmonium interpretation, 
making it a potential tetraquark state with a minimal quark content of $c\bar c u \bar d$.
Meanwhile, numerous states that do not conform to the traditional hadron picture have emerged, forming an exotic particle zoo 
whose structure largely remains a mystery~\cite{Johnson:2024omq}.

To deepen our understanding of exotic states, it is crucial to confirm their existence in independent decay modes.
The decay $\signal$ provides an excellent laboratory for this purpose as its topology enables a 
search for $\Zm$ resonances, observed in $B^0 \to \Zm K^+ \to [\psi(2S) \pi^-] K^+$ decays~\cite{Belle:2009lvn,LHCb-PAPER-2014-014,Belle:2013shl},
and $\Zsp$ resonances, observed in 
\mbox{$B^+ \to \Zsp \phi \to [\jpsi K^+] \phi$} decays~\cite{LHCb-PAPER-2020-044}.
Unlike these three-body \B-meson decay modes, the decay $\signal$ may also receive contributions from $\Xz \to \psi(2S)  \pi^+ \pi^-$ 
or $\Xsz \to \psi(2S) K^+ \pi^-$ states, 
including the possibility of exotic cascade decays, \eg $\Xz \to [\Zm \to \psi(2S) \pi^-] \pi^+$, which have not yet been established.
A better understanding of the rich $K^+ \pi^+ \pi^-$ resonance structure is also
critical for analyses of $B^+ \rightarrow K^+ \pi^+ \pi^- \gamma$ and $B^+ \rightarrow K^+ \pi^+ \pi^- \mu^+ \mu^-$ decays~\cite{Bellee:2019qbt, LHCb-PAPER-2014-030},
which are sensitive to phenomena beyond the Standard Model.

The Belle collaboration conducted the first and (to date) only study of the resonance structure in $\signal$ decays~\cite{BELLE}. 
Due to the limited sample size, with fewer than 1000 signal events, a simplified amplitude fit 
in the three invariant-mass-squared combinations $m^{2}(\Kp \pip \pim)$, $m^{2}(\Kp \pim)$, and $m^{2}(\pip \pim)$ was performed that could not account for contributions from exotic states.
The LHCb collaboration studied the $\psi_2(3823)$ and $\chi_{c1}(3872)$ states in $B^+ \to \jpsi \Kp \pip \pim$ decays 
in one-~\cite{LHCb-PAPER-2020-009} or two-~\cite{LHCb-PAPER-2021-045} dimensional analyses that considered only narrow regions of the phase space.
This paper presents the first four-body amplitude analysis with a vector particle in the final state exploiting the full seven-dimensional phase space of $\signal$ decays. 
The data sample was collected with the LHCb detector in proton-proton ($pp$)
collisions at centre-of-mass energies\footnote{Natural units with $\hbar = c = 1$ are used throughout the paper.}
 of 7, 8 and $13 \tev$, corresponding to an integrated luminosity of $9 \invfb$.

The paper is structured as follows. 
After a description of the LHCb detector and the simulation in Sec.~\ref{sec:Detector},
the event reconstruction and candidate selection are described in Sec.~\ref{sec:Selection}.
The amplitude analysis formalism is discussed in Sec.~\ref{sec:Formalism},
followed by the fit results presented in Sec.~\ref{sec:results}.
Experimental and model-dependent systematic uncertainties are evaluated in
Sec.~\ref{sec:sys} and our conclusions are given in Sec.~\ref{sec:Summary}.

\section{Detector and simulation}
\label{sec:Detector}

The \lhcb detector~\cite{Alves:2008zz,LHCb-DP-2014-002} is a single-arm forward
spectrometer covering the \mbox{pseudorapidity} range $2<\eta <5$,
designed for the study of particles containing \bquark or \cquark
quarks. The detector includes a high-precision tracking system
consisting of a silicon-strip vertex detector (VELO) surrounding the $pp$
interaction region~\cite{LHCb-DP-2014-001}, a large-area silicon-strip detector located
upstream of a dipole magnet with a bending power of about
$4{\mathrm{\,T\,m}}$, and three stations of silicon-strip detectors and straw
drift tubes~\cite{LHCb-DP-2013-003,LHCb-DP-2017-001} placed downstream of the magnet.
The polarity of the dipole magnet is reversed periodically throughout the data-taking process to control systematic asymmetries.
The tracking system provides a measurement of the momentum, \ptot, of charged particles with
a relative uncertainty that varies from 0.5\% at low momentum to 1.0\% at 200\gev.
The minimum distance of a track to a primary vertex (PV), the impact parameter (IP),
is measured with a resolution of $(15+29/\pt)\mum$,
where \pt is the component of the momentum transverse to the beam, in\,\gev.
Different types of charged hadrons are distinguished using information
from two ring-imaging Cherenkov detectors (RICH)~\cite{LHCb-DP-2012-003}.

The online event selection is performed by a trigger~\cite{LHCb-DP-2012-004},
which consists of a hardware stage, based on information from the calorimeter and muon
systems, followed by a software stage, which applies a full event
reconstruction.
At the hardware trigger stage, events are required to have a muon with high \pt or a
hadron, photon or electron with high transverse energy in the calorimeters. For hadrons,
the transverse energy threshold is 3.5\gev.
The software trigger requires a two-, three- or four-track
secondary vertex with a significant displacement from any primary
$pp$ interaction vertex. At least one charged particle
must have a transverse momentum $\pt > 1.6\gev$ and be
inconsistent with originating from a PV\@.
A multivariate algorithm~\cite{BBDT} is used for
the identification of secondary vertices consistent with the decay
of a \bquark hadron.

Simulated events are used to study the detector acceptance and
specific background contributions.
In the simulation, $pp$ collisions are generated using
\pythia~\cite{Sjostrand:2006za,*Sjostrand:2007gs} with a specific \lhcb
configuration~\cite{LHCb-PROC-2010-056}.  Decays of hadrons
are described by \evtgen~\cite{Lange:2001uf}, in which final-state
radiation is generated using \photos~\cite{Golonka:2005pn}.
The simulated signal decays are generated according to a simplified
amplitude model
with an additional pure phase-space component.
The interaction of the generated particles with the detector, and its response,
are implemented using the \geant
toolkit~\cite{Allison:2006ve, *Agostinelli:2002hh} as described in
Ref.~\cite{LHCb-PROC-2011-006}.
The simulation is corrected to match the distributions observed in data of the $\Bu$
kinematics and the number of tracks in an event.

\section{Event reconstruction}
\label{sec:Selection}

The selection of $\signal$ candidates is performed by first reconstructing 
$\psitwos \to \mu^+ \mu^-$  candidates from two oppositely charged particles, identified as muons, that originate from a common vertex displaced from the PV.
Candidate $\psitwos$ mesons with a reconstructed invariant mass within $\pm50\mev$ of the known $\psitwos$ mass~\cite{PDG2022} are
combined with three additional charged tracks to form a $\Bu$ vertex, which must be significantly displaced from any PV.
Hadron candidates are required to be positively identified in the RICH detectors and the hadron candidate most likely to be a kaon is selected as the kaon candidate.
The $\Bu$ proper decay time is required to exceed $0.3\ps$ to suppress most of the prompt combinatorial background.
The reconstructed invariant mass of the $\Bu$ candidate is required to be between $5200\mev$ and $5600\mev$.
Its resolution is improved by performing a kinematic fit~\cite{Hulsbergen:2005pu}
in which the $\Bu$ candidate is constrained to
originate from the associated PV and the reconstructed $\psitwos$ invariant mass is constrained to its known value~\cite{PDG2022}.
The four-momenta of the final-state particles for the amplitude analysis are taken from a kinematic fit that 
additionally constrains the reconstructed $\Bu$ invariant mass to its known value~\cite{PDG2022}.
Kinematic requirements are used
to veto $\Bd$ decays
to $\psitwos K^+ \pim$ or
$\psitwos \phi (\to \Kp \Km)$,
where the $K^-$ is misidentified as a pion,
that are combined with a random pion.
Candidates with an opening angle between any two tracks smaller than 0.5\,mrad are removed to exclude clone tracks, 
\ie tracks which share a large fraction of their hits.

A gradient boosted decision tree (BDT)~\cite{Breiman,AdaBoost} is used to suppress background from random combinations of charged particles.
The multivariate classifier is trained using as signal proxy a background-subtracted~\cite{Pivk:2004ty} data sample 
of $(732.7\pm 1.3) \times 10^3$ \mbox{$\norm$} signal candidates 
selected in the same way as the $\signal$ sample. 
Candidate $\signal$ decays with invariant mass greater than $5400 \mev$ are used as background proxy.
The features used in the BDT classifier are topological variables
related to the vertex separation,
the transverse momentum of the $\Bu$ candidate and its flight distance with respect to the associated PV, 
as well as several variables quantifying the track quality, vertex reconstruction and particle identification.
The selection on the output of the BDT classifier is chosen to optimise the significance of the $\signal$ signal.

\begin{figure}[h]
\centering
\includegraphics[height=!,width=0.49\textwidth]{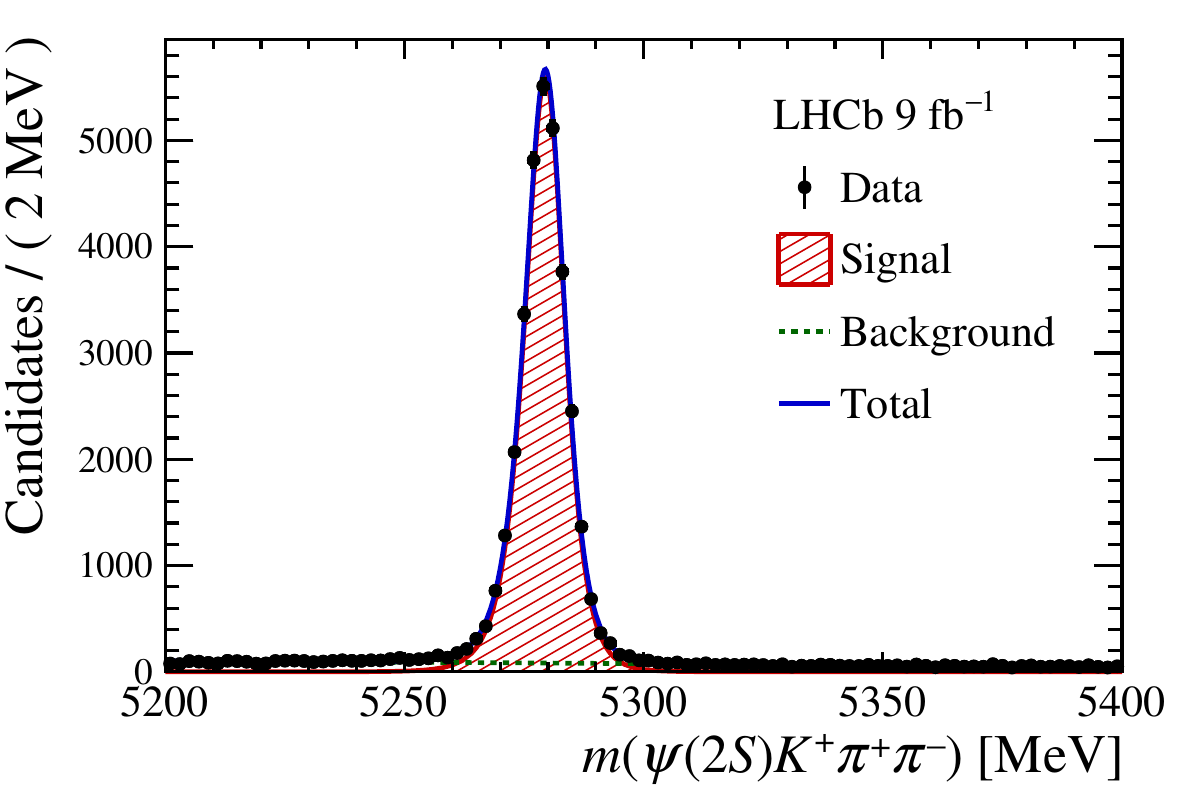}
\includegraphics[height=!,width=0.49\textwidth]{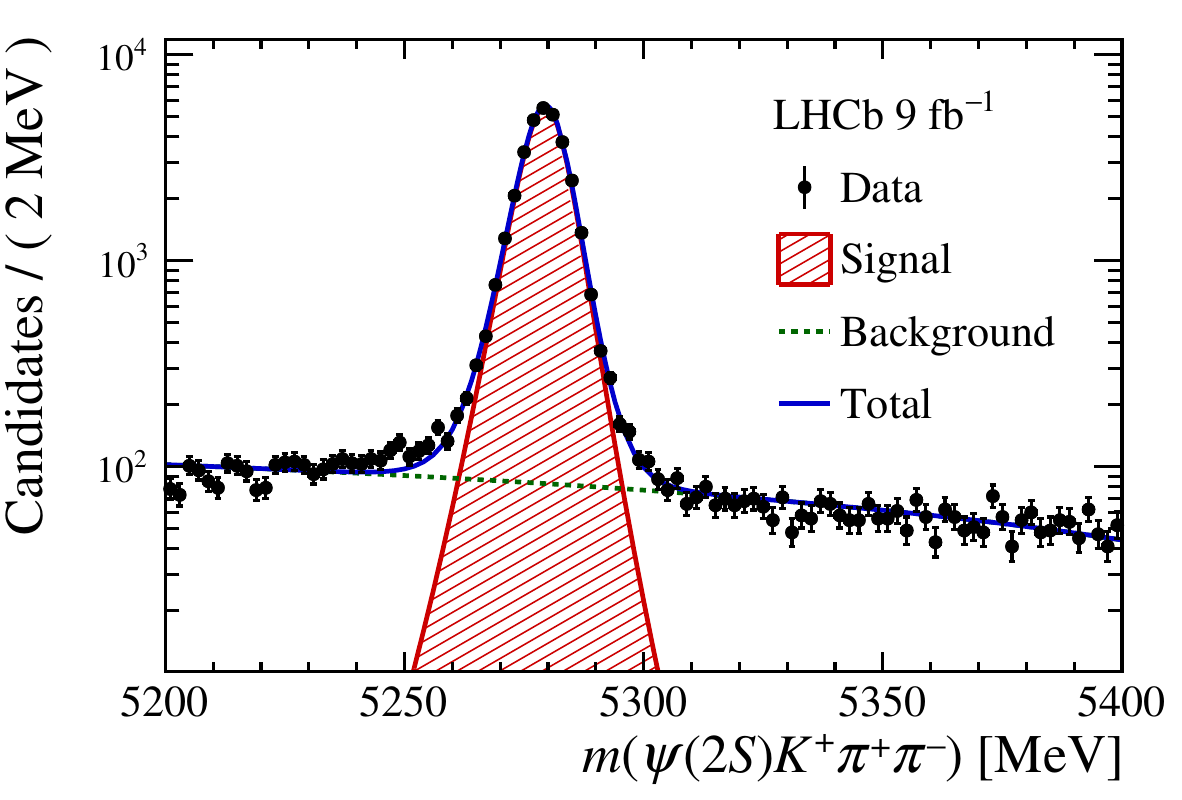}
\caption{Invariant-mass distribution of selected \mbox{\signal} candidates with fit projection overlaid in (left) linear and (right) logarithmic scale.}
\label{fig:massFit}
\end{figure}

Figure~\ref{fig:massFit} displays the $m(\psitwos K^+ \pip \pim)$ distribution of the selected $\Bu$ candidates.
Only candidates within the signal region, defined as $m(\psitwos K^+\pi^+\pi^-)\in[5268,5290] \mev$, are retained for the amplitude analysis.
This range corresponds to approximately 
twice the mass resolution around the known $\Bp$ mass.
An extended, unbinned maximum-likelihood fit is performed to the reconstructed $m(\psitwos K^+ \pip \pim)$ distribution to determine the event yields.
The signal component is modelled using a Johnson $S_U$ function~\cite{10.2307/2332539}, while the combinatorial background is described by a second-order polynomial function.
In the signal region, the signal yield is $30210 \pm 180$ with a background fraction of $f_{\text{B}} = (2.93 \pm 0.05)\%$.

\section{Phenomenology of the decay}
\label{sec:Formalism}

The differential decay rate of a $\Bp$ meson 
with mass $m_{B^+}$,
decaying into three pseudoscalar particles and two muons with four-momenta $p_{i}= (E_{i},\, \vec p_{i}) \, (i \in \{1,2,3,4,5\})$,
is given by
\begin{equation}
	\text{d}\Gamma = \frac{1}{2 \, m_{B^+}} \,  \vert \mathcal A_{B^+}(\phsPoint) \vert^{2} \, \dphs   \, ,
	\label{eq:decayRate}
\end{equation}
where the transition amplitude $\mathcal A_{B^+}(\phsPoint)$ describes the dynamics of the interaction
as a function of the position in $\Bp$ decay phase space, $\phsPoint$,
and 
\dphs\ is the phase-space element \cite{Peskin}.
Each final-state particle contributes three observables,
manifesting in its three-momentum,
summing up to fifteen observables in total.
Four of them are redundant due to four-momentum conservation and
another three are removed due to angular momentum conservation.
Moreover, the dimuon invariant mass is fixed to the known $\psitwos$ mass.
The remaining seven independent degrees of freedom unambiguously determine the kinematics of the decay.
Convenient choices for the kinematic observables
include the invariant-mass combinations of the final-state particles
$m^{2}_{ij} = (p_{i}+p_{j})^{2} , 
m^{2}_{ijk} = (p_{i}+p_{j}+p_{k})^{2},$
with $i,j,k \in \{\psitwos, K^+, \pip, \pim\}$ as well as 
the $\psitwos$ helicity angle and the acoplanarity angle.
These are defined as the angle between the $\mup$ direction
and the $B^+$ direction in the $\psitwos$ rest frame, $\theta$;
and the angle between the $\psitwos$ decay plane 
and the one formed by the $K^+$ and the $(\pi^+\pi^-)$ system, $\chi$~\cite{Beneke:2006hg,LHCb-PAPER-2014-068}.
No particular seven-dimensional
basis is chosen in the amplitude fit, but the full four-vectors are used.
The dimensionality is handled by the phase-space element which can be written in terms of any complete set of 
seven independent kinematic observables, $\phsPoint = (x_1, \ldots, x_7)$, as $\dphs = \phsd(\phsPoint) \, \dphsPoint$,
where $\phi(\phsPoint ) = \left\vert  \partial \phs / \partial(x_1, \ldots x_7) \right\vert$ is the phase-space density~\cite{kinematics}.
In contrast to three-body decays, the phase-space density
function is not uniform in the usual kinematic variables.  

The square of the total amplitude for the 
$\signal$ decay is given by the spin-averaged coherent sum over all 
intermediate-state amplitudes $A_{i}(\phsPoint \vert \lambda_{\mu^+}, \lambda_{\mu^-} )$, 
\begin{equation}
	\vert \mathcal A_{B^+}(\phsPoint ) \vert^2 = \sum_{\lambda_{\mu^+}, \,\lambda_{\mu^-} } \left\vert \sum_{i}  a_{i} \, A_{i}(\phsPoint \vert \lambda_{\mu^+}, \lambda_{\mu^-} )  \right\vert^2  \, ,
\end{equation}
where $\lambda_{\mu^+} (\lambda_{\mu^-})$ denotes the $\mu^+ (\mu^-)$ spin projection.
The complex coefficients $a_{i} = \vert a_{i} \vert \, e^{i \, \phi_i}$ 
are to be determined from data.
To construct the intermediate-state amplitudes $A_{i}(\phsPoint )$,
the isobar approach is used, which 
assumes that
the decay process can be factorised into subsequent two-body decay amplitudes \cite{isobar1,isobar,isobar2}.
This gives rise to two different decay topologies:
quasi two-body decays such as
\mbox{$B^+ \to R_{1} \, R_{2}$}
with 
\mbox{$R_{1} \to h_{1}\,h_{2}$ and $R_{2} \to h_{3}\, \psitwos $} 
or cascade decays such as
\mbox{$B^+ \to \psitwos \, R_{1}$}
with 
\mbox{$R_{1} \to h_{1} \, R_{2}$}
and 
\mbox{$R_{2} \to h_{2} \, h_{3}$},
where the $\psitwos$ meson further decays into a muon pair.
In either case, 
the 
amplitude is parameterised as a product of
orbital angular momentum, $L$, dependent 
form factors $B_{L}$, included for each vertex of the decay tree; 
Breit--Wigner propagators $T_{R}$,  included for each resonance $R$;
and an overall angular distribution represented by a spin factor $S$,
\begin{equation}
	A_{i}(\phsPoint \vert \lambda_{\mu^+}, \lambda_{\mu^-}  ) =  B_{L_{B_s}}(\bold x) \, [B_{L_{R_{1}}}(\phsPoint )  \, T_{R_{1}}(\phsPoint )] \, [B_{L_{R_{2}}}(\phsPoint ) \, T_{R_{2}}(\phsPoint )]  \,  S_{i}(\phsPoint \vert \lambda_{\mu^+}, \lambda_{\mu^-}  )  \, .
	\label{eq:amp4}
\end{equation}
The lineshape of the $\psitwos$ resonance is set to 
a delta function and contributes a constant factor multiplying the final matrix element after integrating over the dimuon invariant mass.

\subsection{Form factors and resonance lineshapes}
\label{sec:lineshapes}

To account for the finite size of the decaying resonances,
the Blatt--Weisskopf penetration factors, 
derived in Ref.~\cite{Bl2}
by assuming a square well interaction potential with radius $r_{\rm BW}$,
are used as form factors, $B_L$.
They depend on
the orbital angular momentum $L$ between the resonance decay products,
and 
the breakup momentum $q$ defined as the magnitude of the three-momentum of one of the decay products in the rest frame of the  resonance.
Their explicit expressions for $L=\{0,1,2\}$ are
\begin{align}
	B_{0}(q)  &= 1,  &
	B_{1}(q)  &= 1 / \sqrt{{1+ (q \, r_{\rm BW})^{2}}} \, \,\, \text{and}  &
	B_{2}(q)  &= 1 / \sqrt{9+3\,(q\, r_{\rm BW})^{2}+(q \, r_{\rm BW})^{4}}.
\end{align}
Resonance lineshapes
are described as functions of the energy-squared, $s$, by Breit--Wigner propagators
\begin{equation}
	T_R(s) = (m_0^{2} - s - i\,m_0 \,\Gamma(s))^{-1},
	\label{eq:BW}
\end{equation}
where the total width, $\Gamma(s)$, is normalised to give the nominal width, $\Gamma_{0}$, when evaluated at the nominal mass $m_{0}$.
For a decay into two stable particles $R \to AB$, the energy dependence of the decay width can be described by 
\begin{equation}
	\Gamma_{R \to AB}(s) = \Gamma_{0} \, \frac{m_{0}}{\sqrt s} \, \left(\frac{q}{q_{0}}\right)^{2L+1} \, \frac{B_{L}(q)^{2}}{B_{L}(q_{0})^{2}}  \, ,
	\label{eq:gamma2}
\end{equation}
where $q_{0}$ is the value of the breakup momentum evaluated at the nominal mass \cite{BW}.
The energy-dependent width for a three-body decay $R \to ABC$
is computed numerically by integrating the transition amplitude-squared over the phase space,
\begin{equation}
	\Gamma_{R \to ABC}(s) =  \frac{1}{m_0} \int \vert A_{R \to ABC} \vert^{2} \, \text{d}\Phi_{3}   ,
	\label{eq:gamma3}
\end{equation}
as described in Ref.~\cite{dArgent:2017gzv}.
The three-body amplitude $A_{R \to ABC}$ is parameterised analogously to the amplitude in Eq.~\ref{eq:amp4}.
To calculate the energy-dependent width for decays of the type $R \to A B (\psitwos \to \mu^+ \mu^-)$, 
we treat them as three-body $R \to AB\psitwos$ decays since the $\psitwos$ is sufficiently narrow to be considered stable.
The running-width distributions for various three-body resonances, calculated from Eq.~\ref{eq:gamma3}, are shown in Appendix~\ref{a:AmpLineShapes}.

Equation~\ref{eq:BW}, with the energy-dependent width from Eq.~\ref{eq:gamma2}, is used by default for resonances decaying into a two-body final state.
For the $\rho(770)^{0}$ resonance, the Gounaris--Sakurai parameterisation is used instead \cite{GS}, see Appendix \ref{a:AmpLineShapes}.
We use the parameterisation from Ref.~\cite{Akhmetshin:2001ig} to include $\rho-\omega$ mixing,
\begin{equation}
	T(s) = T_{GS}(s) \left( 1 + \delta \frac{s}{m_\omega^2} T_\omega(s) \right),
 	\label{eq:rhoOmegaT}
\end{equation}
where $T_\omega$ is the relativistic Breit--Wigner propagator (Eq.~\ref{eq:BW}) of the $\omega$ meson
and $\delta$ is the $\rho-\omega$ mixing parameter.
A fit to the data of the pion vector form factor determines the electromagnetic mixing parameter to be
$\vert \delta_{em} \vert = (1.57 \pm 0.15 \pm 0.05) \times 10^{-3}$~\cite{Akhmetshin:2001ig}.
By relating the electromagnetic current to the clean $d \bar d $ source in $B^0 \to J/\psi \pip \pim$ decays,
it is argued in Ref.~\cite{Daub:2015xja} that the mixing effect can be described by $ \delta_{B^0 \to J/\psi \pi \pi} = -3 \vert\delta_{em} \vert$ in $B^0 \to J/\psi \pip \pim$ decays.
Following a similar argument and assuming a clean $u\bar u$ source for the $\rho$ production in $\signal$ decays, 
we initially fix the mixing parameter to $\delta  = +3 \vert \delta_{em} \vert$.
The parameter $\delta$ is argued to be real in Ref.~\cite{Hanhart:2016pcd} as confirmed experimentally in Refs.~\cite{BaBar:2012bdw,Akhmetshin:2001ig}. 
We therefore fix the phase of $\delta$ to zero.
These assumptions are tested after the amplitude model is built.

The $K^+\pim$ and $\pip\pim$ S-wave contributions, referred to as $[K^+\pim]_{\text S}$ and $[\pip\pim]_{\text S}$ in the following,
are described with the K-matrix formalism~\cite{Chung:1995dx} to preserve coupled-channel unitarity.
The $\pip\pim$ S-wave K-matrix parameterisation is taken from Ref.~\cite{Anisovich:2002ij}, with parameters taken from Ref.~\cite{PhysRevD.78.034023},
and considers the effects of five coupled channels ($\pi\pi, KK, \pi\pi\pi\pi, \eta\eta, \eta\eta^\prime$), 
five poles 
and a nonresonant contribution.
The $K^+\pim$ S-wave couples to two channels $K\pi$ and $K\eta^\prime$, 
and contains only one pole, the $K_0^*(1430)$ resonance.
The isospin state $I=\frac{1}{2}$ contributes to both channels, 
while $I=\frac{3}{2}$ couples to $K\pi$ only.
The parameterisation is taken from Ref.~\cite{FOCUS:2007mcb}.
The couplings to $K^+\pim$ and $\pip\pim$ K-matrix poles and channels
depend on the production mode, which is modelled using the P-vector approach~\cite{Aitchison:1972ay}.
More details about the S-wave models are provided in Appendix~\ref{a:AmpLineShapes}.

Nonresonant contributions in a state of well-defined relative orbital angular momentum are denoted by surrounding the particle system with brackets and indicating the partial-wave state with a subscript;
for example $[\pip \pim]_{\text P}$ refers to a nonresonant dipion P-wave.
The lineshape for nonresonant contributions is set to a constant.

\subsection{Spin densities}

The spin amplitudes are phenomenological descriptions
of decay processes which 
are required to be Lorentz invariant and
compatible with angular momentum conservation and,
where appropriate, parity conservation.
They are constructed in the covariant Zemach (Rarita-Schwinger) tensor formalism
\cite{Zemach,Rarita,helicity3}
and resemble spin-orbit couplings near threshold.
In the following, we briefly introduce 
the fundamental objects of the covariant tensor formalism 
which connect the particle's four-momenta to the spin dynamics of the reaction
and give a general recipe to calculate the spin factors for arbitrary decay trees.
Further details can be found in Refs.~\cite{Zou, Filippini}.

An integer spin-$S$ particle with four-momentum $p$, and spin projection $\lambda$, is represented 
by the polarisation tensor $\epsilon_{(S)}(p,\lambda)$, which is symmetric, traceless and orthogonal to $p$.
These so-called Rarita-Schwinger conditions reduce the $4^{S}$  elements of the rank-$S$ tensor to 
$2S +1$ independent  elements in accordance with the number of degrees of freedom of a spin-$S$ state\cite{Rarita,Zhu}.
The spin projection operator $P^{\mu_{1} \dots \mu_{S} \nu_{1} \dots \nu_{S}}_{(S)}(p_{R})$,  
for a resonance $R$, with spin $S = \{0,1,2\}$, and four-momentum $p_{R}$,
is given by \cite{Filippini}
\begin{align}
	\nonumber
	P_{(0)}^{\mu \nu}(p_{R}) &= 1, \\
	\nonumber
	P_{(1)}^{\mu \nu}(p_{R}) &= \sum_{\lambda_{R}} \eps^{\mu}(p_{R},\lambda_{R}) \, \eps^{*\nu}(p_{R},\lambda_{R}) = 
	- \, g^{\mu \nu} + \frac{p_{R}^{\mu} \, p_{R}^{\nu}}{p_{R}^{2}},    \\
	P_{(2)}^{\mu \nu \alpha \beta}(p_{R})  &=
	 \frac{1}{2} \,  \left[ P_{(1)}^{\mu \alpha}(p_{R})  \, P_{(1)}^{\nu \beta}(p_{R})  + P_{(1)}^{\mu \beta}(p_{R})  \, P_{(1)}^{\nu \alpha}(p_{R}) \right] - \frac{1}{3} \, P_{(1)}^{\mu \nu}(p_{R}) 
	 \, P_{(1)}^{\alpha \beta}(p_{R})    \,,
	\label{eq:pol1}
\end{align}
where $ g^{\mu \nu}$ is the Minkowski metric.
Contracted with an arbitrary tensor, the projection operator selects 
the part of the tensor which satisfies the Rarita-Schwinger conditions.
For a decay process $R \to A B$, with relative orbital angular momentum $L$ between particles $A$ and $B$,
the angular momentum tensor is obtained by 
projecting 
the rank-$L$ tensor $q_R^{\nu_{1}}   \,  q_R^{\nu_{2}}  \dots  \,  q_R^{\nu_{L}} $, constructed from the relative momenta
$q_{R} = p_{A} - p_{B}$,
onto the spin-$L$ subspace,
\begin{equation}
	L_{(L)\mu_{1}  \dots \mu_{L}}(p_{R},q_{R}) = (-1)^{L}  \, P_{(L)\mu_{1}  \dots \mu_{L} \nu_{1}  \dots \nu_{L}}(p_R)  \, 
	 q_R^{\nu_{1}}     \dots  \,  q_R^{\nu_{L}}  .
\end{equation}
Its $\vert \vec q_{R} \vert^{L} $ dependence accounts for the influence of the centrifugal barrier on the transition amplitudes.

Following the isobar approach, the full decay amplitude is described as a product of two-body decay amplitudes.
Initially, one considers the $\psitwos$ meson as a final-state vector particle in the decay $\signal$. 
Then, each sequential two-body decay $R \to A \, B$, 
with relative orbital angular momentum $L_{AB}$, and total intrinsic spin $S_{AB}$,
contributes a term to the overall spin factor given by
\begin{align}
	\nonumber
	S_{R \to A B}(\bold x \vert L_{AB}, S_{AB} ; \lambda_{R}, \lambda_{A}, \lambda_{B})  &=
	 \eps_{(S_{R})}(p_{R},\lambda_{R}) \, \tilde{E}(S_{R},L_{AB},S_{AB}) \,  L_{(L_{AB})}(p_{AB},q_{AB}) \,    \\
          &\times \Omega(\bold x \vert S_{AB} ; \lambda_{A} , \lambda_{B}) ,
          \label{eq:spin1}
\end{align}
where
\begin{align}
	 \Omega(\bold x \vert S_{AB} ; \lambda_{A} , \lambda_{B})  &=  P_{(S_{AB})}(p_R) \, \tilde{E}(S_{AB},S_{A},S_{B})  \, \eps^{*}_{(S_{A})}(p_{A},\lambda_{A})  \, \eps^{*}_{(S_{B})}(p_{B},\lambda_{B})  \,   .
          \label{eq:spin2}
\end{align}
Here, a polarisation vector is assigned to the decaying particle 
and the complex conjugate vectors for each decay product.
The spin and orbital angular momentum couplings are described by the tensors $P_{(S_{AB})}(p_R)$
and $L_{(L_{AB})}(p_{AB},q_{AB})$, respectively.
Firstly, the two spins $S_{A}$ and $S_{B}$ are coupled to a total spin-$S_{AB}$ state, $\Omega(\bold x \vert S_{AB})$,
by projecting the corresponding polarisation vectors  onto the spin-$S_{AB}$
subspace transverse to the momentum of the decaying particle.
Afterwards, the spin and orbital angular momentum tensors are properly contracted with the
polarisation vector of the decaying particle to give a Lorentz scalar.
This requires in some cases to include the tensor $\eps_{\alpha\beta\gamma\delta} \, p_{R}^{\delta}$ via
\begin{equation}
	  \tilde{E}(j_{a},j_{b},j_{c}) = 		
	  \begin{cases}
			  1 & \mbox{if } j_{a} + j_{b} + j_{c} \; {\rm even} \\
			   \eps_{\alpha \beta \gamma \delta} \, p_{R}^{\delta} & \mbox{if } j_{a} + j_{b} + j_{c} \; {\rm odd}
	 \end{cases} \, ,
\end{equation}
where $\eps_{\alpha\beta\gamma\delta}$ is the Levi-Civita symbol and $j$ refers to the arguments of $\tilde{E}$ defined in Eqs.~\ref{eq:spin1}~and~\ref{eq:spin2}.
Its antisymmetric nature ensures the correct parity 
transformation behaviour of the amplitude. 

The spin factor for a given decay chain, for example $B^+ \to \psitwos \, R_{1} $ with $R_{1} \to A \, R_{2}$ and
$R_{2} \to BC$,
is obtained by combining the two-body terms and performing a sum over all unobservable, intermediate spin projections
\begin{equation}
\small
	S^\prime_{i}(\phsPoint | \lambda_\psi)  = \sum_{\lambda_{R_{1}},\lambda_{R_{2}}}   
	S_{B^+ \to \psi R_{1}}(\bold x \vert L_{\psi R_{1}} ; \lambda_{\psi} , \lambda_{R_{1}}) \, 
	S_{R_{1} \to A R_2}(\bold x \vert L_{A R_2} ; \lambda_{R_{1}}  \lambda_{R_{2}}) 
	\, S_{R_{2} \to BC}(\bold x \vert L_{BC} ; \lambda_{R_{2}}) ,
	\label{eq:sf}
\end{equation}
where $\lambda_{B^+} = \lambda_{A} = \lambda_{B}  = \lambda_{C}  = 0$,  
$S_{\psi R_{1}} =  L_{\psi R_{1}}$, $S_{BC} = 0$ and $S_{A R_{2}} =  L_{AR_{2}}$, 
as $B^+$, A, B and C are pseudoscalar particles.
The $\psitwos \to \mu^+ \mu^-$ decay is incorporated by 
the modification
\begin{equation}
S_{i}(\phsPoint \vert \lambda_{\mu^+}, \lambda_{\mu^-}  )  =	\sum_{\lambda_\psi}S^\prime_{i}(\phsPoint  | \lambda_\psi) \,  \epsilon^\alpha_{(1)}(p_\psi,\lambda_\psi)\, \bar u(\lambda_{\mu^-}) \gamma_\alpha \nu(\lambda_{\mup}),
	\label{eq:sf_psi}
\end{equation}
where $\bar u(\lambda_{\mu^-})$ and $\nu(\lambda_{\mup})$ are the Dirac spinors for the fermions $\mu^-$ and $\mu^+$
and the gamma matrix, $\gamma_\alpha$,  describes a vector current.
As a last step, the completeness relations of the polarisation tensors in Eq.~\ref{eq:pol1} are used to simplify the spin-factor expressions.

\subsection{Decay fractions}

The hadronic amplitudes are normalised prior to the amplitude fit such that 
\begin{equation}
	\int   \sum_{\lambda_{\mu^+}, \,\lambda_{\mu^-} }  \left\vert  A_{i}(\phsPoint \vert \lambda_{\mu^+}, \lambda_{\mu^-} ) \right\vert^{2} \, \dphs= 1 .
\end{equation}
This ensures that all amplitude coefficients, $a_i$, are of comparable scale, and makes the fit robust against the choice of starting values for these parameters.
To provide implementation-independent measurements, in addition to the complex coefficients $a_i$, the fit fractions are defined as
\begin{equation}
\label{eq:DefineFitFractions}
	F_{i} \equiv \frac{\int \sum_{\lambda_{\mu^+}, \,\lambda_{\mu^-} }   \left\vert   a_{i} \, A_{i}(\phsPoint  \vert \lambda_{\mu^+}, \lambda_{\mu^-} ) \right\vert^{2} \, \dphs }
	{\int \left\vert  \mathcal A_{B^+}(\phsPoint) \right\vert^{2} \, \dphs}, 
\end{equation}
and are a measure of the relative strength between the different transitions. 
For cascade decays, such as $B^+ \to \psitwos K_1(1270)^+$ with $K_1(1270)^+ \to \{ \Kp \rho(770)^0, K^*(892)^0 \pip, ... \}$,
we also define the 
combined subchannel 
fit fractions by
\begin{equation}
\label{eq:DefineFitFractions2}
	F_{R_i} \equiv \frac{\int \sum_{\lambda_{\mu^+}, \,\lambda_{\mu^-} }   \left\vert \sum_{j \in R_i}  a_{j} \, A_{j}(\phsPoint  \vert \lambda_{\mu^+}, \lambda_{\mu^-} ) \right\vert^{2} \, \dphs }
	{\int \left\vert  \mathcal A_{B^+}(\phsPoint) \right\vert^{2} \, \dphs}, 
\end{equation}
where $R_i$ labels the three-body resonance (\eg $R_i = K_1(1270)^+$)
and $j$ labels the subdecay mode (\eg $j = \{ \Kp \rho(770)^0, K^*(892)^0 \pip, ... \}$).
Then the relative contribution of the subdecay $j$ to the three-body resonance $R_i$ is given by the partial fit fractions
\begin{equation}
\label{eq:DefineFitFractions3}
	f^{R_i}_{j} \equiv \frac{\int \sum_{\lambda_{\mu^+}, \,\lambda_{\mu^-} }   \left\vert a_{j} \, A_{j}(\phsPoint  \vert \lambda_{\mu^+}, \lambda_{\mu^-} ) \right\vert^{2} \, \dphs }
	{\int \sum_{\lambda_{\mu^+}, \,\lambda_{\mu^-} }   \left\vert \sum_{k \in R_i}  a_{k} \, A_{k}(\phsPoint  \vert \lambda_{\mu^+}, \lambda_{\mu^-} ) \right\vert^{2} \, \dphs  }. 
\end{equation}
The interference between amplitude pairs are quantified by the fractions
\begin{equation}
\label{eq:DefineInterferenceFractions}
	I_{ij} \equiv \frac{\int \sum_{\lambda_{\mu^+}, \,\lambda_{\mu^-} }  2\,\Re[a_{i}a^*_{j} \, A_{i}(\phsPoint  \vert \lambda_{\mu^+}, \lambda_{\mu^-} ) A^*_{j}(\phsPoint  \vert \lambda_{\mu^+}, \lambda_{\mu^-} ) ] \, \dphs }
	{\int \left\vert  \mathcal A_{B^+}(\phsPoint) \right\vert^{2} \, \dphs} .
\end{equation}
Constructive interference leads to $I_{ij} > 0$, while destructive interference leads to $I_{ij} < 0$. 
Note that \mbox{$\sum_i F_{i} + \sum_{j<k} I_{j,k} = 1$}.

We ensure that strong decays in the cascade topology have the same pattern regardless of the production mechanism 
by sharing couplings between related subdecays.
For example, given the two $a_i$ parameters required for the S-wave decay
\mbox{\prt{B^+[S] \to \psitwos \, K_1(1270)^+}} with \mbox{\prt{K_1(1270)^+ \to \rho(770)^0 \, \pi^+}}
and \mbox{\prt{K_1(1270)^+ \to K^*(892)^0 \, \pi^+}}, 
the amplitude for the P-wave decay 
\mbox{\prt{B^+[P] \to \psitwos \, K_1(1270)^+}} with 
\mbox{\prt{K_1(1270)^+ \to \rho(770)^0 \, \pi^+}} and
\mbox{\prt{K_1(1270)^+ \to K^*(892)^0 \, \pi^+}} only requires one additional global
complex parameter to represent the different production processes of
\mbox{\prt{B^+[S] \to \psitwos \, K_1(1270)^+}} and 
\mbox{\prt{B^+[P]\to \psitwos \, K_1(1270)^+}},
while the relative magnitude and phase between 
\mbox{\prt{K_1(1270)^+ \to \rho(770)^0 \, \pi^+}} and 
\mbox{\prt{K_1(1270)^+ \to K^*(892)^0 \, \pi^+}} are the same regardless of the production mechanism. 
In the following, if no angular momentum is specified, the lowest angular momentum state compatible
with angular momentum conservation and, where appropriate, parity conservation, is implied.

\section{Amplitude fit}
\label{sec:results}

The total probability density function (PDF) describing the phase-space distribution of
 $\signal$ candidates in the signal region is
composed of  the signal, $\mathcal P_{\text S}(\phsPoint \vert \theta_{\text S})$, and the background, $\mathcal P_{\text B}(\phsPoint \vert \theta_{\text B})$, PDFs,
\begin{equation}
	\mathcal P(\phsPoint \vert \theta) = (1-f_{\text B})  \, \mathcal P_{\text S}(\phsPoint \vert \theta_{\text S}) + f_{\text B} \, \mathcal P_{\text B}(\phsPoint \vert \theta_{\text B}),
\end{equation}
where $f_{\text B}$ is the background fraction determined in Sec.~\ref{sec:Selection} and $\theta = (\theta_{\text S},\theta_{\text B})$ is the total set of fit parameters.
The likelihood function is defined as $\mathcal L(\theta) = \prod_i \mathcal P(\phsPoint_i \vert \theta)$.
Assuming \CP invariance, $B^-$ decays are \CP conjugated and treated as $\signal$ decays.

The signal PDF is built from the total amplitude-squared as 
\begin{equation}
	\mathcal P_S(\phsPoint) = 
	\frac{ \vert \mathcal A_{B^+}(\phsPoint) \vert^2 \,  \epsilon(\phsPoint) \, \phi(\phsPoint) }
	{\int  \vert \mathcal A_{B^+}(\phsPoint) \vert^2 \,  \epsilon(\phsPoint) \,  \dphs  },  
	\label{eg:sigPdf}
\end{equation}
where $\epsilon(\phsPoint)$ describes the efficiency variation across the phase space.
As the efficiency in the numerator leads to an additive constant in the $\log {\mathcal L(\theta)}$ function that does not depend on any fit parameters, it can be neglected in the minimisation procedure. 
The efficiency function still enters via the normalisation integral, 
which is evaluated with the Monte Carlo (MC) integration technique~\cite{dArgent:2017gzv,MarkIIIK3pi,FOCUS4pi,KKpipi} using simulated events that have
been propagated through the full LHCb detector simulation and selection. 
The size of the fully selected MC integration sample is more than 20 times larger than the data sample. 
Since the invariant-mass resolution is much smaller than the decay widths of all considered resonances, resolution effects are neglected in the amplitude fit.

Candidates in the $m(\psitwos K^+ \pim \pip)$ sideband regions, which are defined as \mbox{$[5200, 5240] \, \text{MeV} \cup [5320, 5400] \, \text{MeV}$}, are used to model
the background PDF,
\begin{equation}
	\mathcal P_{\text B}(\phsPoint \vert \theta_{\text B}) 
	= \frac{  B(\phsPoint \vert \theta_{\text B})  \, \epsilon(\bold x)  \, \phsd(\phsPoint) }{\int B(\phsPoint \vert \theta_{\text B}) \, \epsilon(\bold x)  \,  \dphs }.
\label{eq:bkgPDF}
\end{equation} 
The background function $B(\phsPoint \vert \theta_{\text B})$ describes the distribution in phase space
relative to the signal efficiency times phase-space density term, which is ommitted in the numerator as in the signal PDF.
Candidates from the low- and high-mass sidebands are added together such that they have the same relative contribution.
The background function is constructed by training a BDTG algorithm~\cite{Rogozhnikov_2016} to match the phase-space distribution of the MC integration sample 
to the data sidebands.
The value of $B(\phsPoint_i \vert \theta_{\text B})$ for a given data point $\phsPoint_i$ is then obtained by applying 
a pseudo seven-dimensional interpolation method to the weighted MC integration sample~\cite{LHCB-PAPER-2018-041}.

\subsection{Signal model selection}
\label{sec:sigPDF}

The strange-meson spectrum comprises an abundance of resonances potentially contributing to $\signal$ decays
in various topologies and angular-momentum configurations.
Several conventional charmonium states with masses within the phase-space limit also need to be considered.
A multitude of hidden-charm exotic states might contribute as well.
The full list of considered intermediate-state amplitudes can be found in Appendix~\ref{a:decays}.

With the number of possible amplitudes of the order of 100, it is not feasible to fit them all at once
and a reasonably sized subset is chosen to avoid overfitting.
An algorithmic model-building procedure is employed that
starts with a small set of amplitudes (\textit{start} model), which are 
expected to contribute, 
and successively adds amplitudes until a reasonable agreement between the data and fit is achieved.
The quality of the fit in the seven-dimensional phase space is quantified by binning the data and 
calculating the metric
\begin{equation}
	\chi^{2} = \sum_{i=1}^{N_{\text{\rm bins}}} \frac{(N_{i}-\hat N_{i})^{2}}{\sigma^2_{i} + \hat \sigma_{i}^2 },
\end{equation}
where $N_{i}$ is the number of selected candidates in a given bin, 
$\hat N_{i}$ is the event count predicted by the fitted amplitude model
and $N_{\text{\rm bins}}$ is the number of bins.
The uncertainties are given by $\sigma_i^2 = N_i$ and 
$\hat \sigma_i^2 = \sum_{j \in \text{bin} \, i} w^2_j$, 
where the weights $w_j$ match the MC integration sample to the fitted PDF\@.
A robust $\chi^{2}$ calculation is ensured by employing an adaptive binning algorithm \cite{Harnew:2017tlp}.
The phase space is divided such that each bin is populated by at least $25$ candidates ($N_\text{bins}$=1091)
leading to smaller hyper-volumes in regions of high event density.
The $\chi^{2}$ value divided by the number of degrees of freedom,
$\nu = N_{\text{bins}}-1  - N_{\text{par}}$ where $N_{\text{par}}$ is the number of free fit parameters, should be
close to unity.
Individual amplitudes are added one-by-one to the  \textit{start} model and the set of new models is fitted to the data. 
The amplitude that improves the $\chi^{2}/\nu$ value the most is kept and added to the  \textit{start} model. 
The remaining amplitudes are tested again in the subsequent loop until the incremental improvement in fit quality falls below the threshold $\Delta \chi^{2}/\nu = 0.02$~\cite{LHCb-PAPER-2017-040, LHCB-PAPER-2018-041}.

Initially, only conventional hadron states are considered.
The model-building algorithm improves the $\chi^{2}/\nu$ value from 2.54 to 2.05.
While this is a significant improvement with respect to the \textit{start} model, the fit quality is still poor.
The amplitudes included in this \textit{no-exotics} model are detailed in Appendix~\ref{a:modelSel} along with fit projections.
The largest discrepancy in the data description is observed for the $\psitwos \pi^+ \pi^-$ invariant-mass distribution.
The masses and widths of all resonances are fixed to their external values (\cf Appendix~\ref{a:decays});
determining them from the fit marginally improves the fit quality by $\Delta \chi^{2}/\nu = 0.08$ but
does not change the conclusion that nonconventional hadron contributions are needed to describe the data.

As a next step, the model-building procedure is repeated from scratch with the inclusion of the previously observed exotic resonances listed in Appendix~\ref{a:decays},
which could potentially contribute to the $\psitwos \pip \pim$, $\psitwos \pi^\pm$ or $\psitwos K^+$ subsystems.
The $\chi^{2}/\nu$ value is improved to $1.31$ 
and the description of the $\psitwos \pi^+ \pi^-$ invariant-mass distribution is notably better. 
Fit projections and amplitude model details are given in Appendix~\ref{a:modelSel}.
This \textit{known-exotics} model includes four $\Xz \to \psitwos \pi^+ \pi^-$ states
and the \mbox{$\ZsA(4000)^+\to \psitwos K^+$} state observed in $B^+ \to J/\psi \phi K^+ $ decays~\cite{LHCb-PAPER-2020-044}. 
The $\psitwos \pi^+$ system is described primarily by two axial vector contributions, $\ZAone^+$ and $\ZAtwo^+$, with smaller contributions from two vector states, $\ZVone^+$ and $T_{c\bar c}^*(4100)^+$.
This fit also determines the mass and width of the four $\Xz$ states and the two axial vector $\psitwos \pi^+$ states.
Fitting any other resonance's mass and width does not improve the fit result, thus they remain fixed to the external values given in Appendix~\ref{a:decays}.

As some disagreements between fit and data remain, especially in the $m(\psitwos K^+ \pi^-)$ projection,
we explored adding additional $\Xsz \to \psitwos K^+ \pi^-$ exotic contributions of various quantum numbers.
For a given new $\Xsz$ resonance hypothesis, likelihood scans over the allowed $m_0$ mass range are performed with the width fixed to $\Gamma_{0} = \{ 50, 100, 200 \} \mev$.
Local minima in the likelihood profiles are identified and used to
create a pool of potential new states and quantum number assignments.
Subsequent iterations of the model-building algorithm add one of these states until the convergence criterion is reached.
The initial values for the mass and width of new states are set based on the likelihood scans but are floating in the fits.
This procedure adds three new $\Xsz \to \psitwos K^+ \pi^-$ states.
Additional new $\Xz \to \psitwos \pi^+ \pi^-$, $\Xspp \to \psitwos K^+ \pi^+$, $\Zpm \to \psitwos \pi^\pm$ or $\Zsp \to \psitwos K^+$
contributions do not lead to significant improvements in the description of the data,
but are considered in alternative models for studies of systematic uncertainties.
Lastly, we prune the model by removing four insignificant amplitudes.
The resulting model is henceforth referred to as the \textit{baseline} model.

\subsection{Baseline model results}
\label{sec:bestFit}

\begin{figure}[p]
       	 \includegraphics[width=0.329\textwidth,height=!]{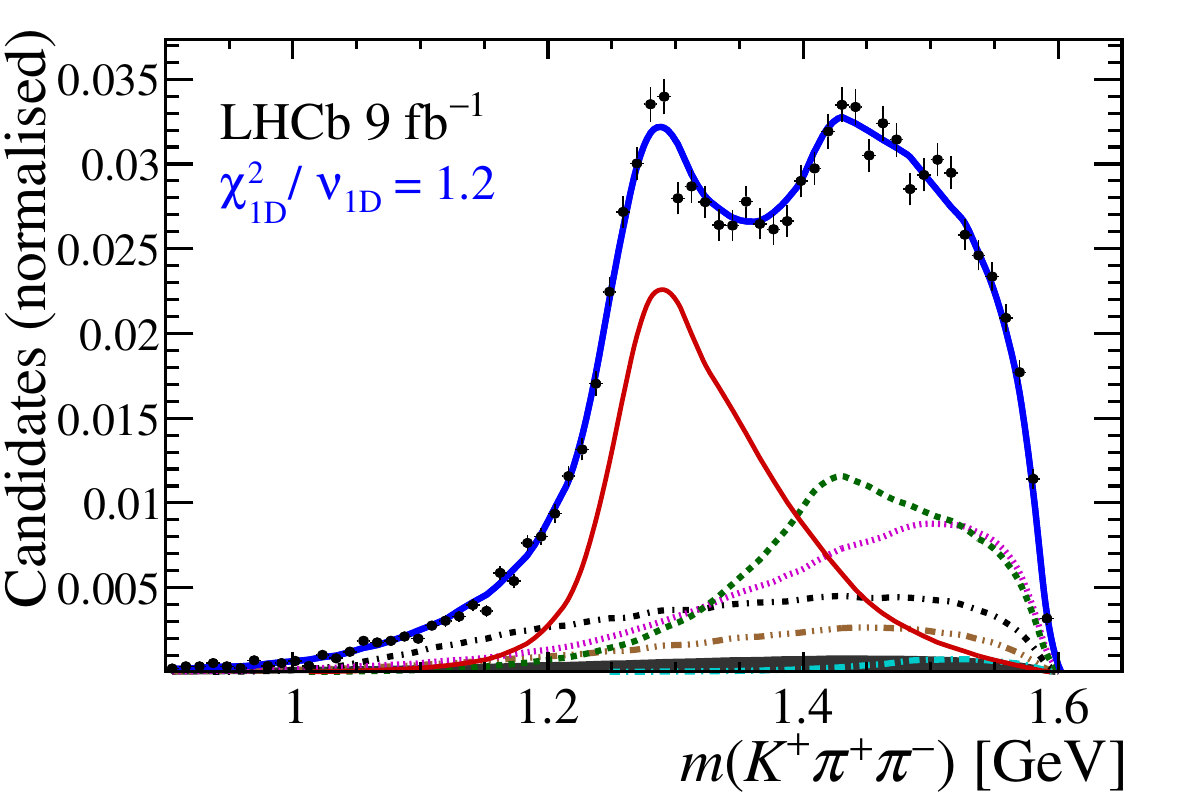}
       	 \includegraphics[width=0.329\textwidth,height=!]{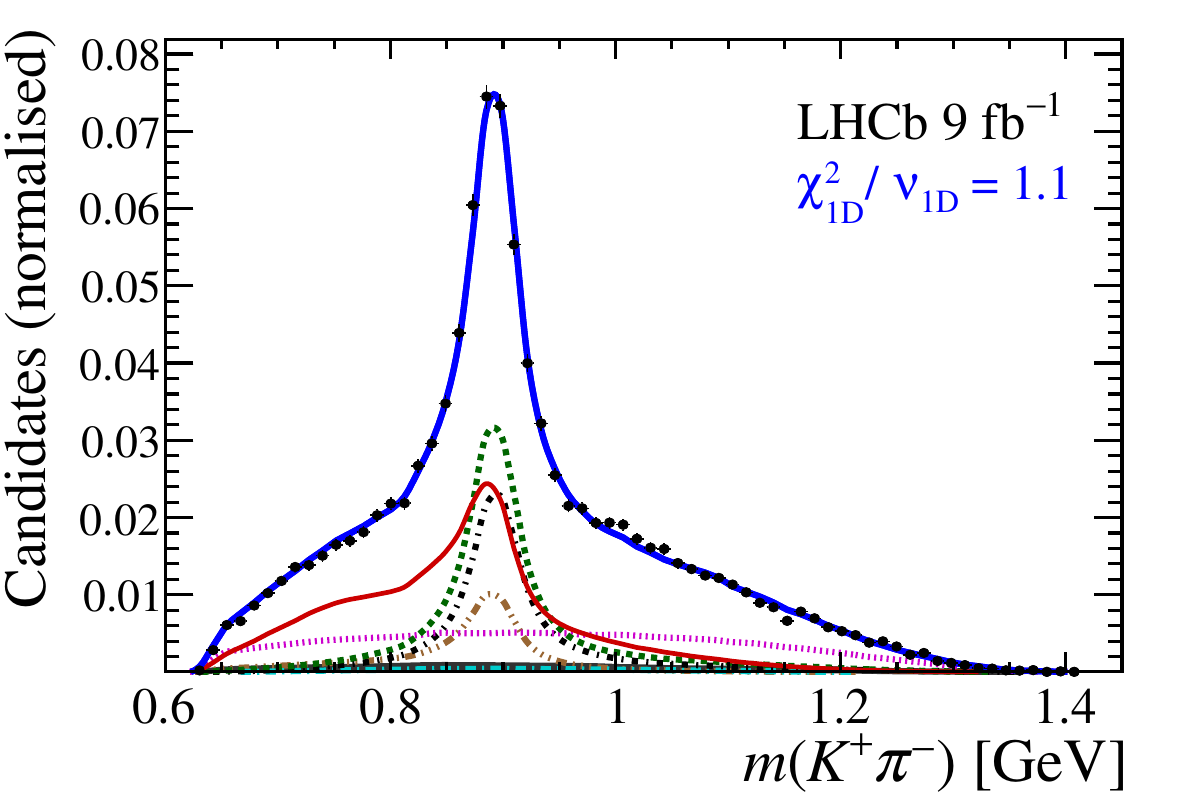}
       	 \includegraphics[width=0.329\textwidth,height=!]{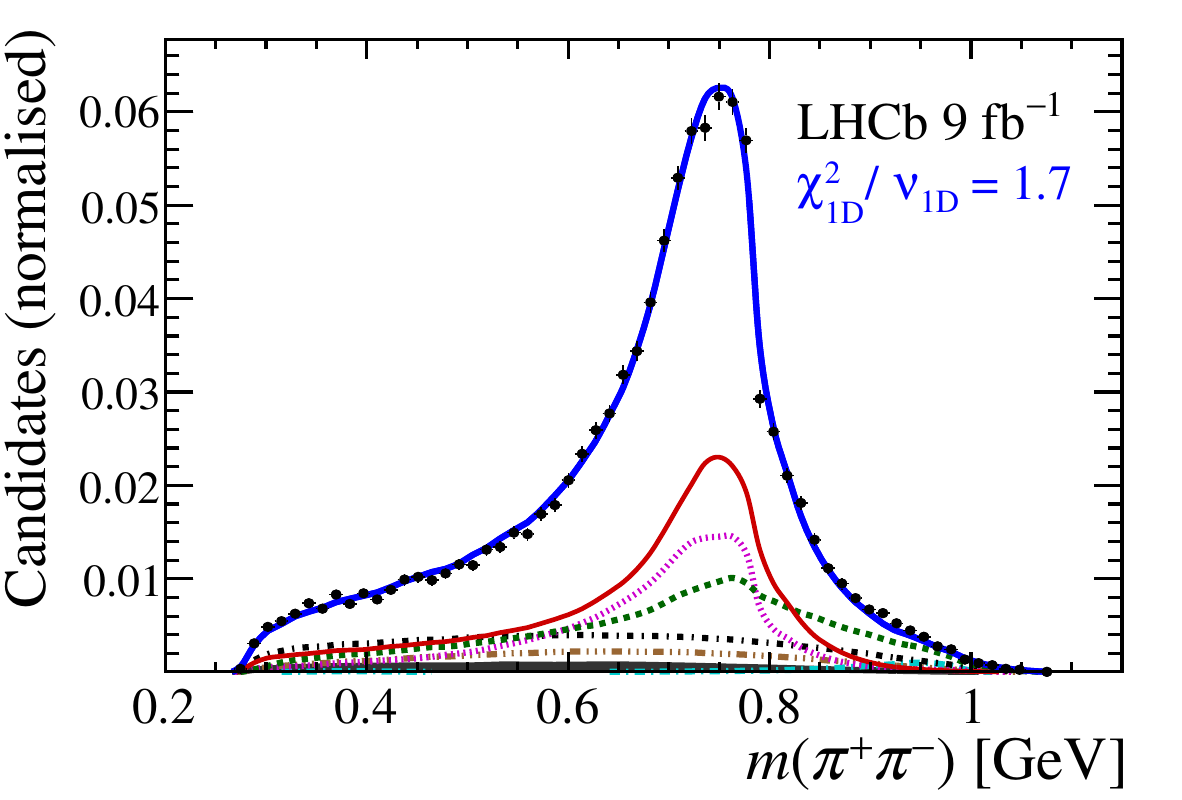}

       	 \includegraphics[width=0.329\textwidth,height=!]{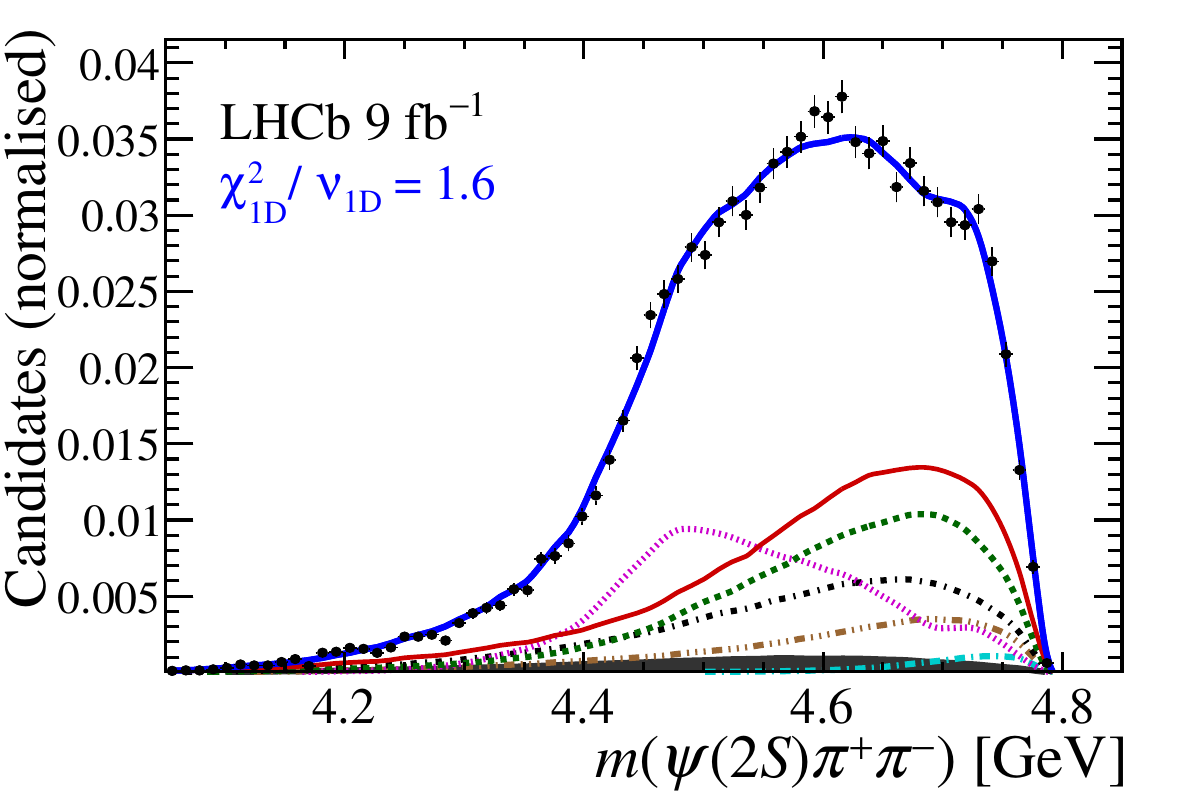}
       	 \includegraphics[width=0.329\textwidth,height=!]{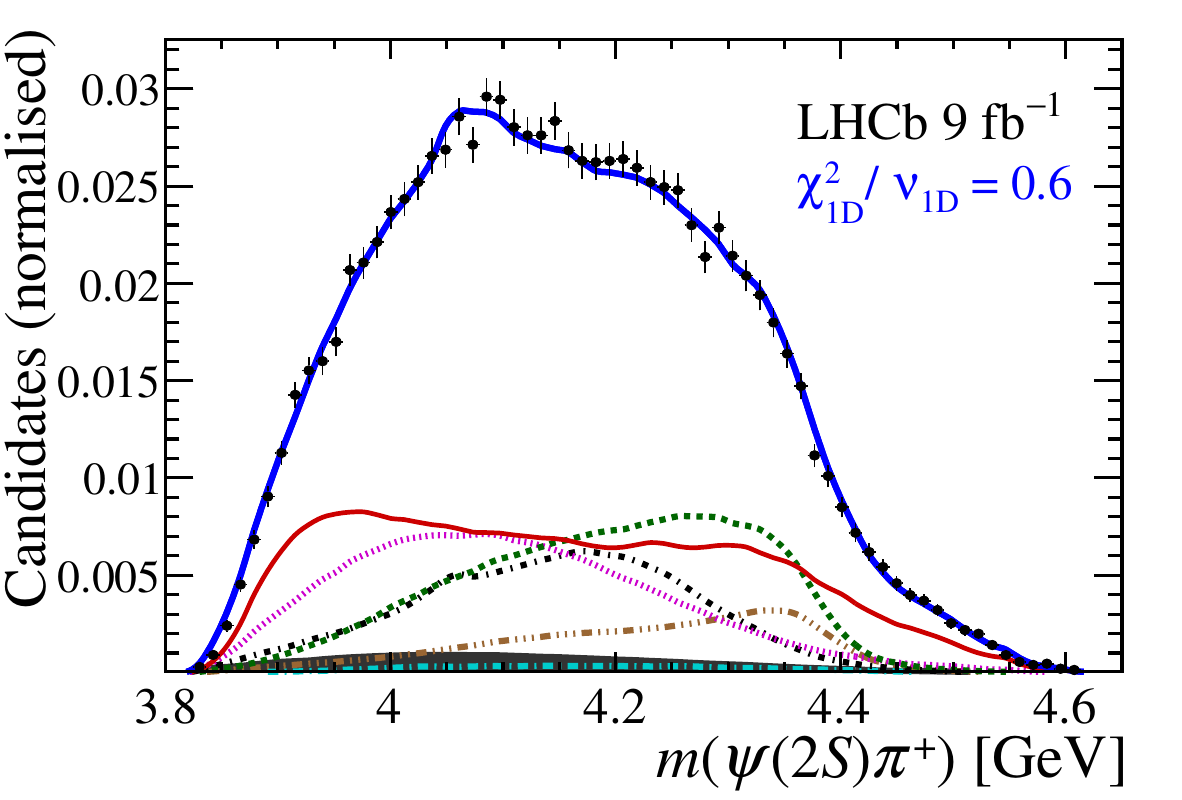}
       	 \includegraphics[width=0.329\textwidth,height=!]{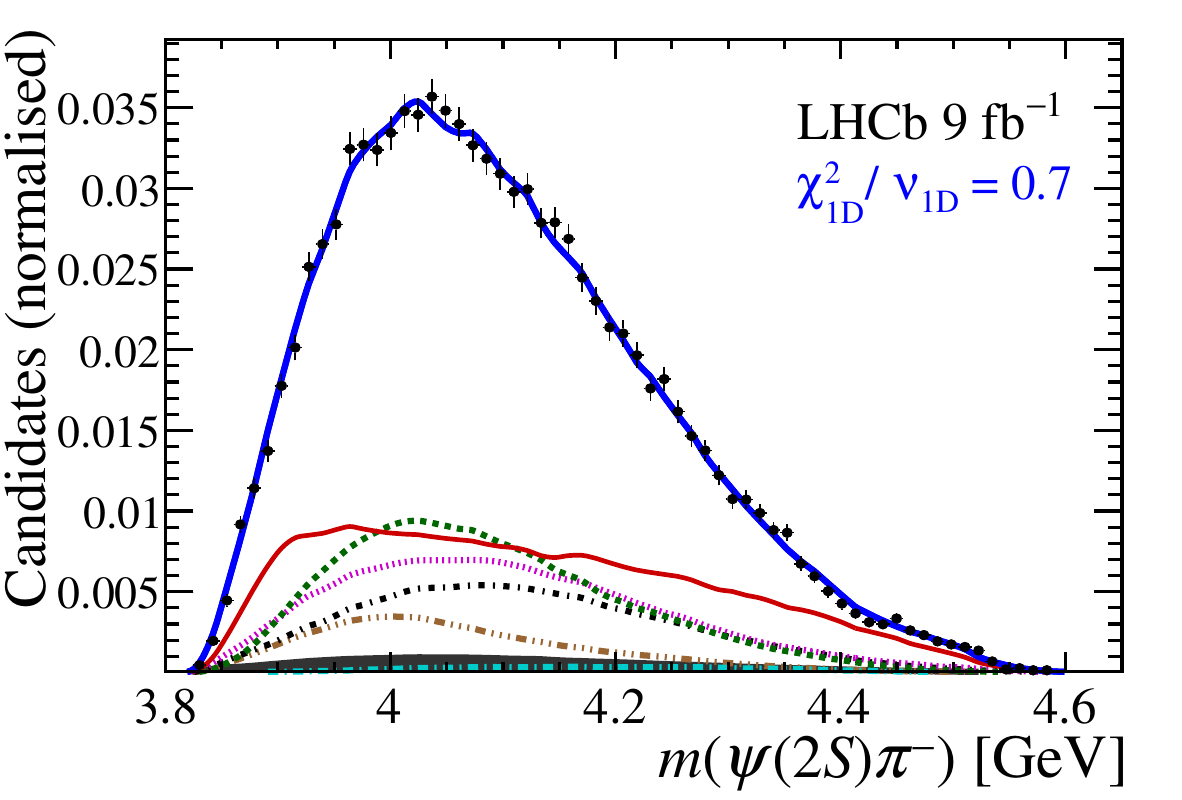}

       	 \includegraphics[width=0.329\textwidth,height=!]{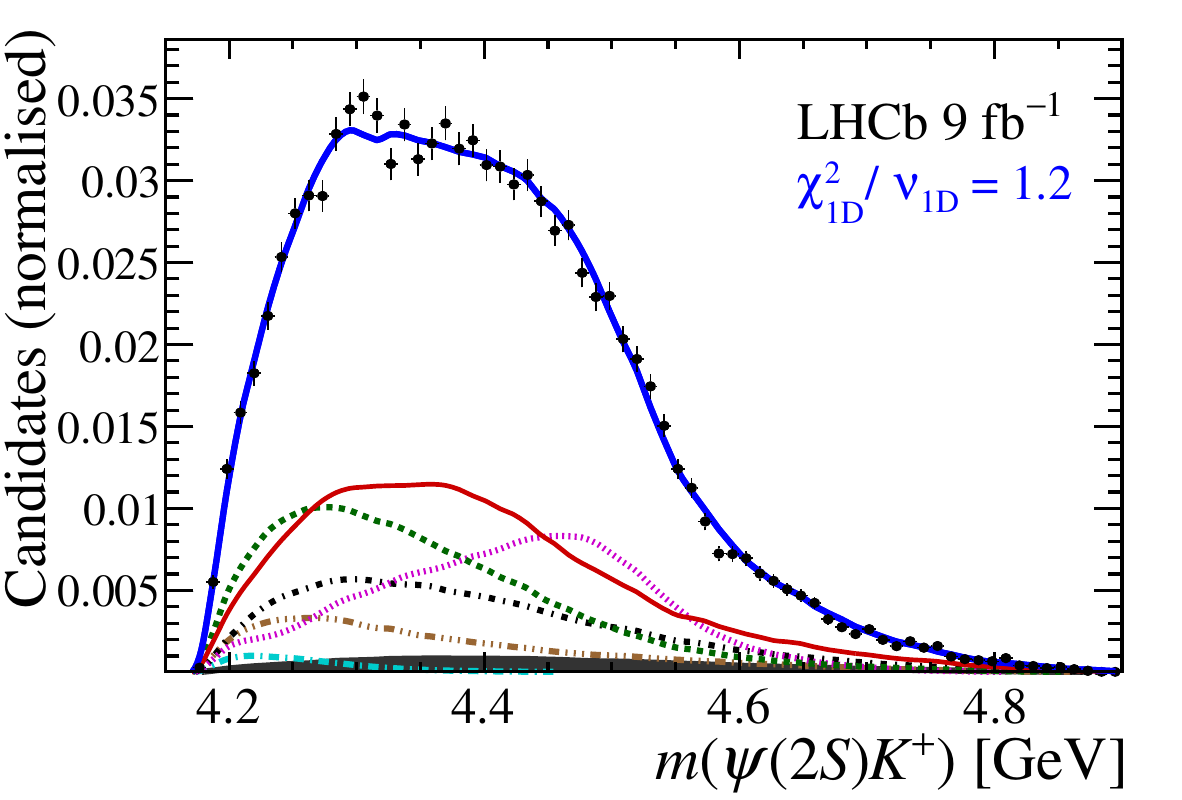}
       	 \includegraphics[width=0.329\textwidth,height=!]{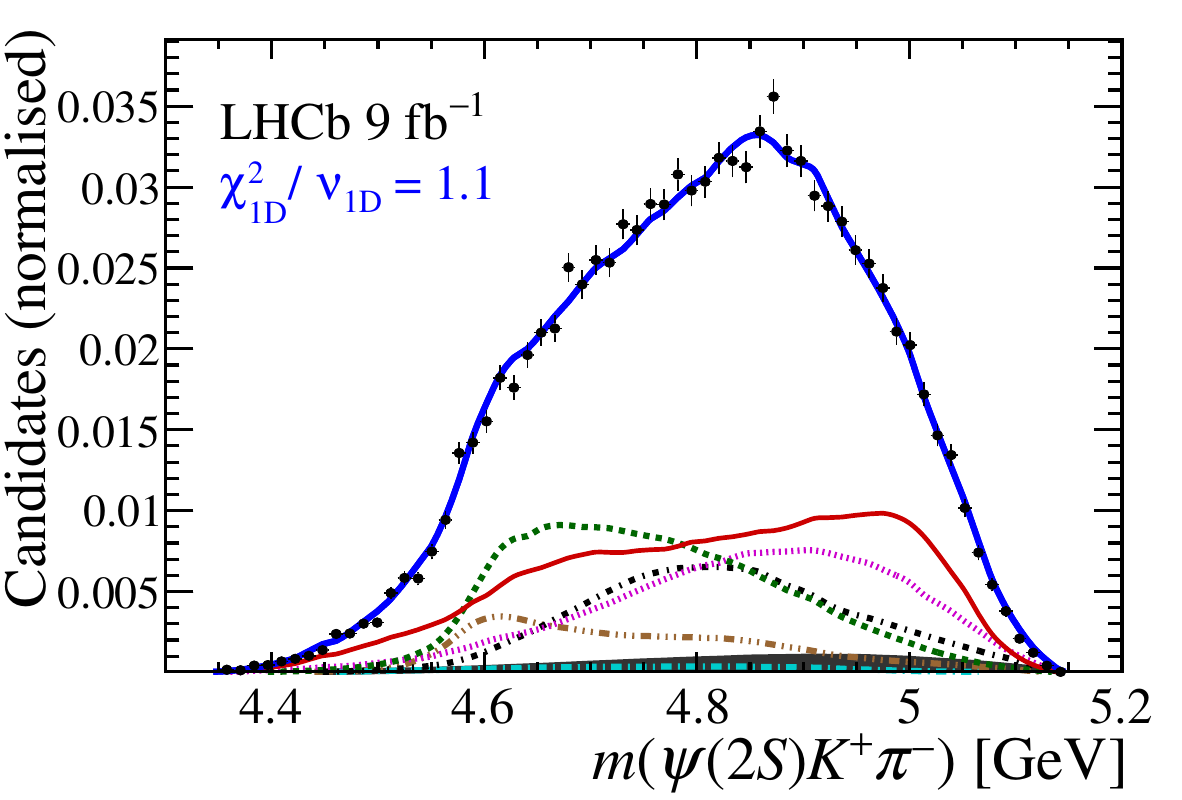}
       	 \includegraphics[width=0.329\textwidth,height=!]{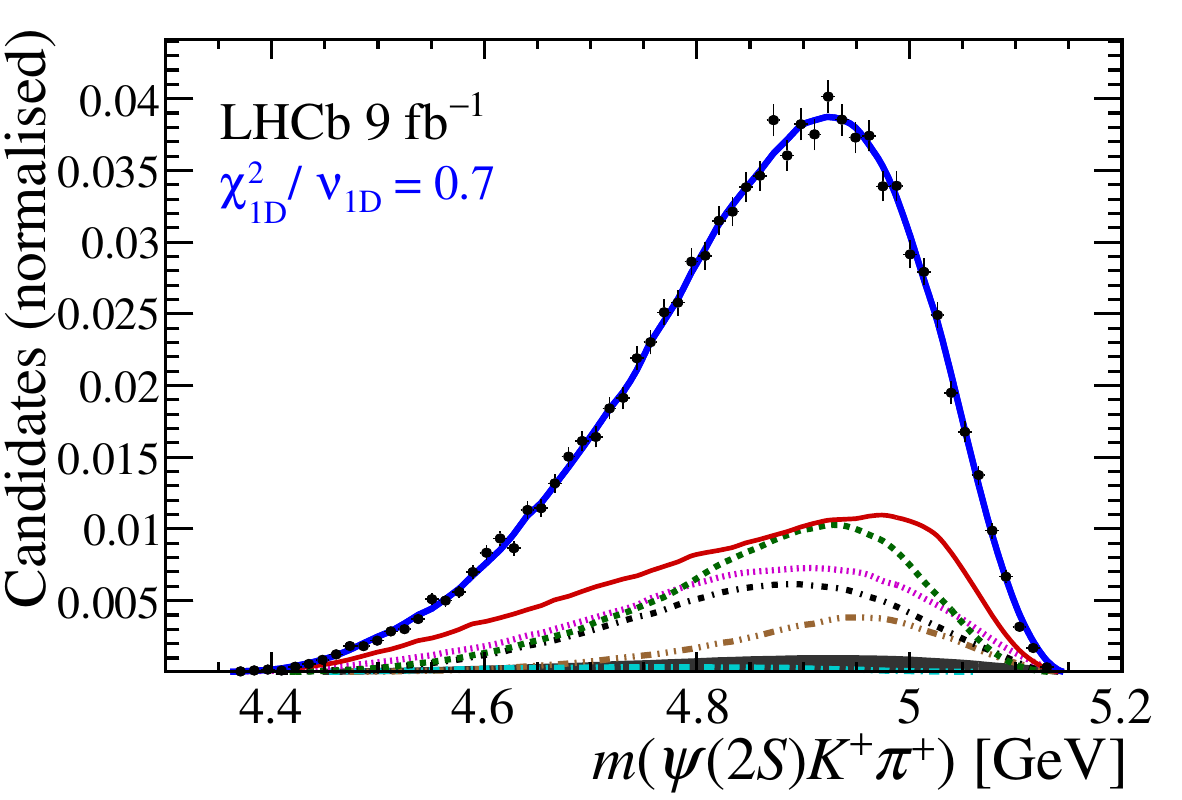}
              
       	 \includegraphics[width=0.329\textwidth,height=!]{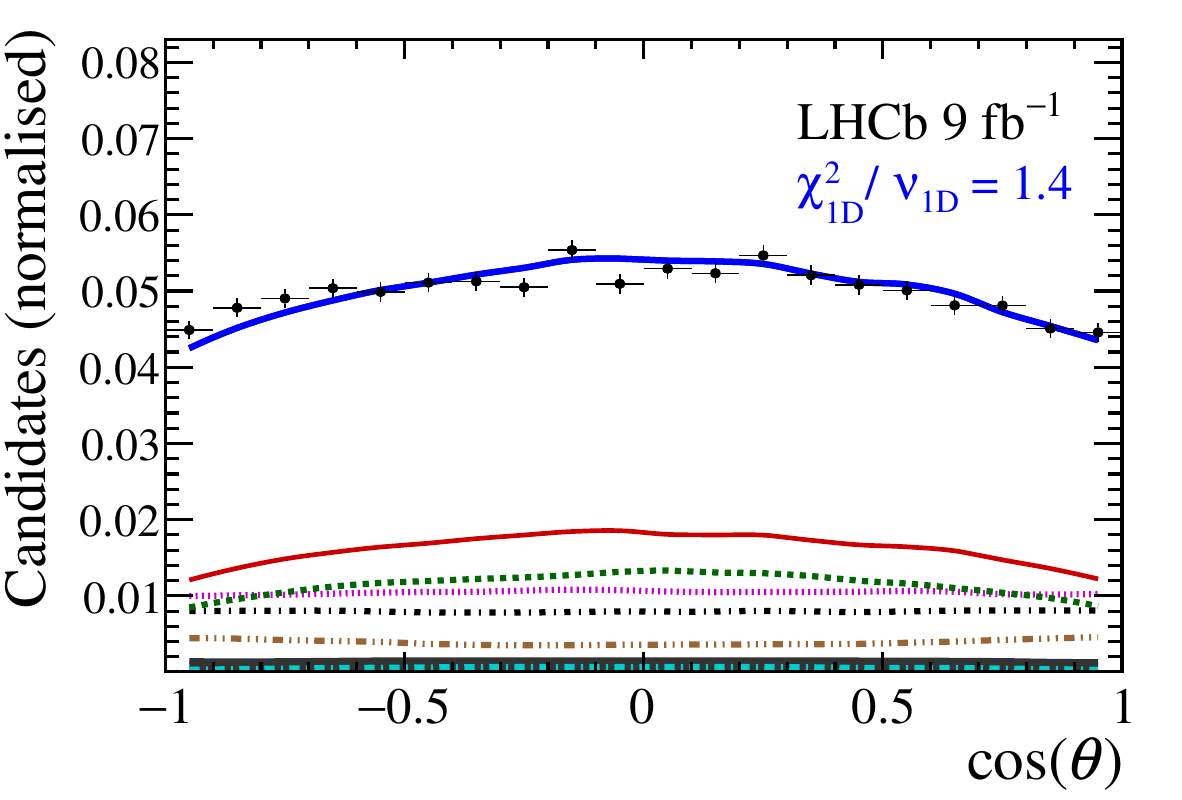}
       	 \includegraphics[width=0.329\textwidth,height=!]{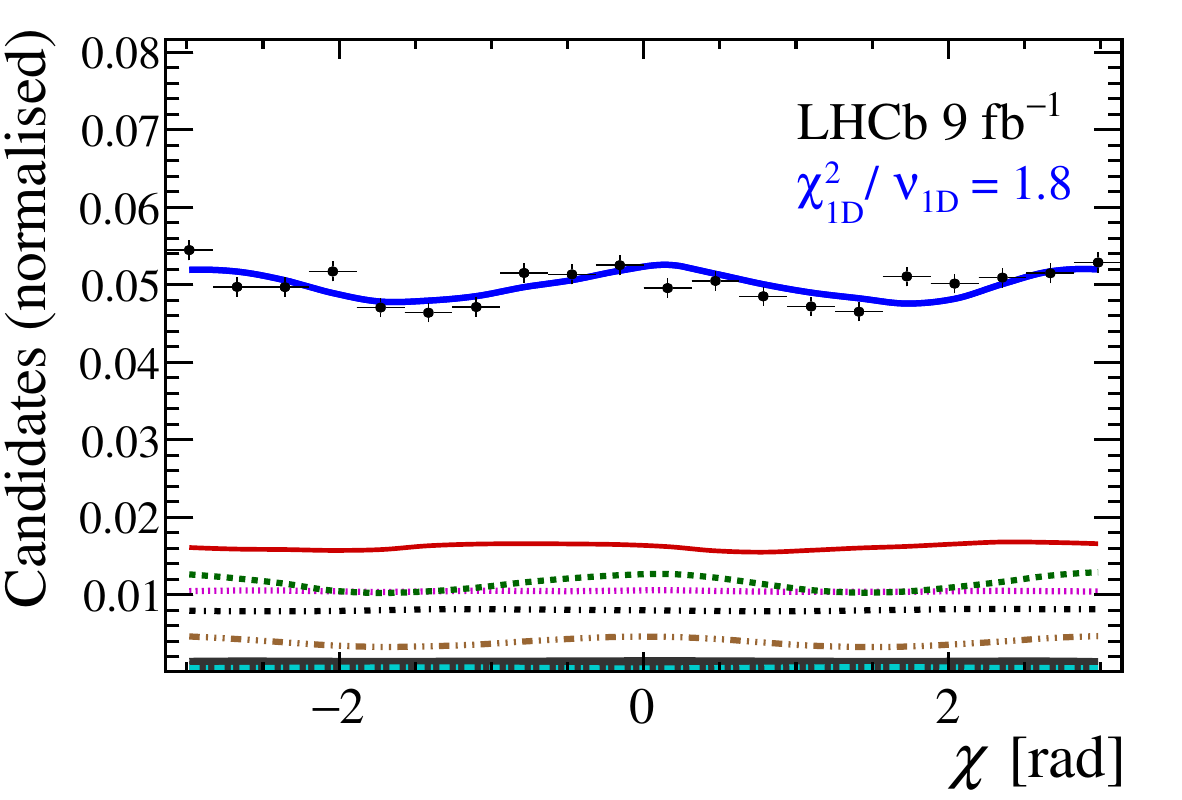}
       	 \includegraphics[width=0.329\textwidth,height=!]{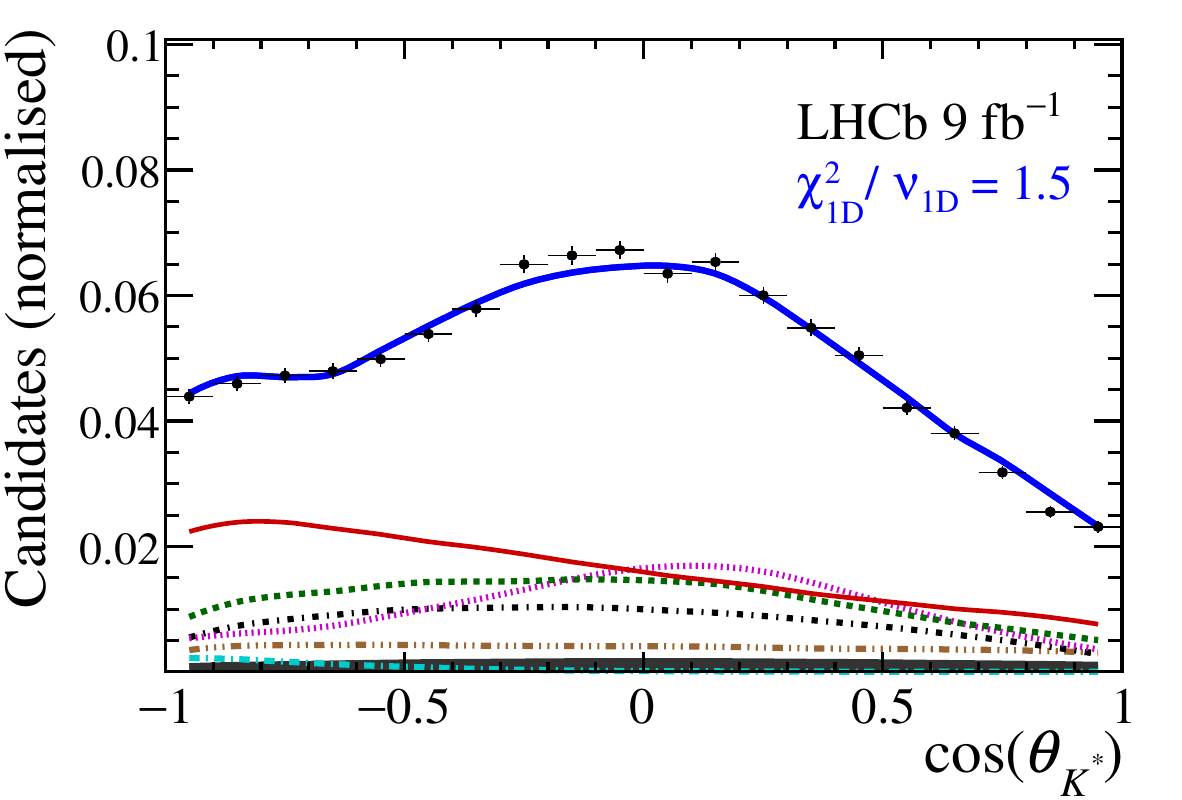}

        \centering
              \includegraphics[width=0.25\textwidth,height=!]{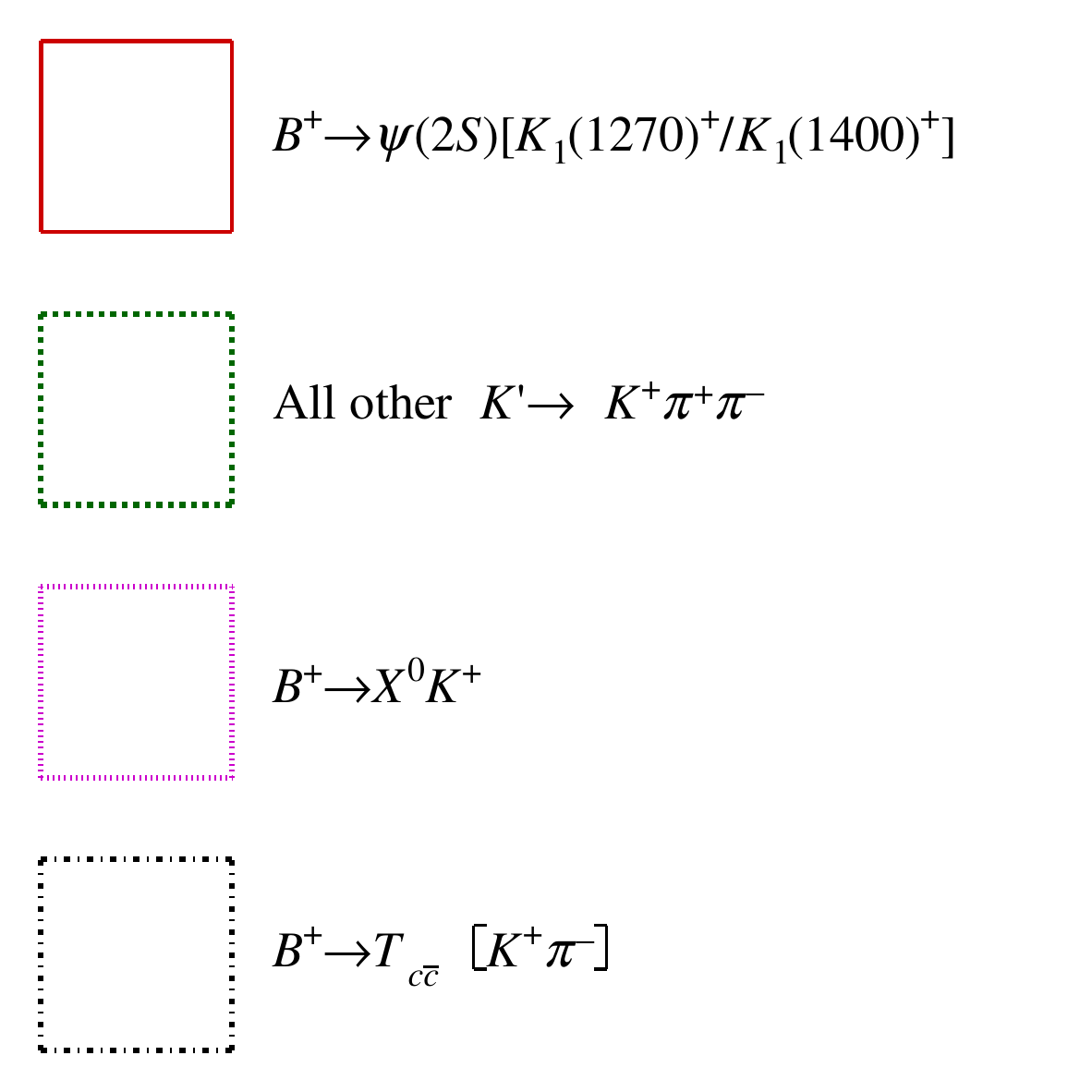}
              \includegraphics[width=0.25\textwidth,height=!]{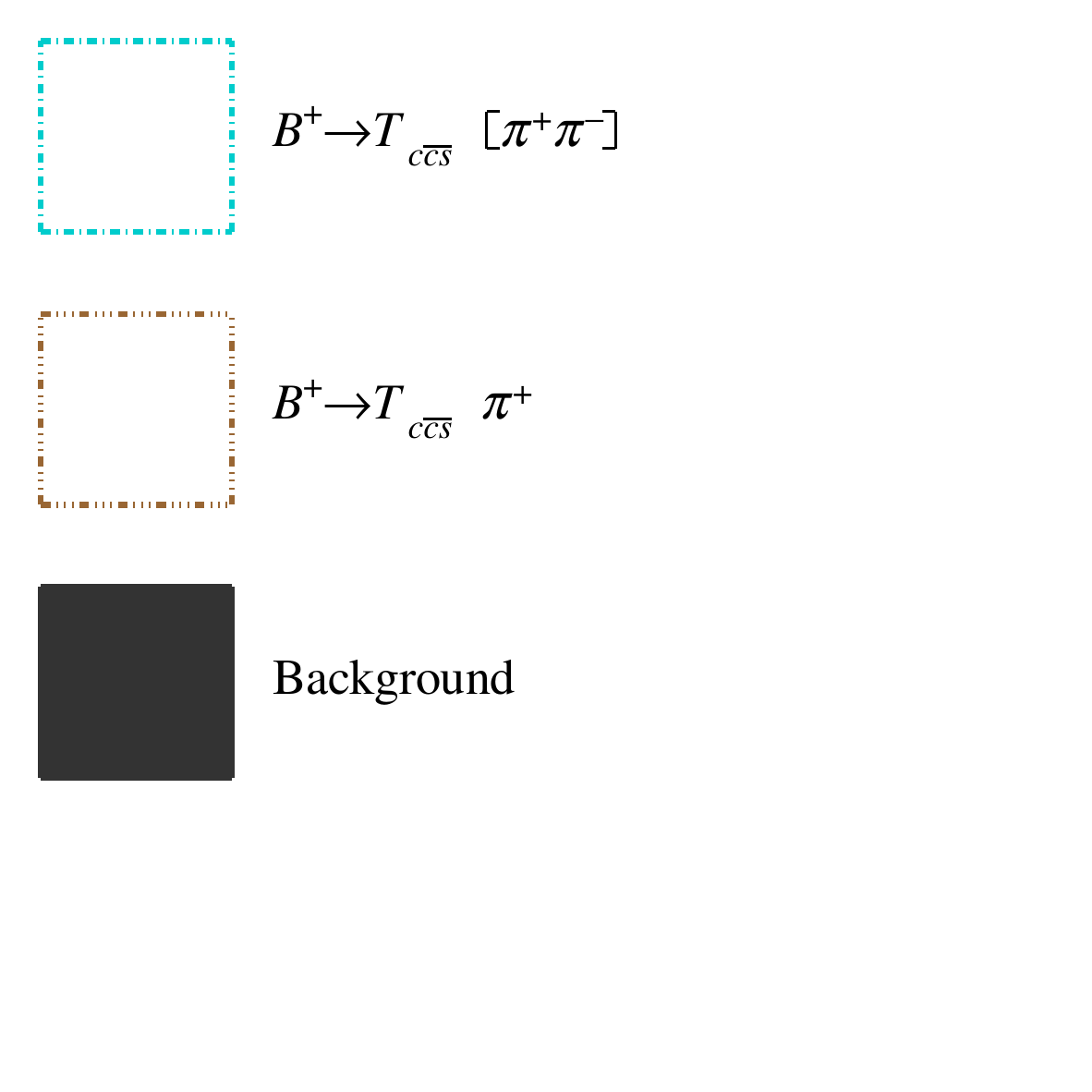}
         
	\caption{Phase-space projections of $\signal$ candidates in the signal region (points with error bars) and fit projections (solid, blue line) for the \textit{baseline} model. 
 The displayed $\chi_{\rm 1D}^2/\nu_{\rm 1D}$ value on each projection gives the sum of squared normalised residuals divided by the number of bins minus one.
 The multidimensional $\chi^2$ value is $\chi^2/\nu= 1.20$ with $\nu=992$.}

         \label{fig:fitBest}

\end{figure}

The \textit{baseline} model consists of 53 amplitudes, has 98 free parameters and achieves a $\chi^{2}/\nu$ ($\chi^{2}/N_{\text{bins}}$) value of $1.21$ $(1.10)$.
From pseudoexperiments, we estimate that this corresponds to a p-value of 0.002 ($3.1\sigma$) before accounting for systematic uncertainties.
Figure~\ref{fig:fitBest} displays the fit projections.
The fit fractions are given in Tables~\ref{tab:fitBest} and~\ref{tab:fitBest2}.
Fit results for the amplitude coefficients and interference fractions can be found in Appendix~\ref{a:models}.
Six $K^{\prime+} \to K^+\pip\pim$ resonances are included that decay via $K^*(892)^0 \pi^+$, $K^+\rho(770)^0$ or S-wave intermediate states.
The largest conventional hadron contribution comes from the $K_1(1270)^+$ resonance produced in S-, P- and D-wave $B^+$ decays.
The extracted branching fractions of the  $K_1(1270)^+$ resonant substructure agrees well with other measurements as detailed in Appendix~\ref{a:k1Results}.
The first subleading component arises from $K^*(1680)^+$ decays. 
The $K_1(1270)^+$ and $K^*(1680)^+$ are the only two resonances included in the $\signal$ analysis by Belle~\cite{BELLE}.
In addition, we observe $K_1(1400)^+,K^*(1410)^+,K_2^*(1430)^+$ and $K(1460)^+$ decays. 
No other established strange resonances exist within the kinematic limit of the decay.

The \textit{baseline} model includes only one conventional charmonium resonance, $\psi(4360)$, with a subpercent fit fraction.
Two other charmonium resonances included in the \textit{no-exotics} model, $\psi(4415)$ and $\psi(4660)$, are no longer significant after considering exotic contributions.
The masses and widths of the conventional hadrons are fixed to external measurements.
Floating these parameters does not significantly improve the fit quality for any resonance ($\Delta \chi^{2}/\nu < 0.01$).
In total, four $\Xz\to\psitwos \pip \pim$, three $\Zpm \to \psitwos \pi^\pm$, 
one $\Zsp \to \psitwos K^+$ and three $\Xsz \to \psitwos K^+ \pi^-$ components are part of the \textit{baseline} model.
Exotic resonance parameters are floated in the fit except for those of the $\ZVone^+$ and $\ZsAone$ states. 
The $\ZVone^+$ state has a small fit fraction with mass and width well-constrained from external measurements~\cite{PDG2022}.
Since the pole mass of the $\ZsAone$ state lies outside of the phase-space boundaries, the fit is not sensitive to its mass and width values. 
Exotic cascade decays of the type $\Xz \to \Zpm \pi^\mp$ are assumed to be charge symmetric, \ie  the magnitudes of the amplitude coefficients are equal for $\Xz \to \Zp \pi^-$ and $\Xz \to \Zm \pi^+$ decays.
For the relative phase, we test $0$ and $\pi$ for each $\Xz$ resonance separately and select the one that gives the larger likelihood value.
Table~\ref{tab:resoParams} lists the measured masses and widths, quantum numbers and significance of all exotic states.

The significance is determined from a likelihood-ratio test.
A given resonance is removed from the model (this might correspond to several amplitudes) and a new fit is performed.
From the change of the likelihood value, $\Delta(-2 \ln \mathcal L)$, we compute the significance as
$n_\sigma = \sqrt{2} \, \text{erfc}^{-1}(\text{Prob}_{\chi^2}(\Delta(-2 \ln \mathcal L), \text{ndf}))$,
where $\text{Prob}_{\chi^2}(\Delta(-2 \ln \mathcal L),\text{ndf}))$ is  the 
chi-squared distribution
and
$\text{erfc}^{-1}$ is the inverse error function converting the p-value into Gaussian standard deviations.
This assumes that $\Delta(-2 \ln \mathcal L)$ follows a chi-squared distribution under the null hypothesis (\ie the resonance is not present in the data).
If the mass and width are fixed, the effective number of degrees of freedom, $\text{ndf}$, is set 
to the number of additional fit parameters under the hypothesis that the resonance exists, $\Delta N_{\rm par}$.
Otherwise, we set it to $\text{ndf} = 2 \Delta N_{\rm par}$~\cite{LHCb-PAPER-2014-014,LHCb-PAPER-2020-044}. 
This method is verified to be conservative through the use of  pseudoexperiments.
The statistical significance of all exotic states is above 8 standard deviations ($\sigma$) and stays above $7\sigma$ when considering systematic uncertainties (see Sec.~\ref{sec:sys}).

To verify the nominal quantum-number assignment of the exotic states,
fits are performed under alternative spin-parity hypotheses with spin up to 2.
The likelihood-ratio test is used to quantify the rejection of these hypotheses.
For all exotic states that are part of the \textit{baseline} model, every alternative spin-parity hypothesis is excluded by more than $8\sigma$.

The largest exotic contribution comes from the scalar resonance $\XSone$, which predominantly decays as $\XSone \to\psitwos \rho(770)^0$.
Its mass agrees well with the $\chi_{c0}(4500)$ state observed in the decay $\chi_{c0}(4500) \to \jpsi \phi$~\cite{LHCb-PAPER-2020-044},
however it is significantly broader. 
Note that the Breit--Wigner parameters are known to be highly model-dependent and the two measurements take
different partial decay widths as the total decay width ($\XSone \to \psitwos \pip \pim$ versus $\chi_{c0}(4500) \to \jpsi \phi$).
This is compounded by the fact that the $\XSone$ or $\chi_{c0}(4500)$ masses are close to the $\psitwos \rho(770)^0$ threshold,
which can significantly distort the lineshapes.
A possible explanation of the disagreeing decay widths might thus be that coupled-channel effects are not accounted for in either measurement.
The $B^+ \to \jpsi \phi K^+$ analysis also includes a huge nonresonant $\jpsi \phi$ $0^+$ component that makes the comparison less straightforward.
Adding an equivalent nonresonant $\psitwos \rho(770)^0$ $0^+$ component to the \textit{baseline} model results in a small fit fraction, \mbox{$(0.48 \pm 0.36)\%$}, 
and decreases the $\XSone$ decay width by $16\mev$.
Assuming isospin conservation in strong decays, we interpret these 
as genuinely different states in the following, 
with the $\XSone$ state having isospin $I=1$ and the $\chi_{c0}(4500)$ state having $I=0$, 
as indicated by their decay modes $\psitwos \rho(770)^0$ and $\jpsi \phi$.
However, if the $\XSone$ state were the $c \bar c (d\bar d - u \bar u)$ isospin partner of
the $\chi_{c0}(4500)$ interpreted as $c \bar c s \bar s$, one would generally expect a larger mass difference of $m_{\chi_{c0}(4500)} - m_{\XSone} \approx +200\mev$.

The $\XAone$ and $\XStwo$ states agree within uncertainties with the $\chi_{c1}(4685)$ and $\chi_{c0}(4700)$ resonances observed in $B^+ \to \jpsi \phi K^+$ decays~\cite{LHCb-PAPER-2020-044}.
Nevertheless, for them to be identical would require a significant isospin violating effect,
which seems implausible.
The mass and the width of the $\XVone$ vector state are not well constrained due to its proximity to the phase-space limit.
Within the large uncertainties, the $\XVone$ resonance parameters are loosely compatible with the $X(4630)^0$ resonance observed in $B^+ \to \jpsi \phi K^+$ decays~\cite{LHCb-PAPER-2020-044}.
However, the quantum numbers of the $X(4630)^0$ resonance are not yet unambiguously determined.
The $J^P = 1^-$ assignment for $X(4630)^0$ is favoured over $J^P = 2^-$ with a significance of $3\sigma$ and other assignments are disfavoured by more than $5\sigma$~\cite{LHCb-PAPER-2020-044}.

A large axial vector component in the $\psitwos \pi^+$ system is described by two resonances, $\ZAone^+$ and $\ZAtwo^+$.
They are predominantly produced in the quasi-two-body topology, $\Bp \to \Zp K^*(892)^0$ and $\Bp \to \Zp [K^+ \pim]_S$,
but are also seen in $\Xz \to \Zpm \pi^\mp$ and $\Xsz \to \Zm K^+$ cascade decays.
This is the first observation of production modes other than
$\Bz \to \Zm \Kp$.
The mass and width of the $\ZAtwo^+$ agree well with those measured in $\Bzb \to \psitwos K^- \pip$ decays~\cite{LHCb-PAPER-2014-014,Belle:2013shl}.
The second, lighter, axial vector state is compatible with the one
reported in the Belle analysis of $\Bzb \to \jpsi K^- \pip$ decays~\cite{Belle:2014nuw},
where $J^P=1^+$ is preferred over $J^P=2^-$ by $4.4\sigma$ and other assignments are excluded.
The LHCb analysis of the $\Bzb \to \psi(2S) K^- \pip$ decay indicated a similar structure in the $0^-$ or $1^+$ $\psitwos \pip$ wave~\cite{LHCb-PAPER-2014-014}.
The $\ZAone^+$ state also has similar mass and width to the $\Z(4250)^+$ state reported in a Belle analysis of $\Bzb \to \chi_{c1}(1P) K^- \pip$ decays~\cite{Belle:2008qeq},
where $J=0$ is the default assumption, though $J=1$ cannot be excluded.
This is the first time that
the $J^P=1^+$ assignment for the $\ZAone^+$
has been unambiguously demonstrated.

A subpercent level contribution from a vector state, $\ZVone^+$, is also observed.
The mass and width of this state are fixed to the parameters of the $\Z(4055)^+$ state seen in the reaction $e^+e^- \to \gamma \psitwos \pip \pim$~\cite{Belle:2014wyt}.
A similar structure has been reported in $\Bzb \to \chi_{c1}(1P) K^- \pip$  ($\Z(4050)^+$~\cite{Belle:2008qeq}) and $\Bzb \to \eta_{c1}(1S) K^- \pip$ ($\Z(4100)^+$~\cite{LHCb-PAPER-2018-034}) decays.  
None of those measurements determined the quantum numbers.
No stable fit results are obtained when floating the $\ZVone^+$ resonance parameters. 
This might be due to the existence of several small $1^-$ $\psitwos \pip$ wave contributions that cannot be resolved with the current statistical precision 
or by residual model deficiencies of the other partial waves being absorbed into the $\ZVone^+$ description.
Thus the high significance of the $\ZVone^+$ component should not be considered as a confirmation of a specific state
but rather an effective description of a generic $1^-$ $\psitwos \pip$ wave contribution.

The baseline model includes the high-mass tail of the $\ZsAone$ resonance observed in $B^+ \to \jpsi \phi K^+$ decays~\cite{LHCb-PAPER-2020-044}.
As alternative models, we replaced the $\ZsAone$ by the $\ZsAtwo$ assuming it to be an axial vector or vector state. 
Both hypotheses are disfavoured by more than $5\sigma$.
Replacing the $\ZsAone$ state by a nonresonant component with an exponential lineshape, 
$\exp{\left(-\alpha \, m^2(\psitwos K^+)\right)}$, 
is also disfavoured by more than $5\sigma$. 
Nonetheless, the results of this analysis should not be considered as an independent confirmation of the $\ZsAone$ state 
as no sufficient sensitivity to the mass and width values is achieved.
All \mbox{$\Xsz \to \psitwos K^+ \pi^-$} states are previously unobserved states.
If interpreted as tetraquark states, their minimal quark content would be $c \bar c \bar s d$.
The $\XsAone$ and $\XsAtwo$
could be radial excitations of the $1^+$ $\ZsA(4000)^0 \to\jpsi K^0_{\text S}$ resonance, for which evidence was found in 
$\Bz \to \jpsi \phi K^0_{\text S}$ decays
\cite{LHCb-PAPER-2022-040}.

The \textit{baseline} model features two broad vector resonances that lie at the edges of the phase space, $\XVone$ and $\XsVone$.
For these states, we perform an additional significance test comparing the Breit--Wigner lineshape hypothesis against a nonresonant hypothesis with exponential lineshape. 
The Breit--Wigner hypothesis is favoured by $5.4\sigma$ for the $\XVone$ and $2.6\sigma$ for the $\XsVone$. 
The proximity to the phase-space boundaries and large uncertainties 
make the interpretation of these components as genuine resonances challenging.
Without independent confirmation of these states from other measurements,
they should be considered an effective description of several $J^P = 1^-$ contributions (within or outside the $\signal$ phase space) that cannot be 
disentangled with the existing data.

\begin{figure}[h]
       	 \includegraphics[width=0.329\textwidth,height=!]{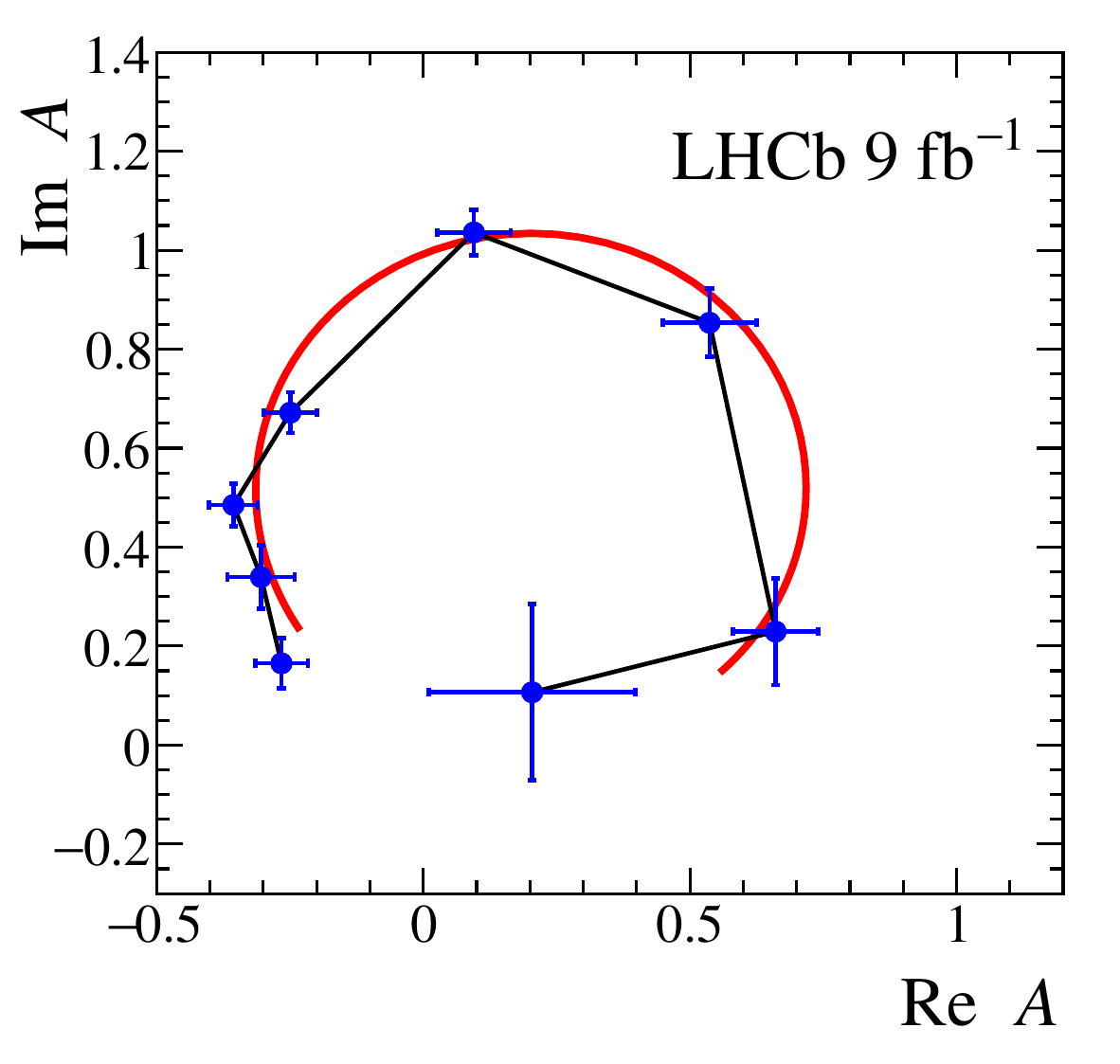}
       	 \includegraphics[width=0.329\textwidth,height=!]{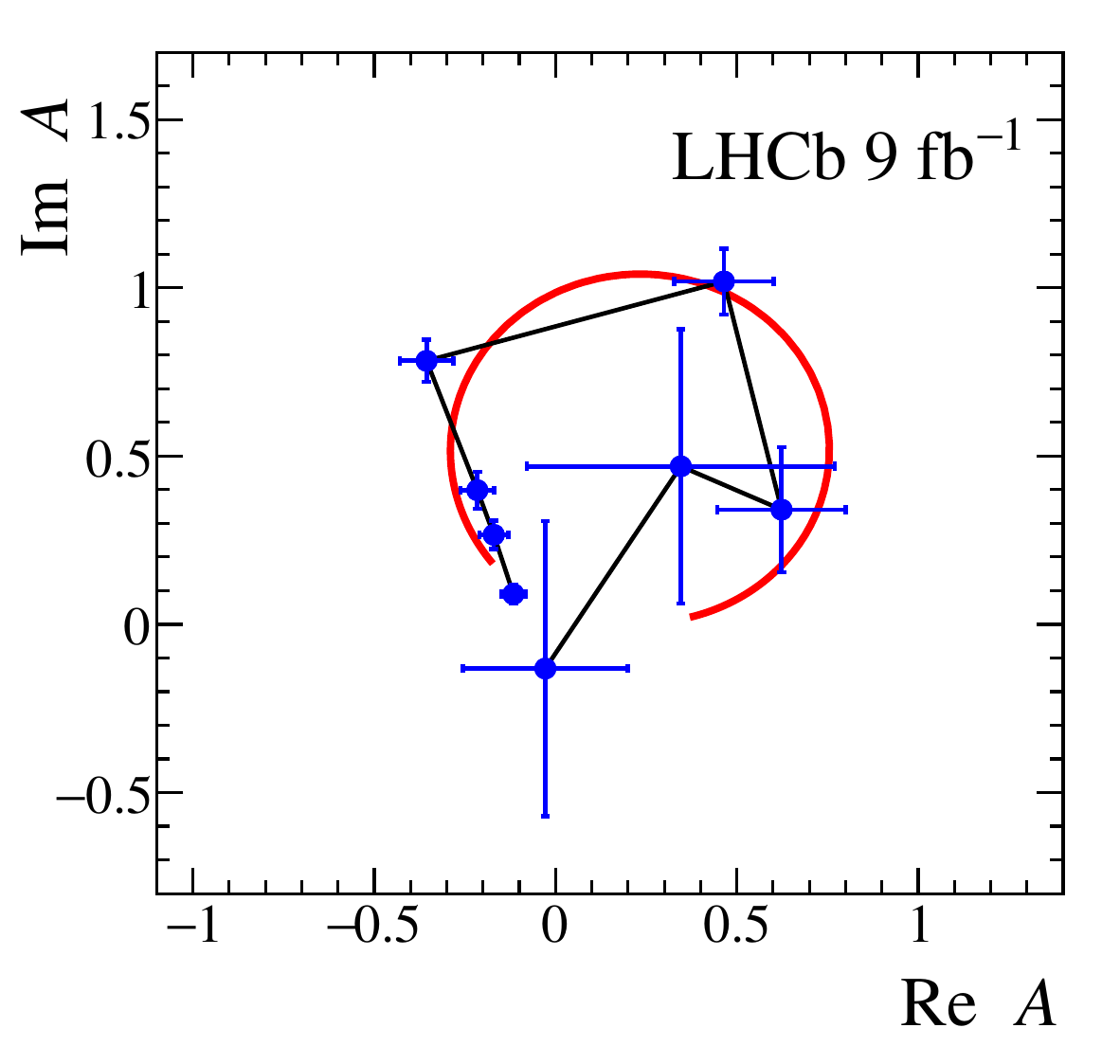}
       	 \includegraphics[width=0.329\textwidth,height=!]{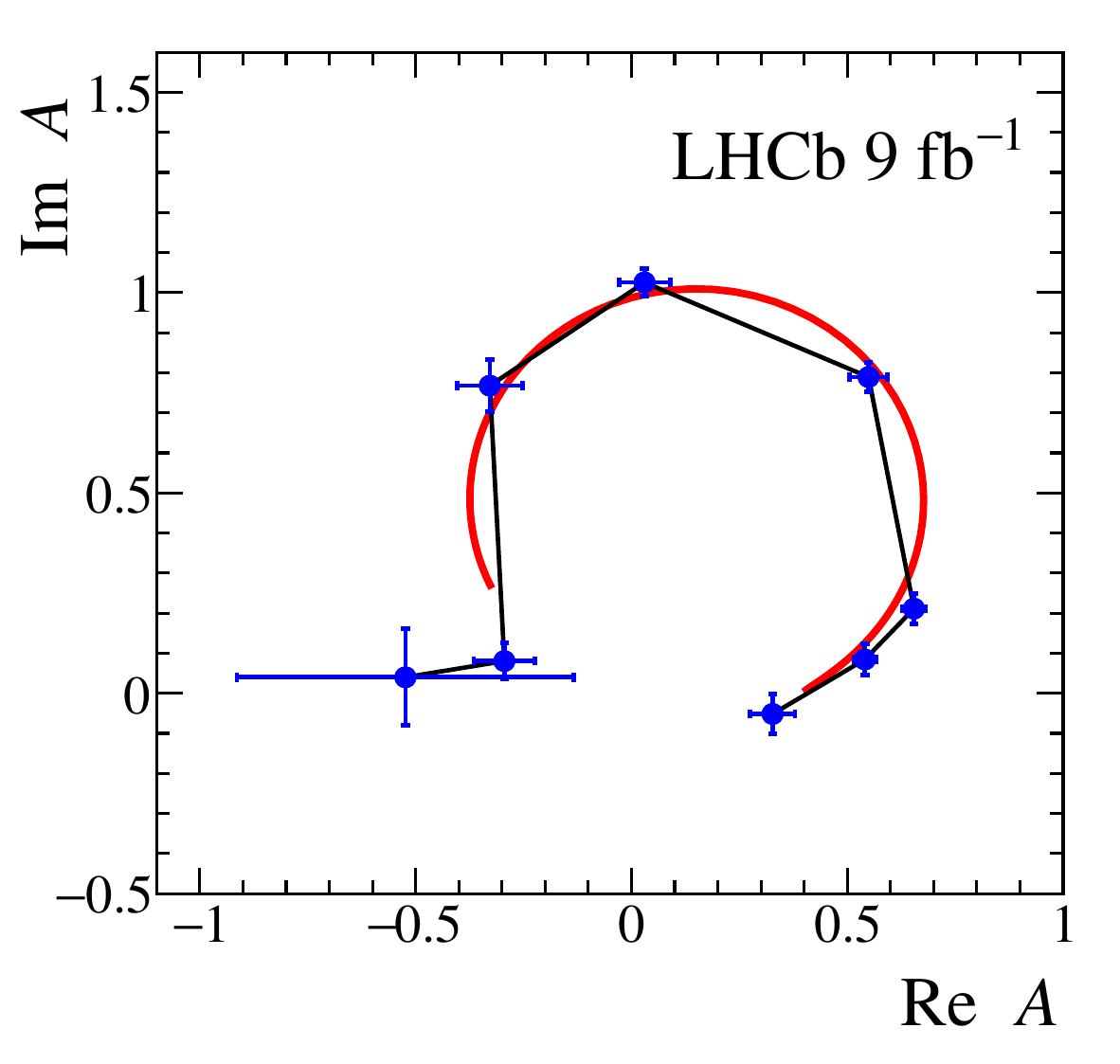}

	\caption{
	Argand diagram for the quasi-model-independent partial-wave analysis for the (left) $\XSone$, (middle) $\XsAone$, and (right) $\ZAone^\pm$ resonances.
	The fitted lineshape knots are displayed as connected points with error bars and the invariant mass increases counterclockwise.
	The Breit--Wigner lineshape with the mass and width from the nominal fit is superimposed (red line). }
	\label{fig:argandMain}
\end{figure}

The resonant nature of the exotic states is probed using a
quasi-model-independent partial-wave analysis (QMIPWA)~\cite{Aitala:2005yh,LHCb-PAPER-2014-014}.
The nominal Breit--Wigner lineshape
of a given resonance is replaced by  a
parameterisation that treats the magnitude and phase of the lineshape at eight discrete
positions in the invariant-mass region around the nominal mass as independent pairs of free parameters to be determined by the fit.
The lineshape is then modelled elsewhere by interpolating between these values using cubic splines~\cite{inbook}. 
The Argand diagrams 
for the largest \mbox{$\Xz \to \psitwos \pip \pim$}, \mbox{$\Xsz \to \psitwos K^+ \pim$} and \mbox{$\Zpm \to \psitwos \pi^\pm$} components 
are shown in Fig.~\ref{fig:argandMain}
and demonstrate a counter-clockwise circular phase motion as expected from a resonance.
More details and QMIPWA studies of the other exotic states are given in Appendix~\ref{a:MIPW}.

\begin{table}[h]
\caption{Fit fractions of the intermediate-state amplitudes contributing to $\signal$ decays for the \textit{baseline} model. The uncertainties are statistical and systematic.
The systematic uncertainties are dominated by the choice of 
amplitude components.
The states $\XVone, \XsVone$, $\ZsAone$ and $\ZVone^+$ should be considered as effective descriptions of generic partial wave contributions due to insufficient sensitivity of the amplitude fit.
}
\centering
\renewcommand{\arraystretch}{1.1}
	 \begin{tabular}{l r} 
\hline
\hline
Decay channel &  Fit fraction $[\%]$  \\ 
\hline
$B^{+}\rightarrow \XSone K^{+}$ & $18.45 \pm 1.31 \pm 2.92$ \\ 
$B^{+}\rightarrow \psi(\text{2S})\,K^{*}(1680)^{+}$ & $8.15 \pm 1.31 \pm 3.51$ \\ 
$B^{+}\rightarrow \psi(\text{2S})\,K_{1}(1270)^{+}$ & $7.60 \pm 0.85 \pm 1.35$ \\ 
$B^{+}[P]\rightarrow \psi(\text{2S})\,K_{1}(1270)^{+}$ & $7.52 \pm 0.60 \pm 1.08$ \\ 
$B^{+}[D]\rightarrow \psi(\text{2S})\,K_{1}(1270)^{+}$ & $6.81 \pm 0.45 \pm 1.18$ \\ 
$B^{+}\rightarrow \psi(\text{2S})\,K_{1}(1400)^{+}$ & $5.78 \pm 0.62 \pm 0.92$ \\ 
$B^{+}\rightarrow \psi(\text{2S})\,K(1460)^{+}$ & $5.26 \pm 0.48 \pm 0.87$ \\ 
$B^{+}[P]\rightarrow \ZAone^{+} \, K^{*}(892)^{0}$ & $4.60 \pm 0.54 \pm 2.17$ \\ 
$B^{+}\rightarrow \XsAone\pi^{+}$ & $4.42 \pm 0.98 \pm 2.17$ \\ 
$B^{+}\rightarrow K_{2}^{*}(1430)^{+}\psi(\text{2S})\,$ & $4.35 \pm 0.29 \pm 0.26$ \\ 
$B^{+}\rightarrow \ZAone^{+} \, K^{*}(892)^{0}$ & $4.02 \pm 0.88 \pm 1.01$ \\ 
$B^{+}\rightarrow \ZAtwo^{+} \, \left[K^{+}\pi^{-}\right]_{\text{S}}$ & $3.41 \pm 0.54 \pm 0.78$ \\ 
$B^{+}\rightarrow \XVone K^{+}$ & $3.24 \pm 0.50 \pm 0.79$ \\ 
$B^{+}\rightarrow \XAone K^{+}$ & $2.89 \pm 0.45 \pm 0.45$ \\ 
$B^{+}[D]\rightarrow \ZAone^{+} \, K^{*}(892)^{0}$ & $2.78 \pm 0.33 \pm 0.61$ \\ 
$B^{+}\rightarrow \XsAtwo \pi^{+}$ & $2.60 \pm 0.66 \pm 1.94$ \\ 
$B^{+}[D]\rightarrow \rho(770)^{0} \, \ZsAone $ & $2.06 \pm 0.22 \pm 0.84$ \\ 
$B^{+}\rightarrow \psi(\text{2S})\,K^{*}(1410)^{+}$ & $1.79 \pm 0.35 \pm 0.74$ \\ 
$B^{+}\rightarrow \XStwo K^{+}$ & $1.73 \pm 0.28 \pm 0.40$ \\ 
$B^{+}\rightarrow \XsVone \pi^{+}$ & $1.59 \pm 0.46 \pm 0.61$ \\ 
$B^{+}\rightarrow \ZsAone \, \left[\pi^{+}\pi^{-}\right]_{\text{S}}$ & $1.24 \pm 0.23 \pm 0.34$ \\ 
$B^{+}\rightarrow \ZAtwo \, K^{*}(892)^{0}$ & $0.75 \pm 0.43 \pm 2.24$ \\ 
$B^{+}\rightarrow \psi(4360)K^{+}$ & $0.64 \pm 0.14 \pm 0.12$ \\ 
$B^{+}\rightarrow \ZVone \, K^{*}(892)^{0}$ & $0.52 \pm 0.10 \pm 0.11$ \\ 
$B^{+}[P]\rightarrow \psi(\text{2S})\,K_{1}(1400)^{+}$ & $0.48 \pm 0.18 \pm 0.40$ \\ \hline
$\text{Sum } B^{+}$ & $102.69 \pm 4.40 \pm 7.41$ \\ 
\hline
\hline
\end{tabular}

 \label{tab:fitBest}
\end{table}

\begin{table}[h]
\caption{Partial fit fractions of the cascade decays contributing to \mbox{$\signal$} decays for the \textit{baseline} model. The uncertainties are statistical and systematic. The systematic uncertainties are dominated by the choice of 
amplitude components.
The states $\XVone, \XsVone$, $\ZsAone$ and $\ZVone^+$ should be considered as effective descriptions of generic partial wave contributions due to insufficient sensitivity of the amplitude fit.
}
\centering
	  \resizebox{0.57\linewidth}{!}{
		\renewcommand{\arraystretch}{1.1}
		\begin{tabular}{l  r@{}r@{}r@{}r@{}r@{}l} 
\hline
\hline
Decay channel &  \multicolumn{6}{c}{Partial fit fraction $f_j^{R_j} [\%]$} \\ 
\hline
$K_{1}(1270)^{+}\rightarrow \rho(770)^{0}K^{+}$ & $50.71 $&$\pm$&$2.18 $&$\pm$&$3.19$  &\\ 
$K_{1}(1270)^{+}\rightarrow K^{*}(892)^{0}\pi^{+}$ & $19.86 $&$\pm$&$1.44 $&$\pm$&$2.05$ &\\ 
$K_{1}(1270)^{+}\rightarrow \left[K^{+}\pi^{-}\right]_{\text{S}}\pi^{+}$ & $11.35 $&$\pm$&$1.45 $&$\pm$&$2.11$ &\\ 
$K_{1}(1270)^{+}[D]\rightarrow K^{*}(892)^{0}\pi^{+}$ & $8.32 $&$\pm$&$0.85 $&$\pm$&$1.54$ &\\ 
$\text{Sum } K_{1}(1270)^{+}$ & $90.24 $&$\pm$&$1.83 $&$\pm$&$3.67$ &\\ \hline
$K_{1}(1400)^{+}\rightarrow K^{*}(892)^{0}\pi^{+}$ & $86.80 $&$\pm$&$3.78 $&$\pm$&$5.34$ &\\ 
$K_{1}(1400)^{+}\rightarrow \rho(770)^{0}K^{+}$ & $22.08 $&$\pm$&$4.40 $&$\pm$&$6.25$ &\\ 
$\text{Sum } K_{1}(1400)^{+}$ & $108.88 $&$\pm$&$0.82 $&$\pm$&$1.84$ &\\  \hline
$K(1460)^{+}\rightarrow \left[\pi^{+}\pi^{-}\right]_{\text{S}}K^{+}$ & $45.13 $&$\pm$&$4.22 $&$\pm$&$10.91$ &\\ 
$K(1460)^{+}\rightarrow K^{*}(892)^{0}\pi^{+}$ & $35.41 $&$\pm$&$4.08 $&$\pm$&$9.72$ &\\ 
$\text{Sum } K(1460)^{+}$ & $80.54 $&$\pm$&$0.67 $&$\pm$&$3.66$ &\\  \hline
$K_{2}^{*}(1430)^{+}\rightarrow K^{*}(892)^{0}\pi^{+}$ & $76.70 $&$\pm$&$3.04 $&$\pm$&$2.43$ &\\ 
$K_{2}^{*}(1430)^{+}\rightarrow \rho(770)^{0}K^{+}$ & $12.71 $&$\pm$&$2.30 $&$\pm$&$1.80$ &\\ 
$\text{Sum } K_{2}^{*}(1430)^{+}$ & $89.41 $&$\pm$&$0.75 $&$\pm$&$0.66$ &\\  \hline
$K^{*}(1410)^{+}\rightarrow K^{*}(892)^{0}\pi^{+}$ & $88.50 $&$\pm$&$8.39 $&$\pm$&$12.65$ &\\ 
$K^{*}(1410)^{+}\rightarrow \rho(770)^{0}K^{+}$ & $38.36 $&$\pm$&$10.46 $&$\pm$&$19.07$ &\\ 
$\text{Sum } K^{*}(1410)^{+}$ & $126.86 $&$\pm$&$4.83 $&$\pm$&$13.38$ &\\  \hline
$K^{*}(1680)^{+}\rightarrow K^{*}(892)^{0}\pi^{+}$ & $49.69 $&$\pm$&$6.72 $&$\pm$&$13.32$ &\\ 
$K^{*}(1680)^{+}\rightarrow \rho(770)^{0}K^{+}$ & $31.16 $&$\pm$&$6.11 $&$\pm$&$11.27$ &\\ 
$\text{Sum } K^{*}(1680)^{+}$ & $80.85 $&$\pm$&$0.64 $&$\pm$&$3.89$ &\\  \hline
$\XSone\rightarrow \rho(770)^{0}\psi(\text{2S})\,$ & $99.04 $&$\pm$&$0.49 $&$\pm$&$1.66$ &\\ 
$\XSone\rightarrow \ZAone^{-} \pi^{+}$ & $0.50 $&$\pm$&$0.25 $&$\pm$&$0.39$ &\\ 
$\XSone\rightarrow \ZAone^{+} \pi^{-}$ & $0.50 $&$\pm$&$0.25 $&$\pm$&$0.39$ &\\ 
$\text{Sum } \XSone$ & $100.03 $&$\pm$&$0.02 $&$\pm$&$1.42$ &\\  \hline
$\XAone\rightarrow \rho(770)^{0}\psi(\text{2S})\,$ & $86.66 $&$\pm$&$7.85 $&$\pm$&$8.49$ &\\ 
$\XAone\rightarrow \ZAone^{-}\pi^{+}$ & $6.62 $&$\pm$&$2.03 $&$\pm$&$2.46$ &\\ 
$\XAone\rightarrow \ZAone^{+}\pi^{-}$ & $6.61 $&$\pm$&$2.03 $&$\pm$&$2.47$ &\\ 
$\text{Sum } \XAone$ & $99.89 $&$\pm$&$7.37 $&$\pm$&$7.56$ &\\  \hline
$\XStwo\rightarrow \rho(770)^{0}\psi(\text{2S})\,$ & $92.35 $&$\pm$&$10.83 $&$\pm$&$15.11$ &\\ 
$\XStwo\rightarrow \ZAtwo^{+}\pi^{-}$ & $17.00 $&$\pm$&$3.82 $&$\pm$&$3.38$ &\\ 
$\XStwo\rightarrow \ZAtwo^{-}\pi^{+}$ & $17.00 $&$\pm$&$3.82 $&$\pm$&$3.37$ &\\ 
$\text{Sum } \XStwo$ & $126.35 $&$\pm$&$10.56 $&$\pm$&$14.40$ &\\  \hline
$\XVone\rightarrow \rho(770)^{0}\psi(\text{2S})\,$ & $41.52 $&$\pm$&$5.19 $&$\pm$&$10.44$ &\\ 
$\XVone\rightarrow \ZAone^{-}\pi^{+}$ & $18.03 $&$\pm$&$5.14 $&$\pm$&$6.59$ &\\ 
$\XVone\rightarrow \ZAone^{+}\pi^{-}$ & $18.03 $&$\pm$&$5.14 $&$\pm$&$6.58$ &\\ 
$\XVone\rightarrow \ZAtwo^{+}\pi^{-}$ & $5.44 $&$\pm$&$2.45 $&$\pm$&$2.29$ &\\ 
$\XVone\rightarrow \ZAtwo^{-}\pi^{+}$ & $5.44 $&$\pm$&$2.45 $&$\pm$&$2.29$ &\\ 
$\text{Sum } \XVone$ & $88.47 $&$\pm$&$11.26 $&$\pm$&$13.07$ &\\  \hline
$\XsAone\rightarrow \psi(\text{2S})\,K^{*}(892)^{0}$ & $50.87 $&$\pm$&$7.79 $&$\pm$&$11.55$ &\\ 
$\XsAone\rightarrow \ZAone^{-}K^{+}$ & $16.53 $&$\pm$&$3.79 $&$\pm$&$12.75$ &\\ 
$\XsAone\rightarrow \ZsAone\pi^{-}$ & $9.84 $&$\pm$&$3.28 $&$\pm$&$5.34$ &\\ 
$\text{Sum } \XsAone$ & $77.23 $&$\pm$&$5.22 $&$\pm$&$17.80$ &\\  \hline
$\XsVone\rightarrow \psi(\text{2S})\,\left[K^{+}\pi^{-}\right]_{\text{S}}$ & $66.28 $&$\pm$&$15.03 $&$\pm$&$17.35$ &\\ 
$\XsVone\rightarrow \ZsAone\pi^{-}$ & $9.37 $&$\pm$&$14.12 $&$\pm$&$13.23$ &\\ 
$\text{Sum } \XsVone$ & $75.65 $&$\pm$&$9.18 $&$\pm$&$13.39$ &\\  \hline

$\XsAtwo\rightarrow \psi(\text{2S})\,K^{*}(892)^{0}$ & \multicolumn{5}{c}{$100$} &\\

$\psi(4430)\rightarrow \psi(\text{2S})\,\left[\pi^{+}\pi^{-}\right]_{\text{S}}$  & \multicolumn{5}{c}{$100$}  &\\

\hline
\hline
\end{tabular}
 	  }
		\label{tab:fitBest2}
\end{table}

\begin{sidewaystable}[h]
\caption{Left: Resonance parameters determined from the fit and the significance of the resonance,  where the uncertainties are statistical and systematic.
The significances are evaluated accounting for statistical (total) uncertainties.
Right: If available, the states listed in the Particle Data Group (PDG)~\cite{PDG2022} with the same $J^P$ quantum numbers and closest mass are listed for comparison. 
The states $\XVone, \XsVone$, $\ZsAone$ and $\ZVone^+$ should be considered as effective descriptions of generic partial wave contributions due to insufficient sensitivity of the amplitude fit.
}
\resizebox{1.0\linewidth}{!}{
		\tiny
		\centering
		\renewcommand{\arraystretch}{1.5}
		\begin{tabular}{l l 
                            r@{}c@{}c@{}c@{}l
                            r@{}c@{}c@{}c@{}l
                            c c c | l 
                            r@{}c@{}l
                            r@{}c@{}l
                            }
		\hline \hline
		Resonance & $J^P$  & \multicolumn{5}{c}{$m_0$ [MeV]} & \multicolumn{5}{c}{$\Gamma_0$ [MeV]}
   & $\Delta(-2 \ln \mathcal L)$ & $\Delta N_{\rm par}$
  & Sign. [$\sigma$] & Res. PDG & \multicolumn{3}{c}{$m_0$ [MeV]} & \multicolumn{3}{c}{$\Gamma_0$ [MeV]}  \\  
		
		\hline

		$\XSone$  & $0^+$ & $4475 $&$\pm$&$ 7 $&$\pm$&$ 12$ & $231 $&$\pm$&$ 19 $&$\pm$&$ 32$ & 675 & 6 & $>20$ $(19)$ & $\chi_{c0}(4500) $&    $4474 $&$\pm$&$ 4 $  & $77 $&$^{+}_{-}$&$ ^{12}_{10}$  \\
		$\XAone$  & $1^+$ & $4653 $&$\pm$&$ 14 $&$\pm$&$ 27$ & $227 $&$\pm$&$ 26 $&$\pm$&$ 22$  & 286 & 6 & 15 (13) & $\chi_{c1}(4685)$ &   $4684$&$^{+}_{-}$&$^{15}_{17} $  &  $126 $&$\pm$&$ 40$ \\
		$\XStwo$ & $0^+$ & $ 4710 $&$\pm$&$ 4 $&$\pm$&$ 5$& $64 $&$\pm$&$ 9 $&$\pm$&$ 10$ & 255 & 6 & 14 (10) & $\chi_{c0}(4700)$&   $4694$&$^{+}_{-}$&$^{16}_{5} $ & $87$&$^{+}_{-}$&$^{18}_{10}$  \\
		$\XVone$  & $1^-$ &  $4785 $&$\pm$&$ 37 $&$\pm$&$ 119$& $457 $&$\pm$&$ 93 $&$\pm$&$ 157$ & 382 & 8 & 17 (12) & $X(4630) $ & $4626$&$^{+}_{-}$&$^{24}_{110} $  & $174$&$^{+}_{-}$&$^{140}_{80}$  \\

		\hline 
		
		$\ZVone^+$  & $1^-$ & \multicolumn{5}{c}{$4054$ (fixed)}  & \multicolumn{5}{c}{$45$ (fixed)}  & 81 & 2 & $8$ (7) & $\Z(4055)^+$ & $4054 $&$\pm$&$ 3.2$ & $45 $&$\pm$&$ 13$  \\
		$\ZAone^+$ & $1^+$ & $4257 $&$\pm$&$ 11 $&$\pm$&$ 17$ & $308 $&$\pm$&$ 20 $&$\pm$&$ 32$  & 842 & 16 & $>20$ ($>20$) & $\ZA(4200)^+$ &  $4196$&$^{+}_{-}$&$^{35}_{32}$ & $370$&$^{+}_{-}$&$^{100}_{150}$  \\
		$\ZAtwo^+$ & $1^+$ & $4468 $&$\pm$&$ 21 $&$\pm$&$ 80$ & $251 $&$\pm$&$ 42 $&$\pm$&$ 82$  & 305 & 10 & $15$ (8)   & $\ZA(4430)^+$ &   $4478$&$^{+}_{-}$&$^{15}_{18}$ & $181 $&$\pm$&$ 31$  \\
		
		\hline
		
		$\XsAone$  & $1^+$ & $4578 $&$\pm$&$ 10 $&$\pm$&$ 18$ & $133 $&$\pm$&$ 28 $&$\pm$&$ 69$  & 287 & 8  & 15 (12) &  \\
		$\XsAtwo$ & $1^+$ & $ 4925 $&$\pm$&$ 22 $&$\pm$&$ 47$& $255 $&$\pm$&$ 55 $&$\pm$&$ 127$ &177  & 4  & 12 (8)  &   \\
		$\XsVone$  & $1^-$ &  $5225 $&$\pm$&$ 86 $&$\pm$&$ 181$& $226 $&$\pm$&$ 76 $&$\pm$&$ 374$ & 149 & 6 & 10 (8) &   \\
		
		\hline

		$\ZsAone$ & $1^+$ & \multicolumn{5}{c}{$4003$  (fixed)} & \multicolumn{5}{c}{$131$  (fixed)} & 597 & 4 & $>20$ (14)  & $\ZsAone$ &  $4003$&$^{+}_{-}$&$^{7}_{15}$ & $131$&$\pm$&$30$   \\

		 \hline \hline
		\end{tabular}
}
\label{tab:resoParams}
\end{sidewaystable}
\clearpage

\subsection{Systematic uncertainties}
\label{sec:sys}

Several sources of systematic uncertainty are considered. 
Experimental issues are discussed first, followed by uncertainties related to the amplitude model and formalism.
The overall fit procedure is tested by generating pseudoexperiments from the \textit{baseline} fit
model using the measured values and subsequently fitting them with the same model.
For each pseudoexperiment and fit parameter, a pull is calculated by dividing the difference
between the fitted and generated values by the statistical uncertainty. The means of the
pull distributions are assigned as systematic uncertainties due to an intrinsic fit bias.

The treatment of the phase-space acceptance relies on simulated data. 
The integration error due to the limited size of the simulated sample used to normalise the signal PDF is estimated by bootstrapping the simulated sample.
The standard deviation of the fit results is assigned as systematic uncertainty.
To assess the uncertainty due to possible data-simulation differences, alternative phase-space acceptances are derived by varying the
selection requirements, for the simulated sample only, on quantities that are expected not to be well described by the simulation.
This includes the trigger~\cite{LHCb-PUB-2014-039} and tracking efficiencies~\cite{LHCb-DP-2013-002}, as well as the particle identification~\cite{LHCb-PUB-2016-021, LHCb-DP-2018-001} and multivariate selection classifier performances.

The uncertainty of the background fraction is propagated to the amplitude fit by varying this parameter within its uncertainty,
which also accounts for alternative signal and background invariant-mass models.
Alternative background phase-space distributions are derived by using additional kinematic observables to train the BDTG algorithm or by
using only candidates from the low- or high-mass sideband.
The amplitude fit is repeated with these alternative background PDFs and the largest deviation to the nominal result is used as systematic uncertainty.

The uncertainties due to fixed masses and widths of resonances are evaluated
by varying them one-by-one within their quoted errors.
The radii of the normalised Blatt--Weisskopf barrier factors are set to 
$r_{\rm BW} = 4  \gev^{-1}$ for the $B^+$, 
$r_{\rm BW} = 1.2  \gev^{-1}$ for all three-body resonances
and $r_{\rm BW} = 2.2  \gev^{-1}$ for all two-body resonances.
Each of these three values is obtained by maximising the likelihood of amplitude fits while fixing the values of the other two. 
To evaluate the associated systematic uncertainties, the radial parameters are varied by $\pm 1\gev^{-1}$.

As argued in Sec.~\ref{sec:lineshapes}, the $\rho-\omega$ mixing parameter $\delta$ is initially fixed to  
\mbox{$\vert \delta \vert = +3 \vert \delta_{em} \vert = (4.71 \pm 0.47) \times 10^{-3}$}~\cite{Akhmetshin:2001ig} 
and $\text{arg}(\delta)=0$
during the model selection.
When floating these parameters using the \textit{baseline} model, we obtain 
$\vert \delta\vert = (3.9 \pm 0.5) \times 10^{-3}$ and $\text{arg}(\delta)=(-11 \pm 7)^\circ$, 
in good agreement with the assumption.
There are multiple topologies with potentially different quark sources for the $\rho$ production to be considered in $\signal$ decays.
Thus we allow for different $\vert \delta \vert$ values for certain channels and obtain:
\begin{align}
\label{eq:rhoOmegaDelta}
\nonumber
K_1(1270)^+ &: \vert \delta \vert = (3.6 \pm 0.9) \times 10^{-3}, \\  \nonumber
\text{all other } K^{\prime+} \to K^+ \, \rho(770)^0 &: \vert \delta \vert = (3.8 \pm 1.3) \times 10^{-3},  \\ \nonumber
\XSone &: \vert \delta \vert = (6.0 \pm 0.8) \times 10^{-3},   \\ \nonumber
\text{all other } \Xz \to \psitwos \, \rho(770)^0 &: \vert \delta \vert = (6.1 \pm 1.2) \times 10^{-3},   \\ 
\Bp \to \Zsp \,\rho(770)^0 &: \vert \delta \vert = (3.9 \pm 2.0) \times 10^{-3}  ,
\end{align}
where the uncertainties are statistical only.
With $\vert \delta \vert \approx 2.3 \vert \delta_{em} \vert$ and $\vert \delta \vert \approx 2.4 \vert \delta_{em} \vert$,
the parameter values for the $K_1(1270)^+$ and the other $K^{\prime+}$ resonances are slightly lower but within uncertainties
consistent with a pure $u\bar u$ source.
Somewhat larger isospin-breaking effects are observed for the exotic $\Xz\to \psitwos \rho(770)^0$ topologies $\vert \delta \vert \approx 3.8 \vert \delta_{em} \vert$.
A large value of  $\vert \delta \vert \approx 11 \vert \delta_{em} \vert$ was also found for $\chi_{c1}(3872) \to \jpsi \rho(770)^0$ decays in Ref.~\cite{LHCb-PAPER-2021-045}. 
In the nominal fit, we fix the parameters to the ones obtained from the dedicated fit displayed in Eq.~\ref{eq:rhoOmegaDelta}. 
A systematic uncertainty is assigned by using $\vert \delta \vert = +3 \vert \delta_{em} \vert$ for all topologies and varying it within the uncertainties of $\vert \delta_{em} \vert$.

The P-vectors for the nominal dipion S-wave model consider the direct coupling to the $\pi\pi$ and $KK$ channels and up to four poles depending on the production mode, see Appendix~\ref{a:AmpLineShapes}.
Additional channels and poles are added to assign a systematic uncertainty. 
A similar procedure is performed for the $K\pi$ S-wave description.

Several alternative lineshape parameterisations are considered as part of the systematic studies.
The Gounaris--Sakurai description for the $\rho(770)^0$ resonance is replaced by a relativistic Breit--Wigner function. 
As an alternative to the $\pi\pi$ S-wave K-matrix parameterisation, the Omn\`es function based on dispersion integrals is used~\cite{Ropertz:2018stk,Garcia-Martin:2011iqs}.
Alternative energy-dependent widths for three-body resonances are derived from Eq.~\ref{eq:gamma3}
taking only the dominant three-body decay mode into account.
For the $\XSone$ state, a coupled-channel decay width is used which includes the $\psitwos \rho(770)^0$ and $\jpsi \phi$ channels,
\begin{equation}
\Gamma(s) = \Gamma_0 \, \frac{g_{\psitwos\rho} \, \rho_{\psitwos\rho}(s) + g_{\jpsi \phi} \, \rho_{\jpsi \phi}(s) }{ g_{\psitwos\rho} \, \rho_{\psitwos\rho}(m_0^2) + g_{\jpsi \phi} \, \rho_{\jpsi \phi}(m_0^2) }.
\end{equation}
We find $g_{\jpsi \phi} / g_{\psitwos\rho} = 0.26 \pm 0.18$ and similar mass and width as in the nominal fit.
The fit is not able to discriminate between the $\jpsi \phi$ and $\jpsi\rho(770)^0$ coupled channels.
For each alteration, an amplitude fit is performed and the standard deviation of the obtained fit results is assigned as a systematic uncertainty. 
The impact of nearby open-charm thresholds is also studied
by using coupled-channel decay widths for the various exotic resonances.
In all cases no sensitivity to the $D^{(*)}_{(s)}D^{(*)}_{(s)}$ open-charm coupling
is observed and no systematic uncertainty is attributed.

A range of modifications to the \textit{baseline} model are examined to assign an additional uncertainty due to the choice of amplitude components.
Fit results for the eleven alternative models can be found in Appendix~\ref{a:models}, Tables~\ref{tab:altModels1_1} to~\ref{tab:altModels2_3}.
The set of alternative models encompasses an ``extended $K^{\prime+}$" model (Model 1), 
where we include additional decay modes and angular momentum configurations of the selected $K^{\prime+} \to K^+\pip\pim$ resonances 
as well as an ``additional $K^{\prime+}$" model (Model 2) that includes additional $K^{\prime+}$ resonances.
In a similar fashion, the ``extended exotic" model (Model 3) 
includes additional decay modes and angular momentum configurations of the selected exotic states.
We searched for new exotic states with the highest significance in each mass dimension to create
``additional exotic"
models.
Model 4 includes additional $1^+ \, \psitwos \pip\pim, 2^- \, \psitwos \pip$ and $0^- \, \psitwos \pip$ states,
while Model 5 includes additional  $0^+ \, \psitwos K^+\pim, 0^+ \, \psitwos K^+ \pip$ and $0^- \, \psitwos K^+$ states. 
The statistical significance of each of those additional states is below $4\sigma$.
An ``additional nonresonant" model (Model 6) that adds several nonresonant amplitudes is also considered.
Another alternative model is derived by repeating the iterative model building with modified binning for the $\chi^2$ calculation (Model 7).
Further models are constructed by adding 25 additional amplitudes to
the \textit{baseline} model and applying a regularisation method~\cite{Guegan:2015mea,dArgent:2017gzv,LHCB-PAPER-2020-030}.
Either a Cauchy~\cite{Gelman_2008} (Model 8)
or a LASSO~\cite{Tibshirani94regressionshrinkage,BIC} regularisation term (Model 9) is used.
Lastly, fits are performed using the 
canonical helicity formalism instead of the covariant tensor formalism.
The definition of S- and D-wave amplitudes differs between the formalisms~\cite{Filippini,Chung:2007nn,Chen:2017gtx}.
There are also differences in the energy-dependent form factors~\cite{JPAC:2017vtd}.
The results are thus not expected to be strictly identical when switching from one formalism to the other,
even if a complete set of partial waves would be used.
The nominal set of partial-wave amplitudes is used for the canonical helicity fit in Model 10, while Model 11 adds additional partial waves.
From this set of alternative models, we compute the sample variance for each observable and take it as the model uncertainty.

The total systematic uncertainty is obtained by summing the components in quadrature.
A full breakdown of the different sources of systematic uncertainty for all fit parameters and observables is given in Appendix~\ref{a:models}.
The total systematic uncertainty is significantly larger than the statistical uncertainty, with the largest contributions coming from the alternative amplitude models.
The smallest significance value found when performing the dominant systematic variations is taken as the significance accounting for systematic uncertainty in Table~\ref{tab:resoParams}.

\section{Summary}
\label{sec:Summary}

The first full amplitude analysis of $\signal$ decays is performed using proton-proton collision data corresponding to $9 \invfb$ recorded with the LHCb detector.
The $\Kp \pip \pim$ spectrum is described by
six $K^{\prime+}$ resonances that decay via $K^*(892)^0 \pi^+$, $K^+\rho(770)^0$ or S-wave intermediate states.
Branching fractions measurements of the $K_1(1270)^+$ resonant substructure are found to be in good agreement with previous
analyses.

A good description of the data in the full seven-dimensional phase space could only be obtained by adding a multitude of exotic hidden-charm components to the model.
Four $\Xz \to \psitwos \pip \pim$ states are identified and the spectrum shows similarities to previously observed $\jpsi \phi$ resonances.
A simultaneous coupled-channel fit of $\signal$ and $\Bp \to \jpsi \phi \Kp$ could provide more insights into the quark-level interpretation of these states.
Further understanding might be gained by examining the structures of $\Xz \to \jpsi \omega$ or $\Xz \to \psitwos \omega$ exotic states.

New production modes of charged charmonium-like states are observed.
The $\ZAtwo^\pm$ resonance is confirmed with a high significance. 
The quantum numbers of the $\ZAone^\pm$ resonance are determined to be $1^+$,
for the first time with a significance exceeding $5\sigma$.
Further studies with more abundant $B^0 \to \psitwos K^+ \pi^-$ and $B^0 \to \jpsi K^+ \pi^-$ samples are required to improve our knowledge of the charged charmonium-like spectrum.
Hidden-charm exotic states that decay to the $\psitwos \Kp \pim$ final state are observed for the first time.
If interpreted as tetraquark states, their minimal quark content would be $c \bar c \bar s d$.
The $\XsAone$ and $\XsAtwo$
might be radial excitations of the $\ZsA(4000)^0$ resonance seen in 
$\Bz \to \ZsA(4000)^0 \phi \to [\jpsi K^0_{\text S}] \phi$ decays.
Despite the high significance of all exotic states, their broad nature and the extremely complex amplitude model with 53 components make the interpretation challenging.
In particular, the broad vector states at the edges of the phase space, $\XVone$ and $\XsVone$, should be considered as effective descriptions rather than genuine resonances at this point.
The same applies to the $\ZsAone$ and $\ZVone^+$ contributions, for which the amplitude fit does not provide sufficient sensitivity to measure their masses and widths.

The $\Kp \pip \pim$ component of the amplitude model provides valuable input 
for studies of new physics effects in $B^+ \rightarrow K^+ \pi^+ \pi^- \gamma$ or $B^+ \rightarrow K^+ \pi^+ \pi^- \mu^+ \mu^-$ decays.
As the first full four-body amplitude analysis featuring a vector particle in the final state, 
the research outlined in this paper establishes a groundwork for subsequent investigations into similar decay modes,
such as  \mbox{$B^+\to \jpsi K^+ \pip \pim$}, \mbox{$B^+\to \chi_{c1}(1P) K^+ \pip \pim$} or \mbox{$B^+\to \chi_{c1}(3872) K^+ \pip \pim$},
that could shed further light on the intricate spectrum of exotic hadrons.

\section*{Acknowledgements}
\noindent We express our gratitude to our colleagues in the CERN
accelerator departments for the excellent performance of the LHC. We
thank the technical and administrative staff at the LHCb
institutes.
We acknowledge support from CERN and from the national agencies:
CAPES, CNPq, FAPERJ and FINEP (Brazil); 
MOST and NSFC (China); 
CNRS/IN2P3 (France); 
BMBF, DFG and MPG (Germany); 
INFN (Italy); 
NWO (Netherlands); 
MNiSW and NCN (Poland); 
MCID/IFA (Romania); 
MICIU and AEI (Spain);
SNSF and SER (Switzerland); 
NASU (Ukraine); 
STFC (United Kingdom); 
DOE NP and NSF (USA).
We acknowledge the computing resources that are provided by CERN, IN2P3
(France), KIT and DESY (Germany), INFN (Italy), SURF (Netherlands),
PIC (Spain), GridPP (United Kingdom), 
CSCS (Switzerland), IFIN-HH (Romania), CBPF (Brazil),
and Polish WLCG (Poland).
We are indebted to the communities behind the multiple open-source
software packages on which we depend.
Individual groups or members have received support from
ARC and ARDC (Australia);
Key Research Program of Frontier Sciences of CAS, CAS PIFI, CAS CCEPP, 
Fundamental Research Funds for the Central Universities, 
and Sci. \& Tech. Program of Guangzhou (China);
Minciencias (Colombia);
EPLANET, Marie Sk\l{}odowska-Curie Actions, ERC and NextGenerationEU (European Union);
A*MIDEX, ANR, IPhU and Labex P2IO, and R\'{e}gion Auvergne-Rh\^{o}ne-Alpes (France);
AvH Foundation (Germany);
ICSC (Italy); 
Severo Ochoa and Mar\'ia de Maeztu Units of Excellence, GVA, XuntaGal, GENCAT, InTalent-Inditex and Prog. ~Atracci\'on Talento CM (Spain);
SRC (Sweden);
the Leverhulme Trust, the Royal Society
 and UKRI (United Kingdom).

\clearpage

\section*{Appendices}

\appendix

\section{Lineshape parameterisations}
\label{a:AmpLineShapes}
\setcounter{figure}{0}
\setcounter{table}{0}
\renewcommand{\thefigure}{A.\arabic{figure}}
\renewcommand{\thetable}{A.\arabic{table}}

\subsubsection*{Model for $\rho(770)^0$ resonance}

	We use the Gounaris--Sakurai parameterisation for the $\rho(770)^{0} \to \pip \pim$ propagator \cite{GS}:
\begin{equation}
    T_{GS}(s) = \frac{1+ f(m_{0}^{2})/ m_{0}^{2}  }
	{m_{0}^{2}+f(s)-s-i\,m_{0}\,\Gamma(s)}  \, ,
\end{equation}
where $\Gamma(s)$ takes on the same form as in Eq.~\ref{eq:gamma2} 
with $r_{\rm BW} = 0 \gev^{-1}$
and the function $f(s)$ is defined as
\begin{align}
	f(s) &= \Gamma_{0} \, \frac{m_{0}^{2}}{q_{0}^{3}} \left[   
	q^{2} \left( h(s) - h(m_{0}^{2}) \right) + \left(s-m_{0}^{2}\right) q_{0}^{2} \, \frac{dh}{d\sqrt{s}} \Bigg{\vert}_{m_{0}}
	\, \right],  \\
	h(s) &= \frac{2}{\pi} \, \frac{q}{\sqrt{s}} \,  \text{ln} \left( \frac{\sqrt{s}+2\,q}{2 \, m_{\pi}}
	\right)     \, .
\end{align}

\subsubsection*{S-wave models}

In the K-matrix formalism, the lineshape is written as
\begin{equation}
	T(s) = \left(  I - i \hat K \hat \rho \right)^{-1} \hat P,
	\label{eq:KmatrixT}
\end{equation}
where $\hat P$ is the process specific production vector,
$\hat \rho$ is the phase space matrix and $\hat K$ is the K matrix.
For the $\pi\pi$ S-wave, the K-matrix parameterisation is taken from Ref.~\cite{PhysRevD.78.034023},
 \begin{equation}
	 \hat K_{ij} = f(s) \, \left(  \sum_\alpha \frac{g_i^\alpha \, g_j^\alpha}{m^2_\alpha - s} +
    	  f_{ij}^{\text{scatt}} \frac{1 \gev^2 - s_0^{\text{scatt}}}{s - s_0^{\text{scatt}}}   \right),
	\label{eq:Kmatrix}
\end{equation}
where the index $i \in \{ \pi\pi, KK, \pi\pi\pi\pi, \eta\eta, \eta\eta^\prime \}$ indicates the channel 
and $\alpha$
indicates the pole.
Slowly varying scattering contributions are described by $f_{ij}^{\text{scatt}}$,
while the function $f(s)$ 
suppresses the kinematic singularity and introduces the Adler zero near the $\pi\pi$ threshold:
$f(s) = \frac{1 \gev^2-s_{A_0}}{s-s_{A_0}} (s - m_\pi^2/2 )$.
The production vector, $\hat P$ has the same pole structure as the K matrix,
 \begin{equation}
	 \hat P_{i} = \left( \sum_\alpha \frac{\beta^\alpha \, g_i^\alpha}{m^2_\alpha - s} +
    	  f_{i}^{\text{prod}} \frac{1 \gev^2 - s_0^{\text{prod}}}{s - s_0^{\text{prod}}}   \right),
	\label{eq:Pmatrix}
\end{equation}
where the complex parameter $\beta^\alpha$ describes the production strength of pole $\alpha$,
the complex parameter $f_{i}^{\text{prod}}$ describes the direct coupling to channel $i$
and $s_0^{\text{prod}}$ is a single real parameter.
As these parameters are process specific, they need to be determined from data.
The $K\pi$ S-wave couples to two channels $K\pi$ and $K\eta^\prime$, 
and contains only one pole, the $K_0^*(1430)$ resonance.
The isospin state $I=\frac{1}{2}$ contributes to both channels, 
while $I=\frac{3}{2}$ couples to $K\pi$ only.
Their parameterisations are
 \begin{align}
	 \hat K^{\frac{1}{2}}_{ij} &=
	 \frac{s-s_0^\frac{1}{2}}{m_K^2+m_\pi^2} \,
	  \left( \frac{g^1_i g^1_j  }{m_1^2-s} +  C_{ij0} + C_{ij1} \tilde s + C_{ij2} \tilde s^2 \right),   \\
	  \hat K^{\frac{3}{2}} &=
	 \frac{s-s_0^\frac{3}{2}}{m_K^2+m_\pi^2} \,
	  \left( D_{110} + D_{111} \tilde s + D_{112} \tilde s^2 \right),
	\label{eq:KmatrixKpi}
\end{align}
where all parameters and definitions are taken from Ref.~\cite{FOCUS:2007mcb}.
The Q-vector approximation is used, which simplifies Eq.~\ref{eq:KmatrixT} to
\begin{equation}
	T(s) = \left(  I - i \hat K \hat \rho\right)^{-1} \hat K \hat \alpha,
	\label{eq:KmatrixQ}
\end{equation}
where $\hat \alpha$ is a diagonal matrix containing a complex parameter for each channel to be determined from data.

\subsubsection*{Running-width functions for three-body resonances}

\begin{figure}[h]
\centering
\includegraphics[height=!,width=0.45\textwidth]{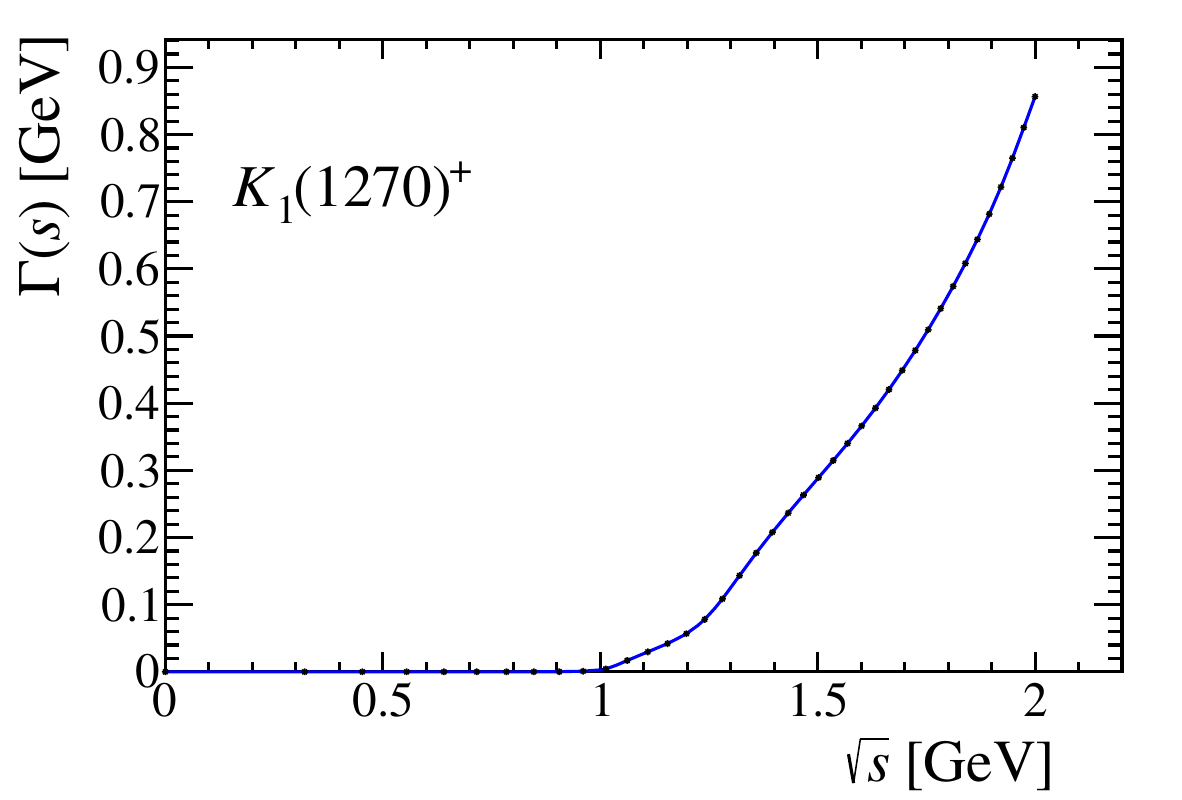}
\includegraphics[height=!,width=0.45\textwidth]{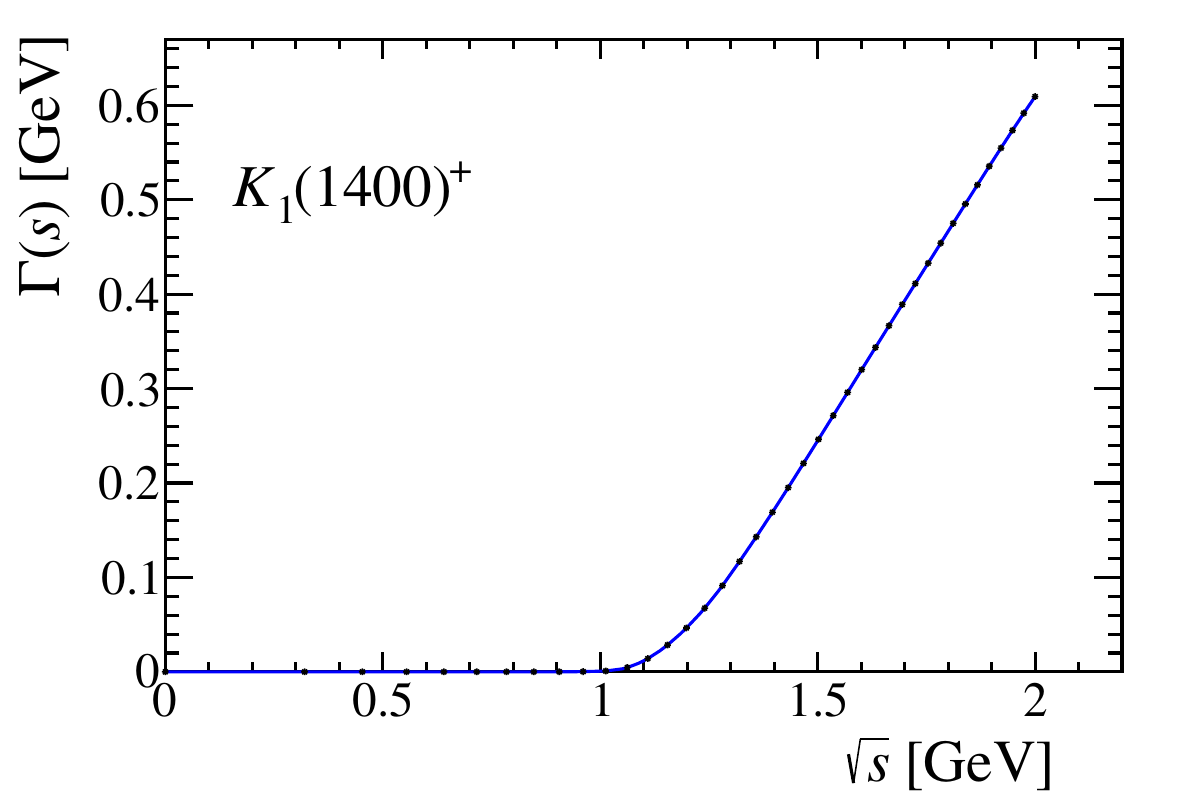}

\includegraphics[height=!,width=0.45\textwidth]{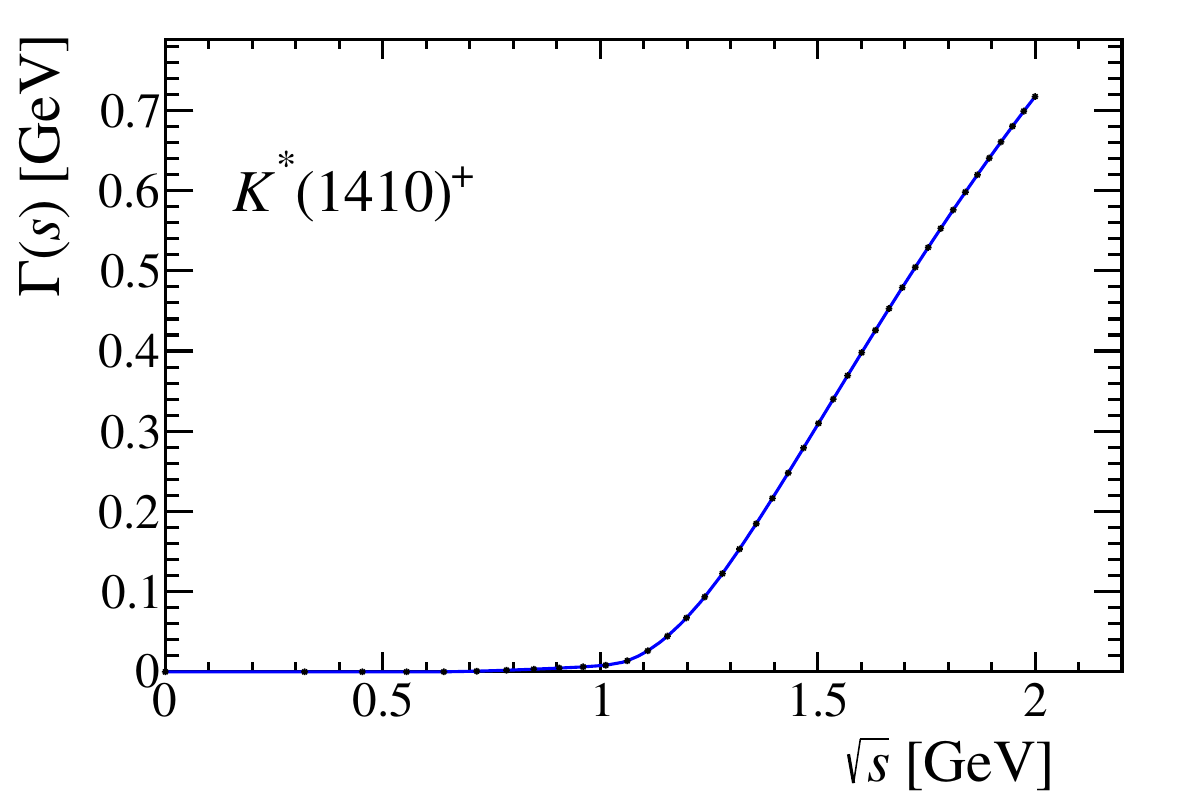}
\includegraphics[height=!,width=0.45\textwidth]{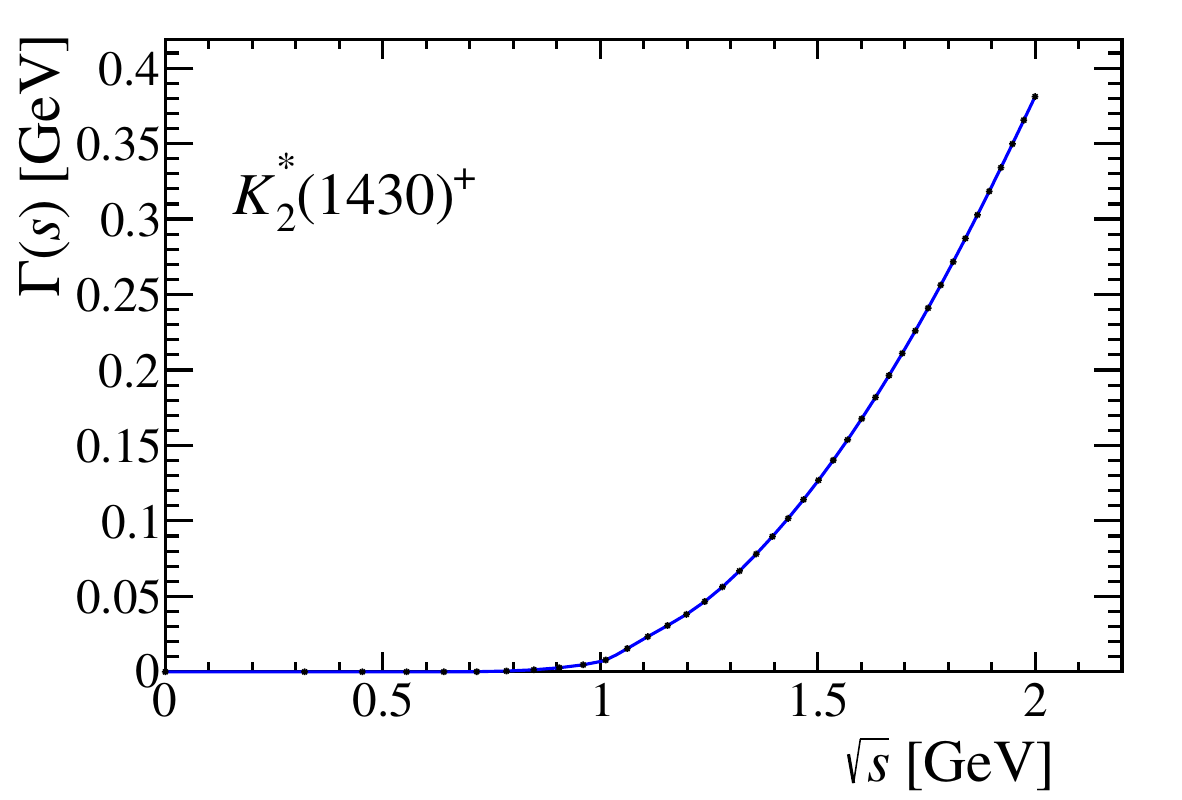}

\includegraphics[height=!,width=0.45\textwidth]{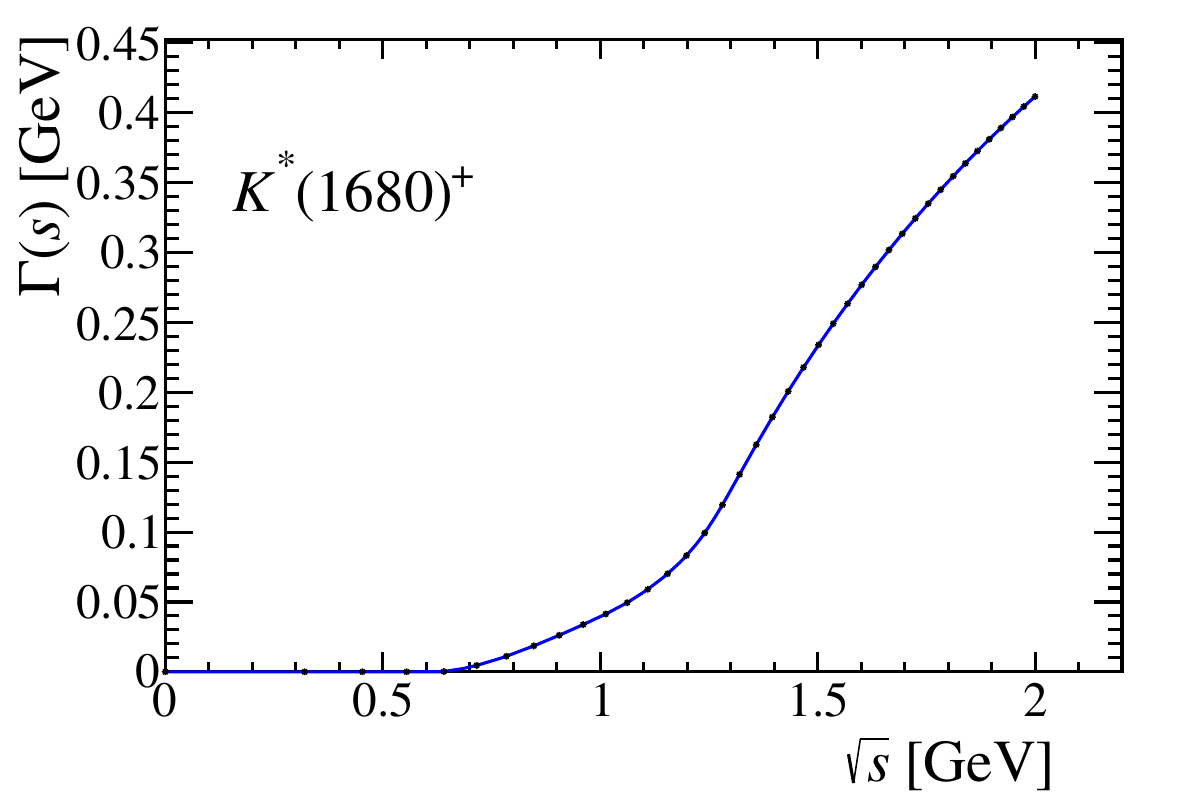}
\includegraphics[height=!,width=0.45\textwidth]{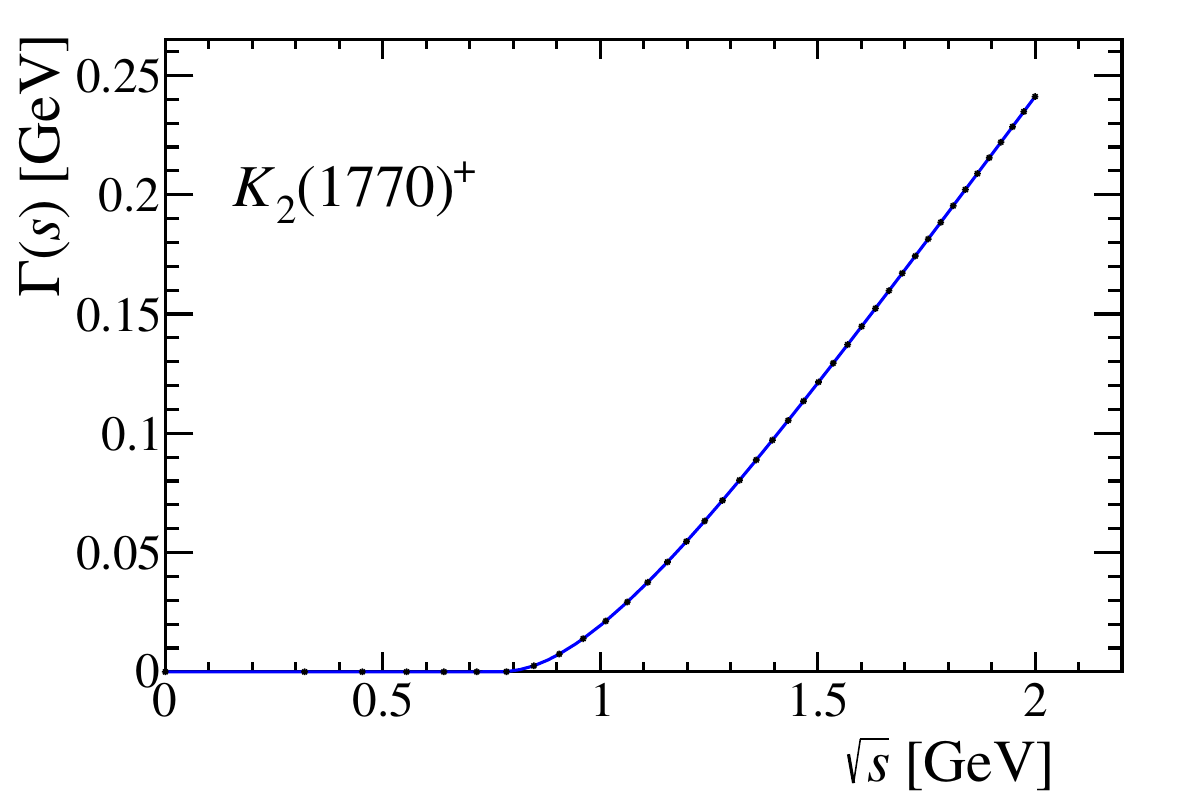}

\includegraphics[height=!,width=0.45\textwidth]{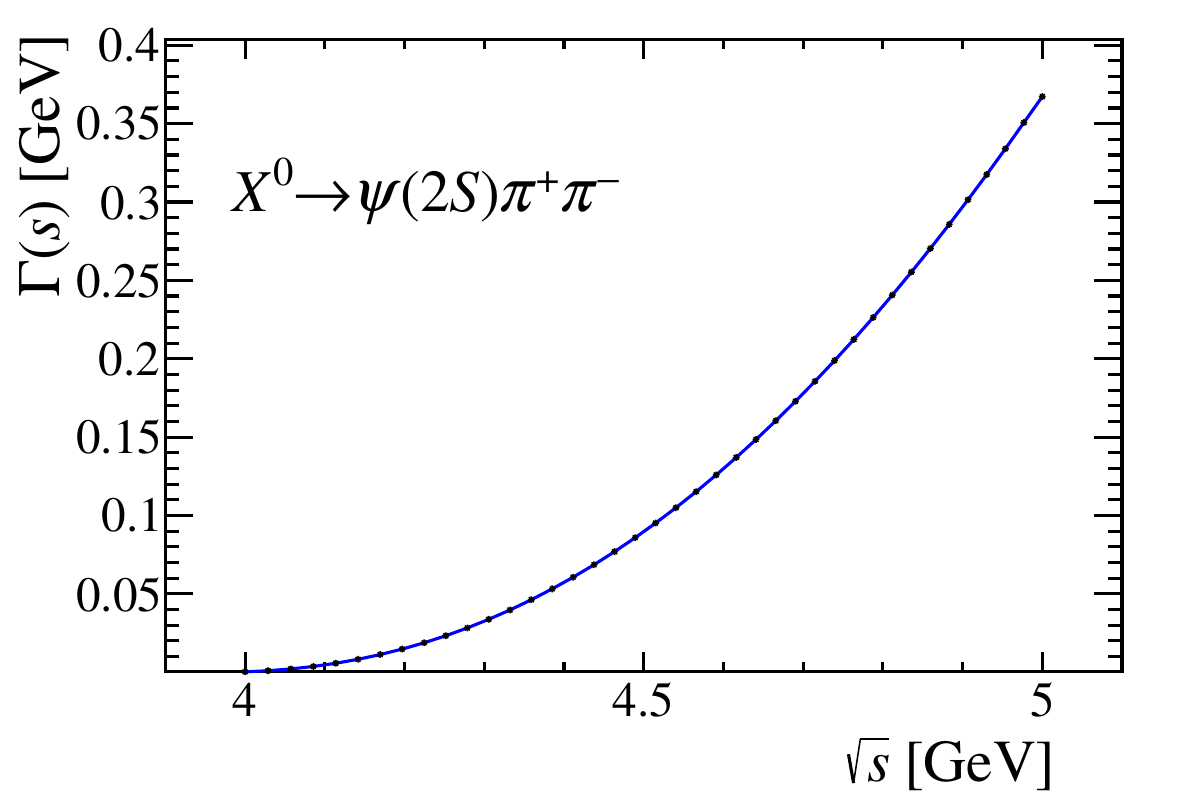}
\includegraphics[height=!,width=0.45\textwidth]{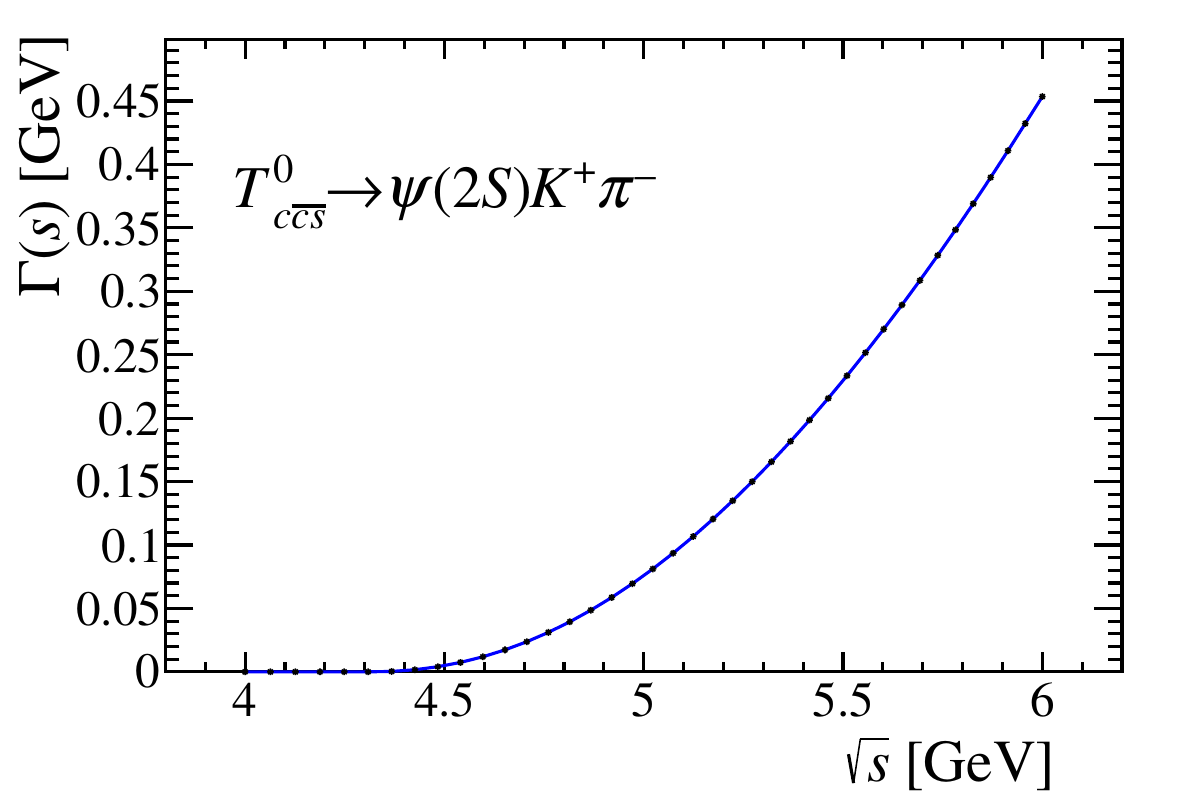}

\caption{Running-width functions of the three-body resonances.
}
\label{fig:rw}
\end{figure}

For the $K(1460)^+$ resonance, the energy-dependent width 
is reproduced from Ref.~\cite{LHCb-PAPER-2017-040}.
We further use the energy-dependent widths for the $K_1(1270)^+$, $K_1(1400)^+$, $K^*(1410)^+$ and $K^*(1680)^+$ mesons from Ref.~\cite{dArgent:2017gzv}. 	
For all other resonances decaying into a three-body final state, an energy-dependent width functions is derived from Eq.~\ref{eq:gamma3} assuming a uniform phase-space population. 
This is straightforward to implement for resonances with limited available information and
has been found to provide reasonable approximations in other studies \cite{dArgent:2017gzv,LHCB-PAPER-2020-030}
The running width functions of the three-body resonances included in the \textit{baseline} model are shown in Fig.~\ref{fig:rw}.

\section{Considered Decay Chains}
\label{a:decays}

\renewcommand{\thefigure}{B.\arabic{figure}}
\renewcommand{\thetable}{B.\arabic{table}}
\setcounter{table}{0}
\setcounter{figure}{0}

Tables~\ref{tab:res1} to \ref{tab:res4} list all previously observed resonances that might contribute to $\signal$ decays.
If quantum numbers are unknown, all possible spin-parity combinations with $J \le 2$ are tested.
We consider the following decay channels 
for all $K^{\prime+} \to \Kp \pip \pim$,
$\Xz \to \psitwos\pip\pim$,
$\Xsz  \to \psitwos \Kp \pim$,
$\Xspp  \to \psitwos \Kp \pip$,
$\Zpm \to \psitwos\pi^\pm$ and
$\Zsp \to \psitwos \Kp$
resonances (unless forbidden by conservation laws):
\begin{itemize}
    \item  $K^{\prime+} \to \Kp \rho(770)^0, \Kp \rho(1450)^0, K^*(892)^0 \pip, K^*(1410)^0 \pip, \Kp [\pi\pi]_S, [K\pi]_S \pip $;
    \item  $\Xz \to \psitwos \rho(770)^0, \psitwos [\pi\pi]_S, \Zpm \pi^\mp $;
    \item  $\Xsz \to \psitwos K^*(892)^0, \psitwos [K\pi]_S, \Zm K^+, Z_s^+ \pi^- $;
    \item  $\Xspp \to \Zp K^+, \Zsp \pi^+$;
    \item  $B^+ \to \Zp \,K^*(892^0), \Zp \,[K\pi]_S$;
    \item  $B^+ \to \Zsp \, \rho(770)^0, \Zsp \, [\pi\pi]_S$.
\end{itemize}
Nonresonant and single resonance amplitudes are obtained from the same topologies by setting \eg the $K^{\prime+}$ lineshape to unity.

\begin{table}[h]
 \caption{Resonances potentially contributing to the $\Kp\pip\pim$ subsystem.}
 \label{tab:res1}
\renewcommand{\arraystretch}{1.}
\footnotesize
\centering
\begin{tabular}{l c 
r@{}c@{}l
r@{}c@{}l
c
}

\hline
\hline
\multicolumn{1}{c}{Name} & \multicolumn{1}{c}{$J^P$} & \multicolumn{3}{c}{$m$[MeV]} & \multicolumn{3}{c}{$\Gamma$[MeV]} & \multicolumn{1}{c}{Source}  \\
\hline

$K_1(1270)^+$ & $1^+$ & 1290 &$\pm$& 2  & 116 &$\pm$& 3  & \cite{LHCb-PAPER-2017-040}\\

$K_1(1400)^+$ & $1^+$ & 1403 &$\pm$& 7  & 174 &$\pm$& 13  & \cite{PDG2022} \\

$K^*(1410)^+$ & $1^-$ & 1414 &$\pm$& 15  & 232 &$\pm$& 21 & \cite{PDG2022} \\

$K_2^*(1430)^+$ & $2^+$ & 1427.3 &$\pm$& 1.5 & 100 &$\pm$& 2.1 & \cite{PDG2022} \\

$K(1460)^+$ & $0^-$ & 1482.4 &$\pm$& 15 & 335.6 &$\pm$& 11 &  \cite{LHCb-PAPER-2017-040} \\

$K_2^*(1580)^+$ & $2^-$ & 1580 & & & 110 & & & \cite{PDG2022} \\

$K(1630)^+$ & $?^?$ & 1629 &$\pm$& 7 & 16 & ${}^{+}_{-}$ & ${}^{19}_{16}$ & \cite{PDG2022} \\

$K_1(1650)^+$ & $1^+$ & 1650 &$\pm$& 50  & 150 &$\pm$& 50 & \cite{PDG2022} \\

$K^*(1680)^+$ & $1^-$ & 1718 &$\pm$& 18 & 322 &$\pm$& 110 & \cite{PDG2022} \\

$K_2(1770)^+$ & $2^-$ & 1773 &$\pm$& 8 & 186 &$\pm$& 14 & \cite{PDG2022} \\

\hline
\hline
\end{tabular}

\caption{Resonances potentially contributing to the $\psitwos\pip\pim$ subsystem and decay channels in which they were seen. All values taken from the PDG~\cite{PDG2022}.}
 \label{tab:res2}
\renewcommand{\arraystretch}{1.}
\footnotesize
\centering
\begin{tabular}{l c 
r@{}c@{}l
r@{}c@{}l
c
}
\hline
\hline
\multicolumn{1}{c}{Name} & \multicolumn{1}{c}{$J^P$} & \multicolumn{3}{c}{$m$[MeV]} & \multicolumn{3}{c}{$\Gamma$[MeV]} & \multicolumn{1}{c}{Decay Channel}  \\
\hline

$\psi(4040)$ & $1^-$ & 4039 &$\pm$& 1 & 80 &$\pm$& 10 & $ DD $ \\

$\chi_{c1}(4140)$ & $1^+$ & 4146.5 &$\pm$& 3.0 & 19 & ${}^{+}_{-}$ & ${}^{7}_{5}$  & $ J/\psi \phi $ \\

$\psi(4160)$ & $1^-$ & 4191 &$\pm$& 5 & 70 &$\pm$& 10 & $ DD $ \\

$X(4160)$ & $?^?$ & 4153& ${}^{+}_{-}$ & ${}^{23}_{21}$ & 136& ${}^{+}_{-}$ & ${}^{60}_{35}$ & $J/\psi \phi, D^*D^* $ \\

$\psi(4230)$ & $1^-$ & 4222.7 &$\pm$& 2.6 & 49 &$\pm$& 8 & $ J/\psi \pi \pi, \psi(2S) \pi \pi$ \\

$\chi_{c1}(4274)$ & $1^+$ & 4286& ${}^{+}_{-}$ & ${}^{8}_{9}$  & 51 &$\pm$& 7 & $ J/\psi \phi $ \\

$X(4350)$ & $?^?$ & 4351 &$\pm$& 5  & 13& ${}^{+}_{-}$ & ${}^{18}_{10}$ & $ J/\psi \phi $ \\

$\psi(4360)$ & $1^-$ & 4372 &$\pm$& 9 & 115 &$\pm$& 13 & $ J/\psi \pi \pi, \psi(2S) \pi \pi, h_c \pi \pi$ \\

$\psi(4415)$ & $1^-$ & 4421 &$\pm$& 4 & 62 &$\pm$& 20 & $ DD $ \\

$\chi_{c0}(4500)$ & $0^+$ & 4474 &$\pm$& 4  & 77& ${}^{+}_{-}$ & ${}^{12}_{10}$ & $ J/\psi \phi $ \\

$X(4630)$ & $1^- (2^-) $ & 4626& ${}^{+}_{-}$ & ${}^{24}_{110}$  & 174& ${}^{+}_{-}$ & ${}^{140}_{80}$ & $ J/\psi \phi $ \\

$\psi(4660)$ & $1^-$ & 4630 &$\pm$& 6 & 72& ${}^{+}_{-}$ & ${}^{14}_{12}$ & $ \psi(2S) \pi \pi $ \\

$\chi_{c1}(4685)$ & $1^+$ & 4684& ${}^{+}_{-}$ & ${}^{15}_{17}$  & 126 &$\pm$& 40 & $ J/\psi \phi $ \\

$\chi_{c0}(4700)$ & $0^+$ & 4694& ${}^{+}_{-}$ & ${}^{16}_{5}$  & 87& ${}^{+}_{-}$ & ${}^{18}_{10}$ & $ J/\psi \phi $ \\

\hline
\hline
\end{tabular}
\caption{Resonances potentially contributing to the $\psitwos\pi^\pm$ subsystem  and decay channels in which they were seen. All values taken from the PDG~\cite{PDG2022}.}
 \label{tab:res3}
\footnotesize
\centering
\begin{tabular}{l c 
r@{}c@{}l
r@{}c@{}l
c
}
\hline
\hline
\multicolumn{1}{c}{Name} & \multicolumn{1}{c}{$J^P$} & \multicolumn{3}{c}{$m$[MeV]} & \multicolumn{3}{c}{$\Gamma$[MeV]} & \multicolumn{1}{c}{Decay Channel}  \\
\hline

$\ZA(3900)^\pm$ & $1^+$ & $3887.1 $&$\pm$&$ 2.6$ & $28.4 $&$\pm$&$ 2.6$ & $ J/\psi \pi $ \\

$\ZV(4020)^\pm$ & $1^-$ & $4024.1 $&$\pm$&$ 1.9$ & $13 $&$\pm$&$ 5$ & $ h_c(1P)\pi, D^*D^* $ \\

$\Z(4050)^\pm$ & $?^?$ & $4051 $&${}^{+}_{-}$&${}^{24}_{40}$ & $82$&${}^{+}_{-}$&${}^{50}_{28}$ & $ \chi_{c_1}(1P) \pi $ \\

$\Z(4055)^\pm$ & $?^?$ & $4054 $&$\pm$&$ 3.2$ & $45 $&$\pm$&$ 13$ & $ \psi(2S) \pi$ \\

$\Z(4100)^\pm$ & $0^+/1^-$ & $4096 $&$\pm$&$ 28$ & $152 $&$\pm$&$ 75$ & $ \eta_c(1S) \pi $ \\

$\ZA(4200)^\pm$ & $1^+ (2^-)$ & $4196$&${}^{+}_{-}$&${}^{35}_{32}$ & $370$&${}^{+}_{-}$&${}^{100}_{150}$ & $ J/\psi \pi $ \\

$\Z(4240)^\pm$ & $0^-(1^+)$ & $4239$&${}^{+}_{-}$&${}^{50}_{21}$ & $220$&${}^{+}_{-}$&${}^{120}_{90}$ & $ \psi(2S) \pi $ \\

$\Z(4250)^\pm$ & $?^?$ & $4248$&${}^{+}_{-}$&${}^{190}_{50}$ & $177$&${}^{+}_{-}$&${}^{320}_{70} $ & $ \chi_{c_1}(1P) \pi$ \\

$\ZA(4430)^\pm$ & $1^+$ & $4478$&${}^{+}_{-}$&${}^{15}_{18}$ & $181 $&$\pm$&$ 31$ & $ J/\psi \pi, \psi(2S) \pi $ \\

\hline
\hline
\end{tabular}
\caption{Resonances potentially contributing to the $\psitwos \Kp$ subsystem  and decay channels in which they were seen. All values taken from the PDG~\cite{PDG2022}.}
 \label{tab:res4}
\renewcommand{\arraystretch}{1.2}

\footnotesize
\centering
\begin{tabular}{l c 
r@{}c@{}l
r@{}c@{}l
c
}
\hline
\hline
\multicolumn{1}{c}{Name} & \multicolumn{1}{c}{$J^P$} & \multicolumn{3}{c}{$m$[MeV]} & \multicolumn{3}{c}{$\Gamma$[MeV]} & \multicolumn{1}{c}{Decay Channel}  \\
\hline

$\ZsA(4000)^+$ & $1^+$ & $4003$&${}^{+}_{-}$&${}^{7}_{15}$ & $131$&$\pm$&$30$ & $ J/\psi K  $ \\
$\ZsA(4220)^+$ & $1^+ (1^-)$ & $4216$&${}^{+}_{-}$&${}^{49}_{38}$ & $233$&${}^{+}_{-}$&${}^{110}_{90}$ & $ J/\psi K  $ \\

\hline
\hline
\end{tabular}
\end{table}

\clearpage
\section{Model selection}
\label{a:modelSel}

\renewcommand{\thefigure}{C.\arabic{figure}}
\renewcommand{\thetable}{C.\arabic{table}}
\setcounter{table}{0}
\setcounter{figure}{0}

\begin{figure}[h]
       	 \includegraphics[width=0.329\textwidth,height=!]{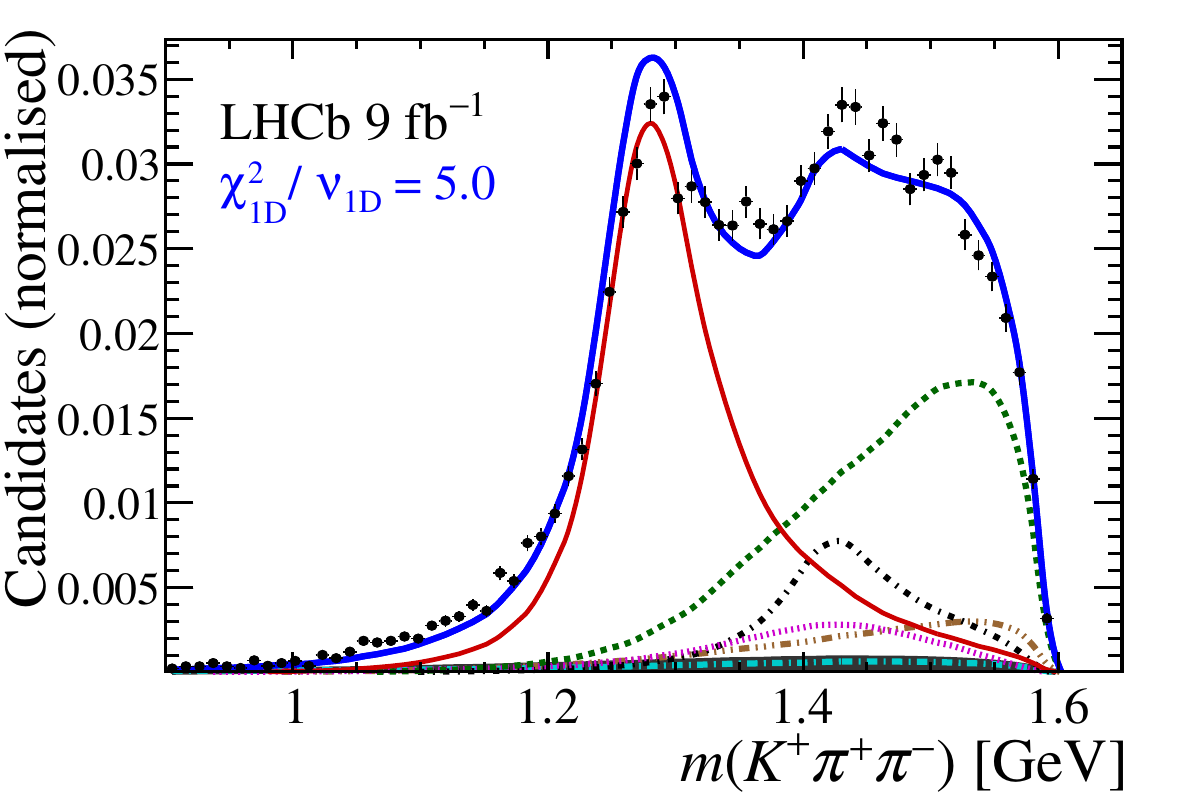}
       	 \includegraphics[width=0.329\textwidth,height=!]{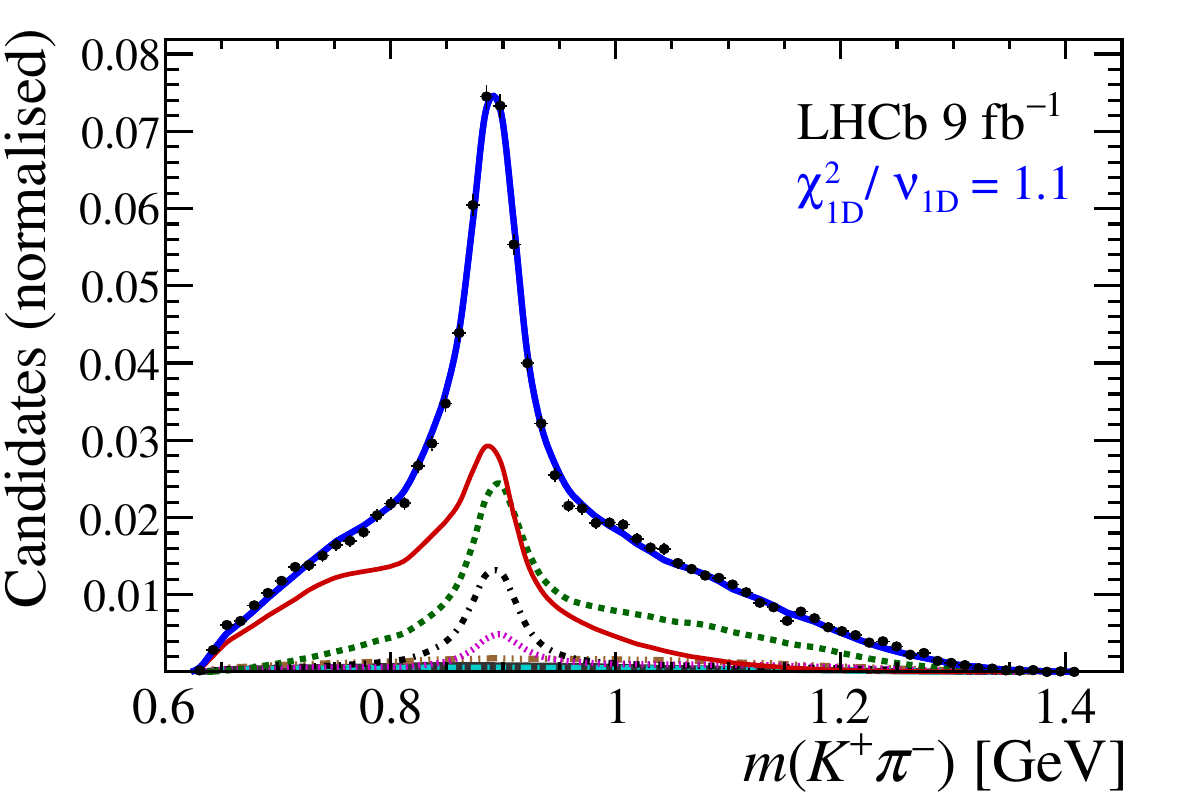}
       	 \includegraphics[width=0.329\textwidth,height=!]{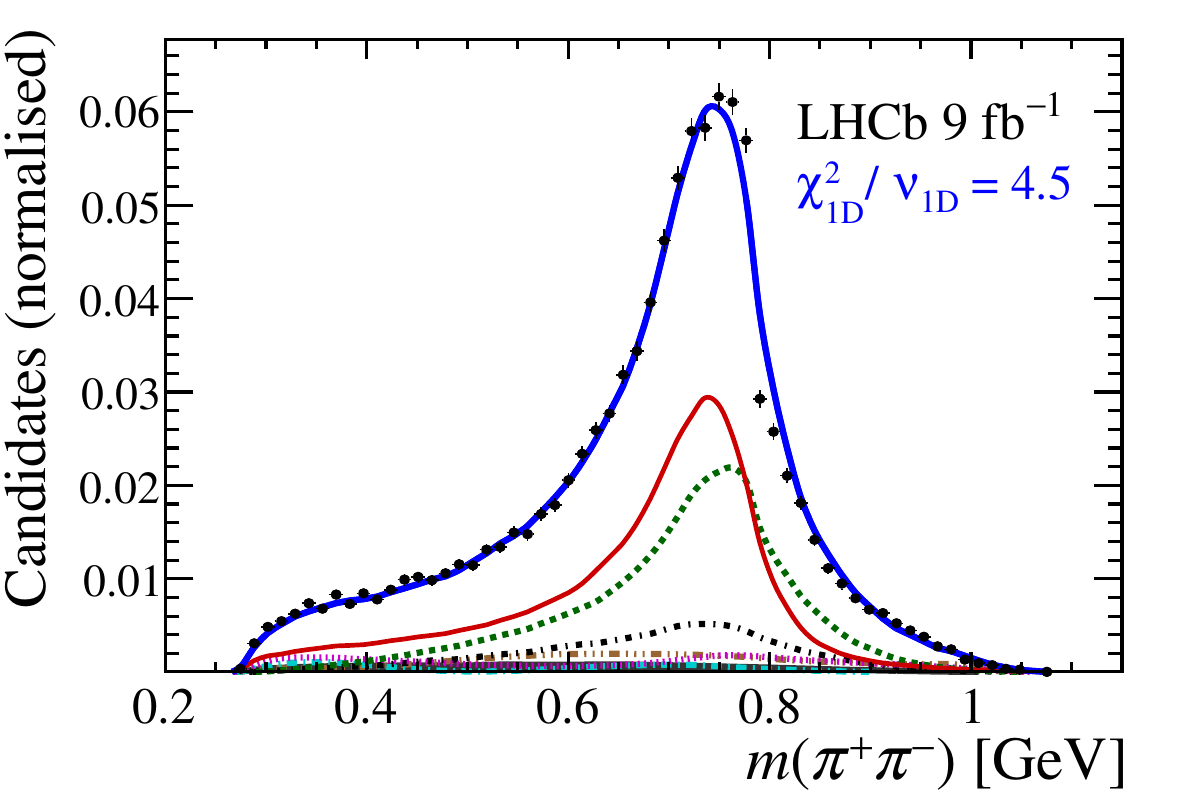}

       	 \includegraphics[width=0.329\textwidth,height=!]{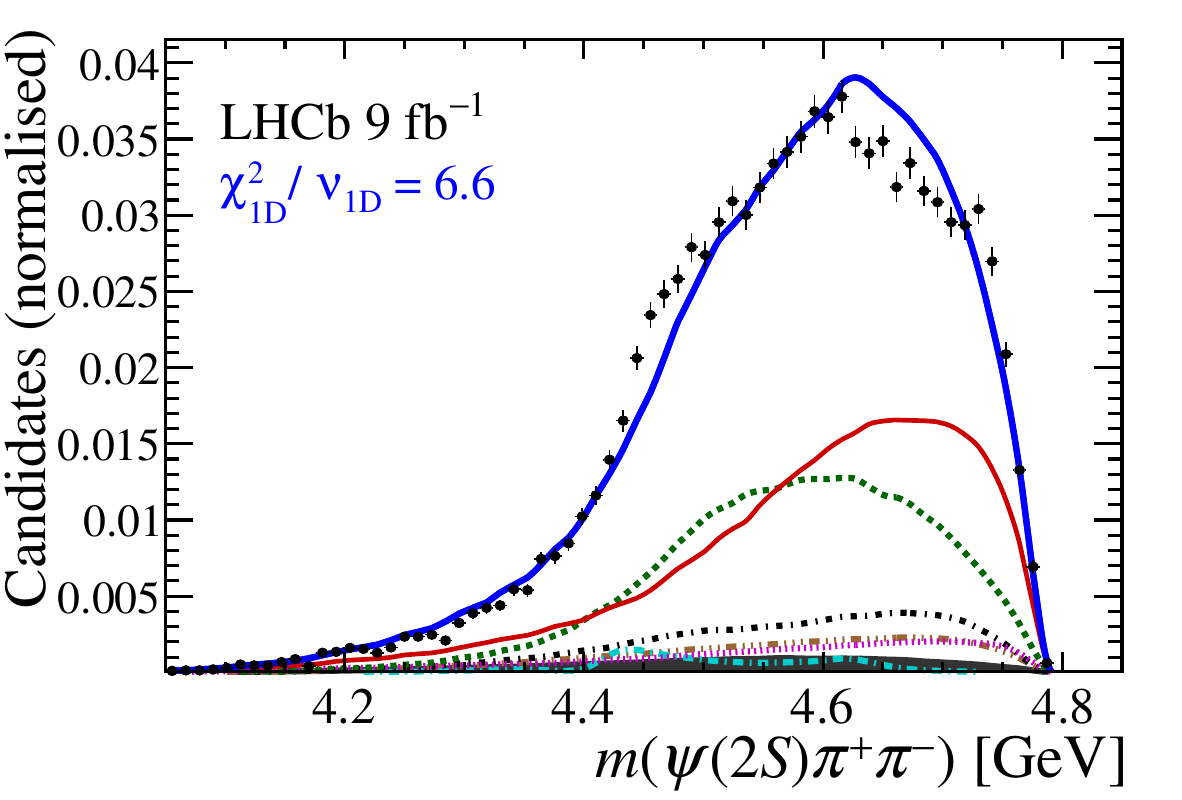}
       	 \includegraphics[width=0.329\textwidth,height=!]{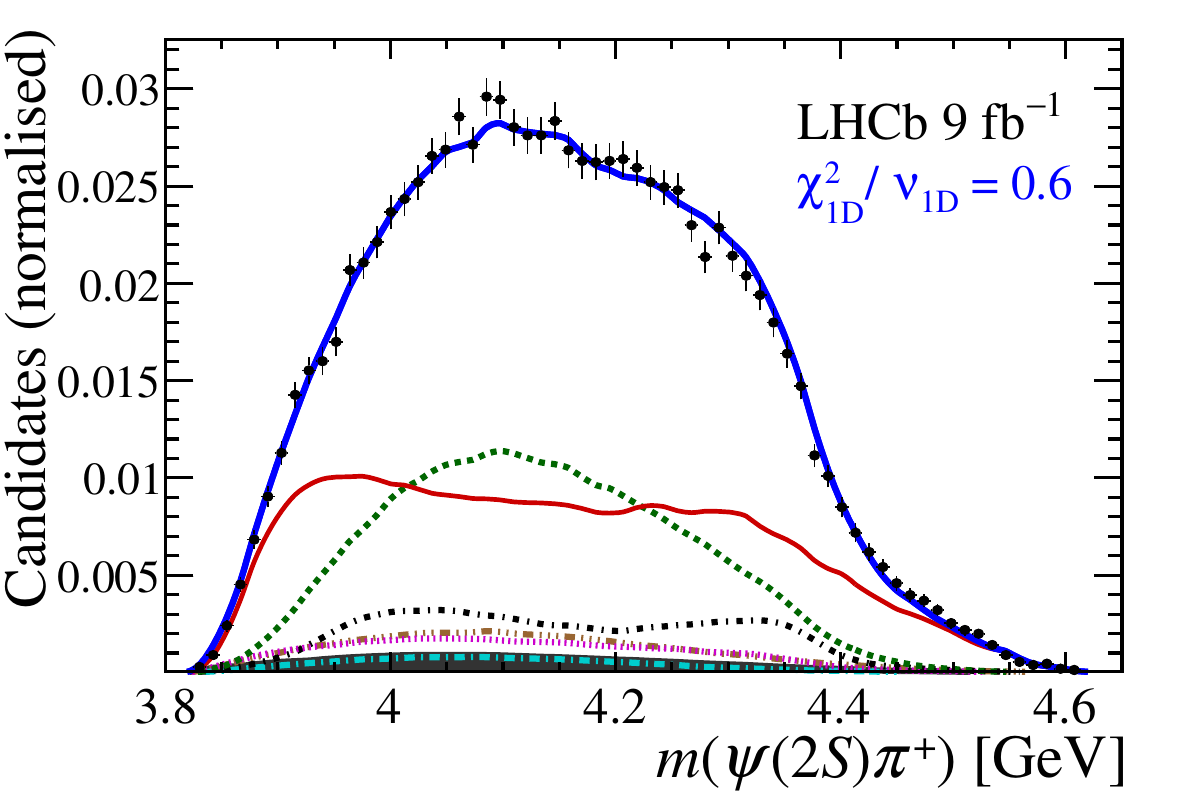}
       	 \includegraphics[width=0.329\textwidth,height=!]{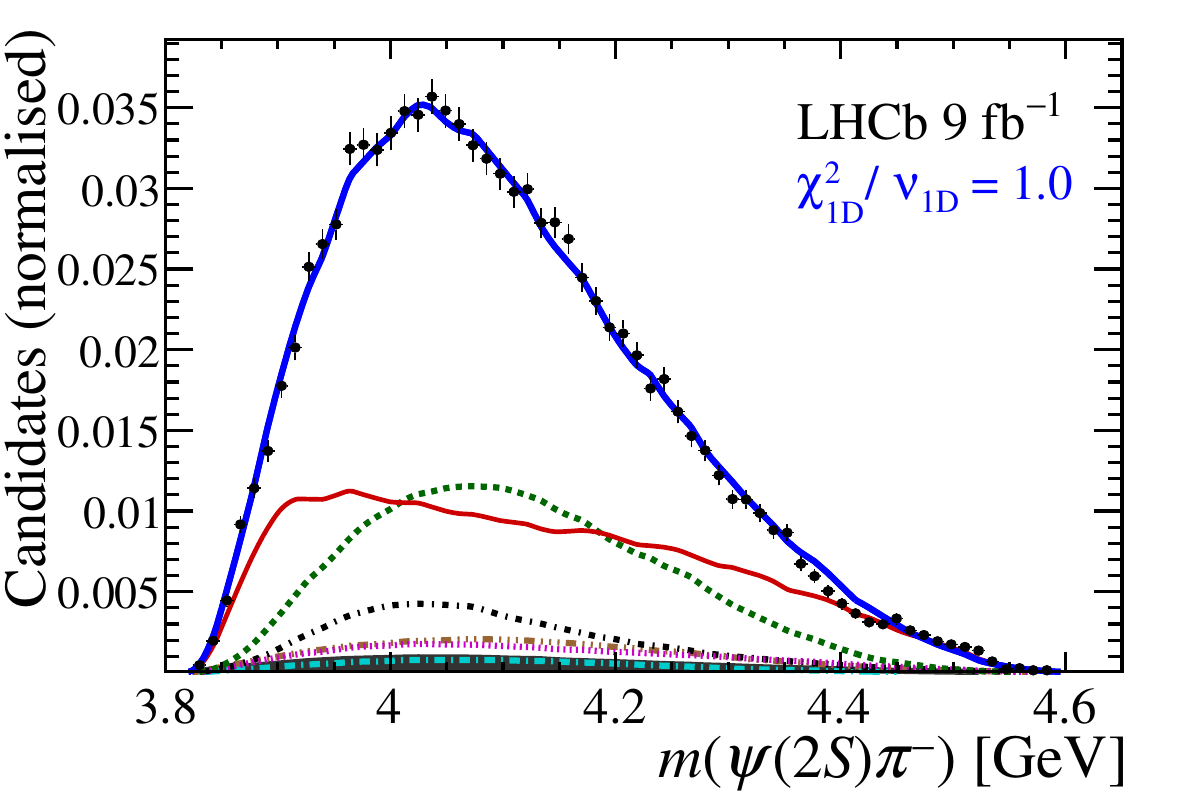}

       	 \includegraphics[width=0.329\textwidth,height=!]{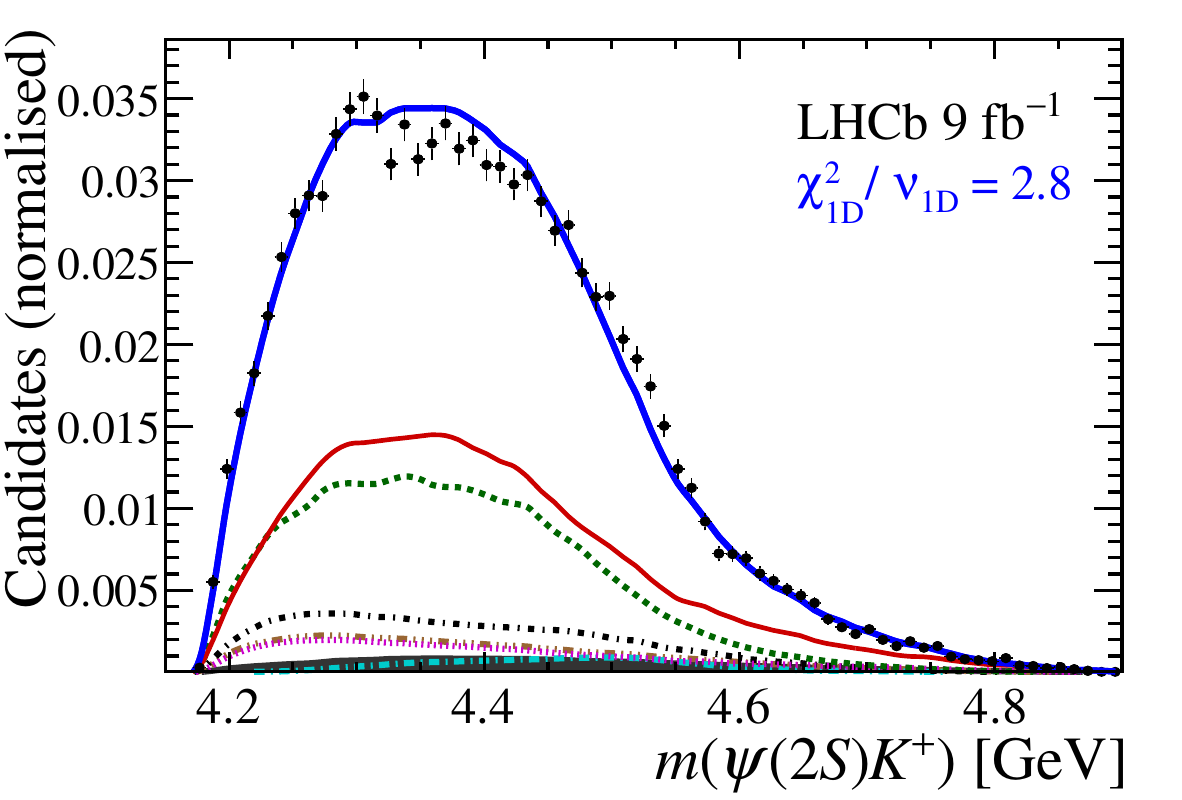}
       	 \includegraphics[width=0.329\textwidth,height=!]{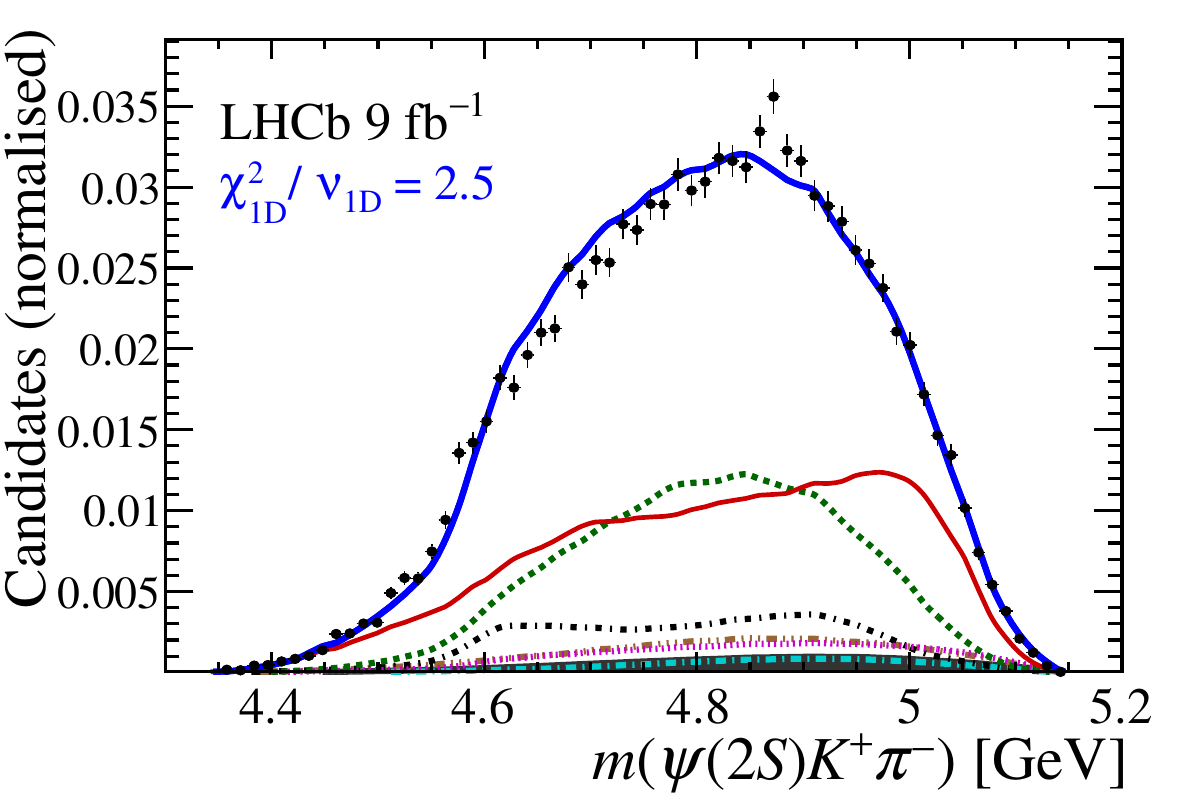}
       	 \includegraphics[width=0.329\textwidth,height=!]{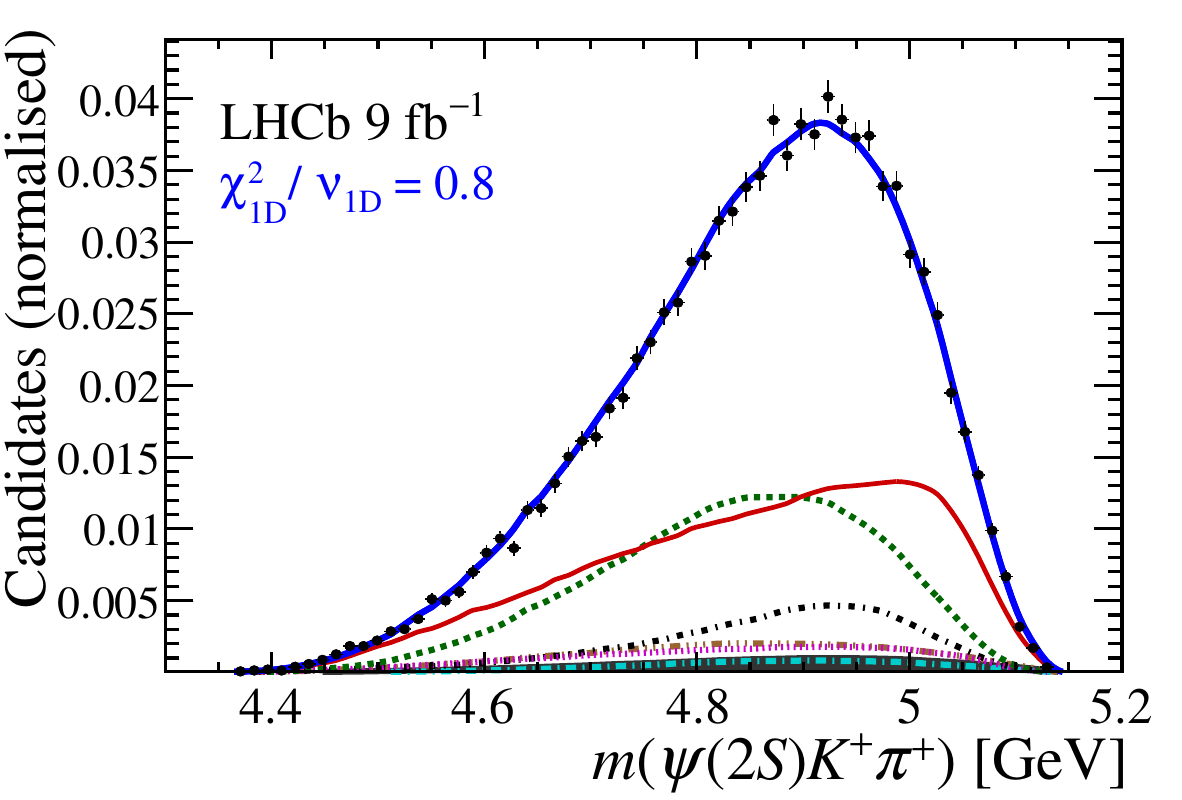}
              
       	 \includegraphics[width=0.329\textwidth,height=!]{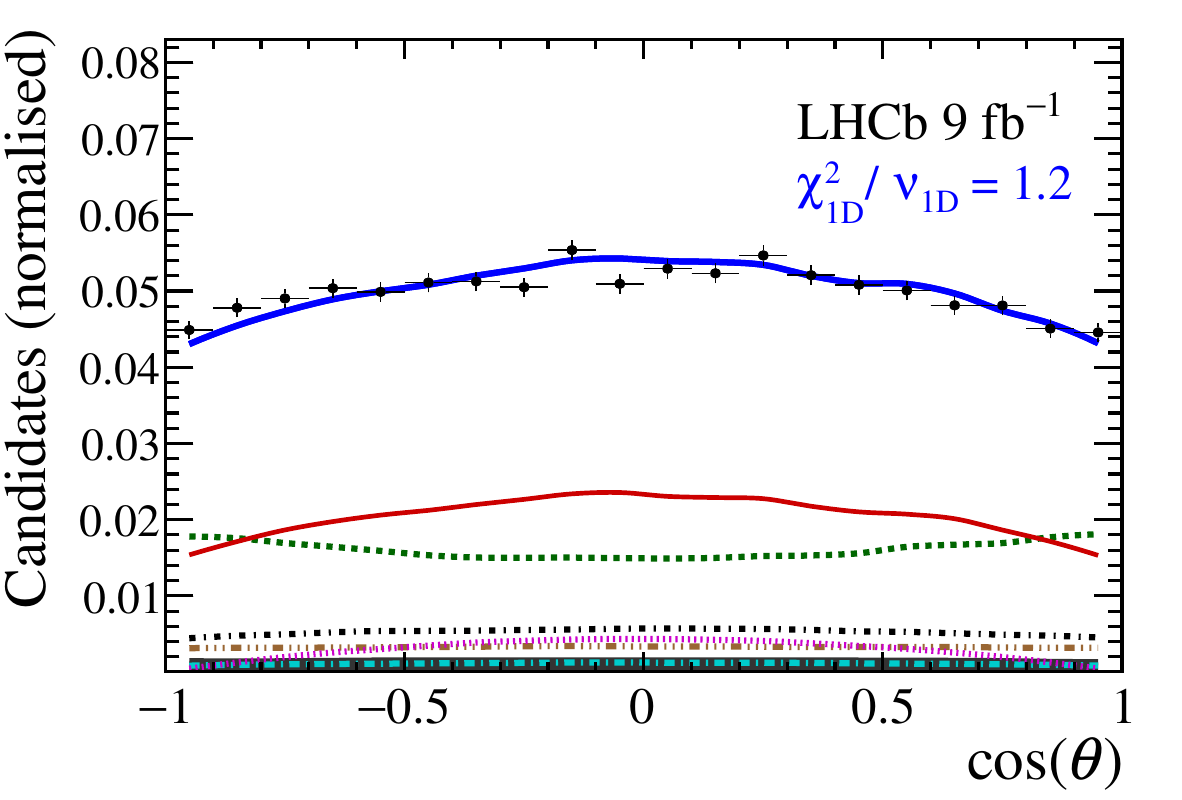}
       	 \includegraphics[width=0.329\textwidth,height=!]{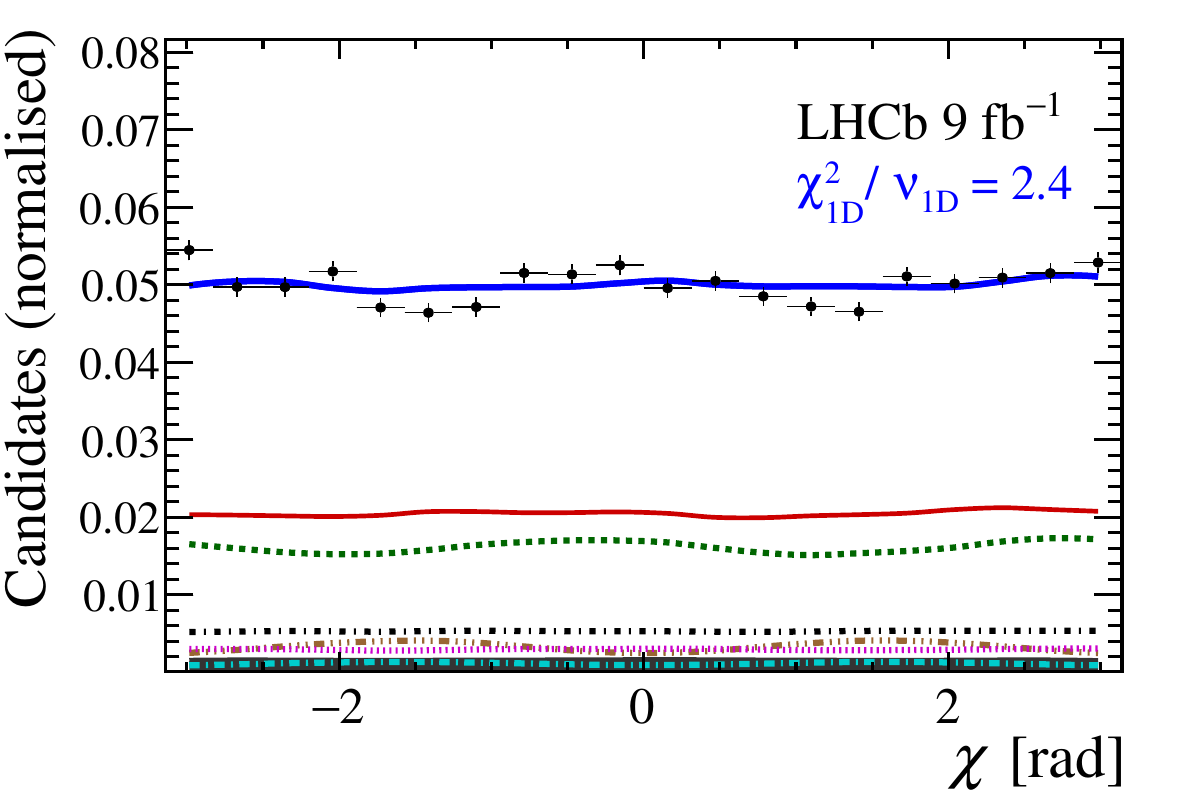}
       	 \includegraphics[width=0.329\textwidth,height=!]{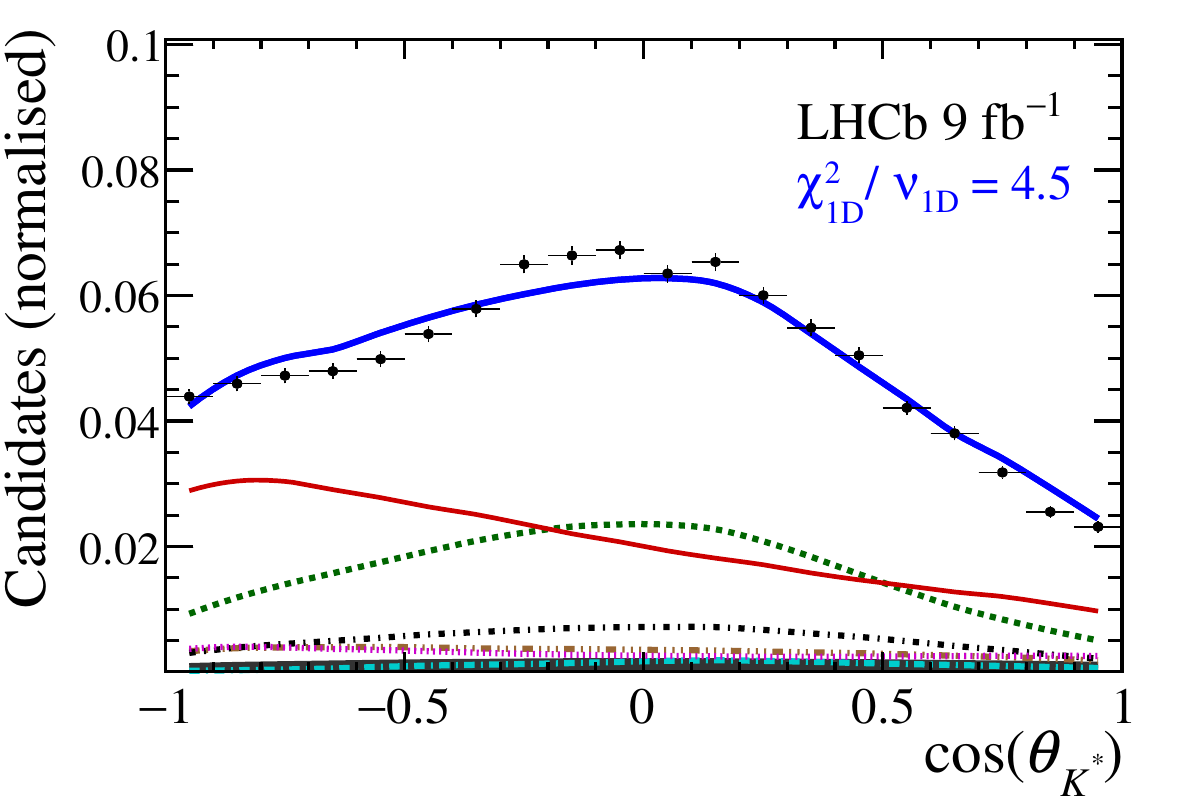}

        \centering
              \includegraphics[width=0.25\textwidth,height=!]{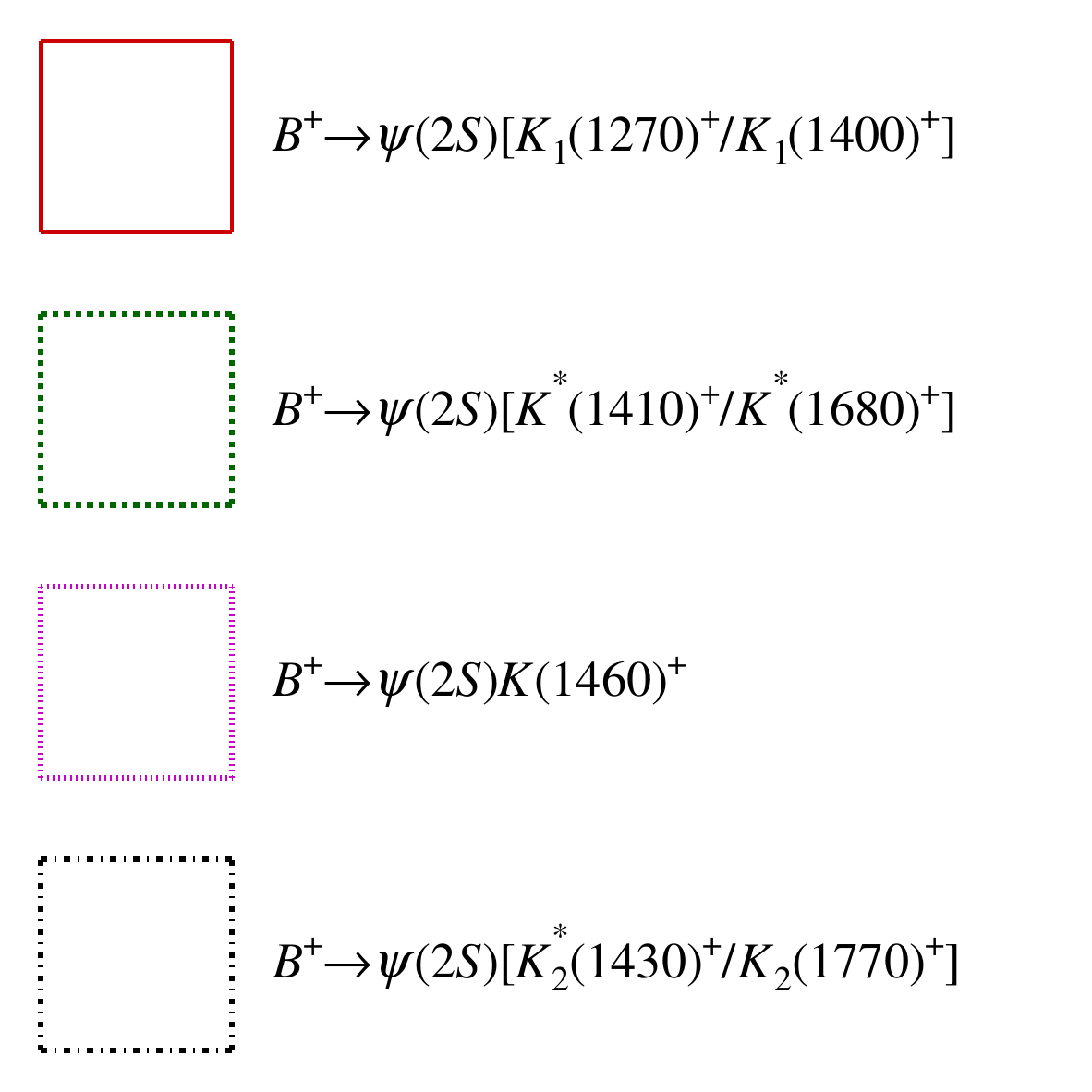}
              \includegraphics[width=0.25\textwidth,height=!]{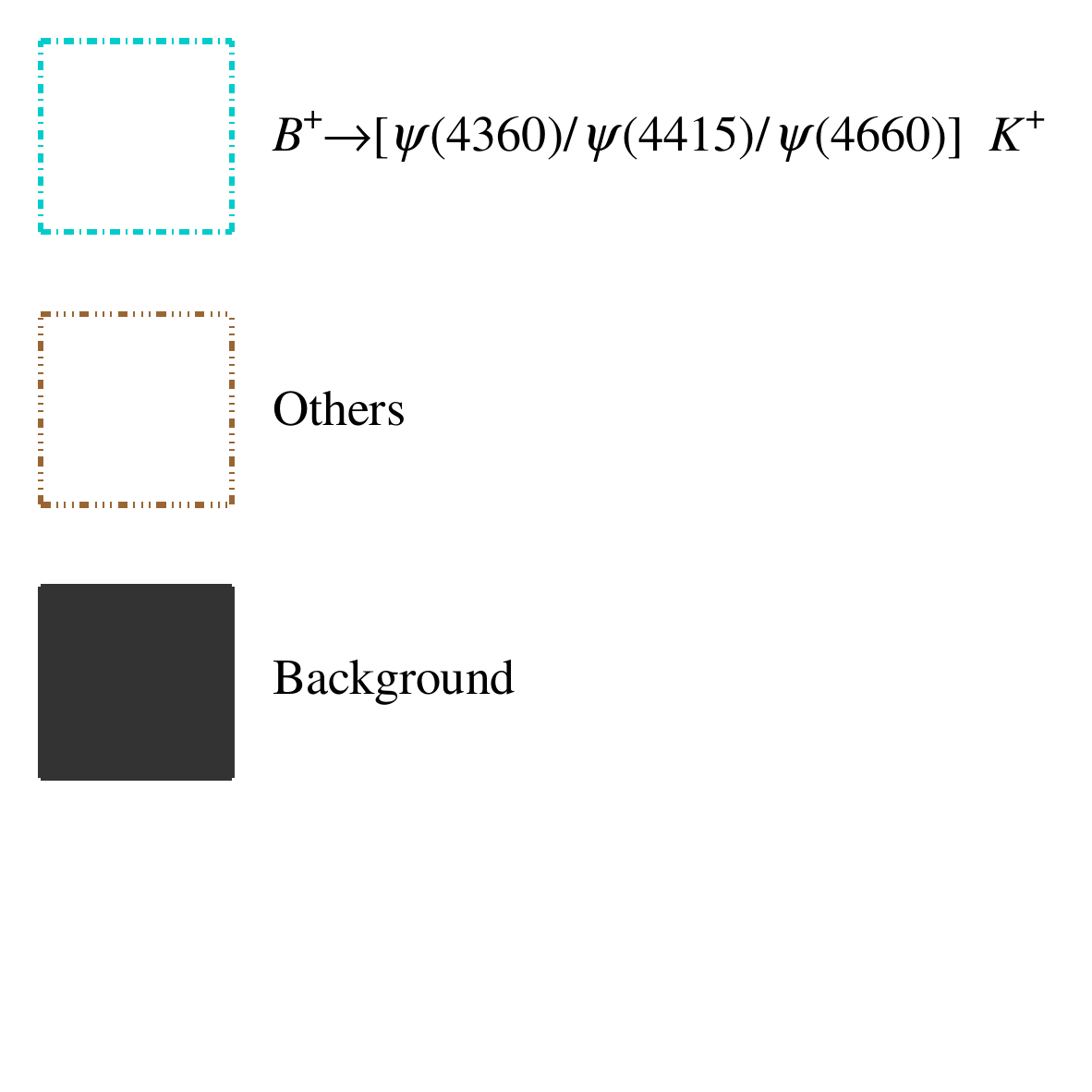}
              
	\caption{Phase-space projections of $\signal$ candidates in the signal region (points with error bars) and fit projections (solid, blue line) for the \textit{no-exotics} model.  The displayed $\chi_{\rm 1D}^2/\nu_{\rm 1D}$ value on each projection gives the sum of squared normalised residuals divided by the number of bins minus one. The multi-dimensional $\chi^2$ value is $\chi^2/\nu= 2.05$ with $\nu=1016$. }
         \label{fig:fitNoExotic}
\end{figure}

\begin{figure}[h]
       	 \includegraphics[width=0.329\textwidth,height=!]{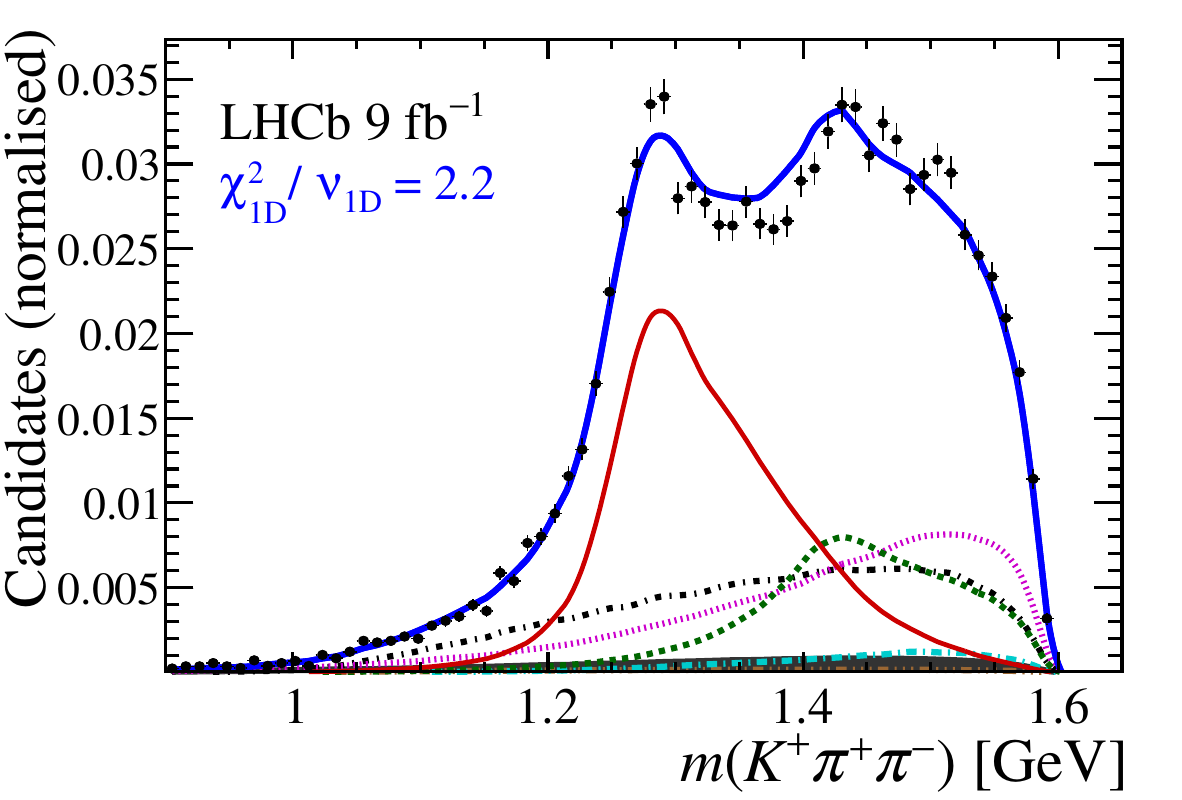}
       	 \includegraphics[width=0.329\textwidth,height=!]{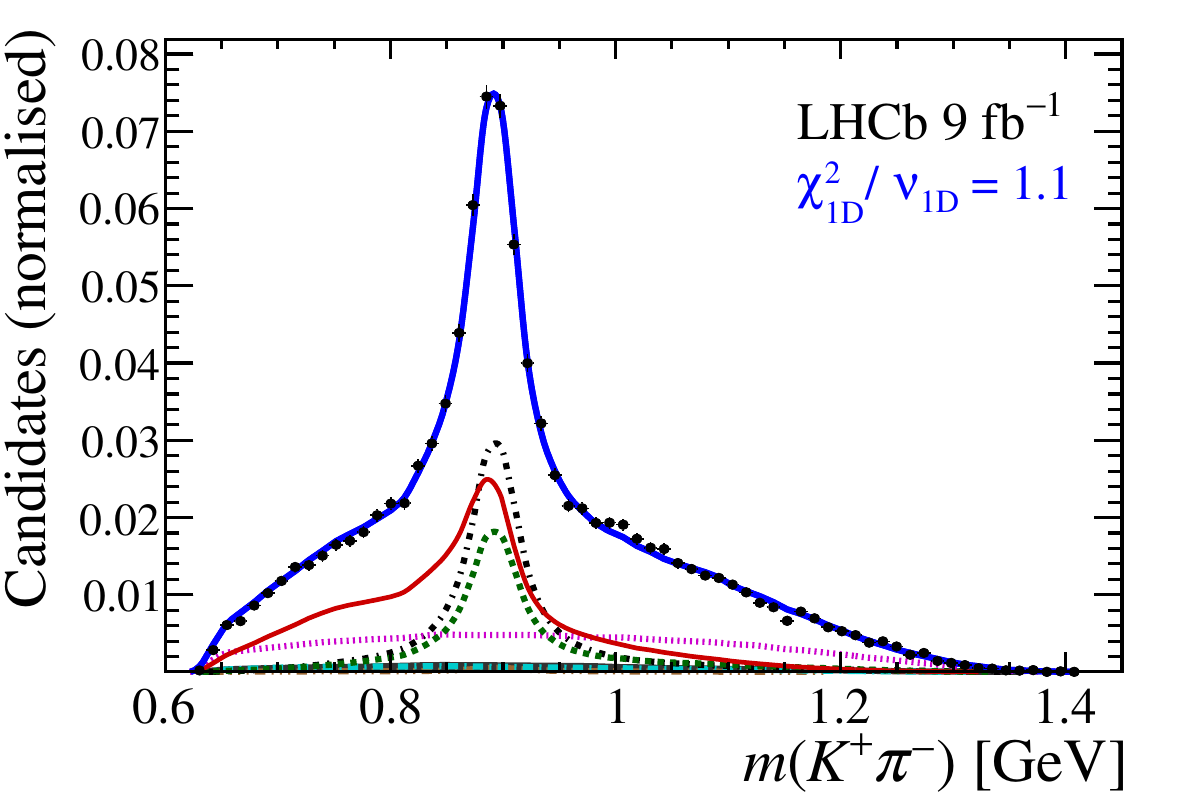}
       	 \includegraphics[width=0.329\textwidth,height=!]{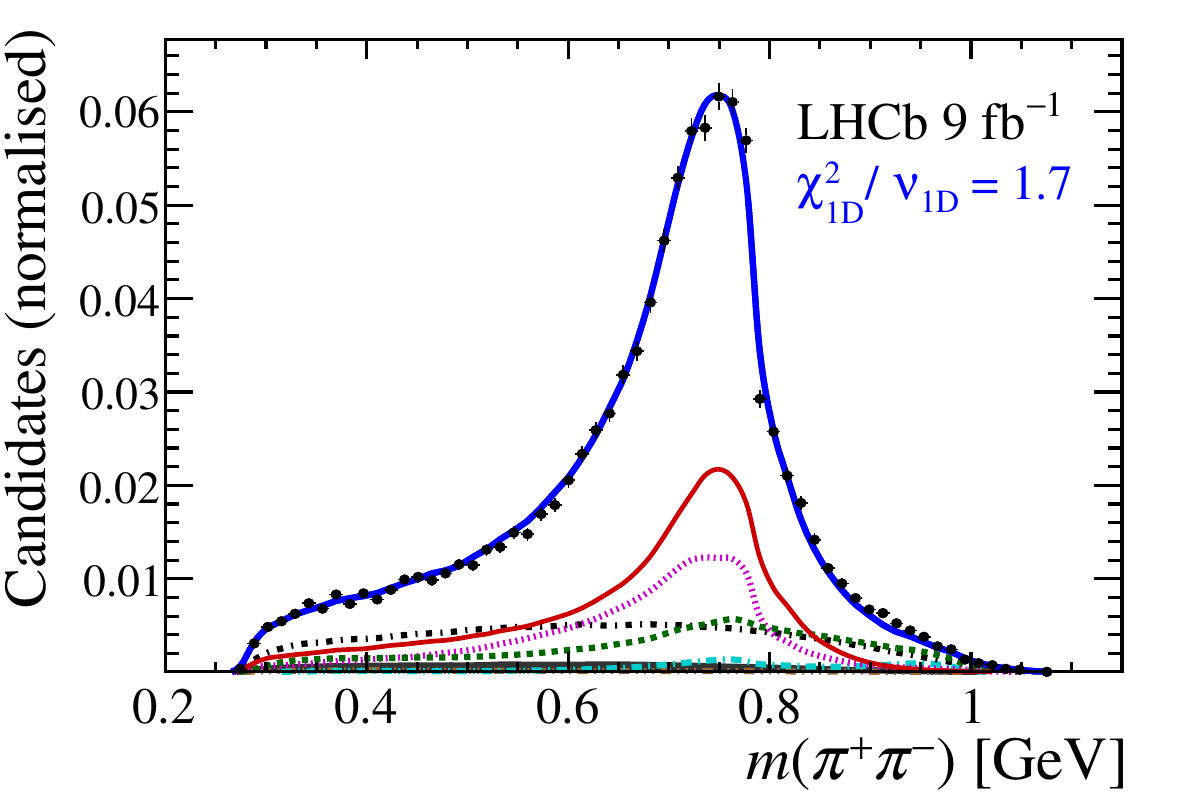}

       	 \includegraphics[width=0.329\textwidth,height=!]{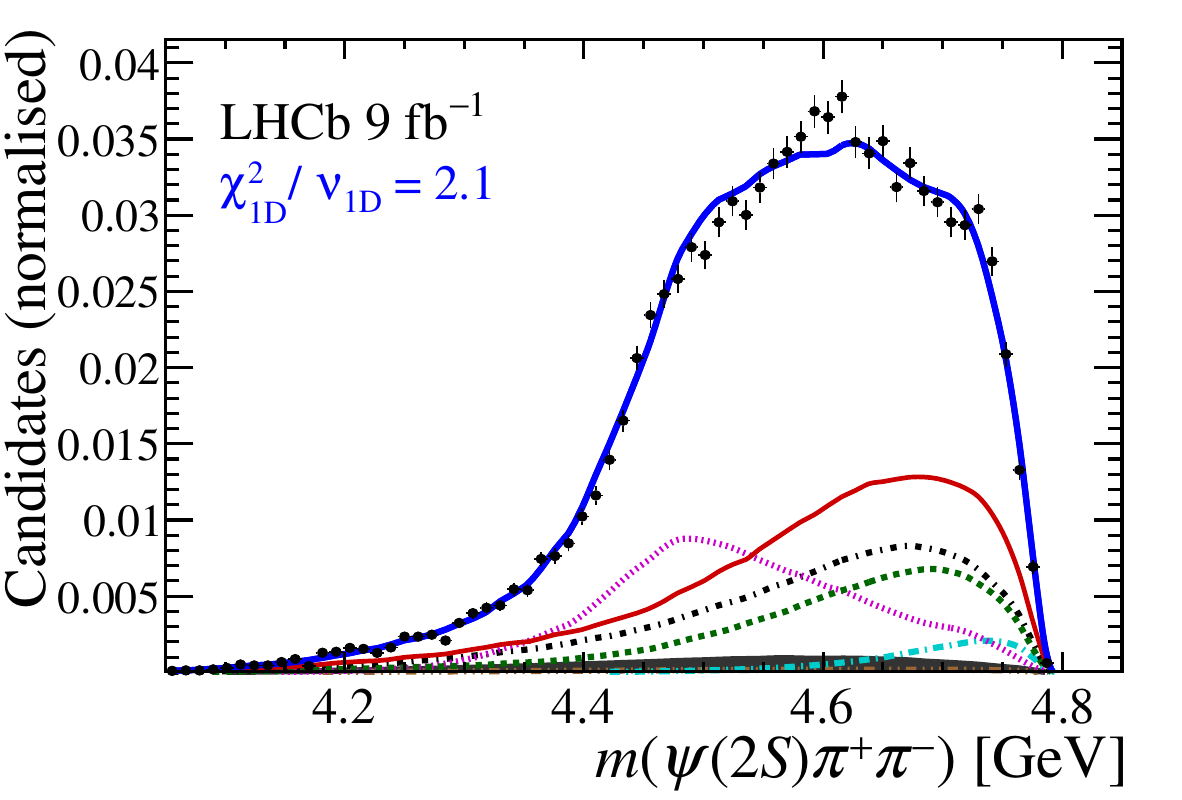}
       	 \includegraphics[width=0.329\textwidth,height=!]{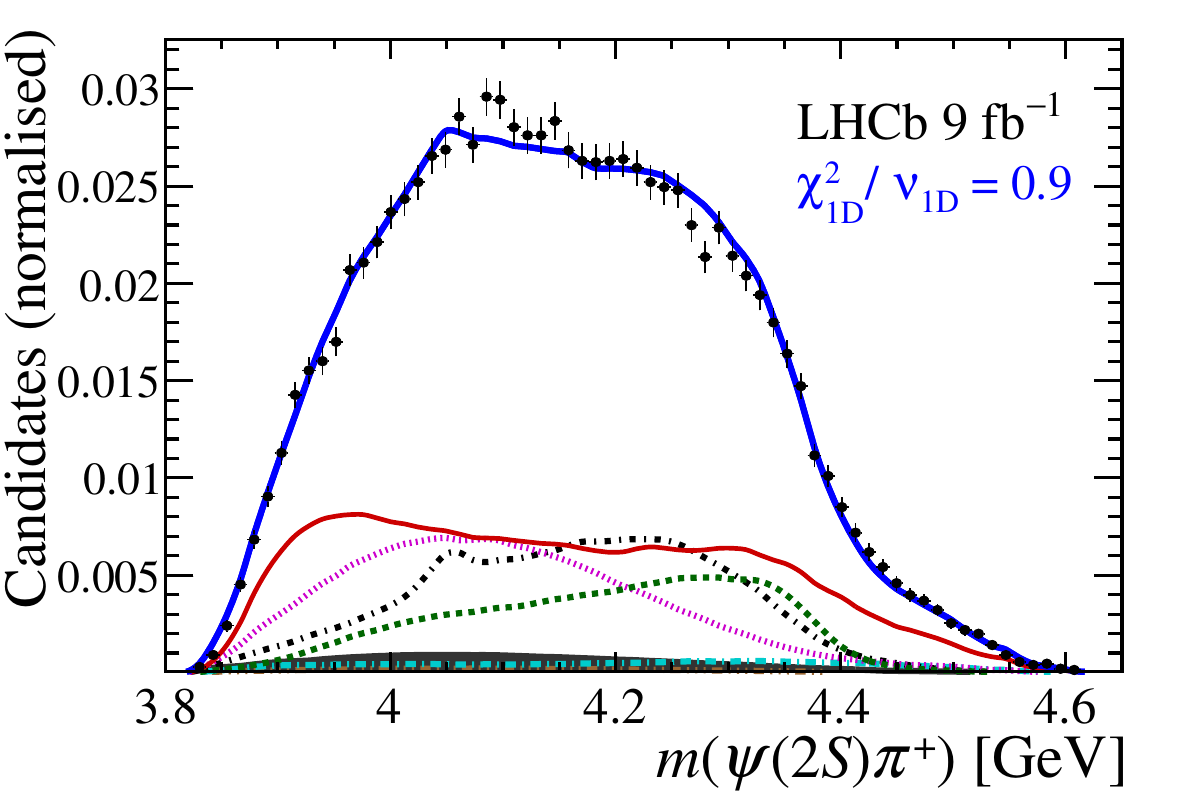}
       	 \includegraphics[width=0.329\textwidth,height=!]{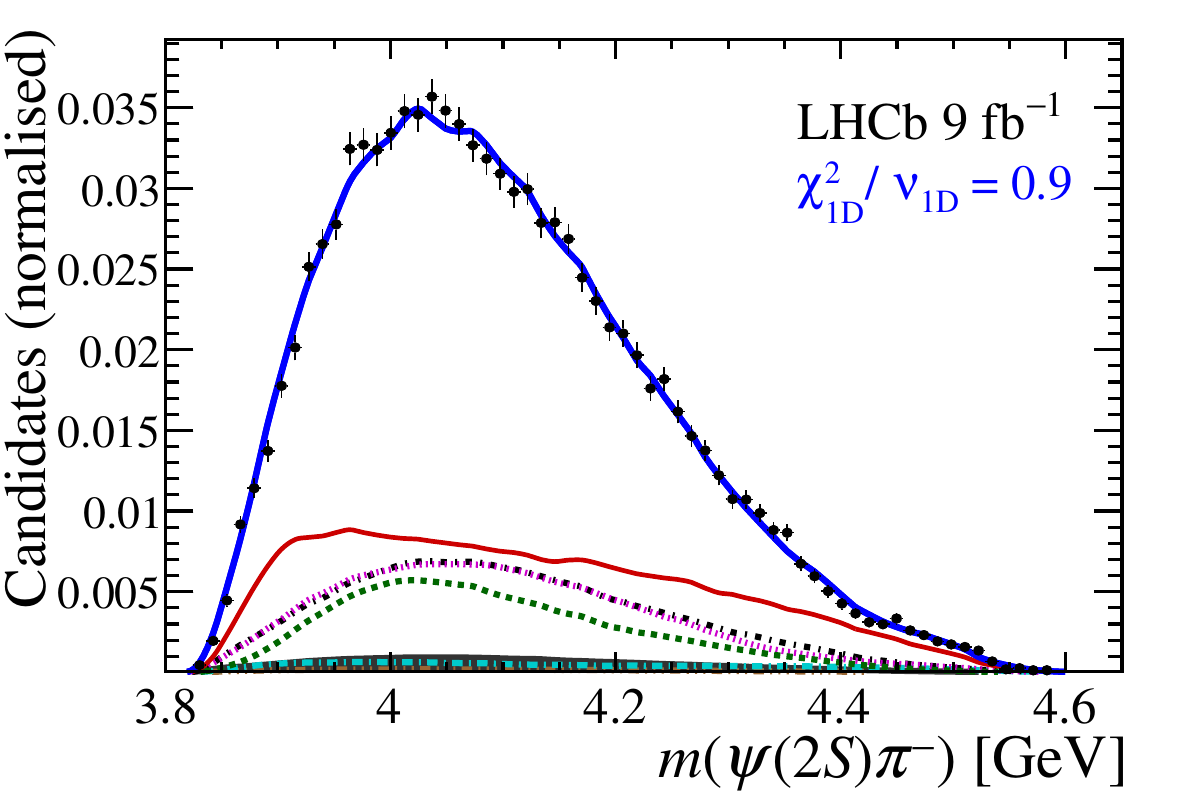}

       	 \includegraphics[width=0.329\textwidth,height=!]{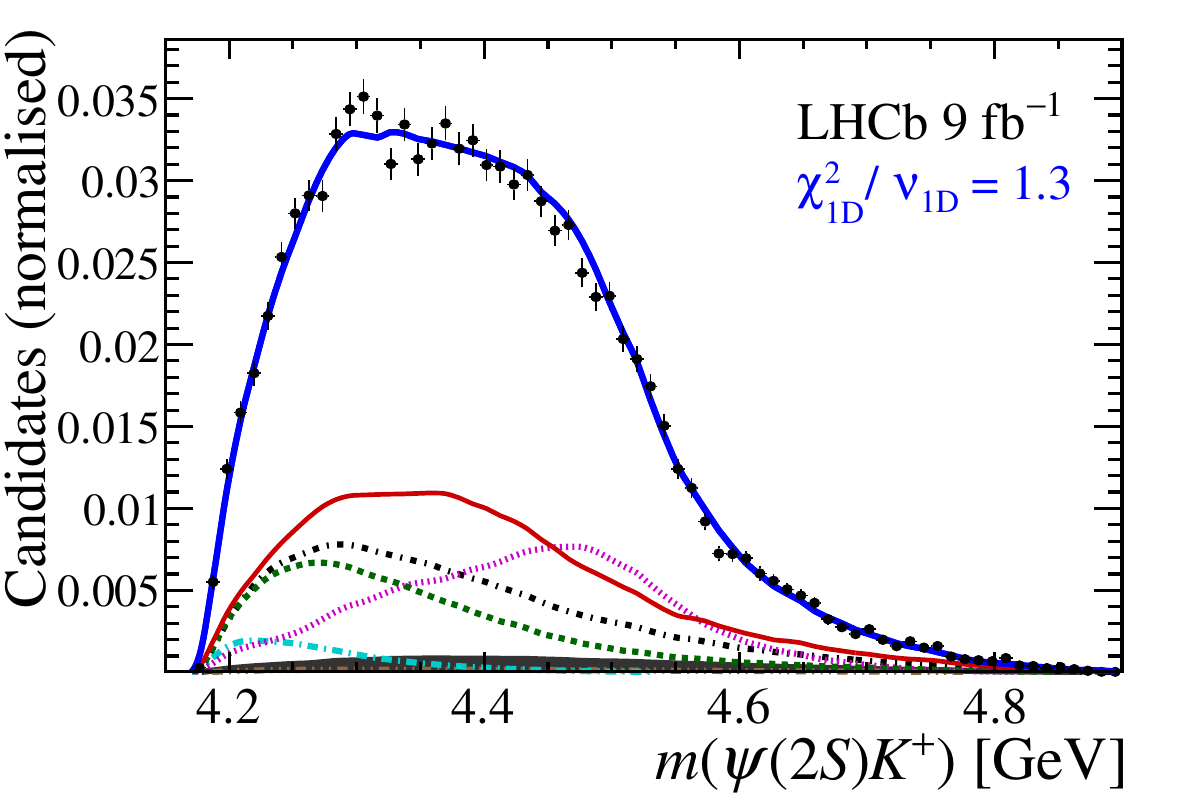}
       	 \includegraphics[width=0.329\textwidth,height=!]{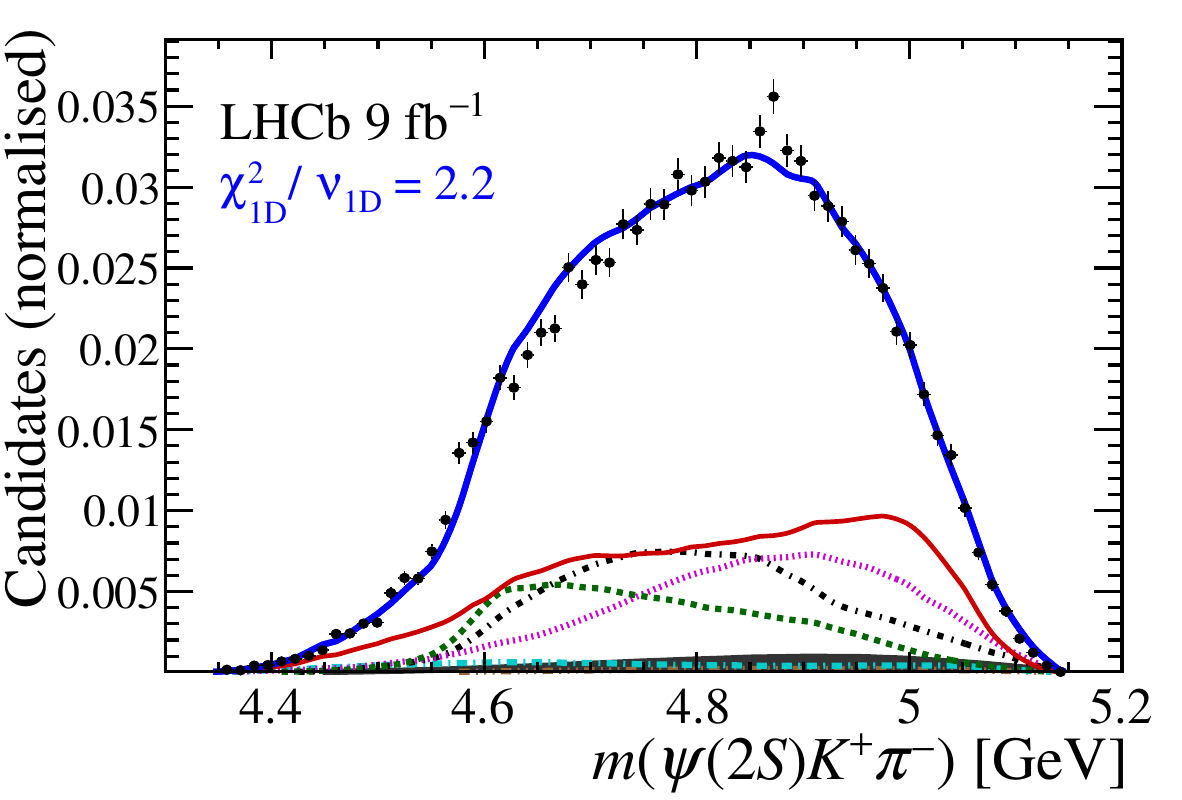}
       	 \includegraphics[width=0.329\textwidth,height=!]{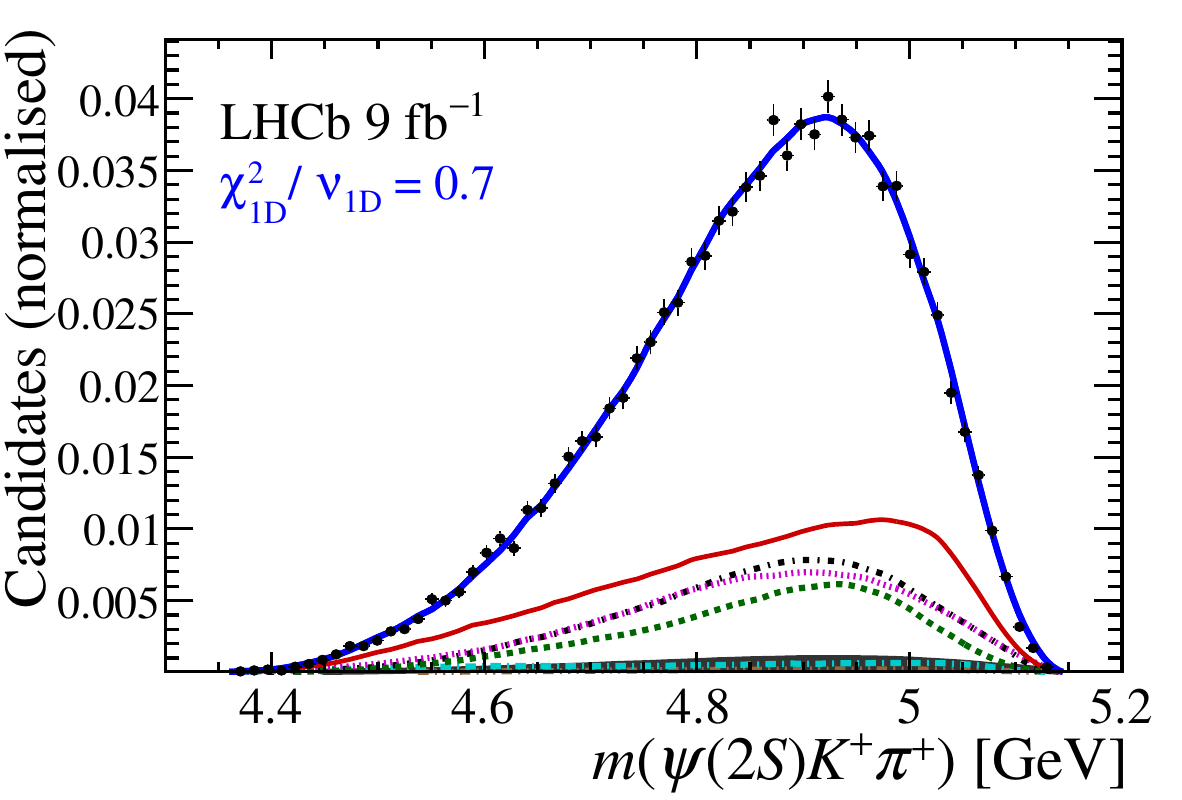}
              
       	 \includegraphics[width=0.329\textwidth,height=!]{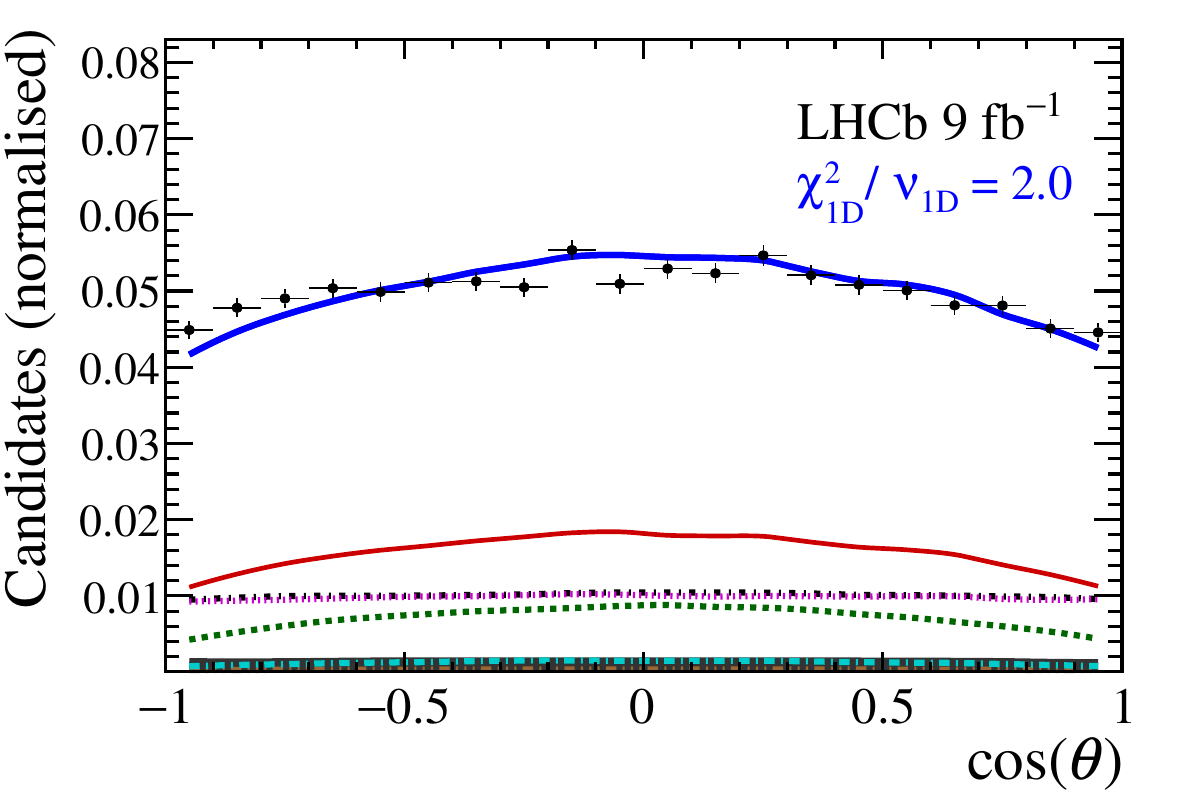}
       	 \includegraphics[width=0.329\textwidth,height=!]{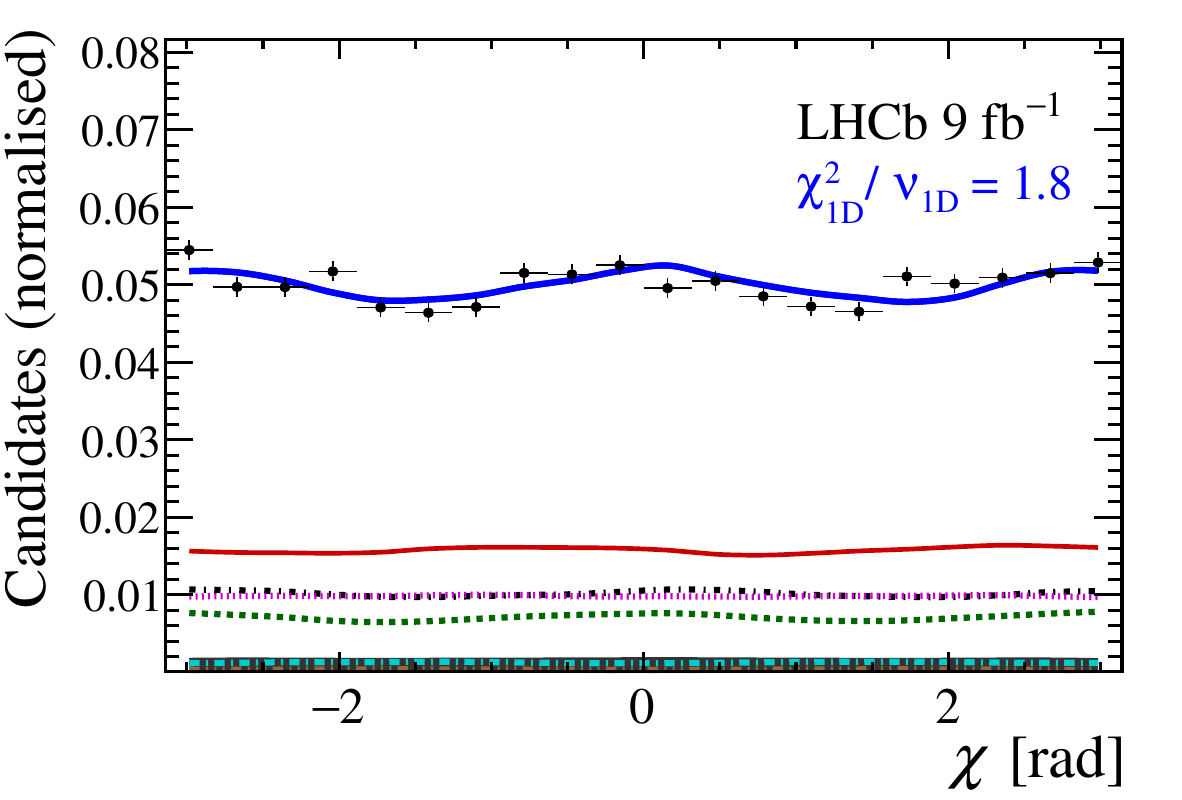}
       	 \includegraphics[width=0.329\textwidth,height=!]{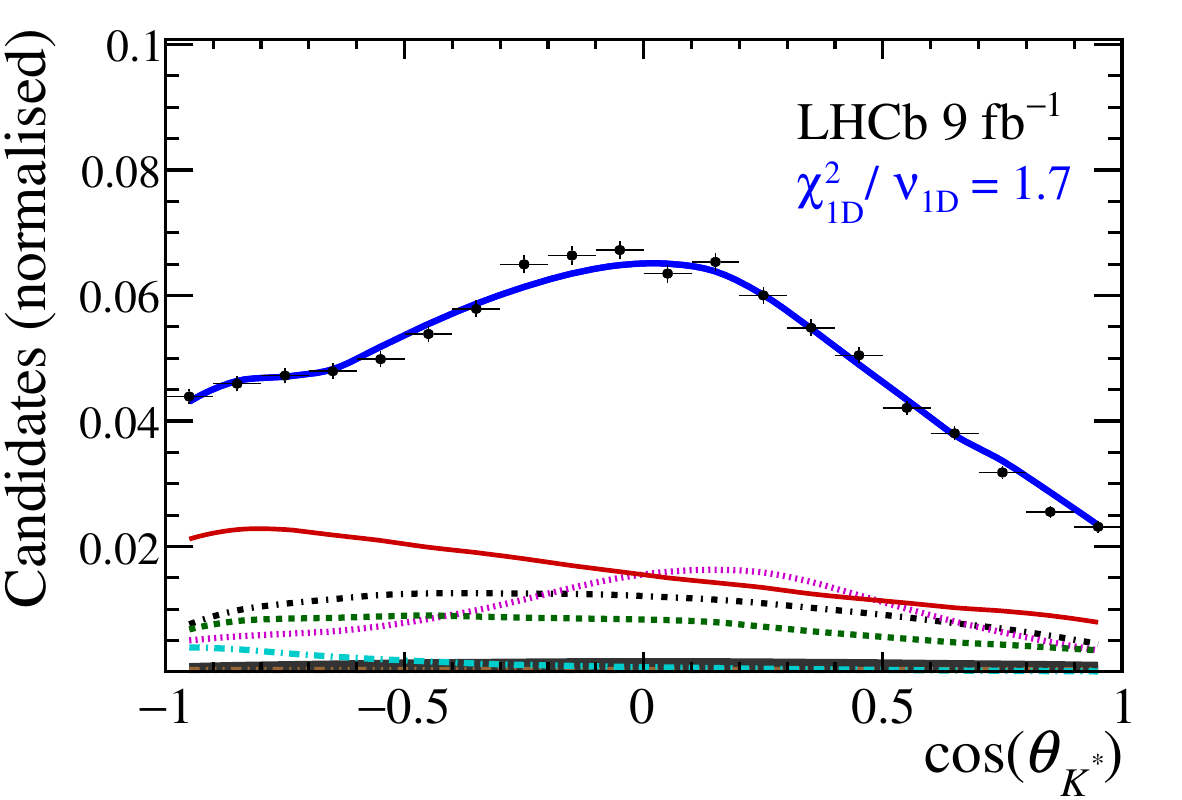}

        \centering
              \includegraphics[width=0.25\textwidth,height=!]{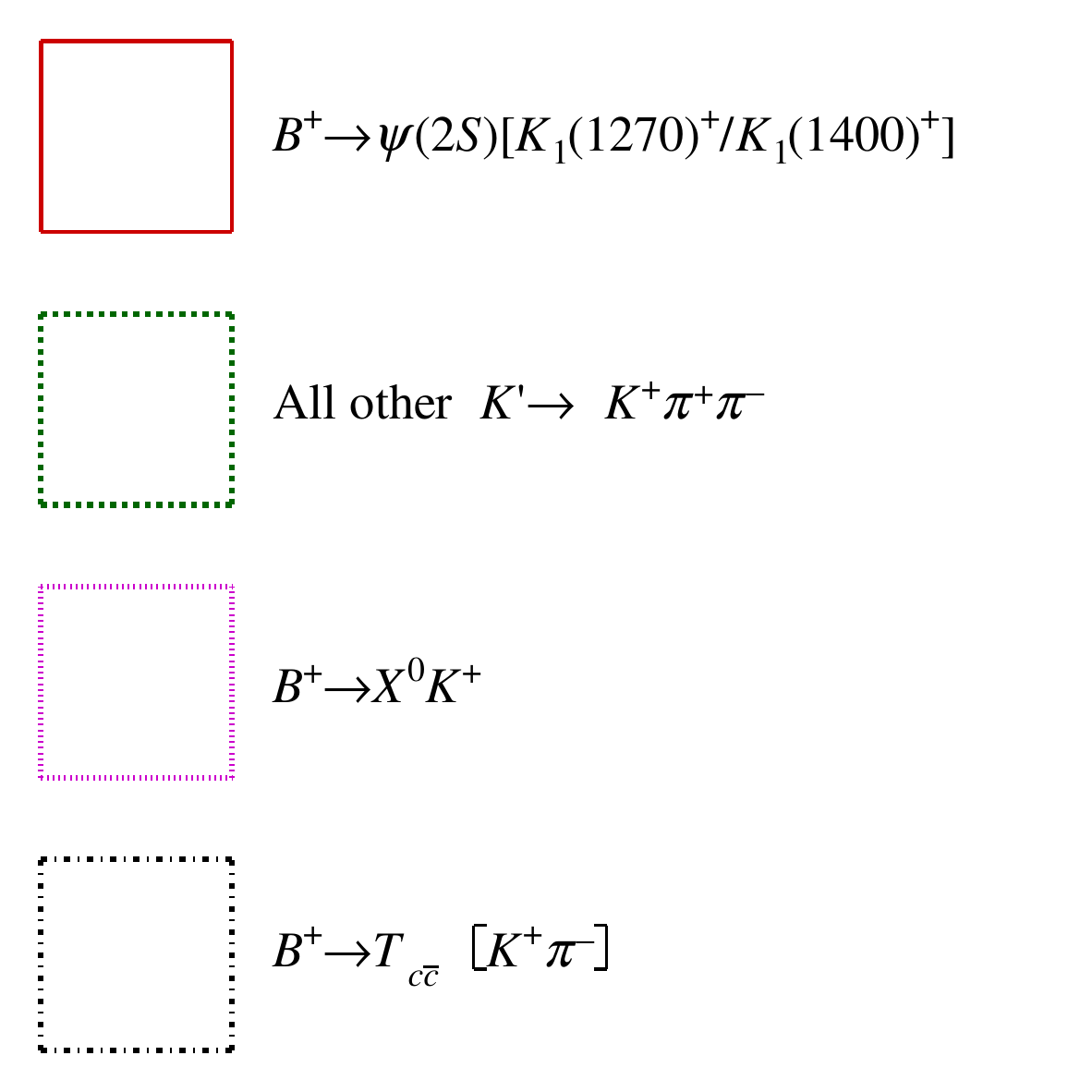}
              \includegraphics[width=0.25\textwidth,height=!]{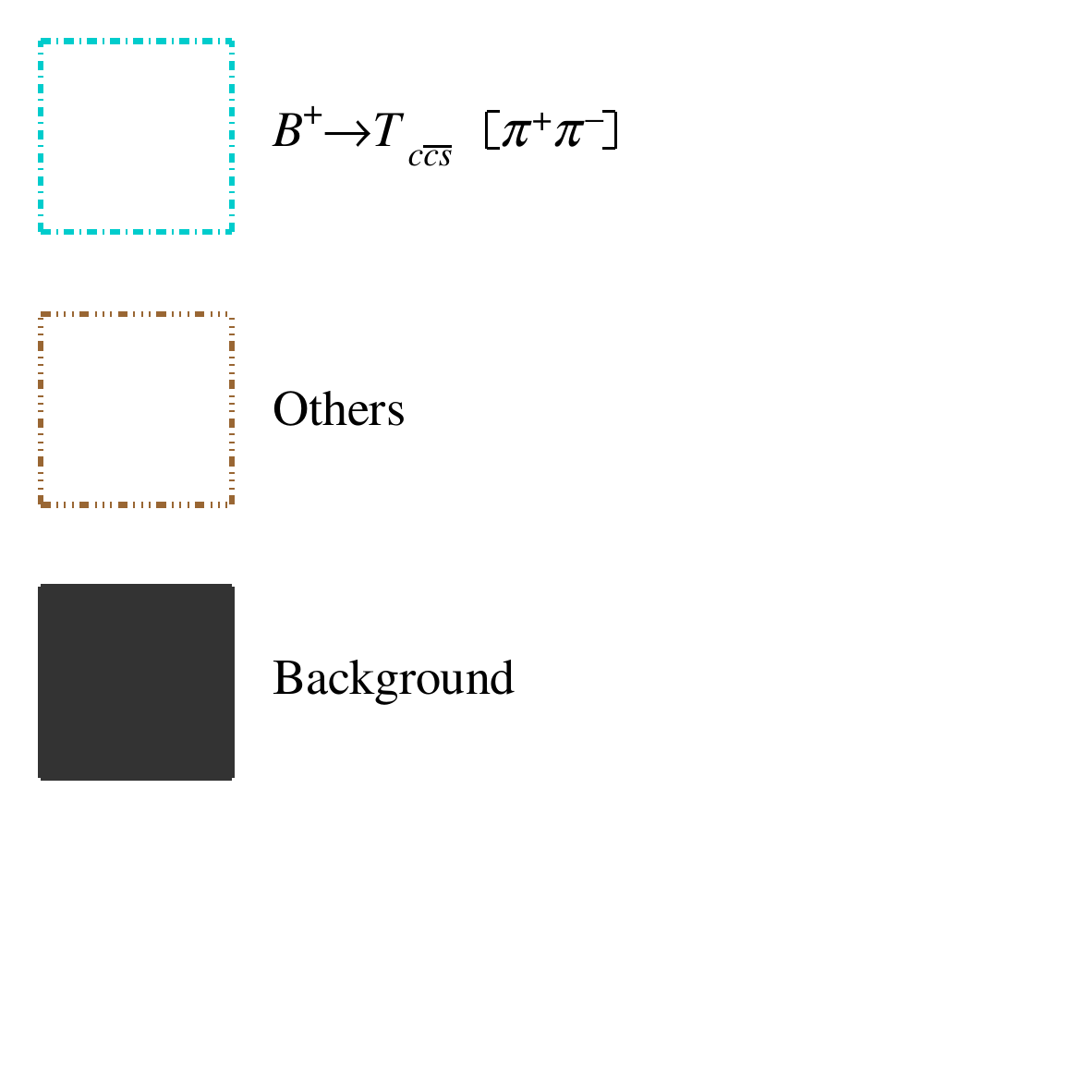}
         
	\caption{Phase-space projections of $\signal$ candidates in the signal region (points with error bars) and fit projections (solid, blue line) for the \textit{known-exotics} model.  The displayed $\chi_{\rm 1D}^2/\nu_{\rm 1D}$ value on each projection gives the sum of squared normalised residuals divided by the number of bins minus one.
	The multi-dimensional $\chi^2$ value is $\chi^2/\nu= 1.31$ with $\nu=1002$. 
	}

         \label{fig:fitExotic}

\end{figure}

Figure~\ref{fig:fitNoExotic} and Figure~\ref{fig:fitExotic} display the fit projections of the  \textit{no-exotics} and \textit{known-exotics} models, respectively.
Fit results of the \textit{start}, \textit{no-exotics}, \textit{known-exotics} and \textit{baseline} models are compared in Tables~\ref{tab:allModels1} and~\ref{tab:allModels2}. 

\begin{table}[h]
\centering
\caption{Fit results of the \textit{start},  \textit{no-exotics}, \textit{known-exotics} and \textit{baseline} models, part 1. Uncertainties are statistical only. }
\resizebox{0.75\linewidth}{!}{
	\renewcommand{\arraystretch}{1.05}
	\begin{tabular}{l r@{}c@{}l r@{}c@{}l r@{}c@{}l r@{}c@{}l }
\hline
\hline
Decay channel $F_{i} [ \% ] $  & \multicolumn{3}{c}{\textit{Start}}  & \multicolumn{3}{c}{\textit{No-exotics}} & \multicolumn{3}{c}{\textit{Known-exotics}} & \multicolumn{3}{c}{\textit{Baseline}} \\
\hline
$B^{+}\rightarrow \psi(\text{2S})\,K_{1}(1270)^{+}$  & $ 52.98 $&$\pm$&$1.13 $  & $ 26.56 $&$\pm$&$0.80 $  & $ 7.50 $&$\pm$&$0.86 $  & $ 7.60 $&$\pm$&$0.85 $  \\ 
$B^{+}\rightarrow \psi(\text{2S})\,K^{*}(1680)^{+}$  & $ 30.48 $&$\pm$&$0.72 $  & $ 22.26 $&$\pm$&$0.77 $  & $ 5.23 $&$\pm$&$0.73 $  & $ 8.15 $&$\pm$&$1.31 $  \\ 
$B^{+}\rightarrow \psi(\text{2S})\,K(1460)^{+}$  & $ 10.69 $&$\pm$&$0.32 $  & $ 6.04 $&$\pm$&$0.52 $  & $ 5.74 $&$\pm$&$0.45 $  & $ 5.26 $&$\pm$&$0.48 $  \\ 
$B^{+}\rightarrow K_{2}^{*}(1430)^{+}\psi(\text{2S})\,$  & $ 6.89 $&$\pm$&$0.47 $  & $ 6.91 $&$\pm$&$0.33 $  & $ 4.25 $&$\pm$&$0.25 $  & $ 4.35 $&$\pm$&$0.29 $  \\ 
$B^{+}\rightarrow \psi(\text{2S})\,K_{1}(1400)^{+}$  & $ 6.43 $&$\pm$&$0.94 $  & $ 2.27 $&$\pm$&$0.32 $  & $ 5.38 $&$\pm$&$0.69 $  & $ 5.78 $&$\pm$&$0.62 $  \\ 
$B^{+}\rightarrow \psi(\text{2S})\,K^{*}(1410)^{+}$  & $ 2.88 $&$\pm$&$0.31 $  & $ 2.61 $&$\pm$&$0.37 $  & $ 2.36 $&$\pm$&$0.43 $  & $ 1.79 $&$\pm$&$0.35 $  \\ 
$B^{+}[P]\rightarrow \psi(\text{2S})\,K_{1}(1270)^{+}$  & \multicolumn{3}{c}{$-$}  & $ 8.50 $&$\pm$&$0.43 $  & $ 4.83 $&$\pm$&$0.46 $  & $ 7.52 $&$\pm$&$0.60 $  \\ 
$B^{+}[D]\rightarrow \psi(\text{2S})\,K_{1}(1270)^{+}$  & \multicolumn{3}{c}{$-$}  & $ 8.46 $&$\pm$&$0.47 $  & $ 7.17 $&$\pm$&$0.50 $  & $ 6.81 $&$\pm$&$0.45 $  \\ 
$B^{+}\rightarrow \psi(\text{2S})\,[\left[\pi^{+}\pi^{-}\right]_{\text{S}}K^{+}]_{\text A}$  & \multicolumn{3}{c}{$-$}  & $ 6.68 $&$\pm$&$0.56 $  & \multicolumn{3}{c}{$-$}  & \multicolumn{3}{c}{$-$}  \\ 
$B^{+}[P]\rightarrow \psi(\text{2S})\,K^{*}(1410)^{+}$  & \multicolumn{3}{c}{$-$}  & $ 4.88 $&$\pm$&$0.42 $  & \multicolumn{3}{c}{$-$}  & \multicolumn{3}{c}{$-$}  \\ 
$B^{+}[P]\rightarrow \psi(\text{2S})\,K^{*}(1680)^{+}$  & \multicolumn{3}{c}{$-$}  & $ 3.96 $&$\pm$&$0.56 $  & \multicolumn{3}{c}{$-$}  & \multicolumn{3}{c}{$-$}  \\ 
$B^{+}\rightarrow K_{2}(1770)^{+}\psi(\text{2S})\,$  & \multicolumn{3}{c}{$-$}  & $ 3.81 $&$\pm$&$0.30 $  & \multicolumn{3}{c}{$-$}  & \multicolumn{3}{c}{$-$}  \\ 
$B^{+}\rightarrow \psi(4415)K^{+}$  & \multicolumn{3}{c}{$-$}  & $ 1.82 $&$\pm$&$0.41 $  & \multicolumn{3}{c}{$-$}  & \multicolumn{3}{c}{$-$}  \\ 
$B^{+}[P]\rightarrow \psi(\text{2S})\,K_{1}(1400)^{+}$  & \multicolumn{3}{c}{$-$}  & $ 1.18 $&$\pm$&$0.17 $  & $ 0.68 $&$\pm$&$0.18 $  & $ 0.48 $&$\pm$&$0.18 $  \\ 
$B^{+}\rightarrow \psi(4360)K^{+}$  & \multicolumn{3}{c}{$-$}  & $ 1.10 $&$\pm$&$0.40 $  & $ 1.11 $&$\pm$&$0.19 $  & $ 0.64 $&$\pm$&$0.14 $  \\ 
$B^{+}\rightarrow \psi(\text{2S})\,K_{1}(1650)^{+}$  & \multicolumn{3}{c}{$-$}  & $ 0.76 $&$\pm$&$0.16 $  & \multicolumn{3}{c}{$-$}  & \multicolumn{3}{c}{$-$}  \\ 
$B^{+}\rightarrow \psi(\text{2S})\,[\rho(770)^{0}K^{+}]_{\text P}$  & \multicolumn{3}{c}{$-$}  & $ 0.36 $&$\pm$&$0.21 $  & \multicolumn{3}{c}{$-$}  & \multicolumn{3}{c}{$-$}  \\ 
$B^{+}\rightarrow \psi(4660)K^{+}$  & \multicolumn{3}{c}{$-$}  & $ 0.32 $&$\pm$&$0.10 $  & \multicolumn{3}{c}{$-$}  & \multicolumn{3}{c}{$-$}  \\ 
$B^{+}\rightarrow \psi(\text{2S})\,[K^{*}(892)^{0}\pi^{+}]_{\text P}$  & \multicolumn{3}{c}{$-$}  & $ 0.06 $&$\pm$&$0.07 $  & \multicolumn{3}{c}{$-$}  & \multicolumn{3}{c}{$-$}  \\ 
$B^{+}\rightarrow \XSone K^{+}$  & \multicolumn{3}{c}{$-$}  & \multicolumn{3}{c}{$-$}  & $ 16.96 $&$\pm$&$1.22 $  & $ 18.45 $&$\pm$&$1.31 $  \\ 
$B^{+}[P]\rightarrow \ZAone^{+} K^{*}(892)^{0}$  & \multicolumn{3}{c}{$-$}  & \multicolumn{3}{c}{$-$}  & $ 7.35 $&$\pm$&$0.53 $  & $ 4.60 $&$\pm$&$0.54 $  \\ 
$B^{+}\rightarrow \XVone K^{+}$  & \multicolumn{3}{c}{$-$}  & \multicolumn{3}{c}{$-$}  & $ 4.32 $&$\pm$&$0.52 $  & $ 3.24 $&$\pm$&$0.50 $  \\ 
$B^{+}\rightarrow \ZAtwo^{+} \left[K^{+}\pi^{-}\right]_{\text{S}}$  & \multicolumn{3}{c}{$-$}  & \multicolumn{3}{c}{$-$}  & $ 3.85 $&$\pm$&$0.52 $  & $ 3.41 $&$\pm$&$0.54 $  \\ 
$B^{+}\rightarrow \XStwo K^{+}$  & \multicolumn{3}{c}{$-$}  & \multicolumn{3}{c}{$-$}  & $ 3.03 $&$\pm$&$0.42 $  & $ 1.73 $&$\pm$&$0.28 $  \\ 
$B^{+}\rightarrow \XAone K^{+}$  & \multicolumn{3}{c}{$-$}  & \multicolumn{3}{c}{$-$}  & $ 2.31 $&$\pm$&$0.40 $  & $ 2.89 $&$\pm$&$0.45 $  \\ 
$B^{+}[D]\rightarrow \ZAone^{+} K^{*}(892)^{0}$  & \multicolumn{3}{c}{$-$}  & \multicolumn{3}{c}{$-$}  & $ 2.05 $&$\pm$&$0.31 $  & $ 2.78 $&$\pm$&$0.33 $  \\ 
$B^{+}\rightarrow \ZAtwo^{+} K^{*}(892)^{0}$  & \multicolumn{3}{c}{$-$}  & \multicolumn{3}{c}{$-$}  & $ 1.92 $&$\pm$&$1.30 $  & $ 0.75 $&$\pm$&$0.43 $  \\ 
$B^{+}[D]\rightarrow \rho(770)^{0}\ZsAone $  & \multicolumn{3}{c}{$-$}  & \multicolumn{3}{c}{$-$}  & $ 1.84 $&$\pm$&$0.20 $  & $ 2.06 $&$\pm$&$0.22 $  \\ 
$B^{+}\rightarrow \ZAone^{+} K^{*}(892)^{0}$  & \multicolumn{3}{c}{$-$}  & \multicolumn{3}{c}{$-$}  & $ 1.36 $&$\pm$&$0.85 $  & $ 4.02 $&$\pm$&$0.88 $  \\ 
$B^{+}\rightarrow \ZsAone \left[\pi^{+}\pi^{-}\right]_{\text{S}}$  & \multicolumn{3}{c}{$-$}  & \multicolumn{3}{c}{$-$}  & $ 1.12 $&$\pm$&$0.21 $  & $ 1.24 $&$\pm$&$0.23 $  \\ 
$B^{+}\rightarrow \ZVone K^{*}(892)^{0}$  & \multicolumn{3}{c}{$-$}  & \multicolumn{3}{c}{$-$}  & $ 1.04 $&$\pm$&$0.15 $  & $ 0.52 $&$\pm$&$0.10 $  \\ 
$B^{+}\rightarrow [\psi(\text{2S})\,K^{*}(892)^{0}]_{\text{S}}\pi^{+}$  & \multicolumn{3}{c}{$-$}  & \multicolumn{3}{c}{$-$}  & $ 0.41 $&$\pm$&$0.51 $  & \multicolumn{3}{c}{$-$}  \\ 
$B^{+}\rightarrow [\psi(\text{2S})\,K^{+}]_{\text{A}}\left[\pi^{+}\pi^{-}\right]_{\text{S}}$  & \multicolumn{3}{c}{$-$}  & \multicolumn{3}{c}{$-$}  & $ 0.32 $&$\pm$&$0.13 $  & \multicolumn{3}{c}{$-$}  \\ 
$B^{+}[P]\rightarrow \Z(4100)^{+}K^{*}(892)^{0}$  & \multicolumn{3}{c}{$-$}  & \multicolumn{3}{c}{$-$}  & $ 0.31 $&$\pm$&$0.13 $  & \multicolumn{3}{c}{$-$}  \\ 
$B^{+}[P]\rightarrow [\psi(\text{2S})\,K^{+}]_{\text{V}}\rho(770)^{0}$  & \multicolumn{3}{c}{$-$}  & \multicolumn{3}{c}{$-$}  & $ 0.07 $&$\pm$&$0.05 $  & \multicolumn{3}{c}{$-$}  \\ 

$B^{+}\rightarrow \XsAone\pi^{+}$  & \multicolumn{3}{c}{$-$}  & \multicolumn{3}{c}{$-$}  & \multicolumn{3}{c}{$-$}  & $ 4.42 $&$\pm$&$0.98 $  \\ 
$B^{+}\rightarrow \XsAtwo\pi^{+}$  & \multicolumn{3}{c}{$-$}  & \multicolumn{3}{c}{$-$}  & \multicolumn{3}{c}{$-$}  & $ 2.60 $&$\pm$&$0.66 $  \\ 
$B^{+}\rightarrow \XsVone\pi^{+}$  & \multicolumn{3}{c}{$-$}  & \multicolumn{3}{c}{$-$}  & \multicolumn{3}{c}{$-$}  & $ 1.59 $&$\pm$&$0.46 $  \\  \hline
$\text{Sum } B^{+}$  & $ 110.34 $&$\pm$&$1.99 $  & $ 108.56 $&$\pm$&$1.21 $  & $ 92.50 $&$\pm$&$2.15 $  & $ 102.69 $&$\pm$&$4.40 $  \\ 
\hline
$\XSone \text{ mass [MeV]}$  & \multicolumn{3}{c}{$-$}  & \multicolumn{3}{c}{$-$}  & $ 4467.1 $&$\pm$&$7.6 $  & $ 4474.5 $&$\pm$&$6.9 $  \\ 
$\XSone \text{ width [MeV]}$  & \multicolumn{3}{c}{$-$}  & \multicolumn{3}{c}{$-$}  & $ 259.2 $&$\pm$&$22.6 $  & $ 230.8 $&$\pm$&$19.5 $  \\ 
$\XAone \text{ mass [MeV]}$  & \multicolumn{3}{c}{$-$}  & \multicolumn{3}{c}{$-$}  & $ 4506.1 $&$\pm$&$15.4 $  & $ 4652.5 $&$\pm$&$14.4 $  \\ 
$\XAone \text{ width [MeV]}$  & \multicolumn{3}{c}{$-$}  & \multicolumn{3}{c}{$-$}  & $ 250.0 $&$\pm$&$25.2 $  & $ 227.0 $&$\pm$&$26.3 $  \\ 
$\XStwo \text{ mass [MeV]}$  & \multicolumn{3}{c}{$-$}  & \multicolumn{3}{c}{$-$}  & $ 4696.4 $&$\pm$&$5.6 $  & $ 4710.2 $&$\pm$&$4.4 $  \\ 
$\XStwo \text{ width [MeV]}$  & \multicolumn{3}{c}{$-$}  & \multicolumn{3}{c}{$-$}  & $ 98.8 $&$\pm$&$10.5 $  & $ 63.6 $&$\pm$&$8.7 $  \\ 
$\XVone \text{ mass [MeV]}$  & \multicolumn{3}{c}{$-$}  & \multicolumn{3}{c}{$-$}  & $ 4748.0 $&$\pm$&$33.1 $  & $ 4784.6 $&$\pm$&$37.4 $  \\ 
$\XVone \text{ width [MeV]}$  & \multicolumn{3}{c}{$-$}  & \multicolumn{3}{c}{$-$}  & $ 444.2 $&$\pm$&$78.9 $  & $ 457.1 $&$\pm$&$93.4 $  \\ 
$\ZAone^{+} \text{ mass [MeV]}$  & \multicolumn{3}{c}{$-$}  & \multicolumn{3}{c}{$-$}  & $ 4311.0 $&$\pm$&$14.5 $  & $ 4257.3 $&$\pm$&$10.8 $  \\ 
$\ZAone^{+} \text{ width [MeV]}$  & \multicolumn{3}{c}{$-$}  & \multicolumn{3}{c}{$-$}  & $ 343.3 $&$\pm$&$18.8 $  & $ 308.2 $&$\pm$&$20.4 $  \\ 
$\ZAtwo^{+} \text{ mass [MeV]}$  & \multicolumn{3}{c}{$-$}  & \multicolumn{3}{c}{$-$}  & $ 4481.7 $&$\pm$&$13.1 $  & $ 4468.2 $&$\pm$&$21.0 $  \\ 
$\ZAtwo^{+} \text{ width [MeV]}$  & \multicolumn{3}{c}{$-$}  & \multicolumn{3}{c}{$-$}  & $ 180.5 $&$\pm$&$23.5 $  & $ 250.9 $&$\pm$&$41.6 $  \\ 
$\XsAone \text{ mass [MeV]}$  & \multicolumn{3}{c}{$-$}  & \multicolumn{3}{c}{$-$}  & \multicolumn{3}{c}{$-$}  & $ 4577.8 $&$\pm$&$9.5 $  \\ 
$\XsAone \text{ width [MeV]}$  & \multicolumn{3}{c}{$-$}  & \multicolumn{3}{c}{$-$}  & \multicolumn{3}{c}{$-$}  & $ 133.0 $&$\pm$&$28.4 $  \\ 
$\XsAtwo \text{ mass [MeV]}$  & \multicolumn{3}{c}{$-$}  & \multicolumn{3}{c}{$-$}  & \multicolumn{3}{c}{$-$}  & $ 4924.9 $&$\pm$&$22.0 $  \\ 
$\XsAtwo \text{ width [MeV]}$  & \multicolumn{3}{c}{$-$}  & \multicolumn{3}{c}{$-$}  & \multicolumn{3}{c}{$-$}  & $ 255.0 $&$\pm$&$54.6 $  \\ 
$\XsVone \text{ mass [MeV]}$  & \multicolumn{3}{c}{$-$}  & \multicolumn{3}{c}{$-$}  & \multicolumn{3}{c}{$-$}  & $ 5225.1 $&$\pm$&$85.7 $  \\ 
$\XsVone \text{ width [MeV]}$  & \multicolumn{3}{c}{$-$}  & \multicolumn{3}{c}{$-$}  & \multicolumn{3}{c}{$-$}  & $ 226.2 $&$\pm$&$75.5 $  \\ 
\hline
$\Delta(-2 \ln \mathcal L)$  & \multicolumn{3}{c}{$ 0.0 $}  & \multicolumn{3}{c}{$ -6214.2  $}  & \multicolumn{3}{c}{$ -11038.0  $}  & \multicolumn{3}{c}{$ -12021.3  $}  \\ 
$ N_{\rm par} $  &  \multicolumn{3}{c}{$18 $}  & \multicolumn{3}{c}{$74 $}  & \multicolumn{3}{c}{$ 88 $}  & \multicolumn{3}{c}{$ 98 $}  \\ 
$ \chi^2/\nu $  & \multicolumn{3}{c}{$ 2.54 $}  & \multicolumn{3}{c}{$ 2.05 $}  & \multicolumn{3}{c}{$ 1.31 $}  & \multicolumn{3}{c}{$ 1.21 $}  \\ 
$ \chi^2/N_{\rm bins} $  & \multicolumn{3}{c}{$ 2.50 $}  & \multicolumn{3}{c}{$ 1.91 $}  & \multicolumn{3}{c}{$ 1.20 $}  & \multicolumn{3}{c}{$ 1.10 $}  \\ 
\hline
\hline
\end{tabular} }
\label{tab:allModels1}
\end{table}

\begin{table}[h]
\centering
\caption{Fit results of the \textit{start}, \textit{no-exotics}, \textit{known-exotics} and \textit{baseline} models, part 2. Uncertainties are statistical only. }
\resizebox{0.82\linewidth}{!}{
	\renewcommand{\arraystretch}{1.05}
	\begin{tabular}{l r@{}c@{}l r@{}c@{}l r@{}c@{}l r@{}c@{}l }
\hline
\hline
Decay channel $F_{i} [ \% ] $  & \multicolumn{3}{c}{\textit{Start}}  & \multicolumn{3}{c}{\textit{No-exotics}} & \multicolumn{3}{c}{\textit{Known-exotics}} & \multicolumn{3}{c}{\textit{Baseline}} \\
\hline

$K_{1}(1270)^{+}\rightarrow \rho(770)^{0}K^{+}$  & $ 71.49 $&$\pm$&$1.50 $  & $ 45.70 $&$\pm$&$1.87 $  & $ 46.52 $&$\pm$&$2.33 $  & $ 50.71 $&$\pm$&$2.18 $  \\ 
$K_{1}(1270)^{+}\rightarrow K^{*}(892)^{0}\pi^{+}$  & $ 33.40 $&$\pm$&$1.24 $  & $ 26.16 $&$\pm$&$1.06 $  & $ 21.63 $&$\pm$&$1.77 $  & $ 19.86 $&$\pm$&$1.44 $  \\ 
$K_{1}(1270)^{+}\rightarrow \left[K^{+}\pi^{-}\right]_{\text{S}}\pi^{+}$  & $0.00 $&$\pm$&$0.00$  & $6.25 $&$\pm$&$0.98$  & $ 9.85 $&$\pm$&$1.28 $  & $ 11.35 $&$\pm$&$1.45 $  \\ 
$K_{1}(1270)^{+}[D]\rightarrow K^{*}(892)^{0}\pi^{+}$  & \multicolumn{3}{c}{$-$}  & $ 6.75 $&$\pm$&$0.58 $  & $ 9.35 $&$\pm$&$1.09 $  & $ 8.32 $&$\pm$&$0.85 $  \\ 
$K_{1}(1270)^{+}\rightarrow \rho(1450)^{0}K^{+}$  & \multicolumn{3}{c}{$-$}  & $ 7.12 $&$\pm$&$1.60 $  & \multicolumn{3}{c}{$-$}  & \multicolumn{3}{c}{$-$}  \\ 
$K_{1}(1270)^{+}\rightarrow K^{*}(1410)^{0}\pi^{+}$  & \multicolumn{3}{c}{$-$}  & $ 6.45 $&$\pm$&$1.05 $  & \multicolumn{3}{c}{$-$}  & \multicolumn{3}{c}{$-$}  \\ 
$\text{Sum } K_{1}(1270)^{+}$  & $ 104.89 $&$\pm$&$0.77 $  & $ 98.42 $&$\pm$&$3.03 $  & $ 87.35 $&$\pm$&$1.59 $  & $ 90.24 $&$\pm$&$1.83 $  \\  \hline

$K_{1}(1400)^{+}\rightarrow K^{*}(892)^{0}\pi^{+}$  & \multicolumn{3}{c}{$100$}  & \multicolumn{3}{c}{$100$}  & $ 80.56 $&$\pm$&$4.15 $  & $ 86.80 $&$\pm$&$3.78 $  \\ 
$K_{1}(1400)^{+}\rightarrow \rho(770)^{0}K^{+}$  & \multicolumn{3}{c}{$-$}  & \multicolumn{3}{c}{$-$}  & $ 29.32 $&$\pm$&$4.62 $  & $ 22.08 $&$\pm$&$4.40 $  \\ 
$\text{Sum } K_{1}(1400)^{+}$  & \multicolumn{3}{c}{$100$}  & \multicolumn{3}{c}{$100$}  & $ 109.88 $&$\pm$&$0.70 $  & $ 108.88 $&$\pm$&$0.82 $  \\  \hline

$K^{*}(1410)^{+}\rightarrow K^{*}(892)^{0}\pi^{+}$  & \multicolumn{3}{c}{$100$}  & $ 86.92 $&$\pm$&$5.04 $  & $ 71.42 $&$\pm$&$8.38 $  & $ 88.50 $&$\pm$&$8.39 $  \\ 
$K^{*}(1410)^{+}\rightarrow \rho(770)^{0}K^{+}$  & \multicolumn{3}{c}{$-$}  & $ 50.09 $&$\pm$&$6.23 $  & $ 57.57 $&$\pm$&$9.24 $  & $ 38.36 $&$\pm$&$10.46 $  \\ 
$\text{Sum } K^{*}(1410)^{+}$  & \multicolumn{3}{c}{$100$}  & $ 137.01 $&$\pm$&$1.27 $  & $ 128.99 $&$\pm$&$3.96 $  & $ 126.86 $&$\pm$&$4.83 $  \\  \hline

$K_{2}^{*}(1430)^{+}\rightarrow K^{*}(892)^{0}\pi^{+}$  & $ 80.70 $&$\pm$&$0.38 $  & $ 77.88 $&$\pm$&$2.48 $  & $ 78.90 $&$\pm$&$2.92 $  & $ 76.70 $&$\pm$&$3.04 $  \\ 
$K_{2}^{*}(1430)^{+}\rightarrow \rho(770)^{0}K^{+}$  & $ 36.15 $&$\pm$&$0.17 $  & $ 11.74 $&$\pm$&$1.83 $  & $ 11.31 $&$\pm$&$2.16 $  & $ 12.71 $&$\pm$&$2.30 $  \\ 
$\text{Sum } K_{2}^{*}(1430)^{+}$  & $ 116.86 $&$\pm$&$0.54 $  & $ 89.62 $&$\pm$&$0.65 $  & $ 90.22 $&$\pm$&$0.80 $  & $ 89.41 $&$\pm$&$0.75 $  \\  \hline

$K(1460)^{+}\rightarrow \left[\pi^{+}\pi^{-}\right]_{\text{S}}K^{+}$  & $ 47.43 $&$\pm$&$2.14 $  & $ 27.78 $&$\pm$&$3.04 $  & $ 41.60 $&$\pm$&$3.72 $  & $ 45.13 $&$\pm$&$4.22 $  \\ 
$K(1460)^{+}\rightarrow K^{*}(892)^{0}\pi^{+}$  & $ 36.40 $&$\pm$&$2.05 $  & $ 31.89 $&$\pm$&$3.72 $  & $ 38.26 $&$\pm$&$3.61 $  & $ 35.41 $&$\pm$&$4.08 $  \\ 
$K(1460)^{+}\rightarrow \rho(770)^{0}K^{+}$  & \multicolumn{3}{c}{$-$}  & $ 24.56 $&$\pm$&$3.12 $  & \multicolumn{3}{c}{$-$}  & \multicolumn{3}{c}{$-$}  \\ 
$\text{Sum } K(1460)^{+}$  & $ 83.83 $&$\pm$&$0.09 $  & $ 84.22 $&$\pm$&$1.07 $  & $ 79.86 $&$\pm$&$0.42 $  & $ 80.54 $&$\pm$&$0.67 $  \\  \hline

$K_{1}(1650)^{+}[D]\rightarrow K^{*}(892)^{0}\pi^{+}$  & \multicolumn{3}{c}{$-$}  & \multicolumn{3}{c}{$100$}  & \multicolumn{3}{c}{$-$}  & \multicolumn{3}{c}{$-$}  \\ 

$K^{*}(1680)^{+}\rightarrow \rho(770)^{0}K^{+}$  & $ 77.92 $&$\pm$&$2.00 $  & $ 80.86 $&$\pm$&$1.63 $  & $ 81.58 $&$\pm$&$5.43 $  & $ 31.16 $&$\pm$&$6.11 $  \\ 
$K^{*}(1680)^{+}\rightarrow K^{*}(892)^{0}\pi^{+}$  & $ 11.32 $&$\pm$&$1.47 $  & $ 8.03 $&$\pm$&$1.01 $  & $ 16.62 $&$\pm$&$5.05 $  & $ 49.69 $&$\pm$&$6.72 $  \\ 
$\text{Sum } K^{*}(1680)^{+}$  & $ 89.24 $&$\pm$&$0.67 $  & $ 88.89 $&$\pm$&$0.72 $  & $ 98.20 $&$\pm$&$3.14 $  & $ 80.85 $&$\pm$&$0.64 $  \\  \hline

$K_{2}(1770)^{+}\rightarrow \rho(770)^{0}K^{+}$  & \multicolumn{3}{c}{$-$}  & $ 68.79 $&$\pm$&$3.59 $  & \multicolumn{3}{c}{$-$}  & \multicolumn{3}{c}{$-$}  \\ 
$K_{2}(1770)^{+}\rightarrow K^{*}(892)^{0}\pi^{+}$  & \multicolumn{3}{c}{$-$}  & $ 23.21 $&$\pm$&$3.18 $  & \multicolumn{3}{c}{$-$}  & \multicolumn{3}{c}{$-$}  \\ 
$\text{Sum } K_{2}(1770)^{+}$  & \multicolumn{3}{c}{$-$}  & $ 92.00 $&$\pm$&$0.81 $  & \multicolumn{3}{c}{$-$}  & \multicolumn{3}{c}{$-$}  \\  \hline

$\XSone\rightarrow \rho(770)^{0}\psi(\text{2S})\,$  & \multicolumn{3}{c}{$-$}  & \multicolumn{3}{c}{$-$}  & $ 97.56 $&$\pm$&$0.71 $  & $ 99.04 $&$\pm$&$0.49 $  \\ 
$\XSone\rightarrow \ZAone^{-} \pi^{+}$  & \multicolumn{3}{c}{$-$}  & \multicolumn{3}{c}{$-$}  & $ 1.28 $&$\pm$&$0.38 $  & $ 0.50 $&$\pm$&$0.25 $  \\ 
$\XSone\rightarrow \ZAone^{+} \pi^{-}$  & \multicolumn{3}{c}{$-$}  & \multicolumn{3}{c}{$-$}  & $ 1.28 $&$\pm$&$0.38 $  & $ 0.50 $&$\pm$&$0.25 $  \\ 
$\text{Sum } \XSone$  & \multicolumn{3}{c}{$-$}  & \multicolumn{3}{c}{$-$}  & $ 100.13 $&$\pm$&$0.04 $  & $ 100.03 $&$\pm$&$0.02 $  \\  \hline

$\XAone\rightarrow \rho(770)^{0}\psi(\text{2S})\,$  & \multicolumn{3}{c}{$-$}  & \multicolumn{3}{c}{$-$}  & $ 73.56 $&$\pm$&$10.54 $  & $ 86.66 $&$\pm$&$7.85 $  \\ 
$\XAone\rightarrow \ZAone^{-} \pi^{+}$  & \multicolumn{3}{c}{$-$}  & \multicolumn{3}{c}{$-$}  & $ 3.78 $&$\pm$&$1.94 $  & $ 6.62 $&$\pm$&$2.03 $  \\ 
$\XAone\rightarrow \ZAone^{+} \pi^{-}$  & \multicolumn{3}{c}{$-$}  & \multicolumn{3}{c}{$-$}  & $ 3.77 $&$\pm$&$1.94 $  & $ 6.61 $&$\pm$&$2.03 $  \\ 
$\text{Sum } \XAone$  & \multicolumn{3}{c}{$-$}  & \multicolumn{3}{c}{$-$}  & $ 81.11 $&$\pm$&$8.55 $  & $ 99.89 $&$\pm$&$7.37 $  \\  \hline

$\XStwo\rightarrow \rho(770)^{0}\psi(\text{2S})\,$  & \multicolumn{3}{c}{$-$}  & \multicolumn{3}{c}{$-$}  & $ 67.06 $&$\pm$&$8.29 $  & $ 92.35 $&$\pm$&$10.83 $  \\ 
$\XStwo\rightarrow \ZAtwo^{+} \pi^{-}$  & \multicolumn{3}{c}{$-$}  & \multicolumn{3}{c}{$-$}  & $ 22.61 $&$\pm$&$2.97 $  & $ 17.00 $&$\pm$&$3.82 $  \\ 
$\XStwo\rightarrow \ZAtwo^{-} \pi^{+}$  & \multicolumn{3}{c}{$-$}  & \multicolumn{3}{c}{$-$}  & $ 22.53 $&$\pm$&$2.96 $  & $ 17.00 $&$\pm$&$3.82 $  \\ 
$\text{Sum } \XStwo$  & \multicolumn{3}{c}{$-$}  & \multicolumn{3}{c}{$-$}  & $ 112.20 $&$\pm$&$7.48 $  & $ 126.35 $&$\pm$&$10.56 $  \\  \hline

$\XVone\rightarrow \rho(770)^{0}\psi(\text{2S})\,$  & \multicolumn{3}{c}{$-$}  & \multicolumn{3}{c}{$-$}  & $ 42.65 $&$\pm$&$4.41 $  & $ 41.52 $&$\pm$&$5.19 $  \\ 
$\XVone\rightarrow \ZAone^{+} \pi^{-}$  & \multicolumn{3}{c}{$-$}  & \multicolumn{3}{c}{$-$}  & $ 21.09 $&$\pm$&$4.17 $  & $ 18.03 $&$\pm$&$5.14 $  \\ 
$\XVone\rightarrow \ZAone^{-} \pi^{+}$  & \multicolumn{3}{c}{$-$}  & \multicolumn{3}{c}{$-$}  & $ 21.07 $&$\pm$&$4.17 $  & $ 18.03 $&$\pm$&$5.14 $  \\ 
$\XVone\rightarrow \ZAtwo^{+} \pi^{-}$  & \multicolumn{3}{c}{$-$}  & \multicolumn{3}{c}{$-$}  & $ 4.46 $&$\pm$&$2.01 $  & $ 5.44 $&$\pm$&$2.45 $  \\ 
$\XVone\rightarrow \ZAtwo^{-} \pi^{+}$  & \multicolumn{3}{c}{$-$}  & \multicolumn{3}{c}{$-$}  & $ 4.44 $&$\pm$&$2.00 $  & $ 5.44 $&$\pm$&$2.45 $  \\ 
$\text{Sum } \XVone$  & \multicolumn{3}{c}{$-$}  & \multicolumn{3}{c}{$-$}  & $ 93.71 $&$\pm$&$9.74 $  & $ 88.47 $&$\pm$&$11.26 $  \\  \hline

$\XsAone\rightarrow \psi(\text{2S})\,K^{*}(892)^{0}$  & \multicolumn{3}{c}{$-$}  & \multicolumn{3}{c}{$-$}  & \multicolumn{3}{c}{$-$}  & $ 50.87 $&$\pm$&$7.79 $  \\ 
$\XsAone\rightarrow \ZAone^{-} K^{+}$  & \multicolumn{3}{c}{$-$}  & \multicolumn{3}{c}{$-$}  & \multicolumn{3}{c}{$-$}  & $ 16.53 $&$\pm$&$3.79 $  \\ 
$\XsAone\rightarrow \ZsAone \pi^{-}$  & \multicolumn{3}{c}{$-$}  & \multicolumn{3}{c}{$-$}  & \multicolumn{3}{c}{$-$}  & $ 9.84 $&$\pm$&$3.28 $  \\ 
$\text{Sum } \XsAone$  & \multicolumn{3}{c}{$-$}  & \multicolumn{3}{c}{$-$}  & \multicolumn{3}{c}{$-$}  & $ 77.23 $&$\pm$&$5.22 $  \\  \hline

$\XsVone\rightarrow \psi(\text{2S})\,\left[K^{+}\pi^{-}\right]_{\text{S}}$  & \multicolumn{3}{c}{$-$}  & \multicolumn{3}{c}{$-$}  & \multicolumn{3}{c}{$-$}  & $ 66.28 $&$\pm$&$15.03 $  \\ 
$\XsVone\rightarrow \ZsAone\pi^{-}$  & \multicolumn{3}{c}{$-$}  & \multicolumn{3}{c}{$-$}  & \multicolumn{3}{c}{$-$}  & $ 9.37 $&$\pm$&$14.12 $  \\ 
$\text{Sum } \XsVone$  & \multicolumn{3}{c}{$-$}  & \multicolumn{3}{c}{$-$}  & \multicolumn{3}{c}{$-$}  & $ 75.65 $&$\pm$&$9.18 $  \\  \hline
$\psi(4430)\rightarrow \psi(\text{2S})\,\left[\pi^{+}\pi^{-}\right]_{\text{S}}$  & \multicolumn{3}{c}{$-$}  & \multicolumn{3}{c}{$100$}  & \multicolumn{3}{c}{$100$}  & \multicolumn{3}{c}{$100$}  \\ 

$\psi(4415)\rightarrow \psi(\text{2S})\,\left[\pi^{+}\pi^{-}\right]_{\text{S}}$  & \multicolumn{3}{c}{$-$}  & \multicolumn{3}{c}{$100$}  & \multicolumn{3}{c}{$-$}  & $ - $  \\ 

$\psi(4660)\rightarrow \psi(\text{2S})\,\left[\pi^{+}\pi^{-}\right]_{\text{S}}$  & \multicolumn{3}{c}{$-$}  & \multicolumn{3}{c}{$100$}  & \multicolumn{3}{c}{$-$}  & $ - $  \\ 

$\XsAtwo\rightarrow \psi(\text{2S})\,K^{*}(892)^{0}$  & \multicolumn{3}{c}{$-$}  & \multicolumn{3}{c}{$-$}  & \multicolumn{3}{c}{$-$}  & \multicolumn{3}{c}{$100$}  \\

\hline
\hline
\end{tabular} }
\label{tab:allModels2}
\end{table}

\clearpage
\section{Amplitude models}
\label{a:models}

\renewcommand{\thefigure}{D.\arabic{figure}}
\renewcommand{\thetable}{D.\arabic{table}}
\setcounter{table}{0}
\setcounter{figure}{0}

\begin{figure}[h]
       	 \includegraphics[width=0.329\textwidth,height=!]{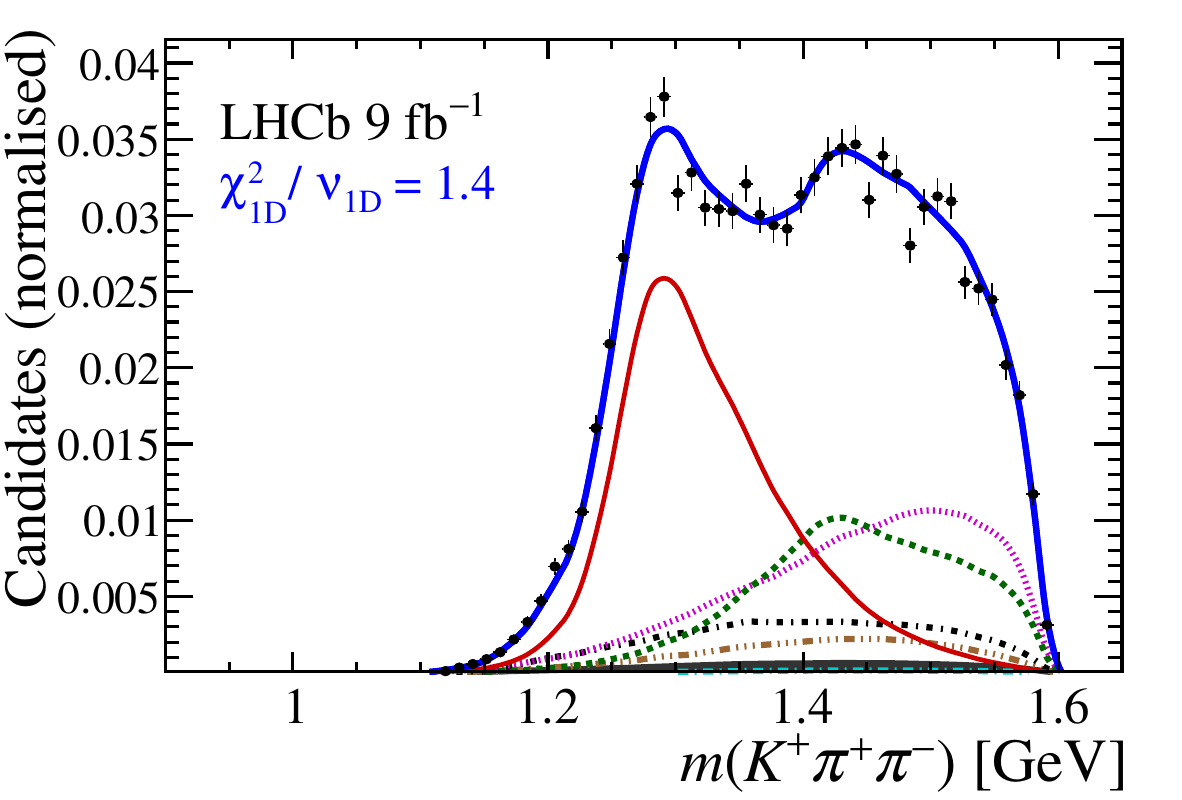}
       	 \includegraphics[width=0.329\textwidth,height=!]{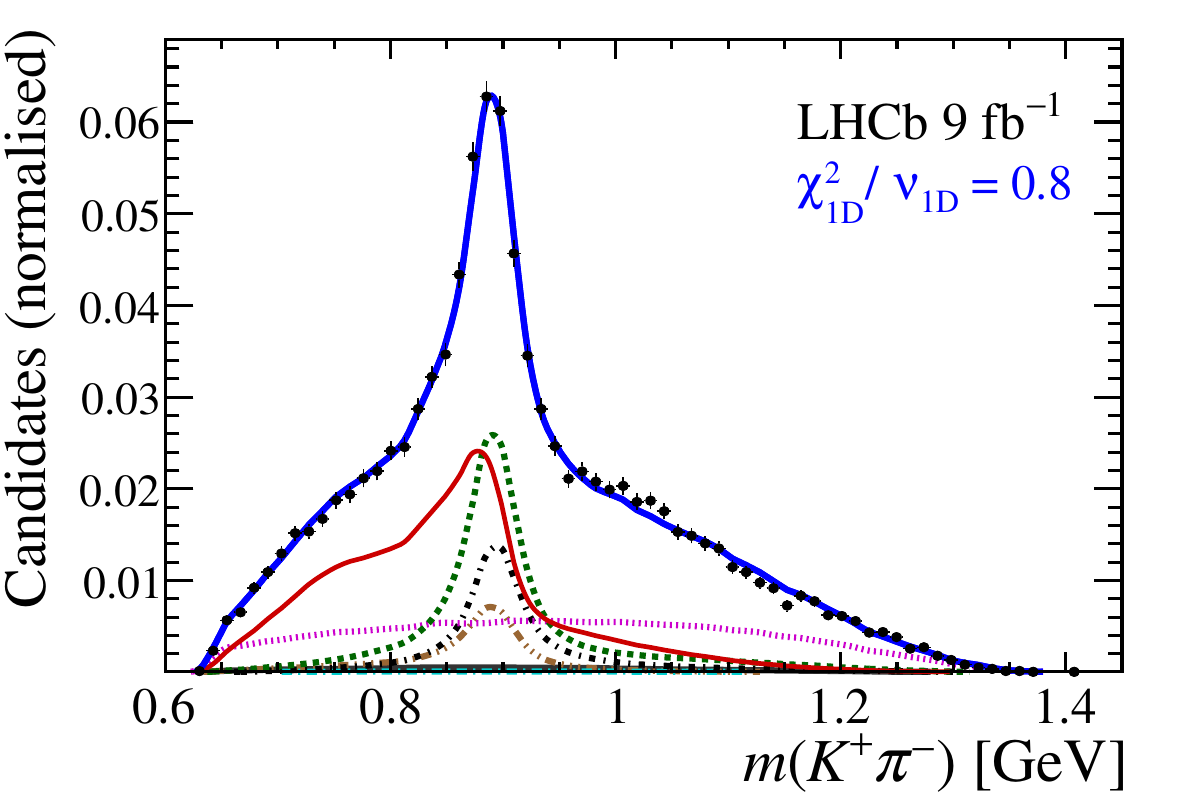}
       	 \includegraphics[width=0.329\textwidth,height=!]{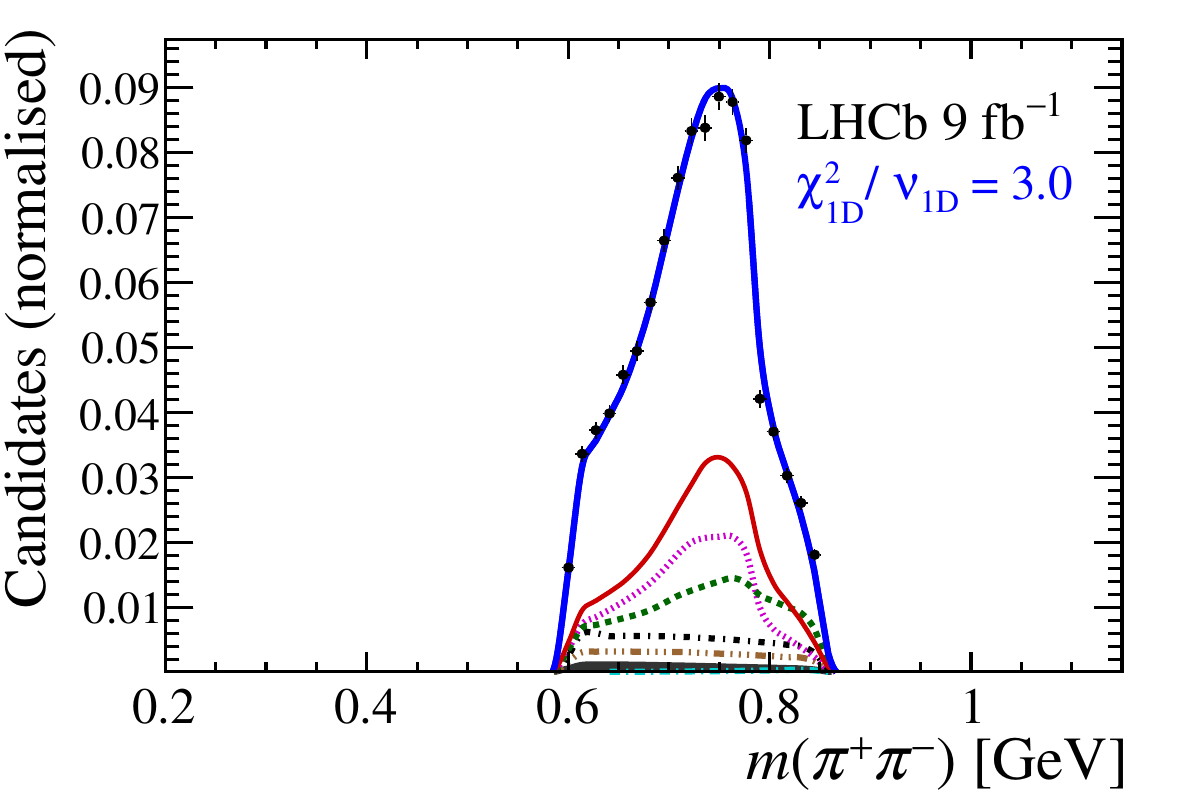}

       	 \includegraphics[width=0.329\textwidth,height=!]{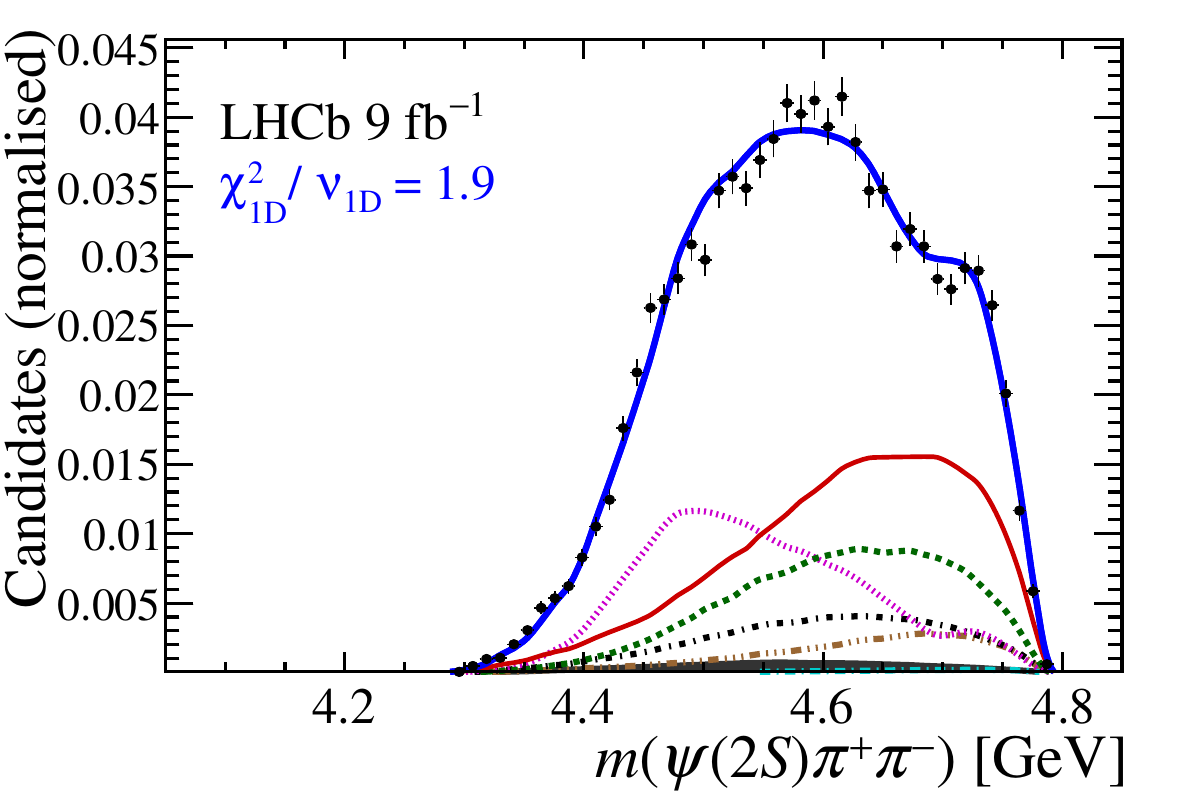}
       	 \includegraphics[width=0.329\textwidth,height=!]{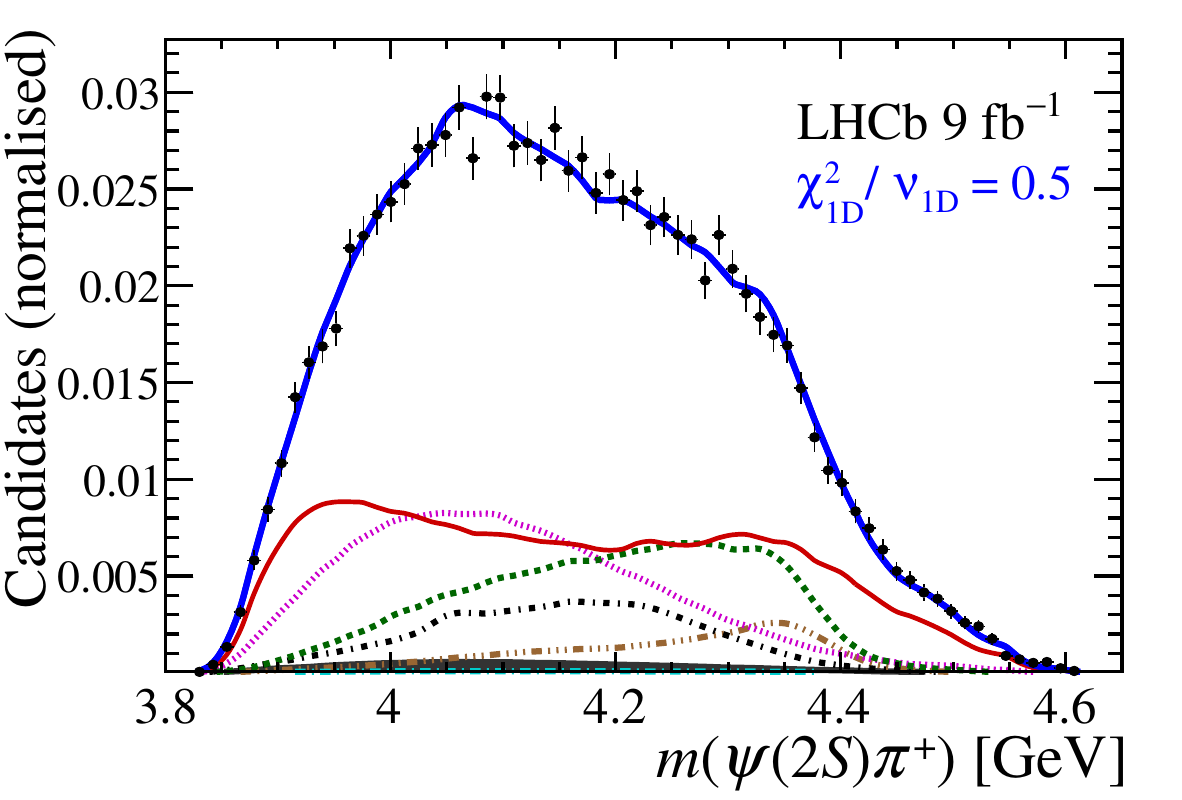}
       	 \includegraphics[width=0.329\textwidth,height=!]{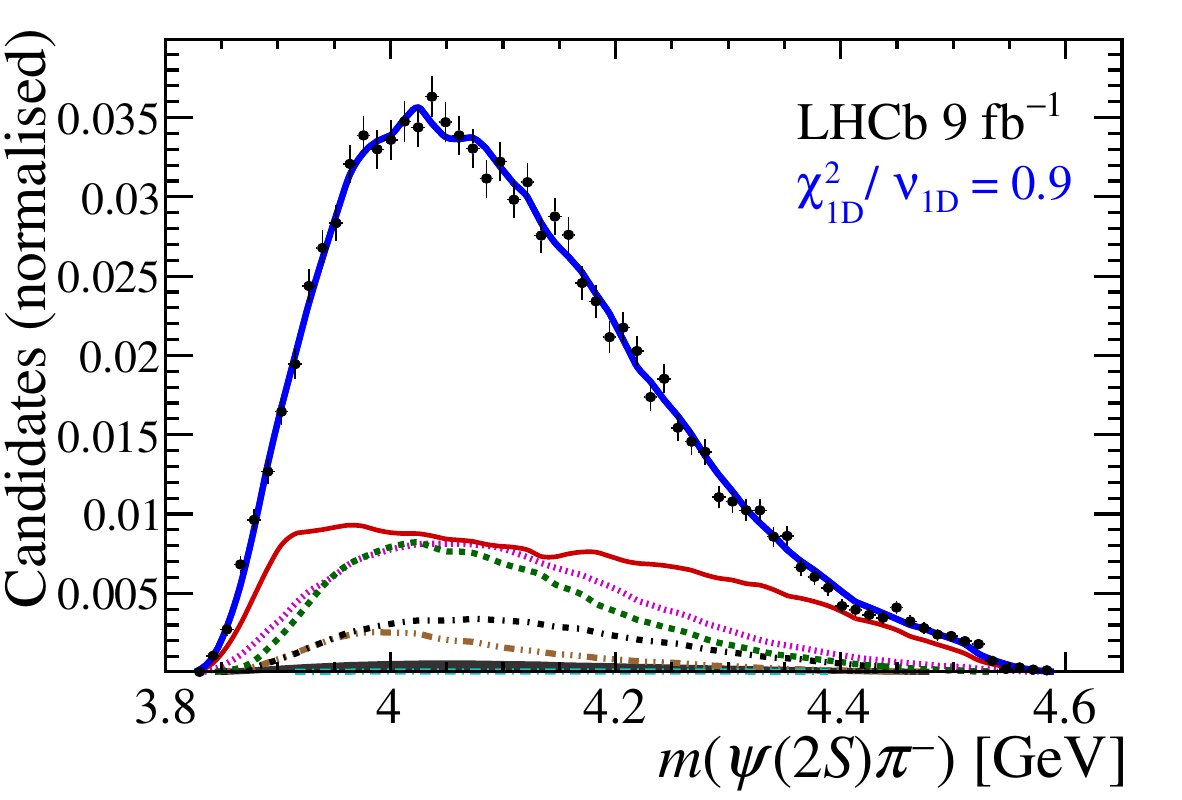}

       	 \includegraphics[width=0.329\textwidth,height=!]{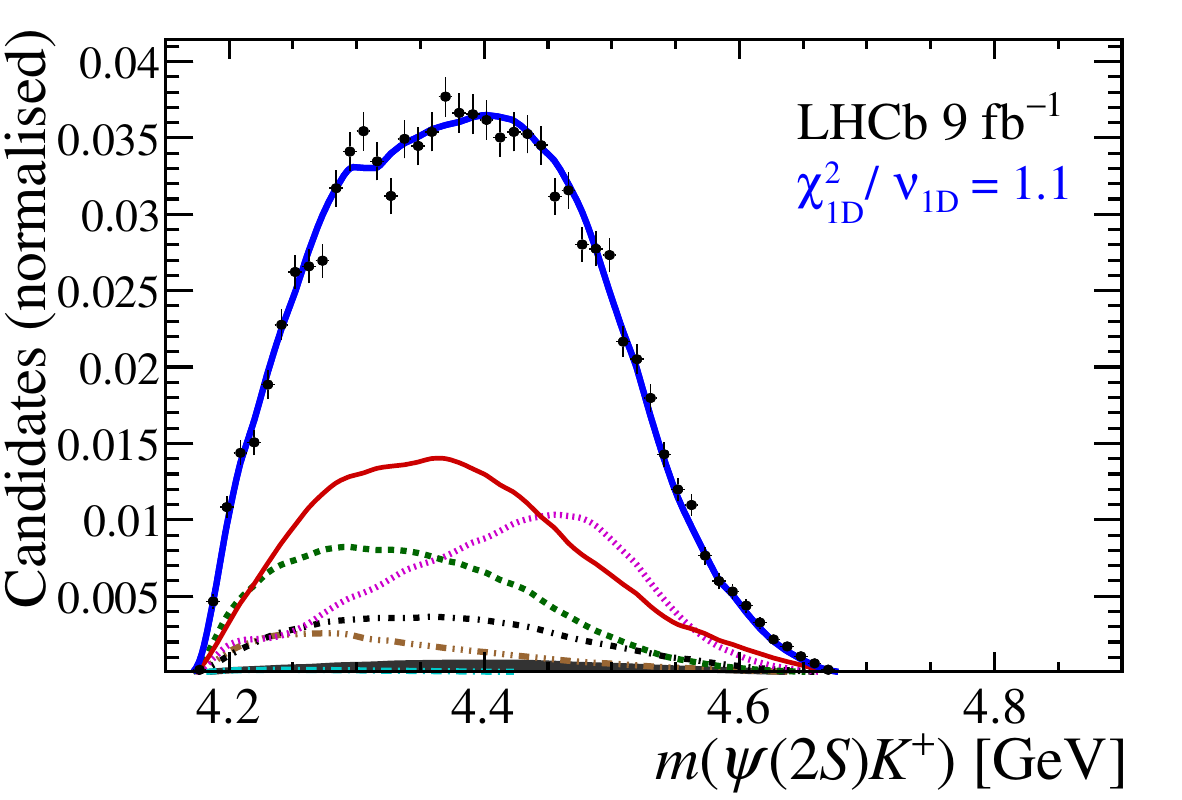}
       	 \includegraphics[width=0.329\textwidth,height=!]{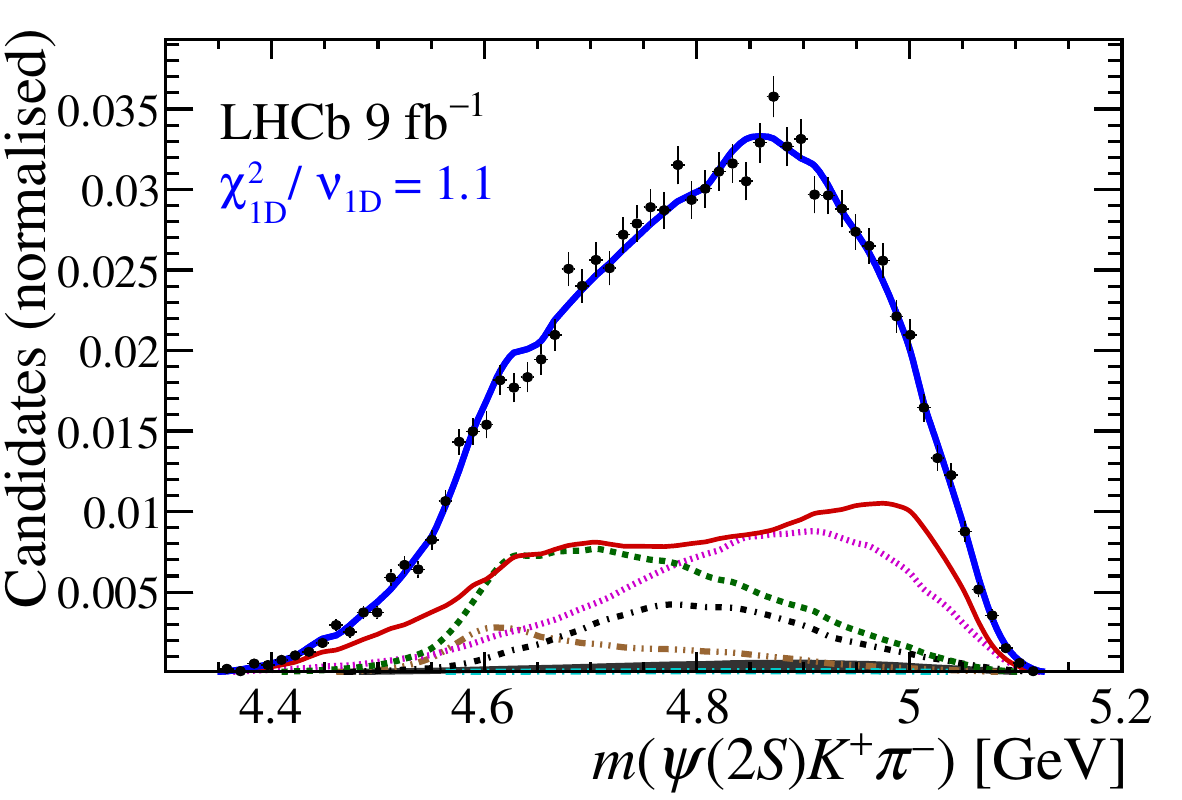}
       	 \includegraphics[width=0.329\textwidth,height=!]{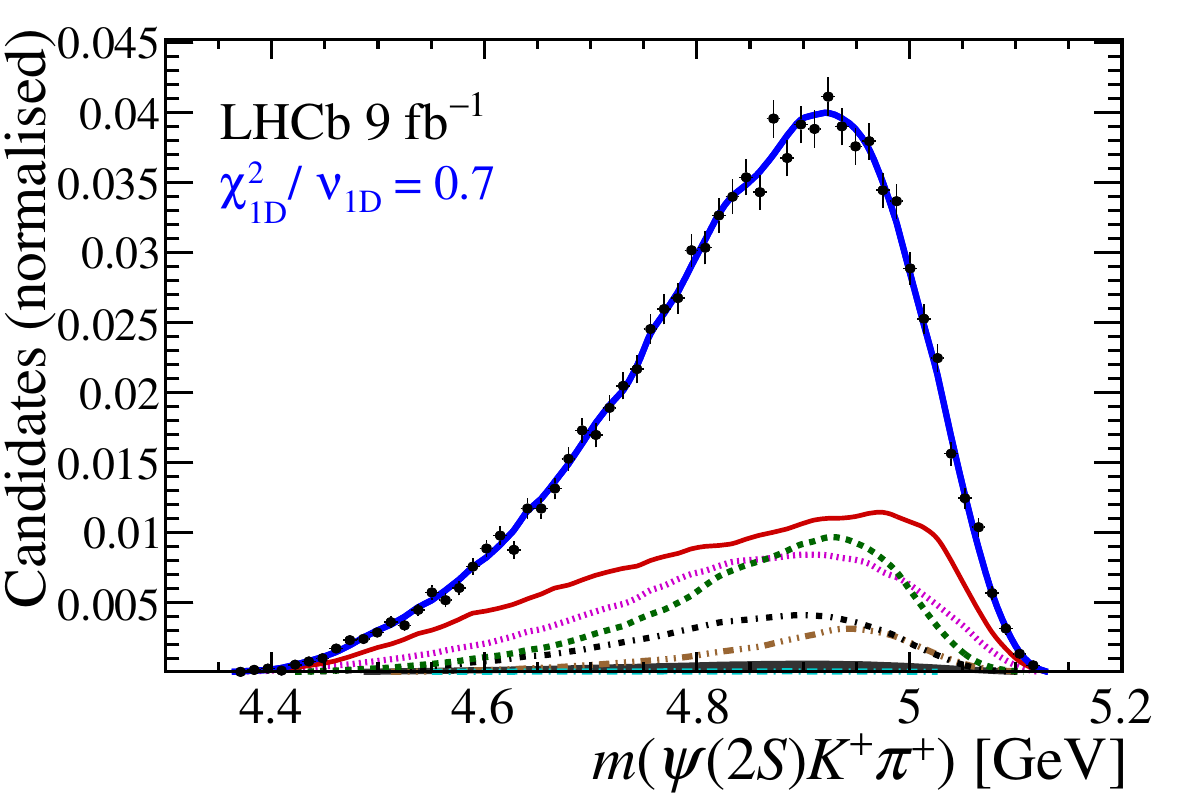}
              
       	 \includegraphics[width=0.329\textwidth,height=!]{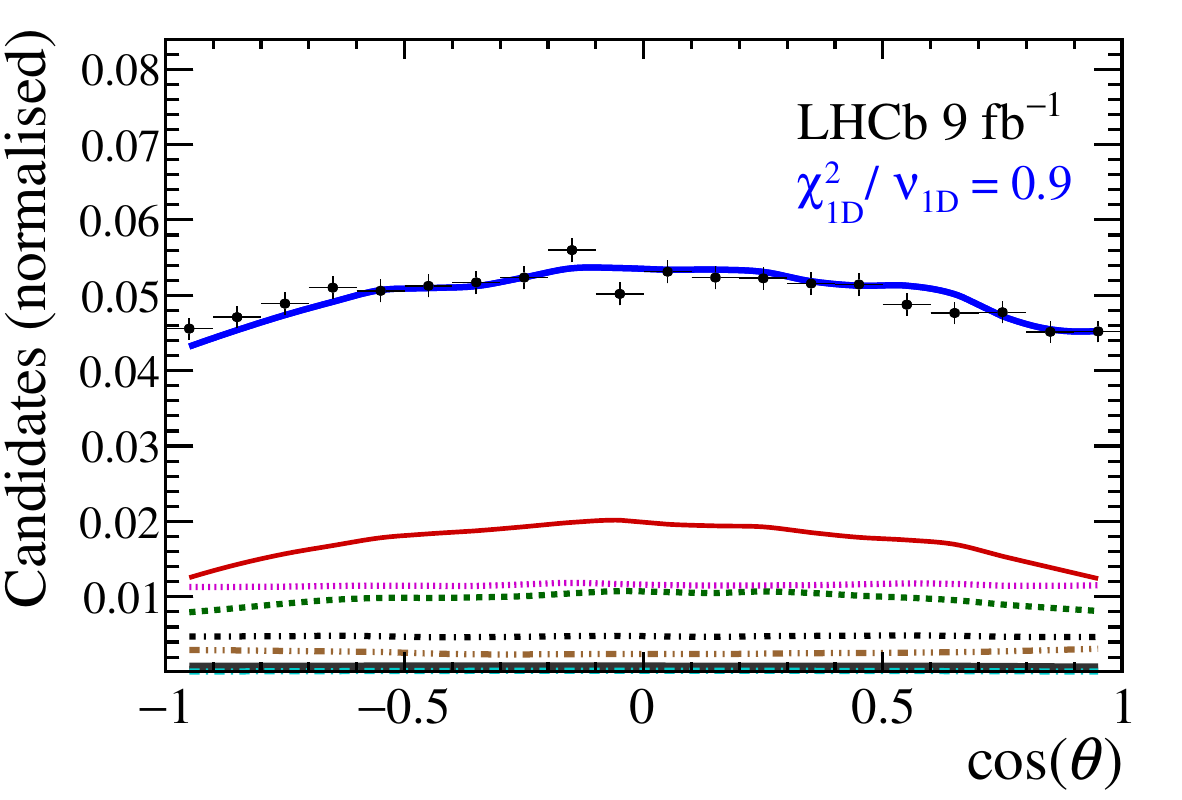}
       	 \includegraphics[width=0.329\textwidth,height=!]{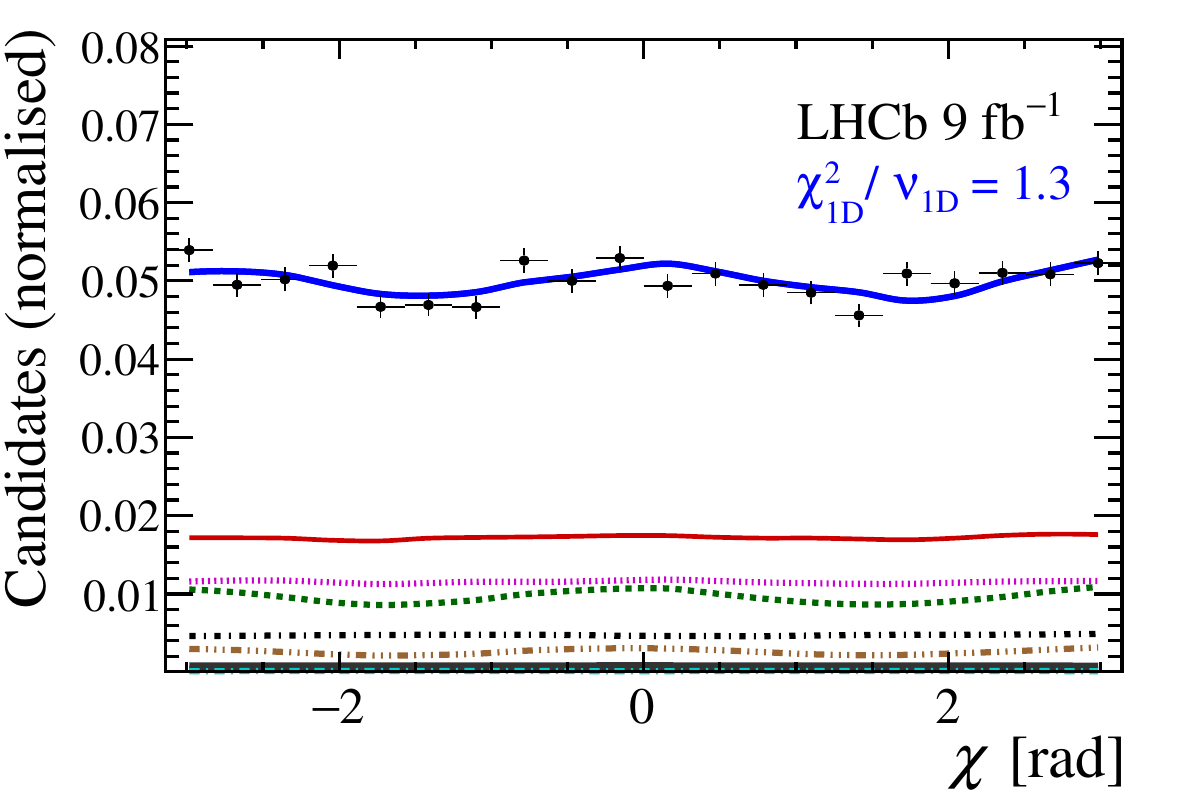}
       	 \includegraphics[width=0.329\textwidth,height=!]{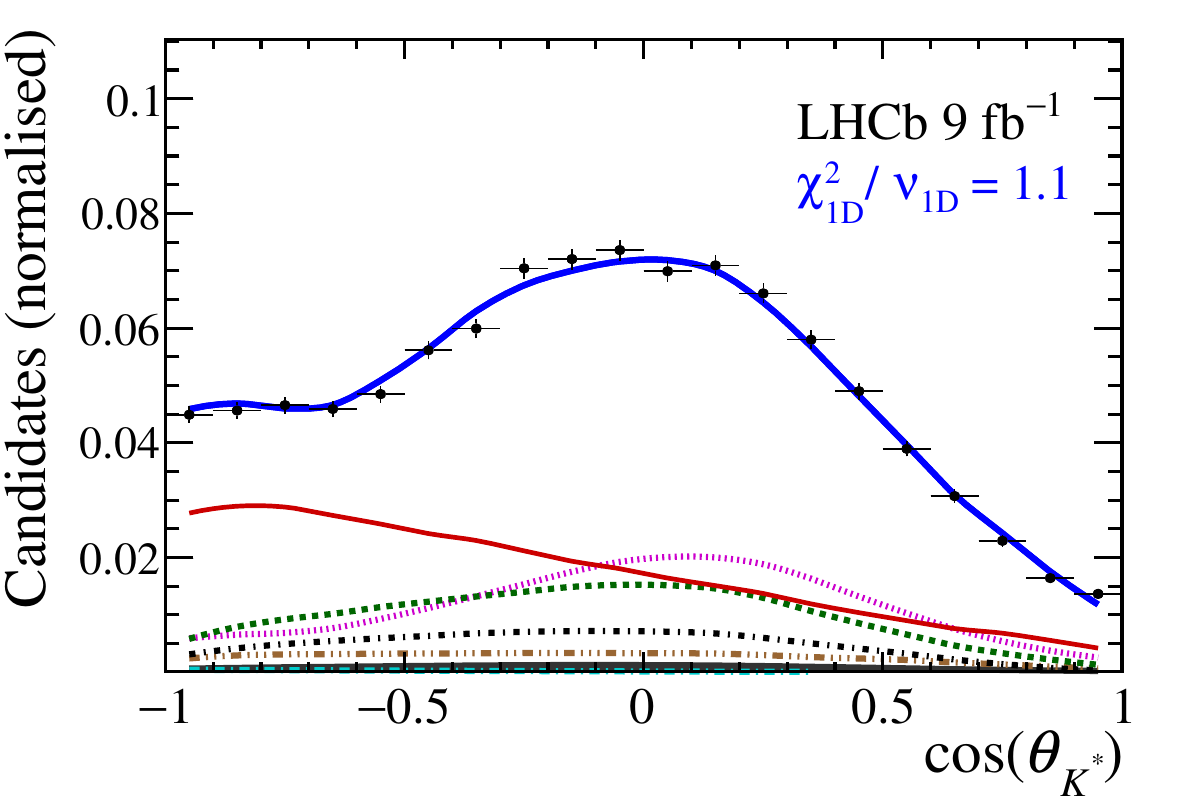}

        \centering
              \includegraphics[width=0.25\textwidth,height=!]{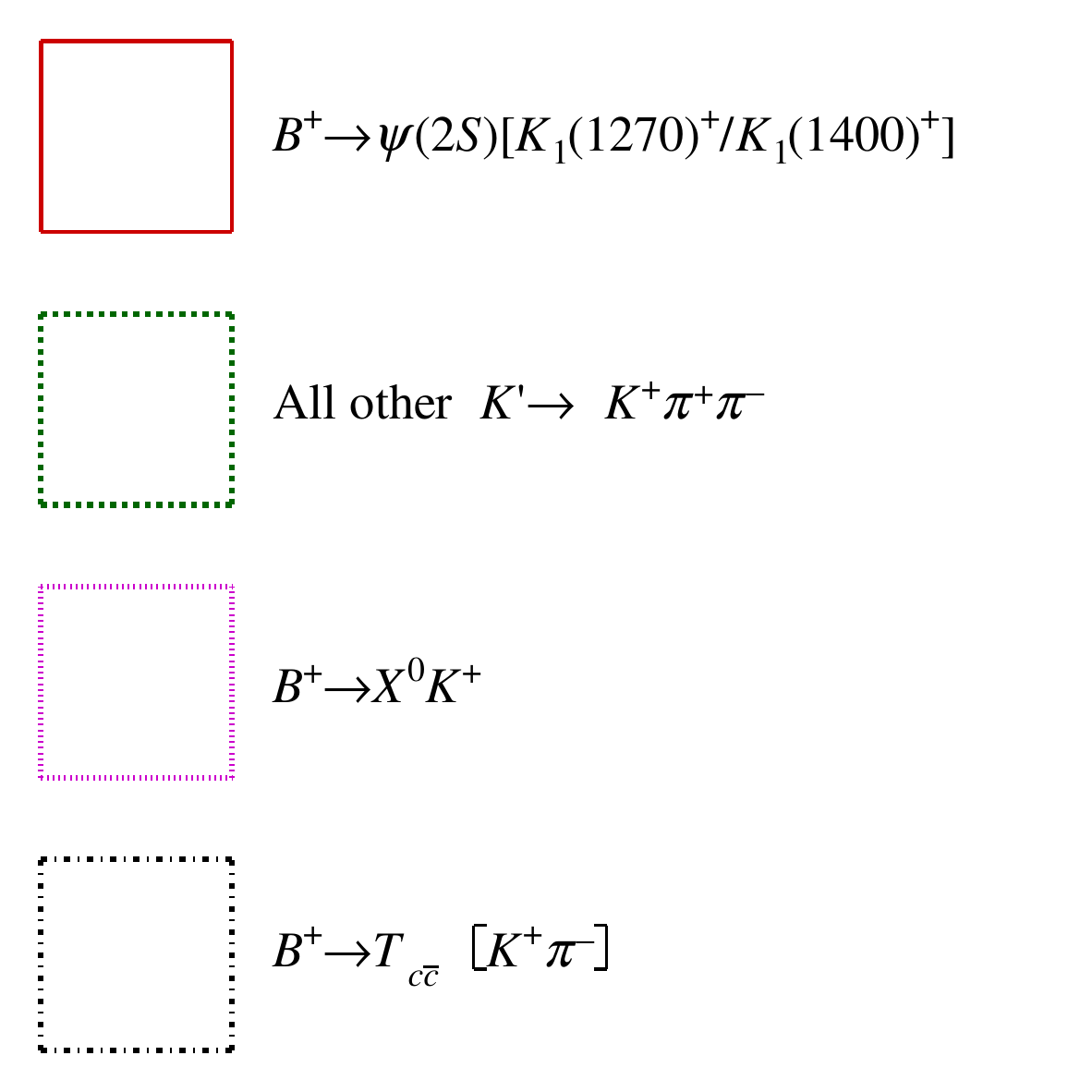}
              \includegraphics[width=0.25\textwidth,height=!]{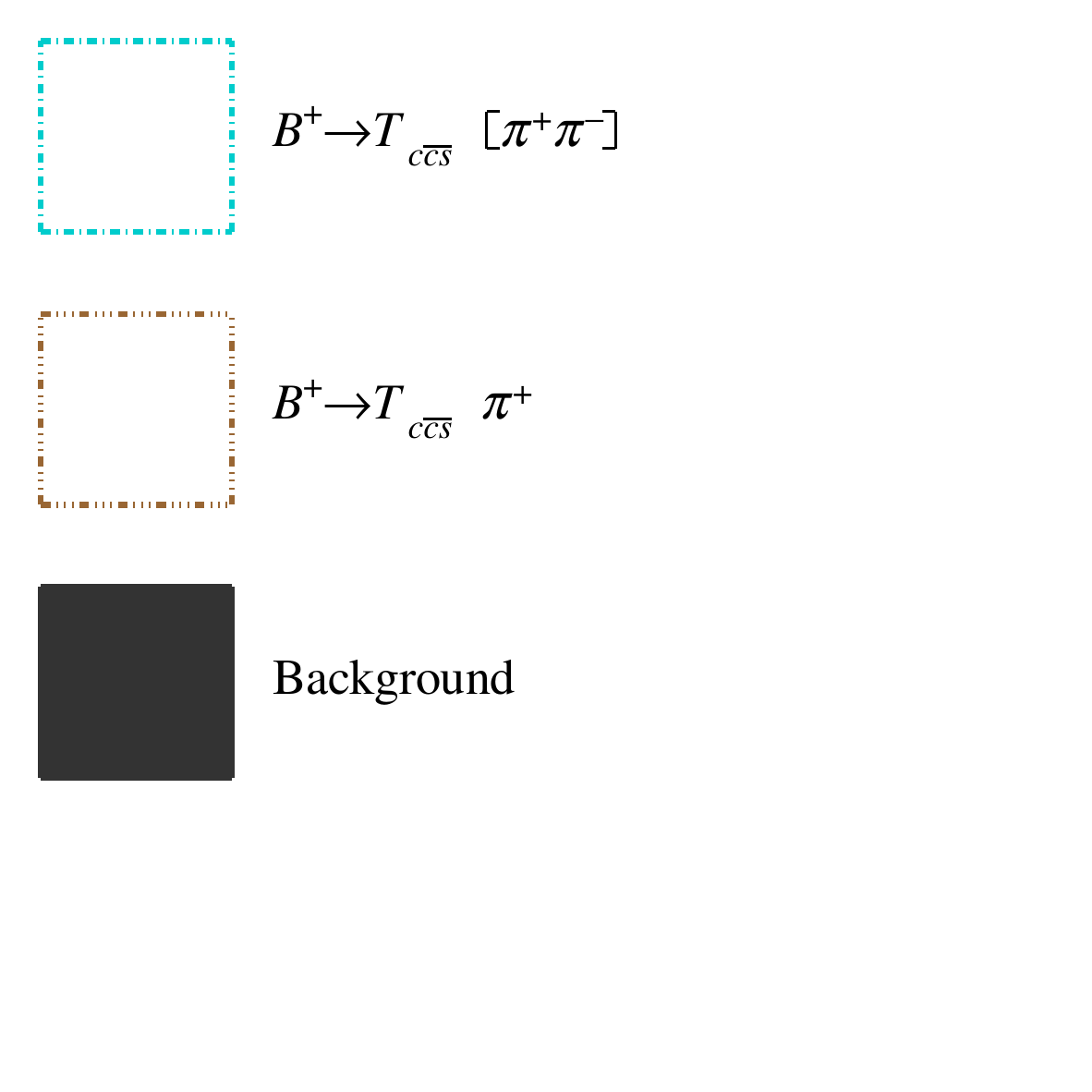}
              
	\caption{Phase-space projections of $\signal$ candidates in the $m(\pip\pim)$ region around the $\rho(770)$ resonance (points with error bars) and fit projections (solid, blue line) for the \textit{baseline} model.  The displayed $\chi_{\rm 1D}^2/\nu_{\rm 1D}$ value on each projection gives the sum of squared normalised residuals divided by the number of bins minus one.
	The multi-dimensional $\chi^2$ value is $\chi^2/\nu= 1.19$ with $\nu=737$.}

         \label{fig:fitBest2}

\end{figure}

\begin{figure}[h]
       	 \includegraphics[width=0.329\textwidth,height=!]{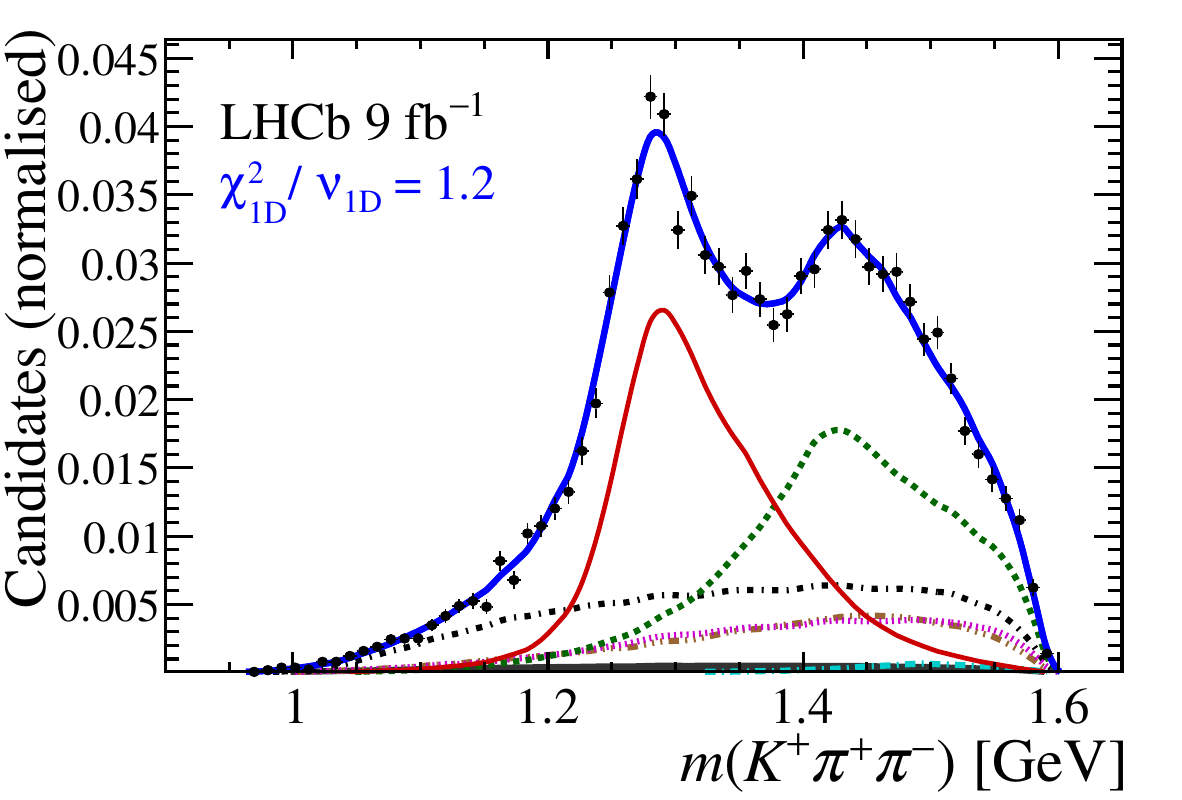}
       	 \includegraphics[width=0.329\textwidth,height=!]{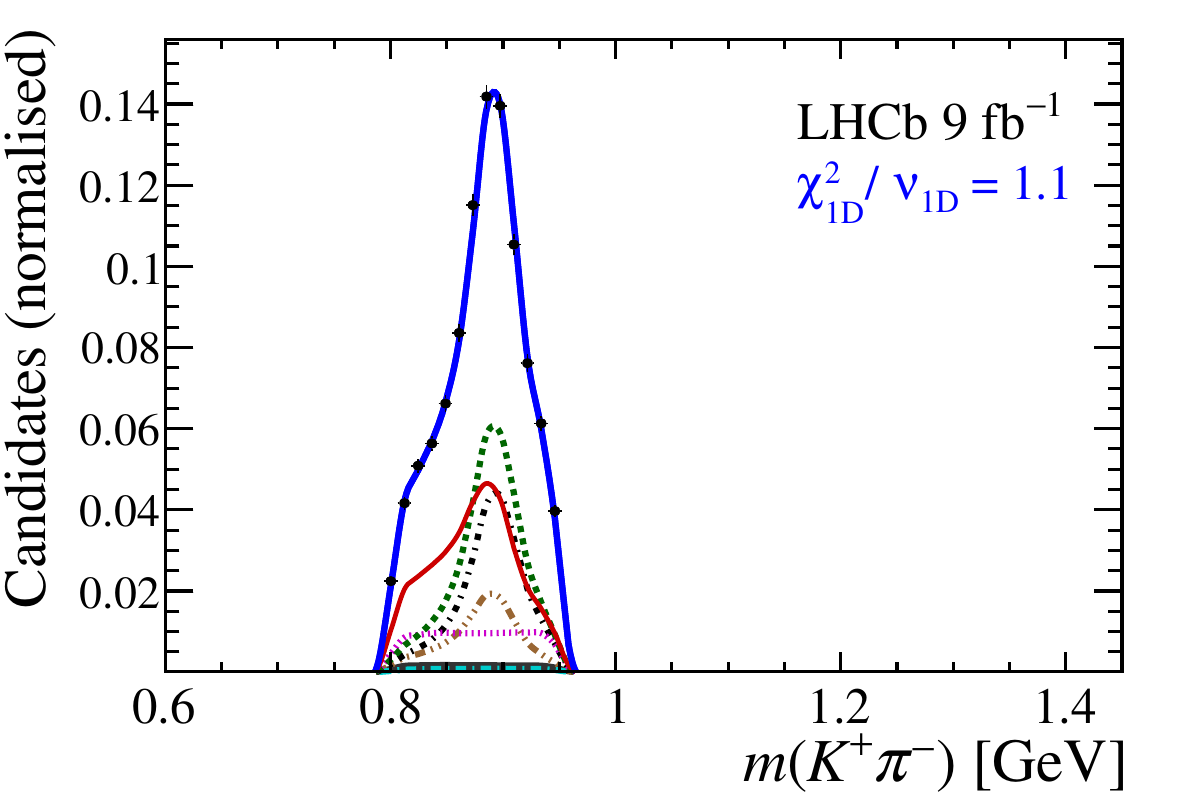}
       	 \includegraphics[width=0.329\textwidth,height=!]{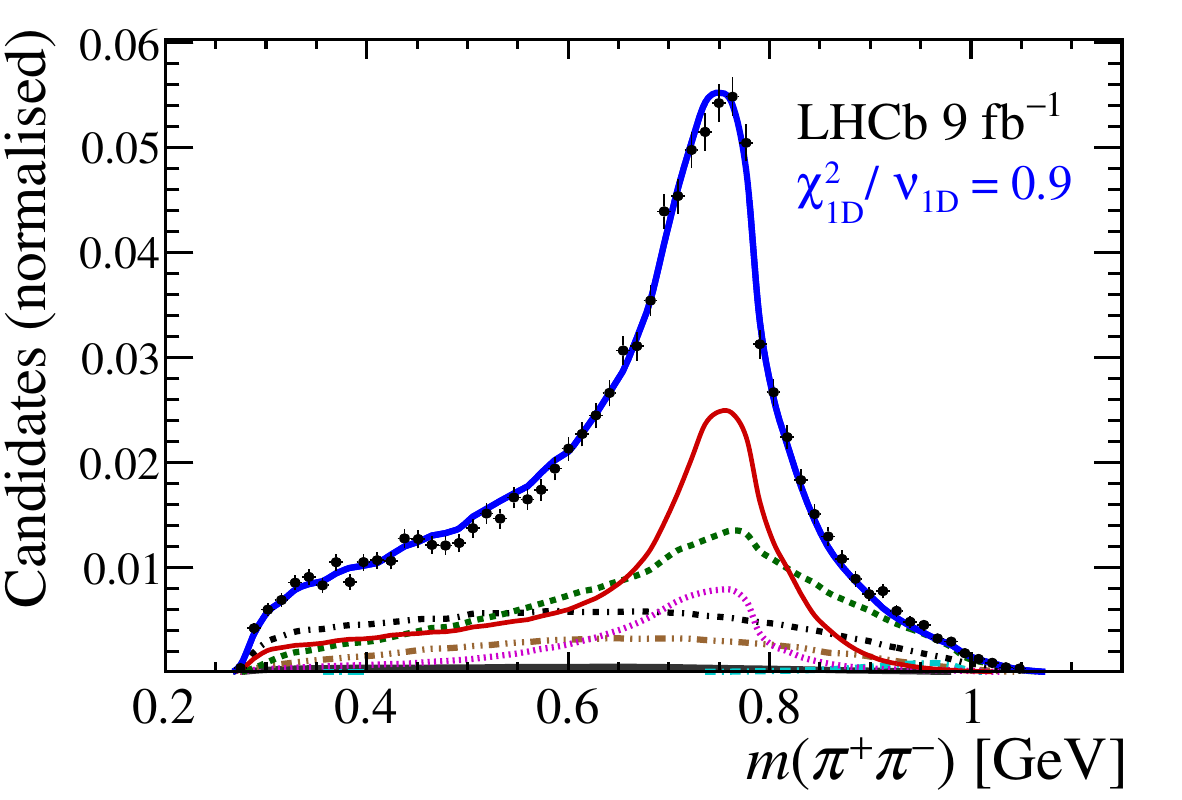}

       	 \includegraphics[width=0.329\textwidth,height=!]{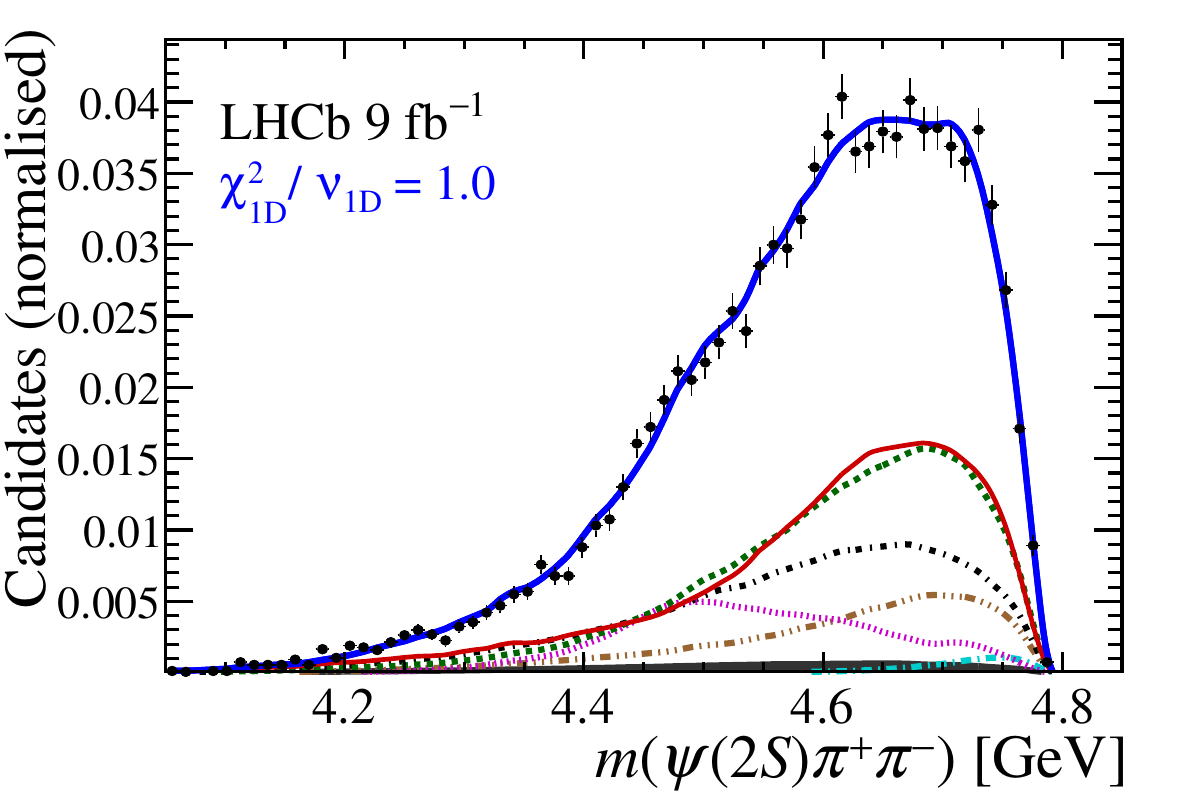}
       	 \includegraphics[width=0.329\textwidth,height=!]{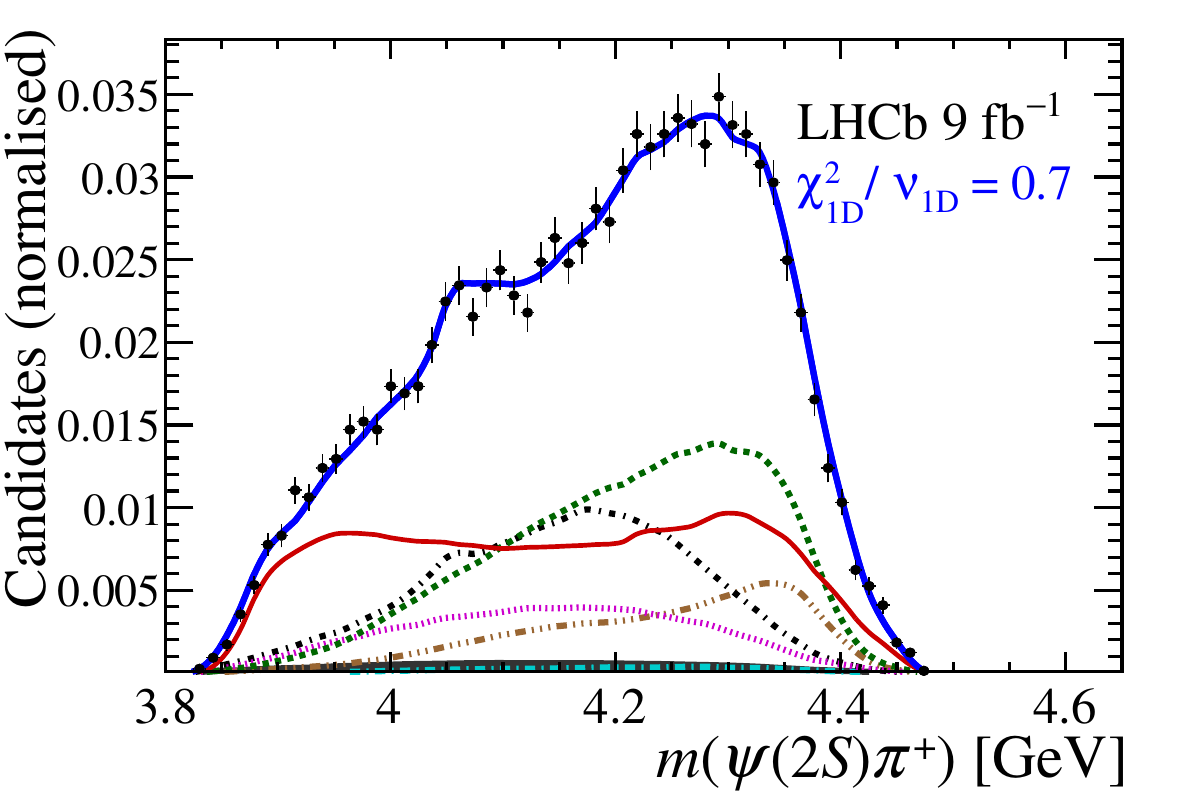}
       	 \includegraphics[width=0.329\textwidth,height=!]{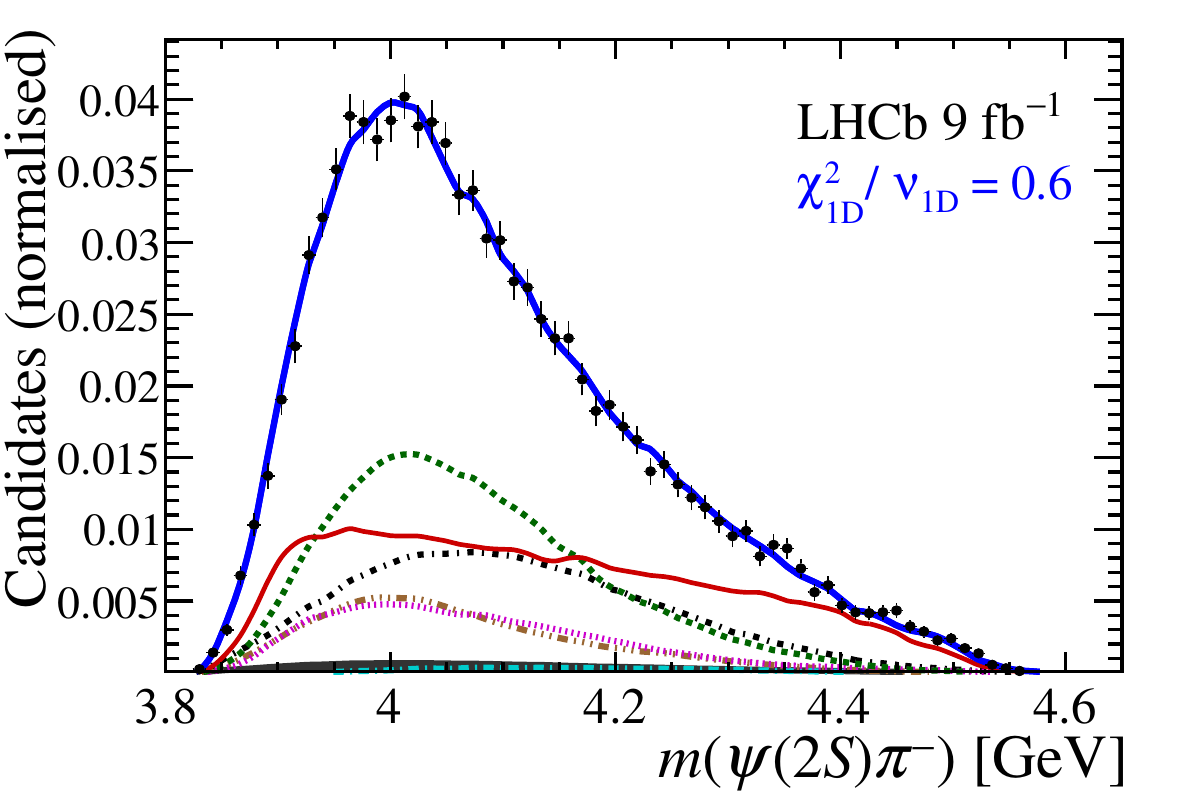}

       	 \includegraphics[width=0.329\textwidth,height=!]{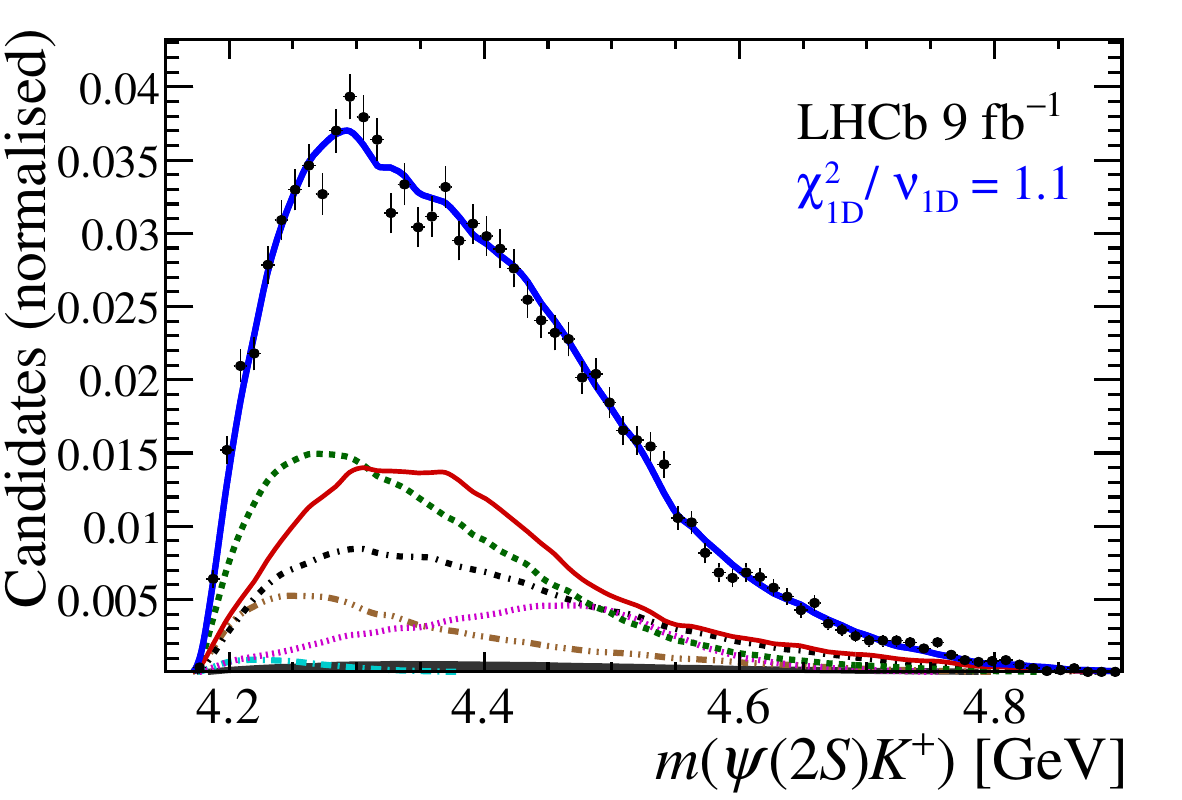}
       	 \includegraphics[width=0.329\textwidth,height=!]{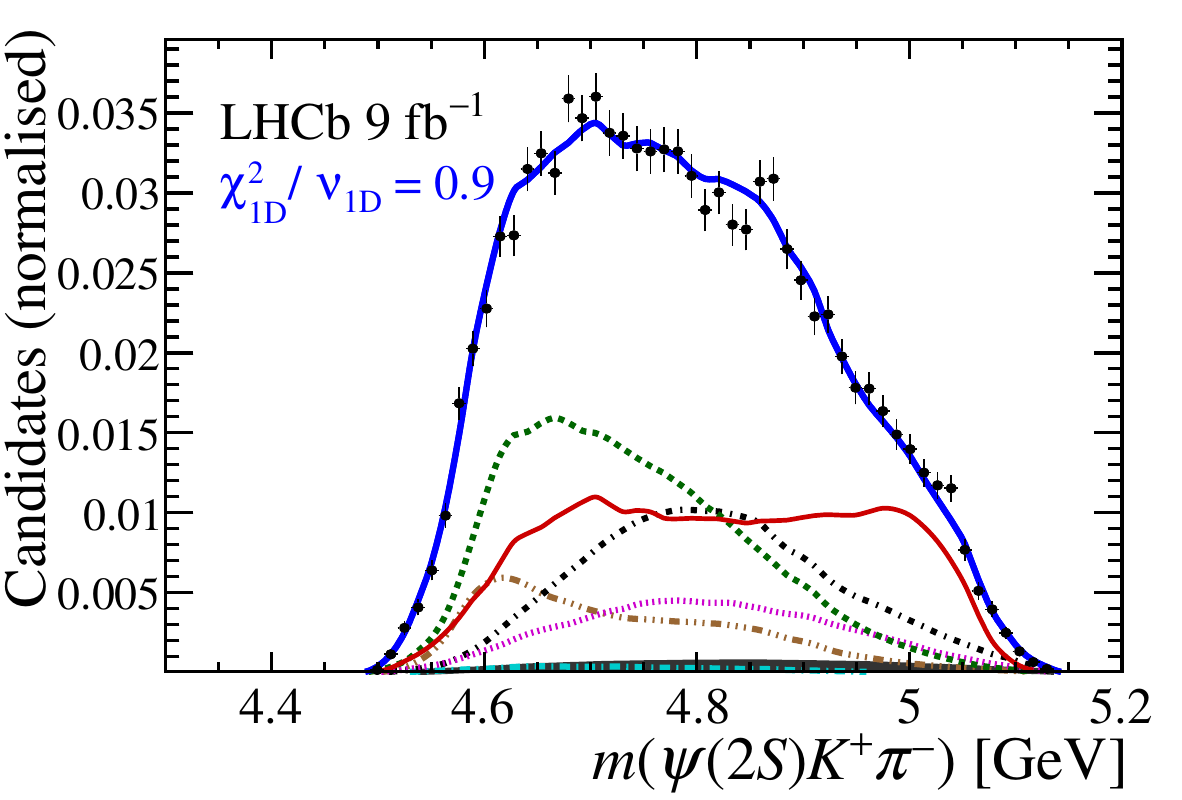}
       	 \includegraphics[width=0.329\textwidth,height=!]{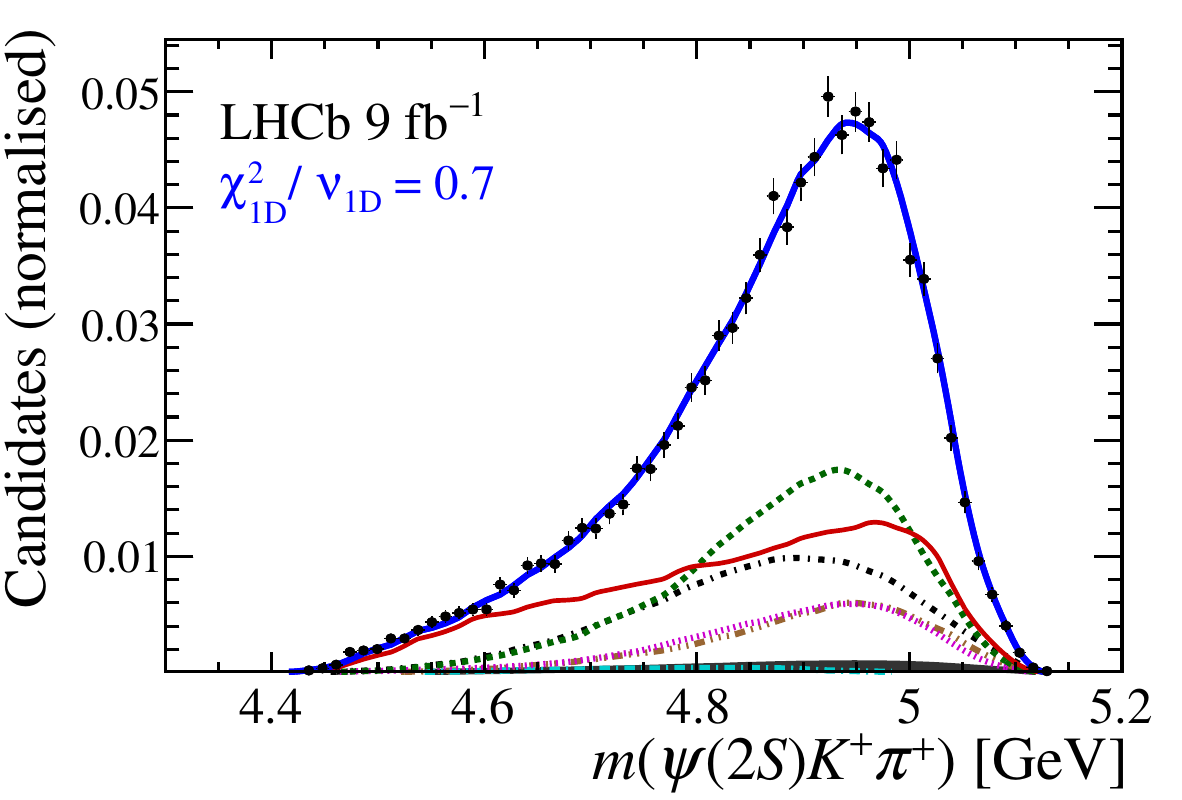}
              
       	 \includegraphics[width=0.329\textwidth,height=!]{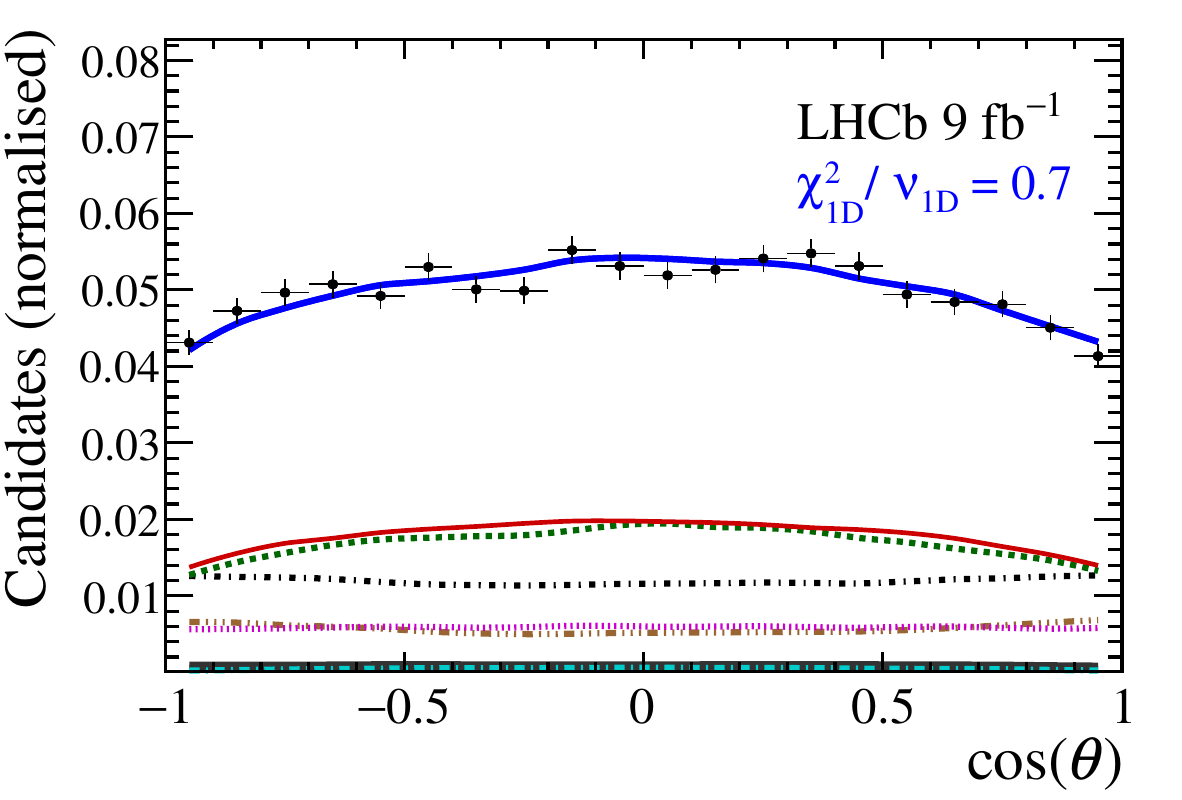}
       	 \includegraphics[width=0.329\textwidth,height=!]{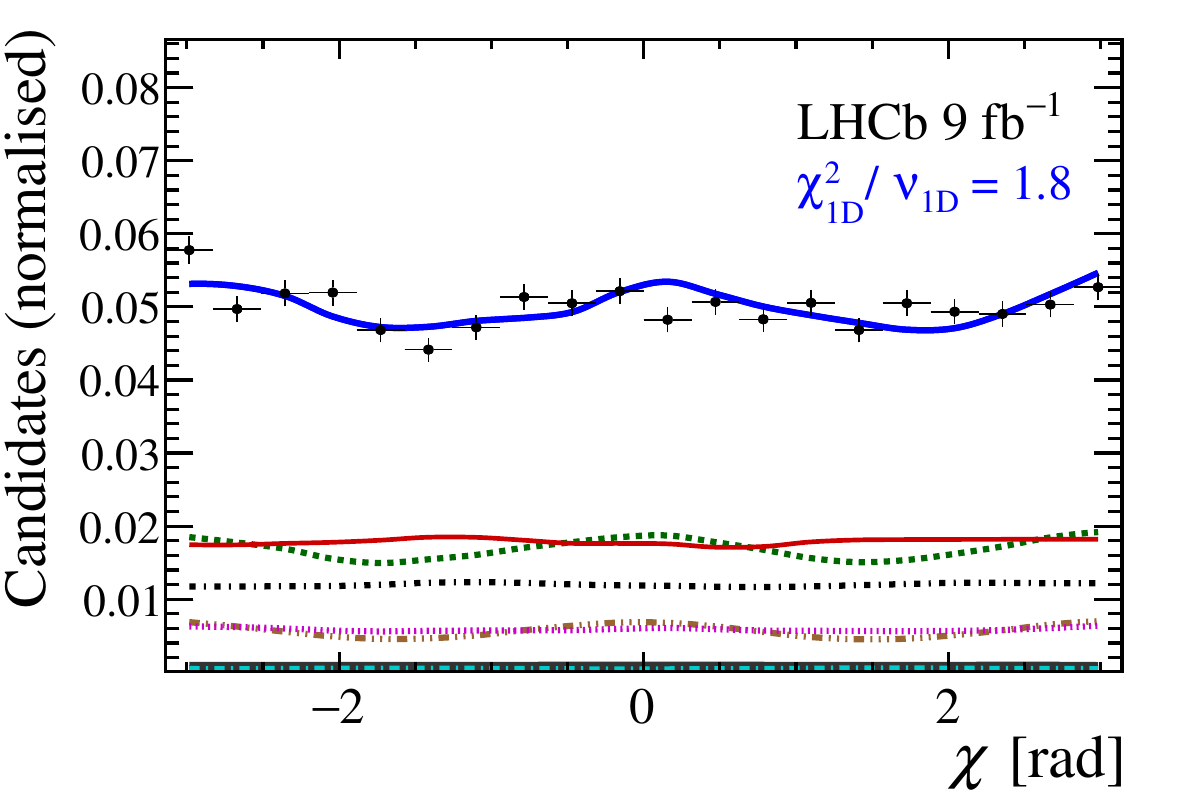}
       	 \includegraphics[width=0.329\textwidth,height=!]{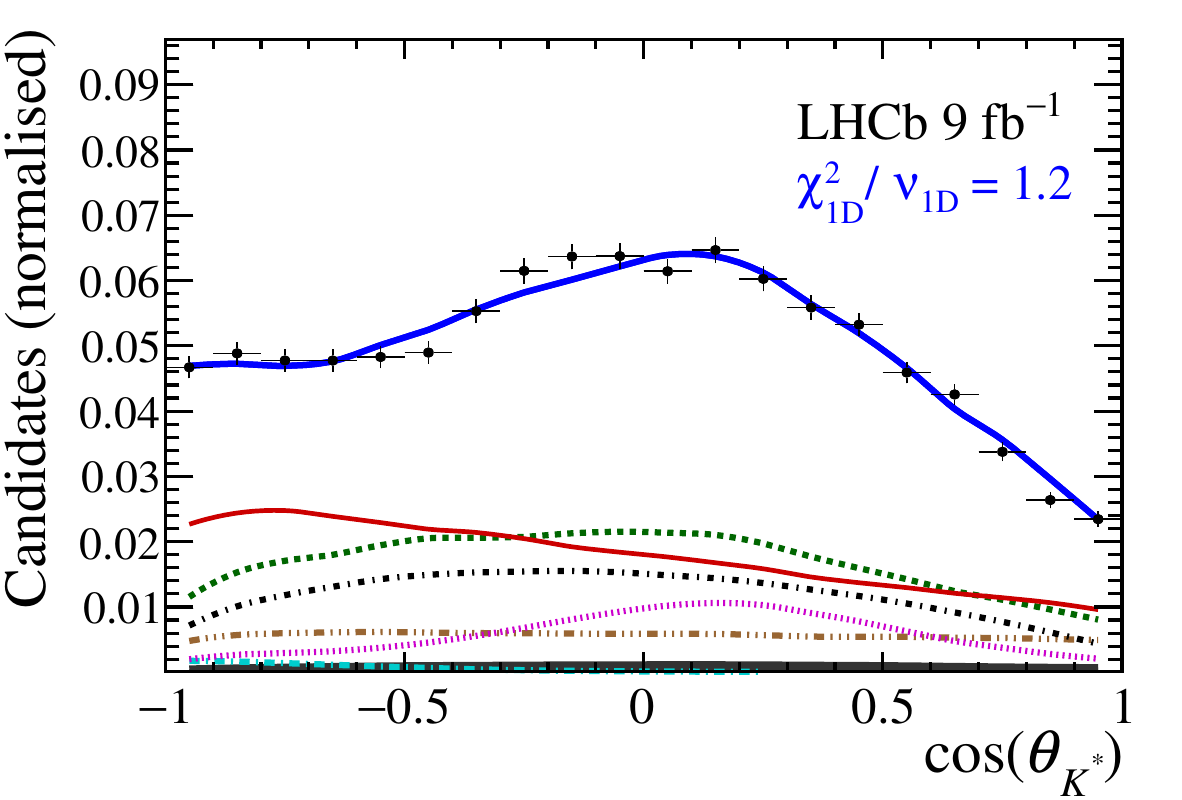}

        \centering
              \includegraphics[width=0.25\textwidth,height=!]{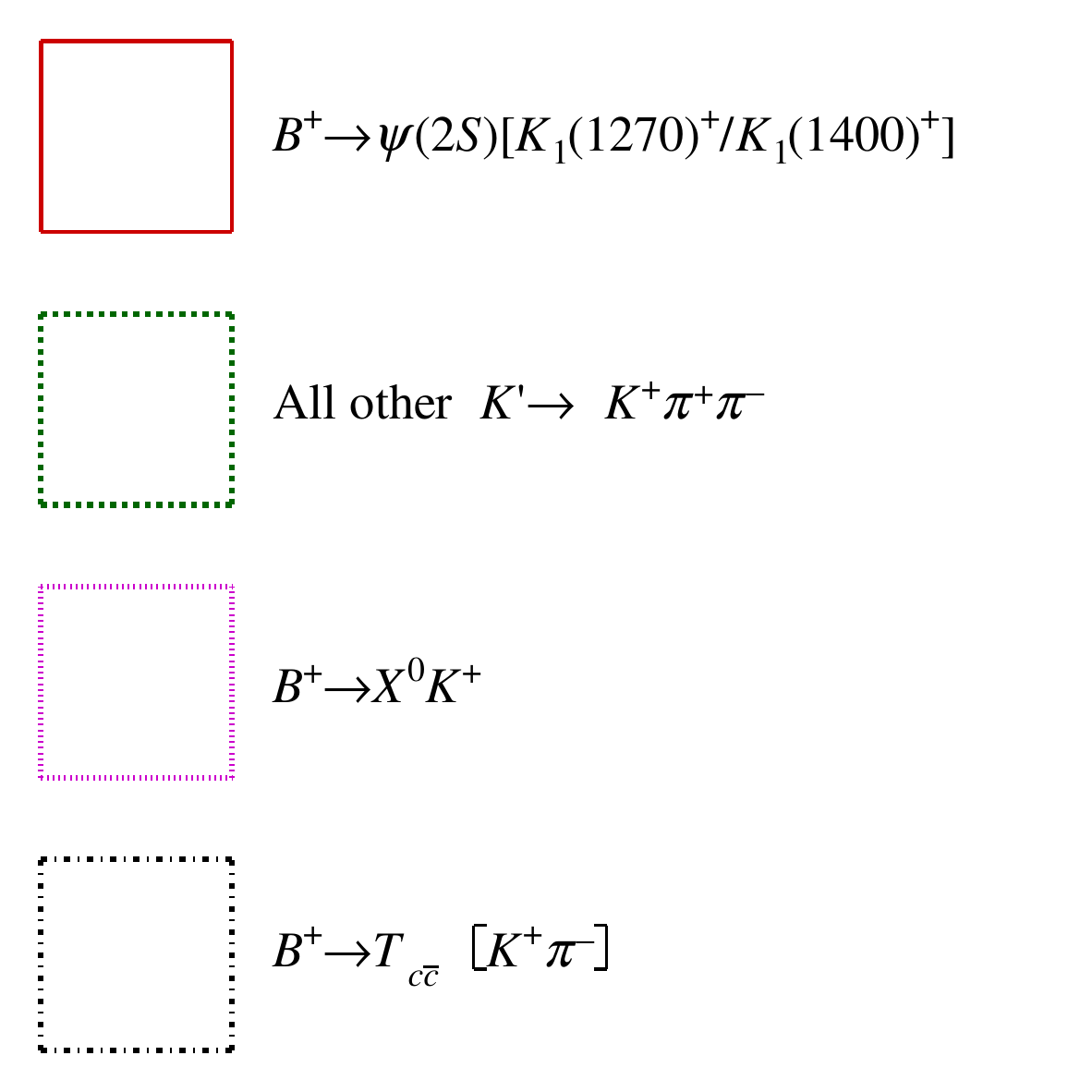}
              \includegraphics[width=0.25\textwidth,height=!]{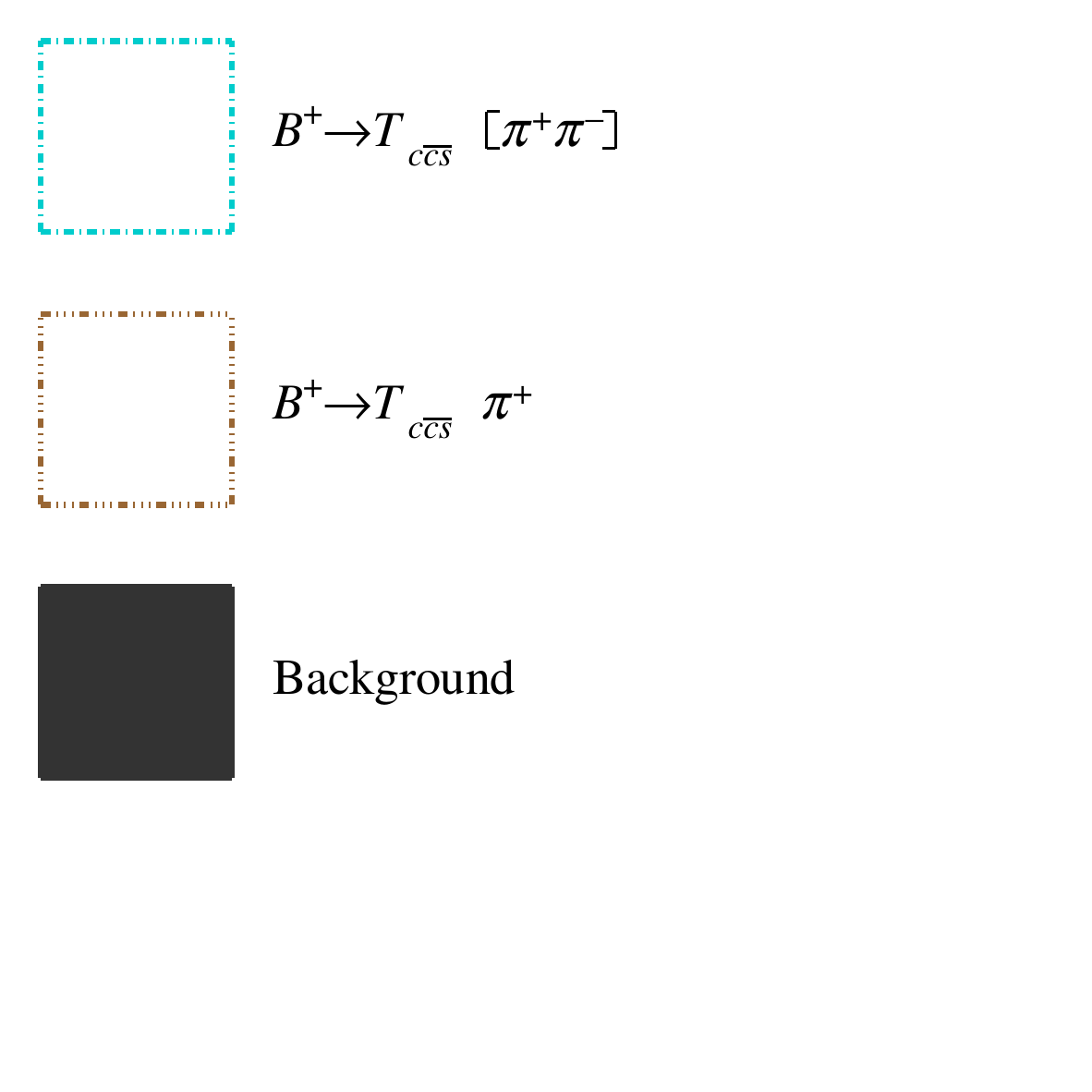}
              
	\caption{Phase-space projections of $\signal$ candidates in the $m(\Kp\pim)$ region around the $K^*(892)$ resonance (points with error bars) and fit projections (solid, blue line) for the \textit{baseline} model.  The displayed $\chi_{\rm 1D}^2/\nu_{\rm 1D}$ value on each projection gives the sum of squared normalised residuals divided by the number of bins minus one. The multi-dimensional $\chi^2$ value is $\chi^2/\nu= 1.09$  with $\nu=577$.}

         \label{fig:fitBest3}

\end{figure}

\begin{figure}[h]
       	 \includegraphics[width=0.329\textwidth,height=!]{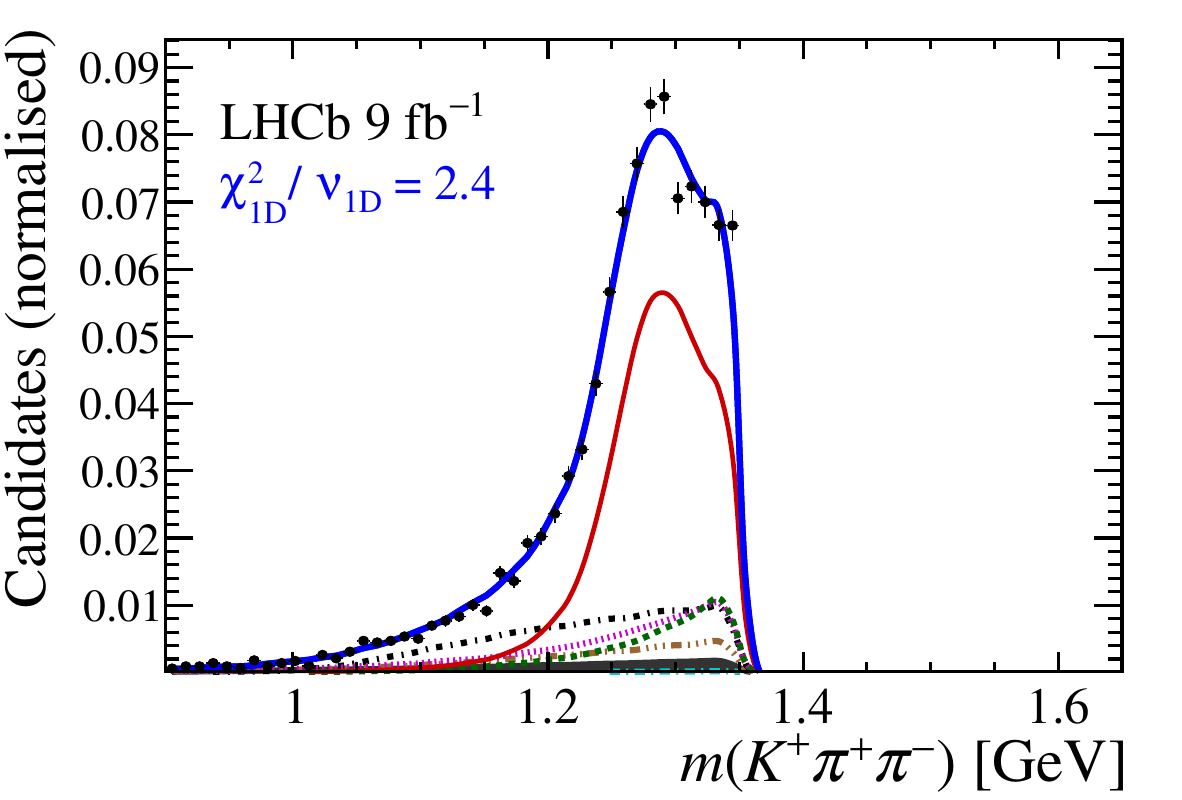}
       	 \includegraphics[width=0.329\textwidth,height=!]{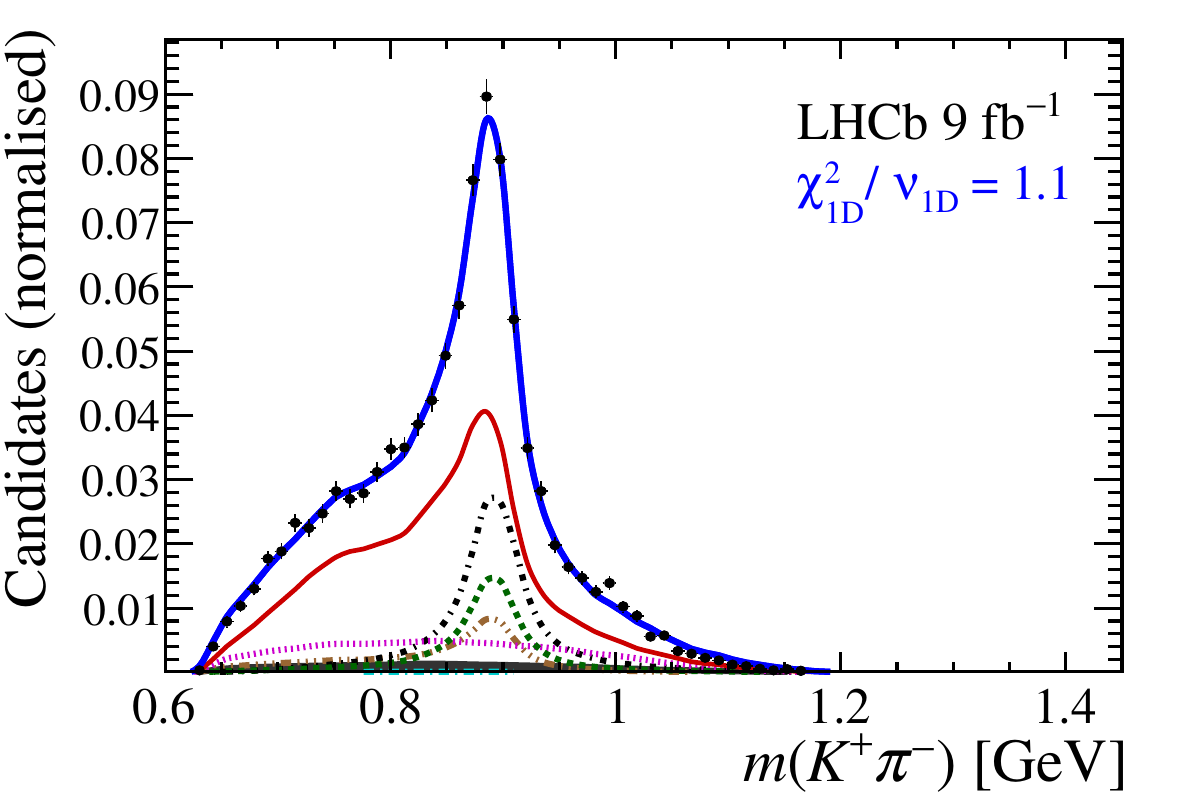}
       	 \includegraphics[width=0.329\textwidth,height=!]{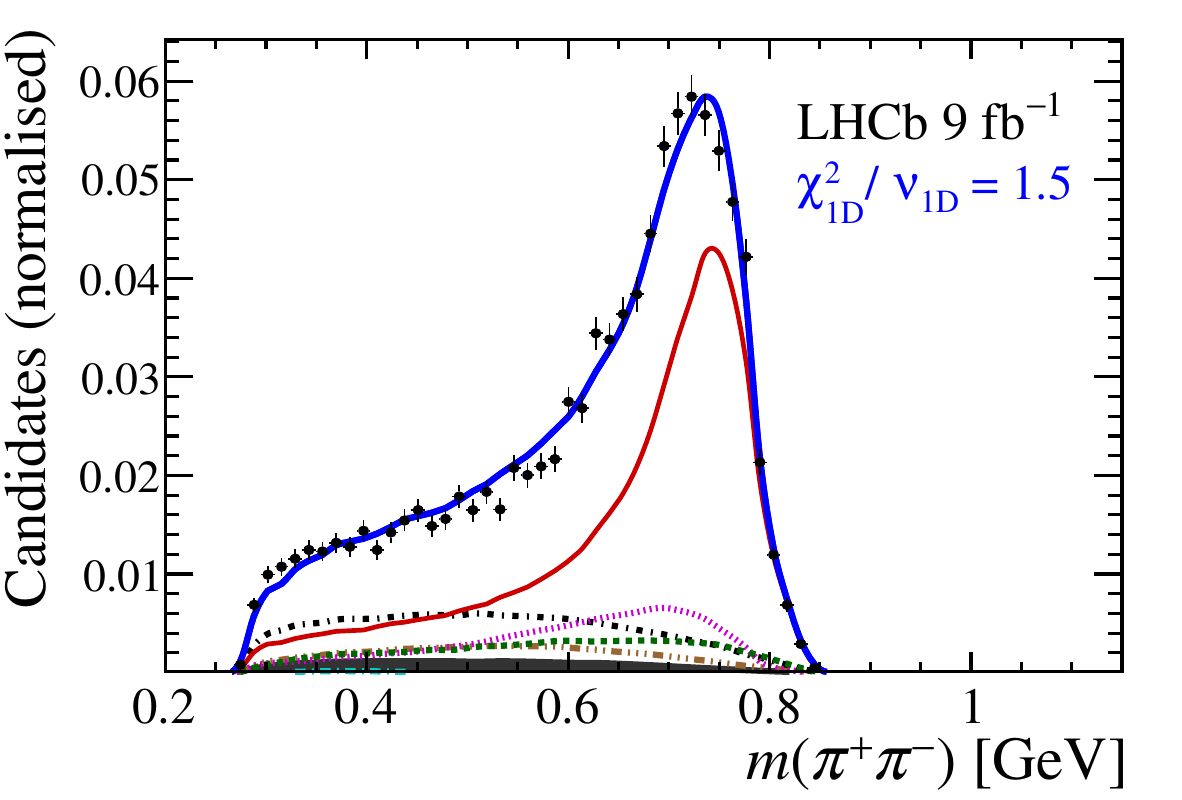}

       	 \includegraphics[width=0.329\textwidth,height=!]{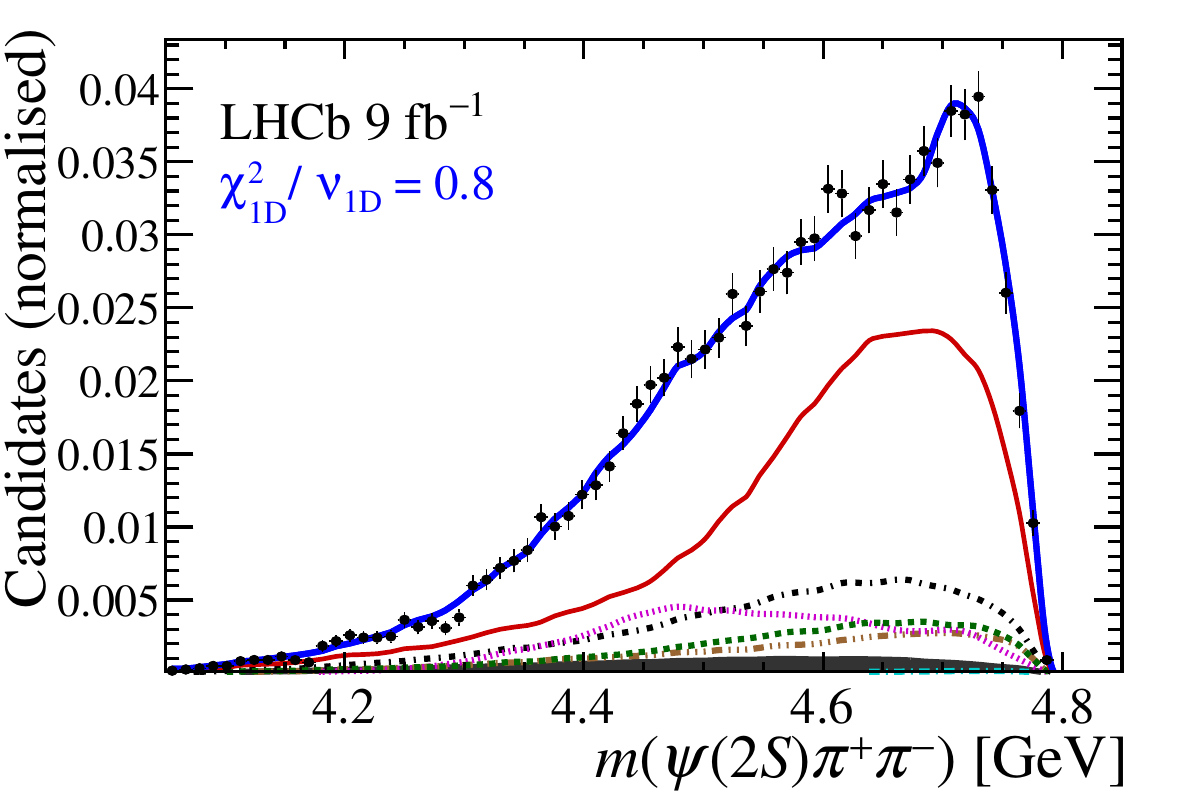}
       	 \includegraphics[width=0.329\textwidth,height=!]{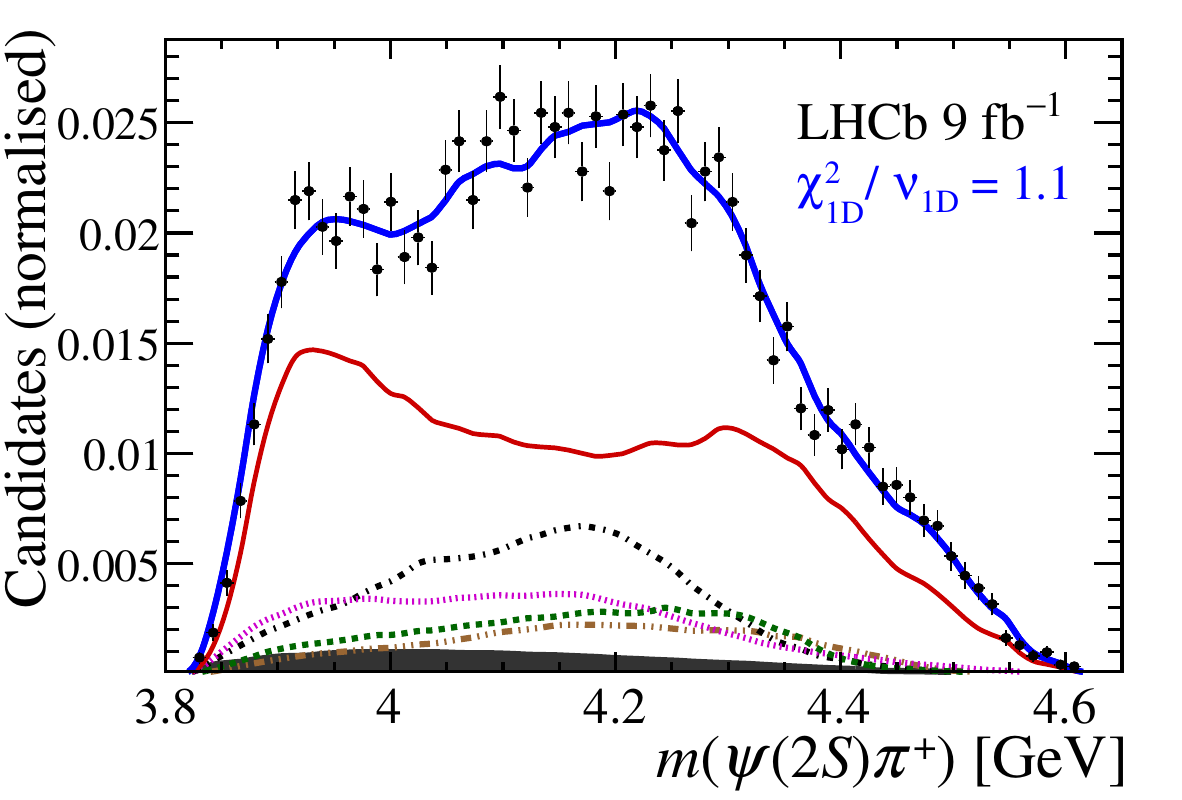}
       	 \includegraphics[width=0.329\textwidth,height=!]{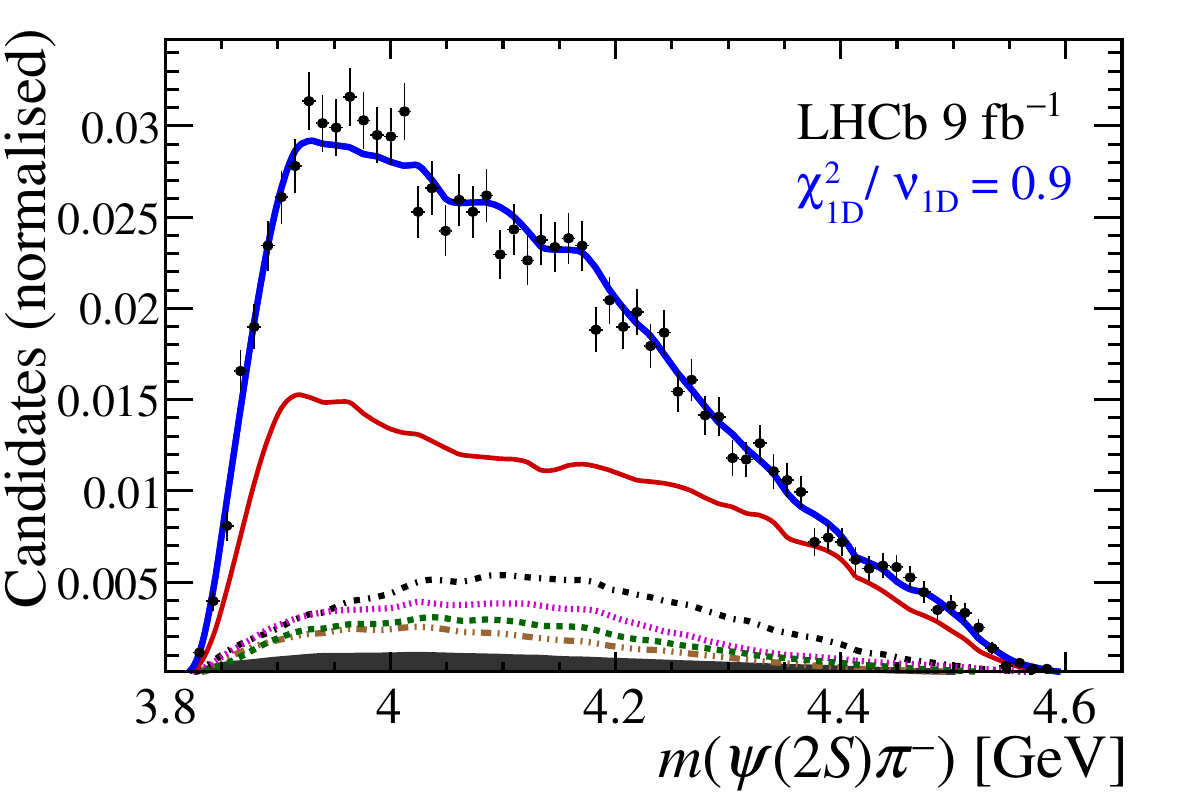}

       	 \includegraphics[width=0.329\textwidth,height=!]{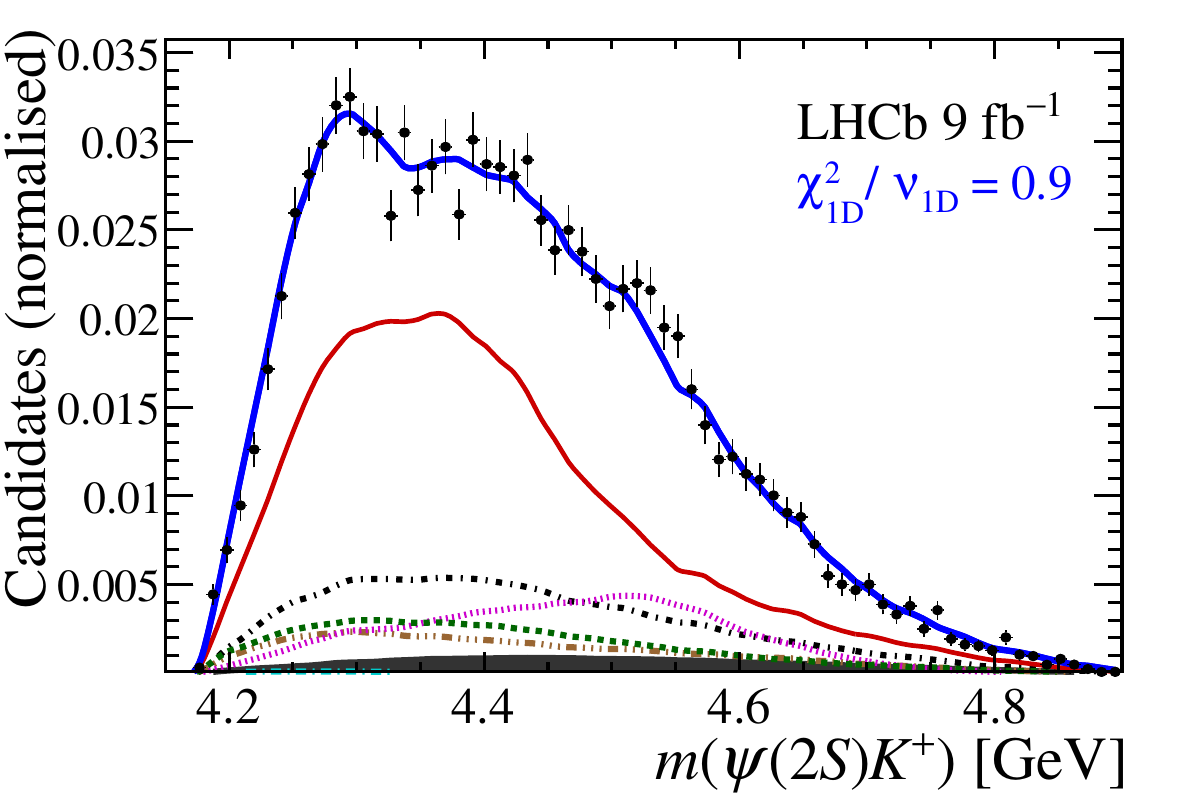}
       	 \includegraphics[width=0.329\textwidth,height=!]{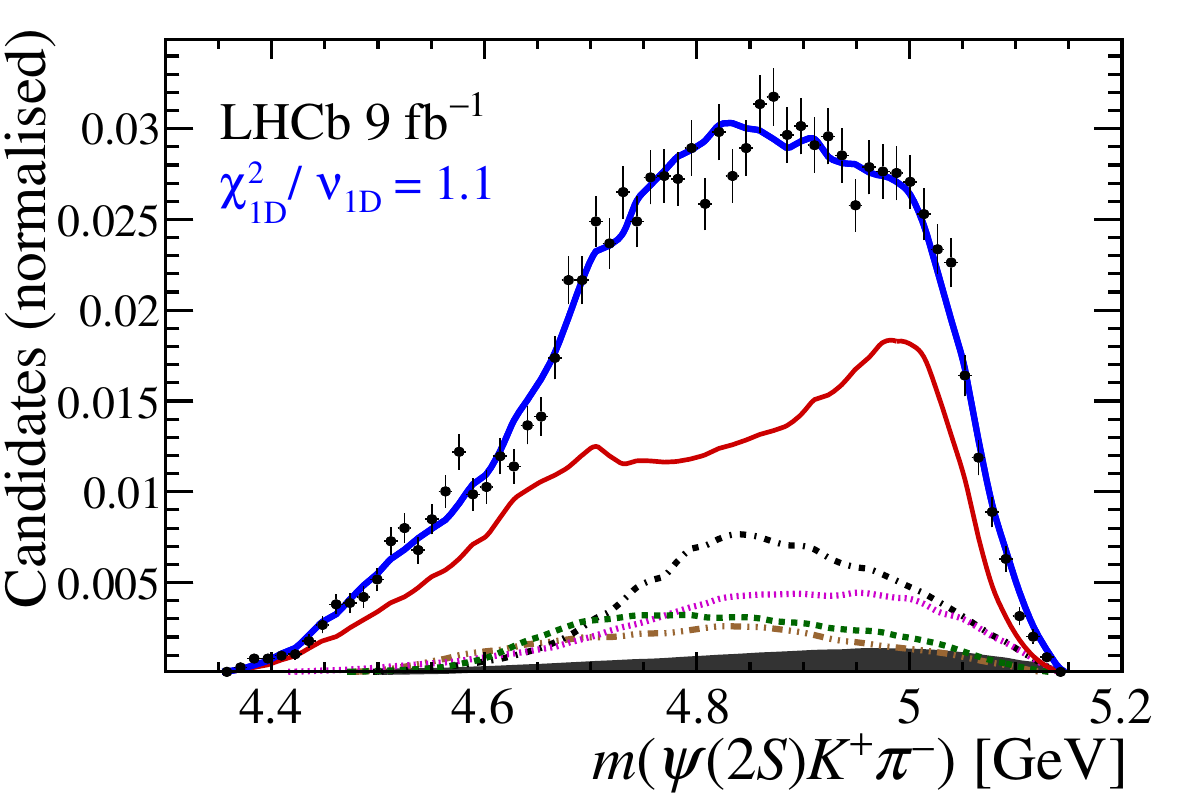}
       	 \includegraphics[width=0.329\textwidth,height=!]{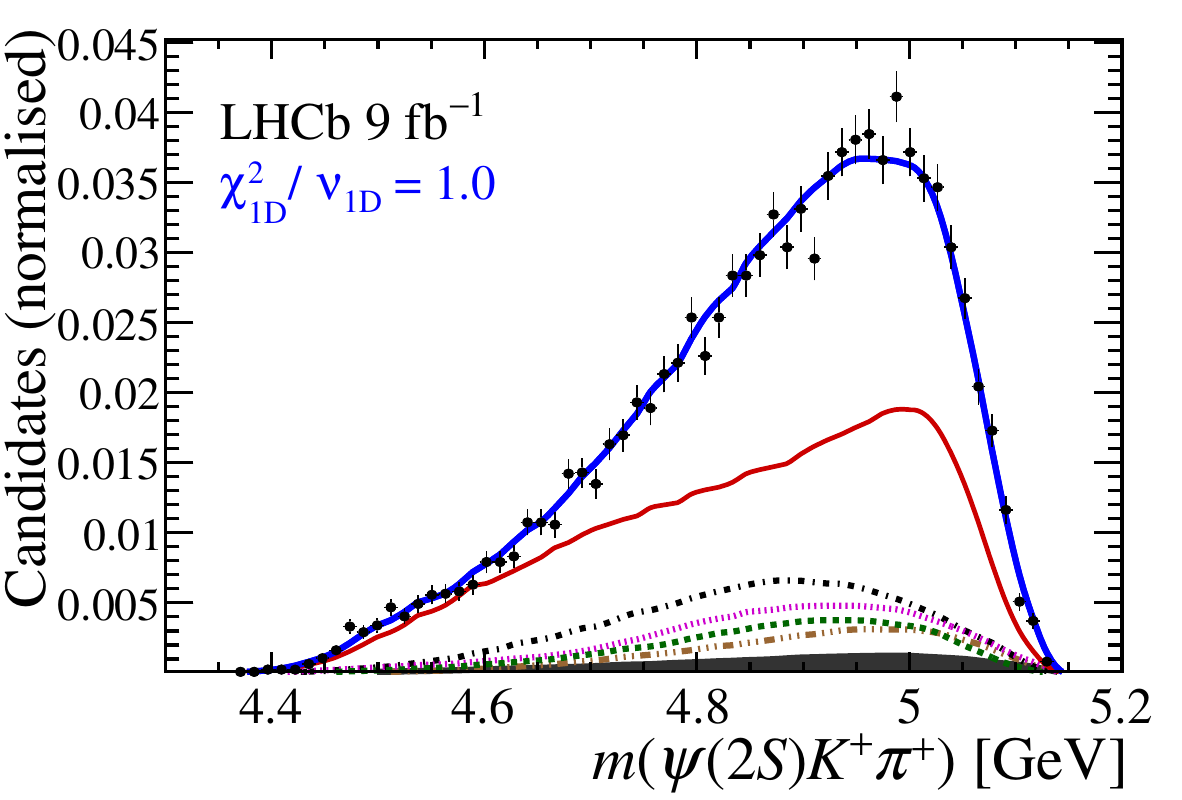}
              
       	 \includegraphics[width=0.329\textwidth,height=!]{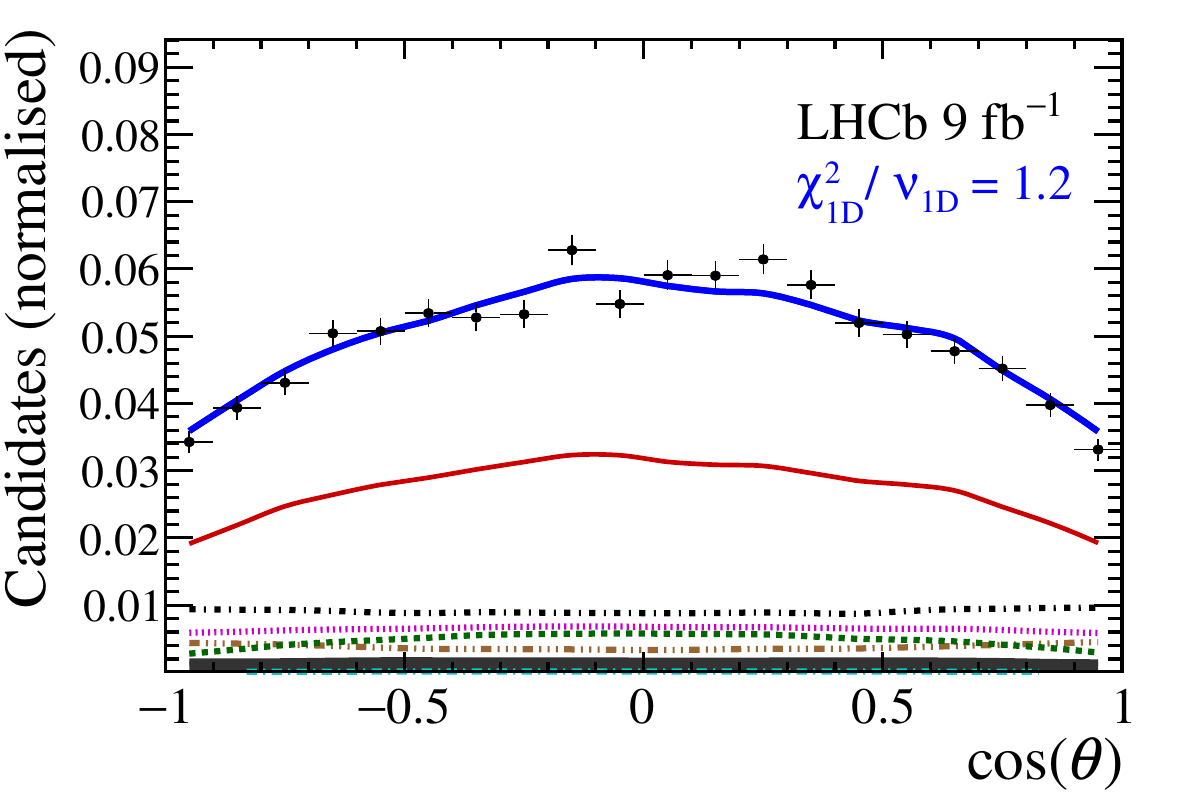}
       	 \includegraphics[width=0.329\textwidth,height=!]{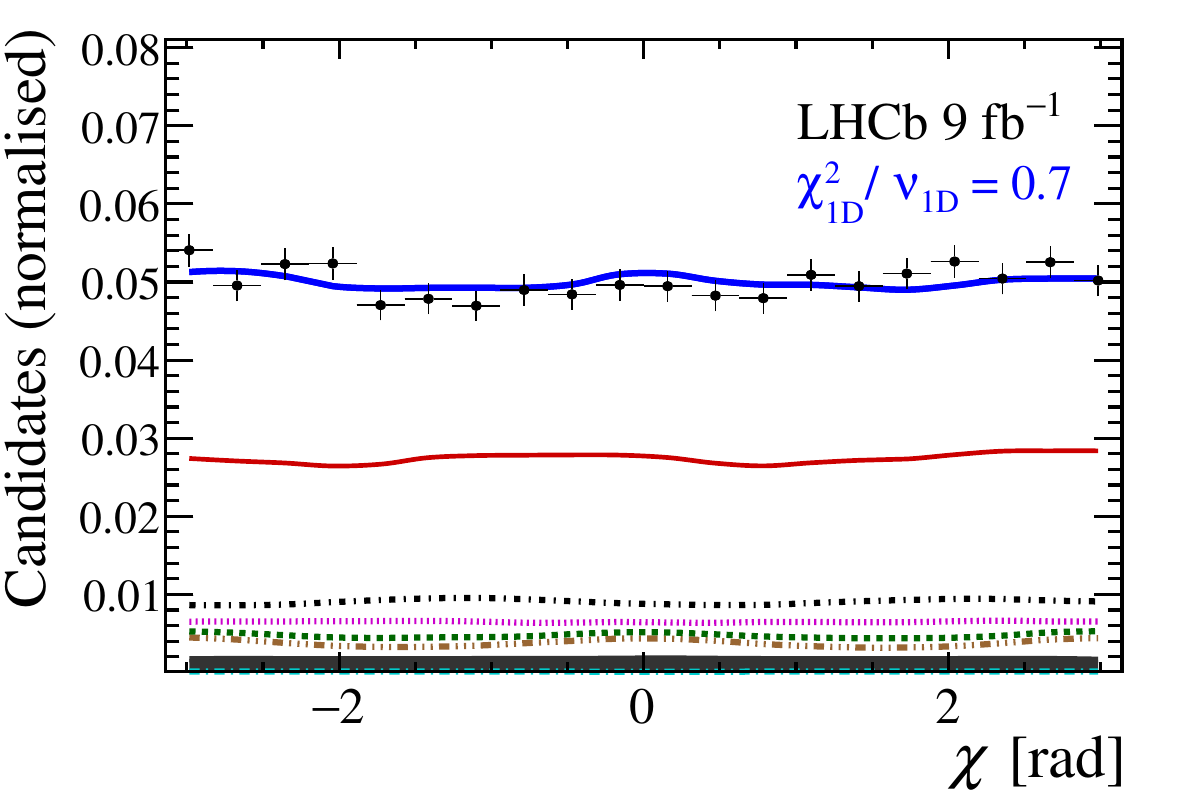}
       	 \includegraphics[width=0.329\textwidth,height=!]{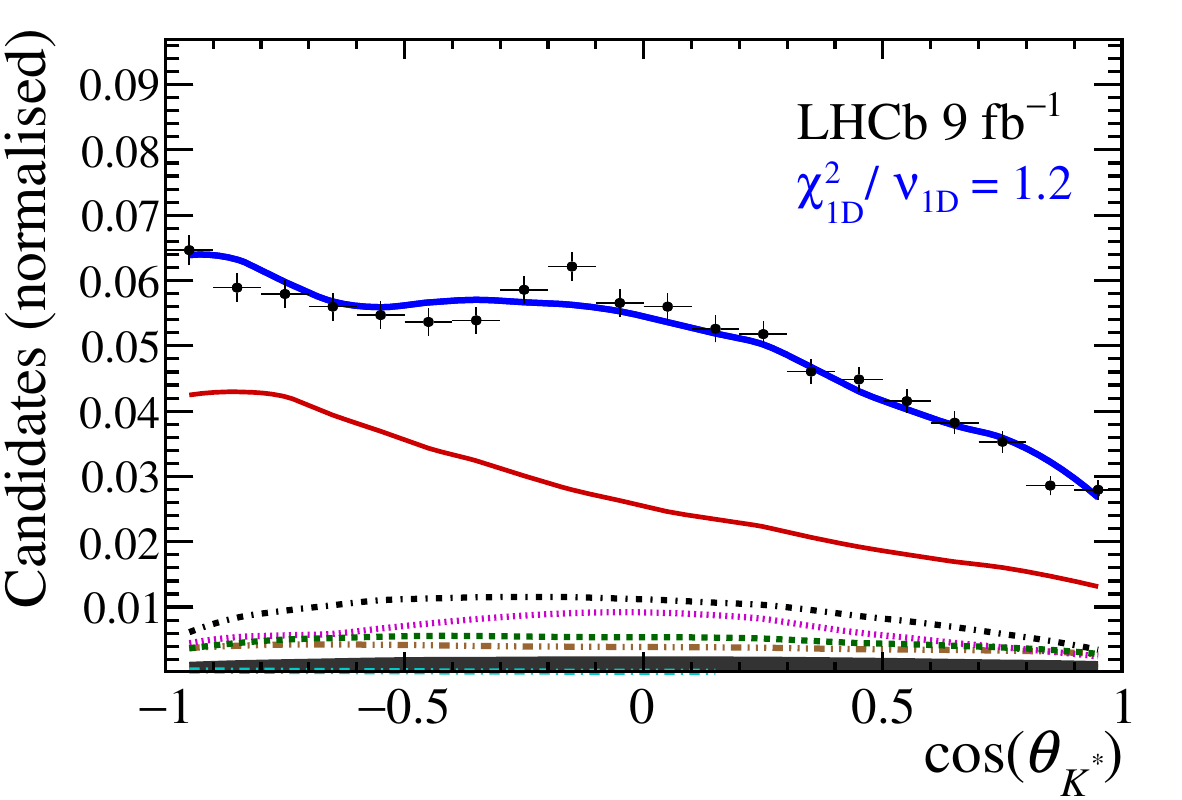}

        \centering
              \includegraphics[width=0.25\textwidth,height=!]{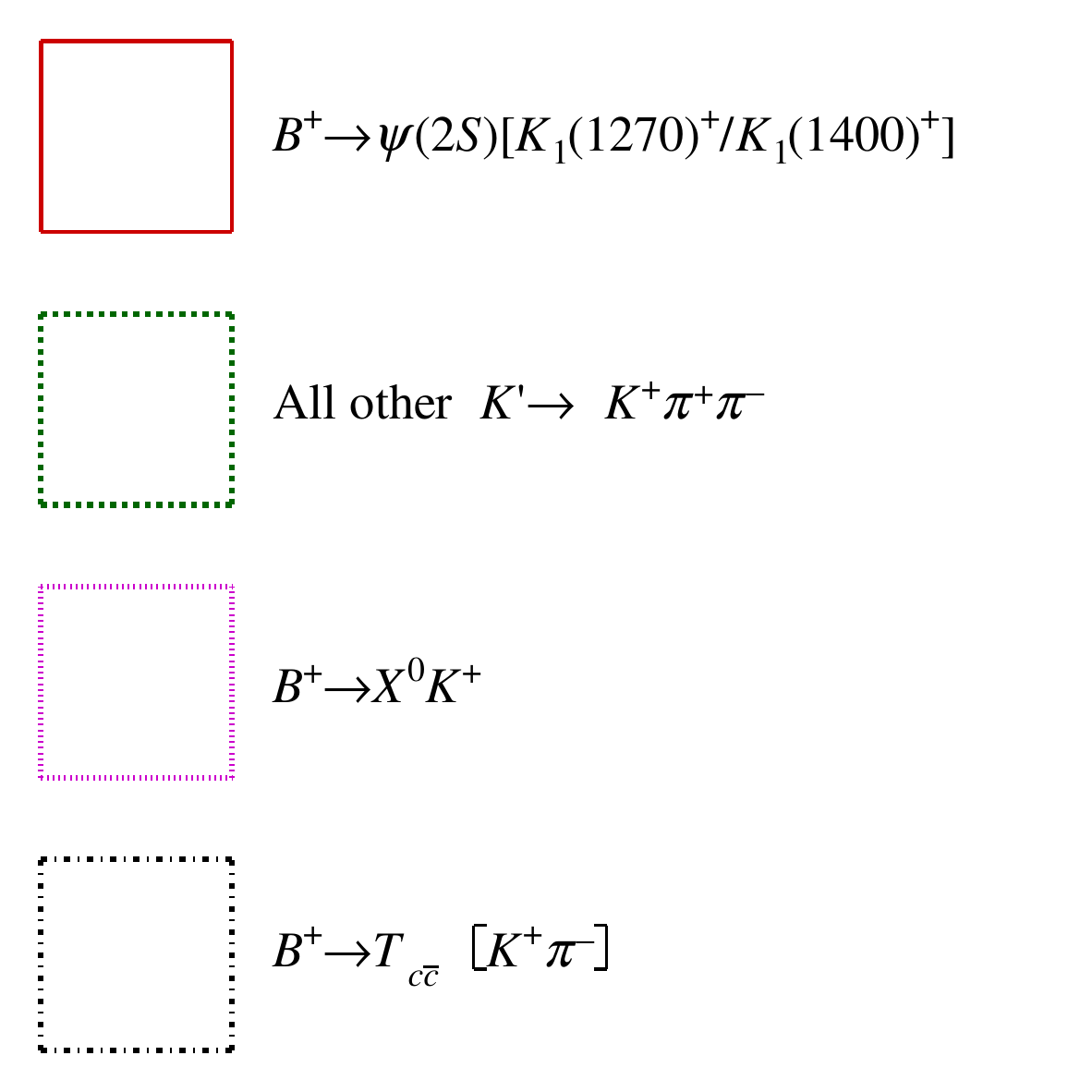}
              \includegraphics[width=0.25\textwidth,height=!]{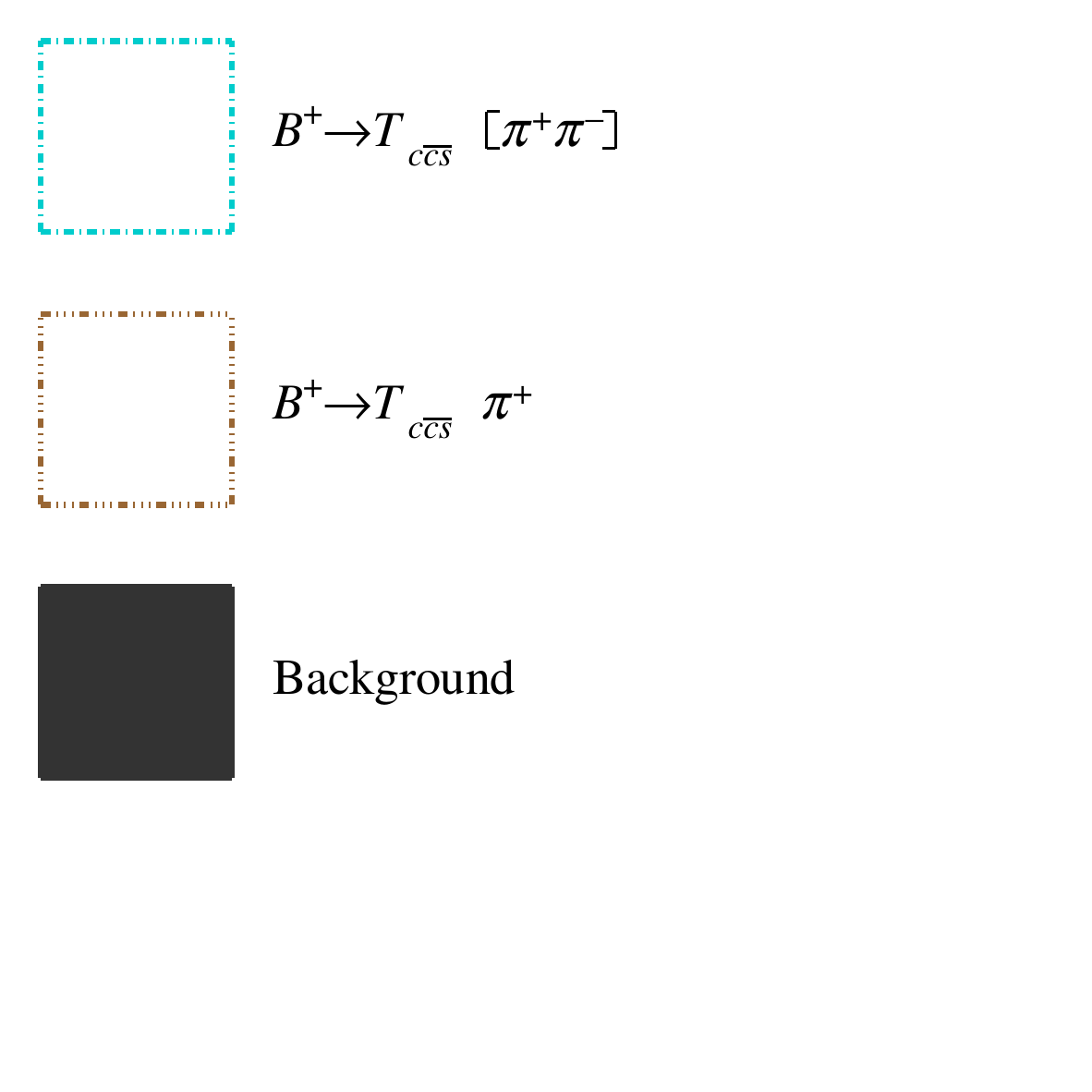}

	\caption{Phase-space projections of $\signal$ candidates in the in low $m(\Kp\pip\pim)$ region (points with error bars) and fit projections (solid, blue line) for the \textit{baseline} model.  The displayed $\chi_{\rm 1D}^2/\nu_{\rm 1D}$ value on each projection gives the sum of squared normalised residuals divided by the number of bins minus one. The multi-dimensional $\chi^2$ value is $\chi^2/\nu= 1.09$  with $\nu=652$.}

         \label{fig:fitBest4}

\end{figure}

\begin{figure}[h]
       	 \includegraphics[width=0.329\textwidth,height=!]{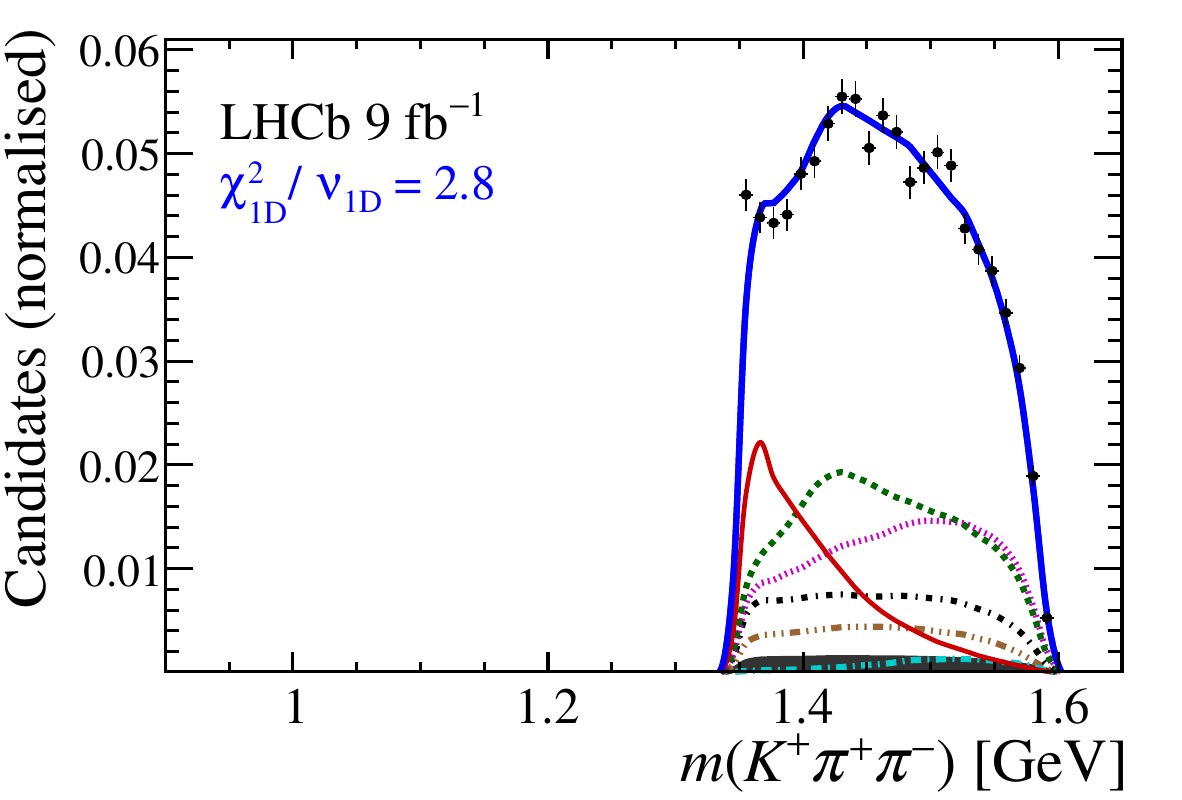}
       	 \includegraphics[width=0.329\textwidth,height=!]{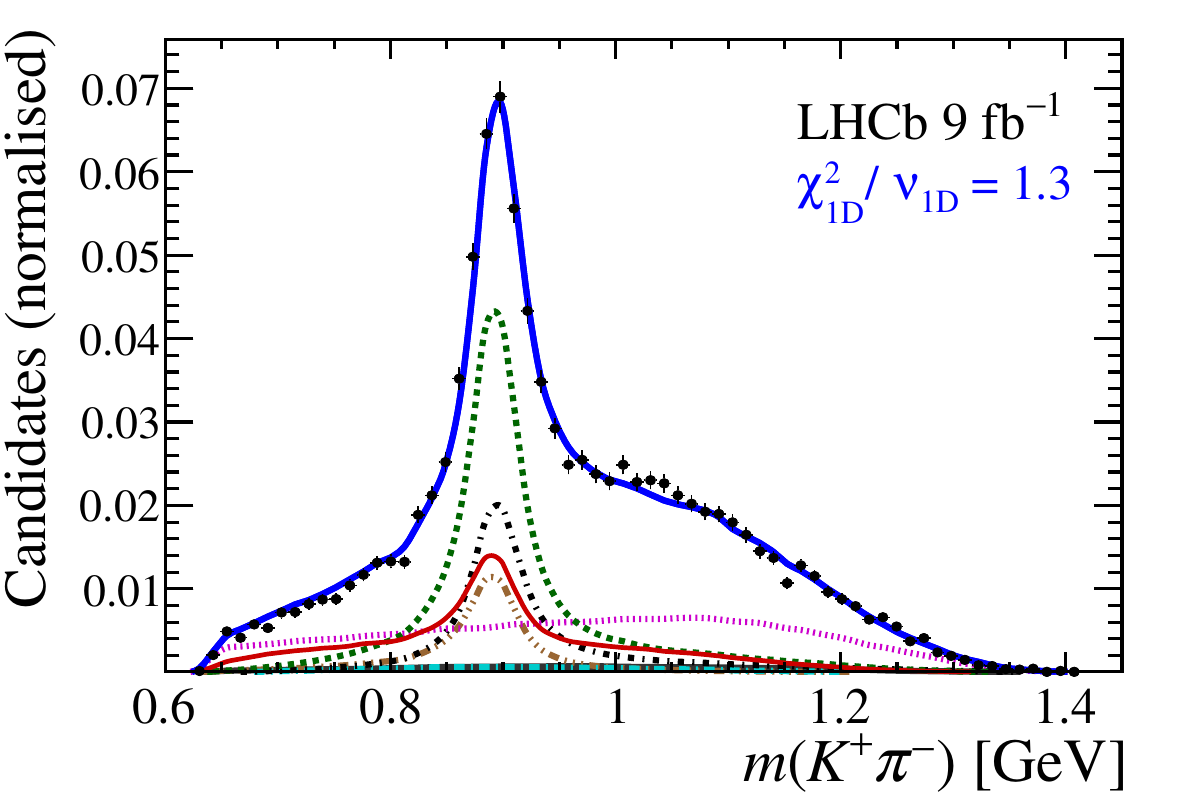}
       	 \includegraphics[width=0.329\textwidth,height=!]{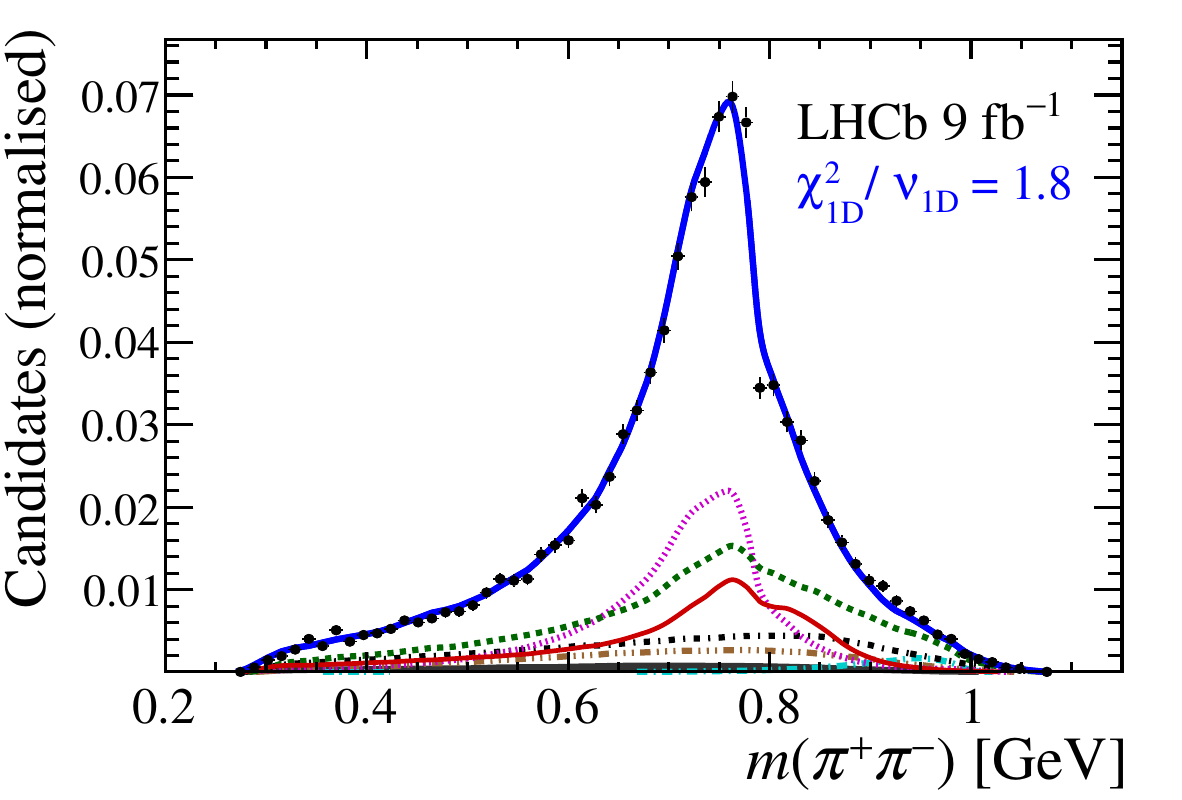}

       	 \includegraphics[width=0.329\textwidth,height=!]{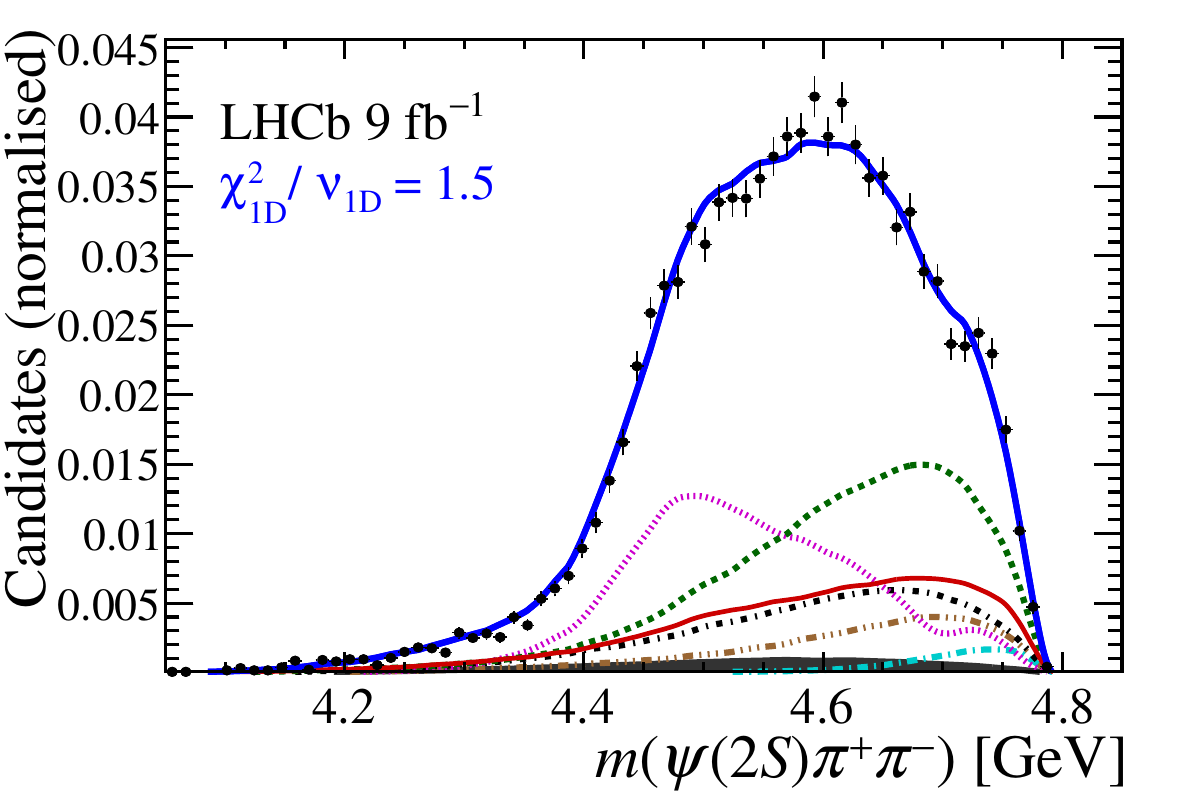}
       	 \includegraphics[width=0.329\textwidth,height=!]{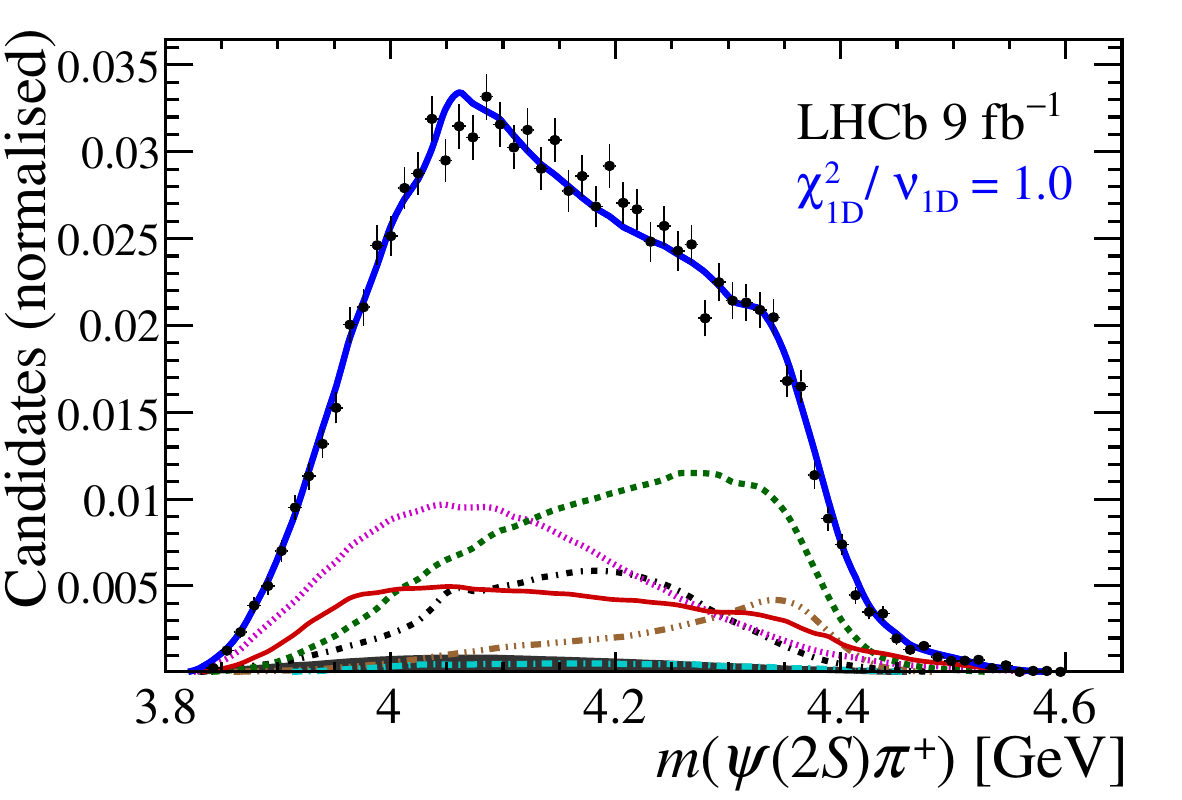}
       	 \includegraphics[width=0.329\textwidth,height=!]{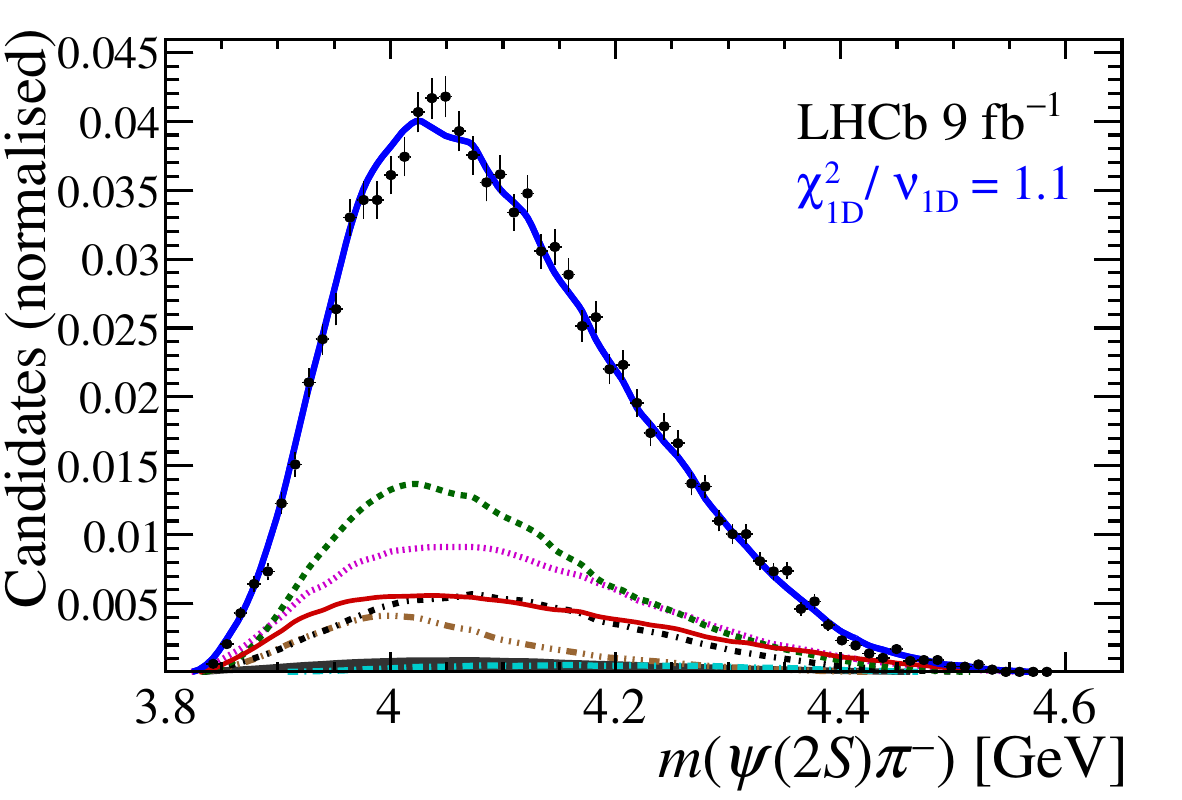}

       	 \includegraphics[width=0.329\textwidth,height=!]{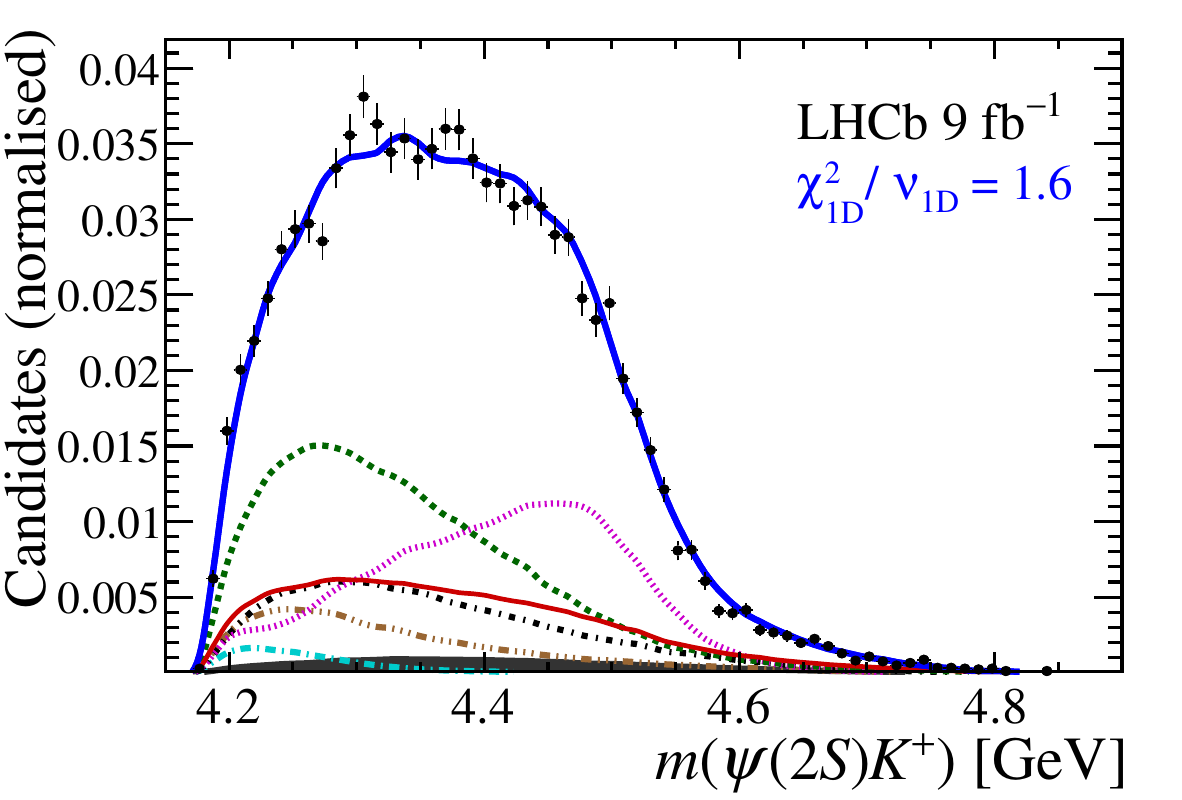}
       	 \includegraphics[width=0.329\textwidth,height=!]{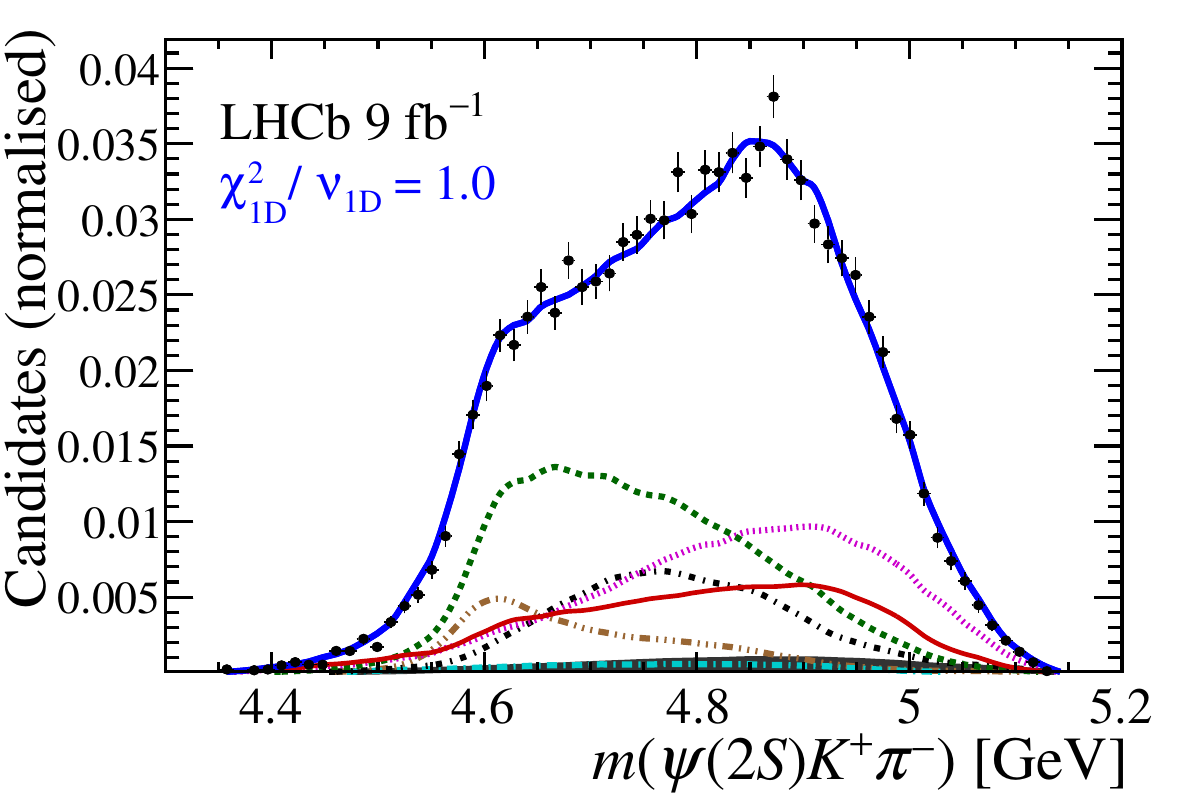}
       	 \includegraphics[width=0.329\textwidth,height=!]{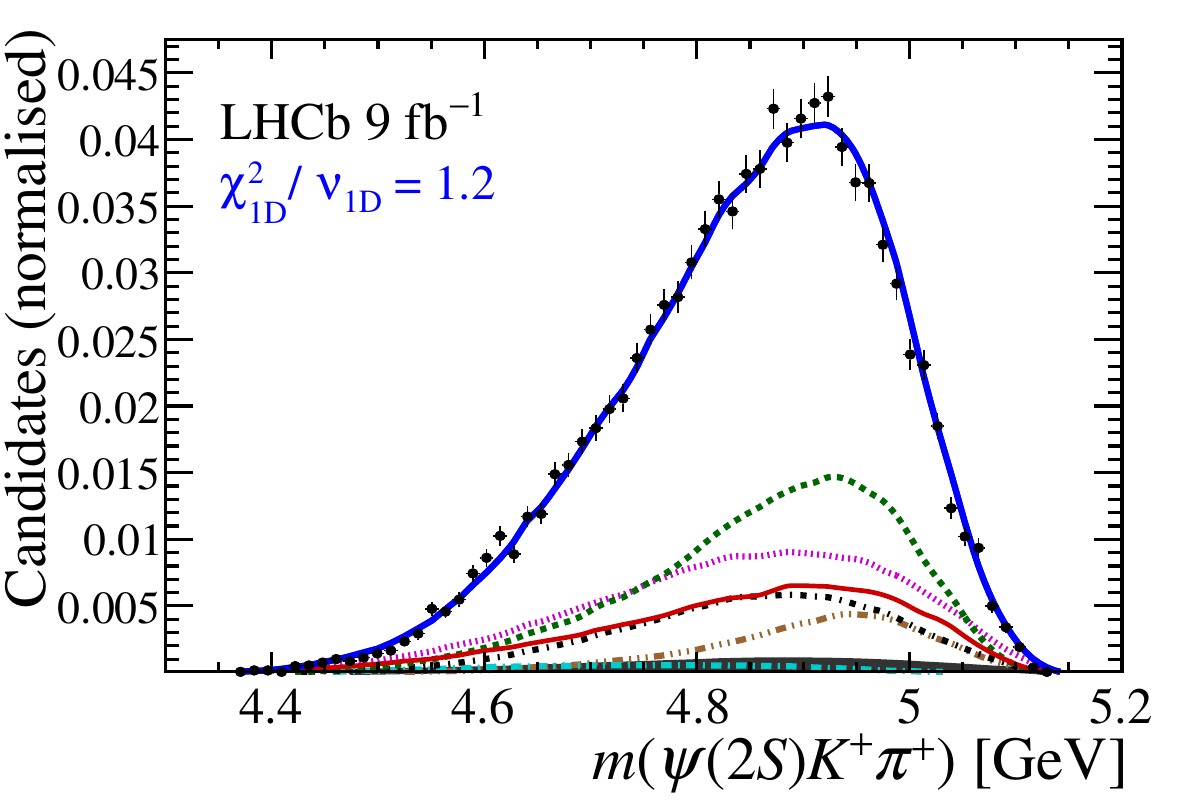}
              
       	 \includegraphics[width=0.329\textwidth,height=!]{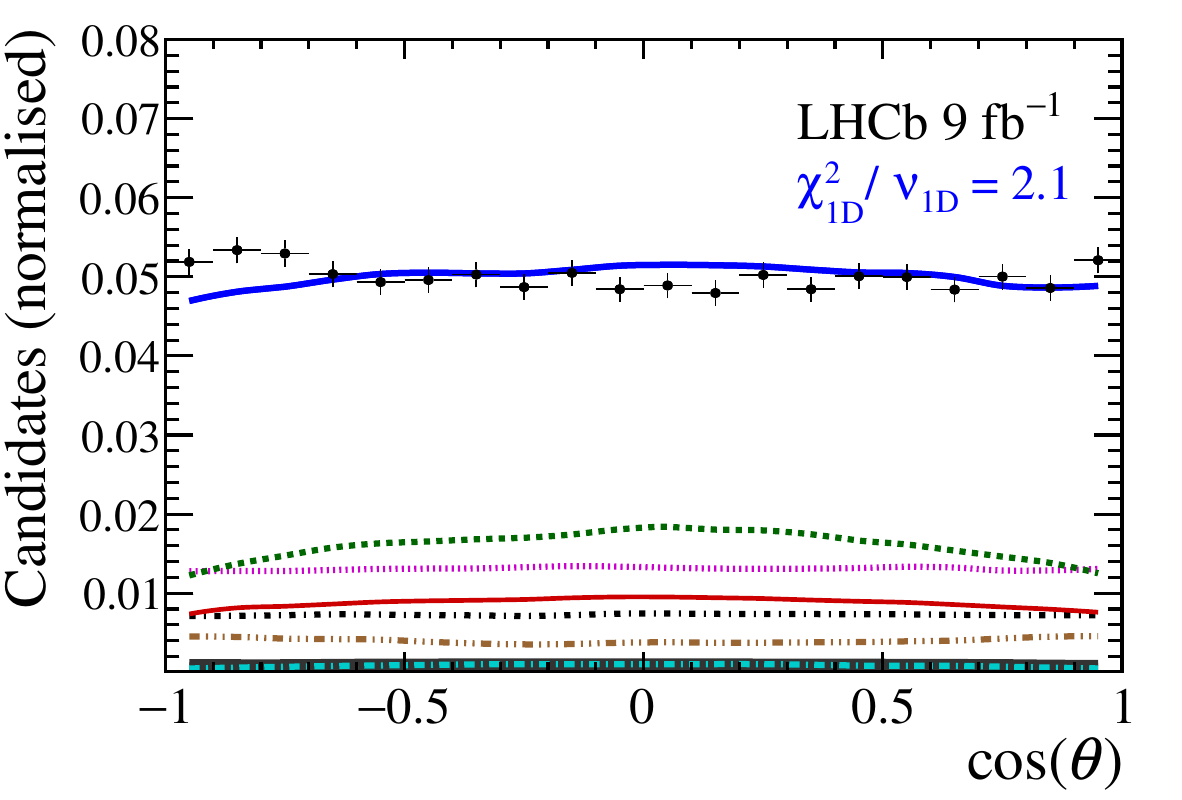}
       	 \includegraphics[width=0.329\textwidth,height=!]{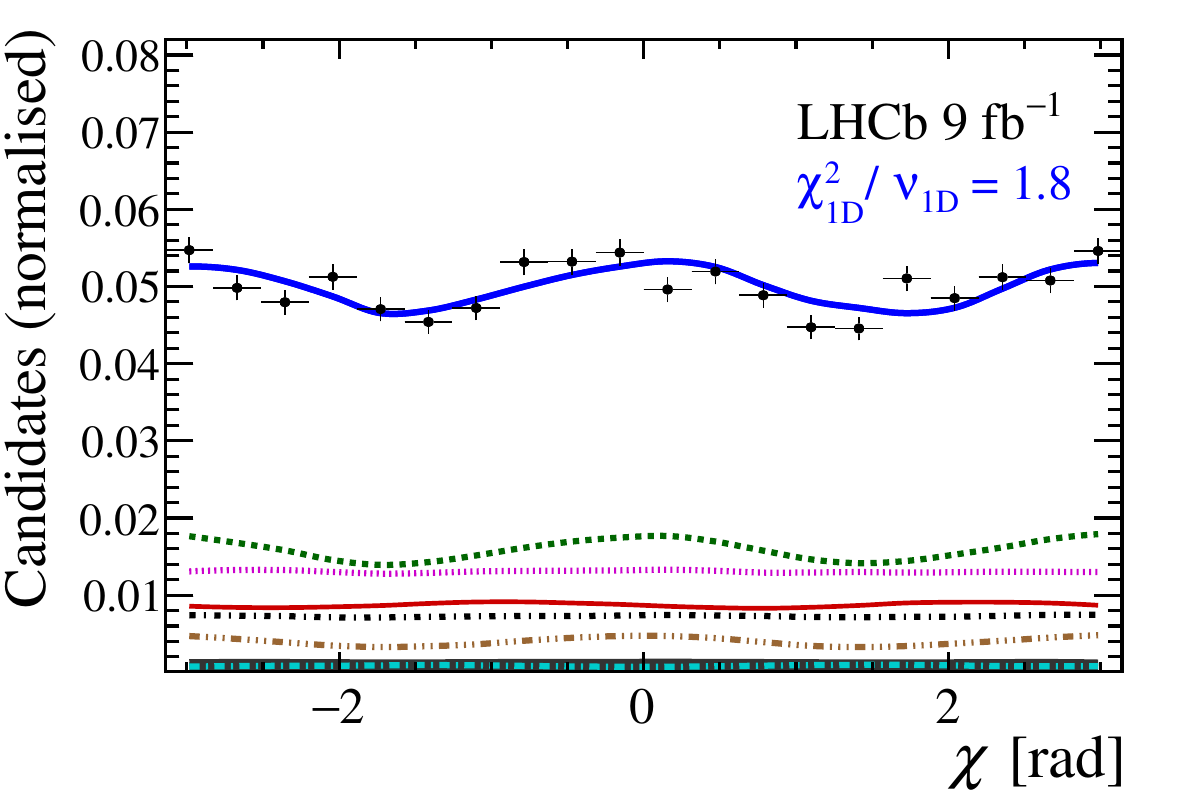}
       	 \includegraphics[width=0.329\textwidth,height=!]{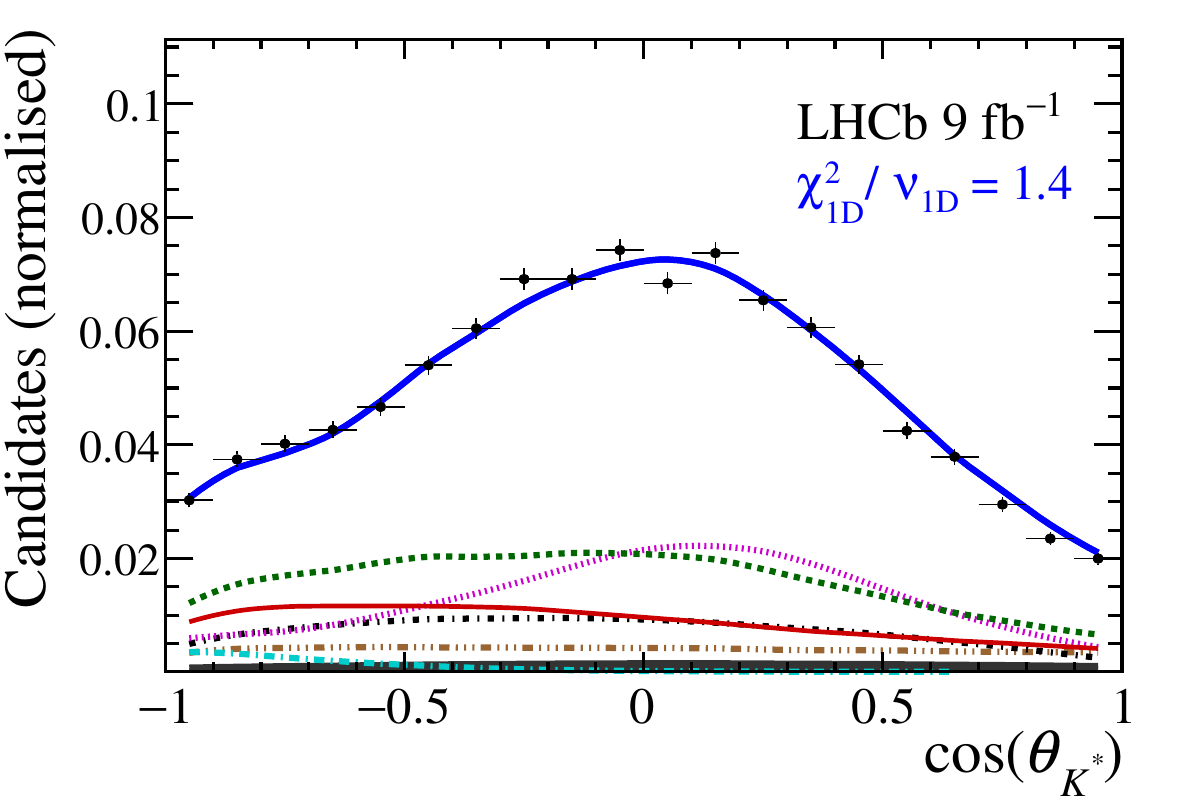}

        \centering
              \includegraphics[width=0.25\textwidth,height=!]{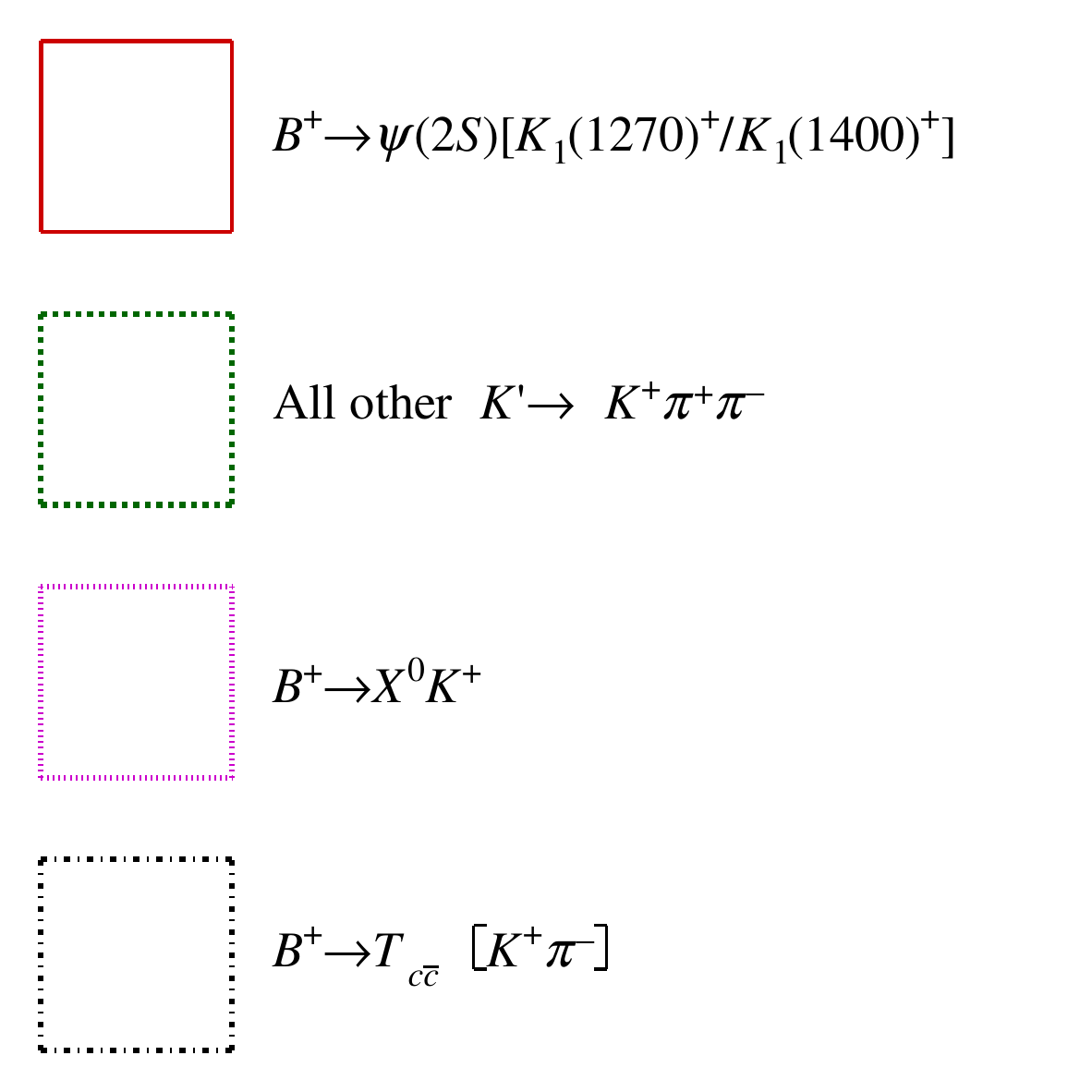}
              \includegraphics[width=0.25\textwidth,height=!]{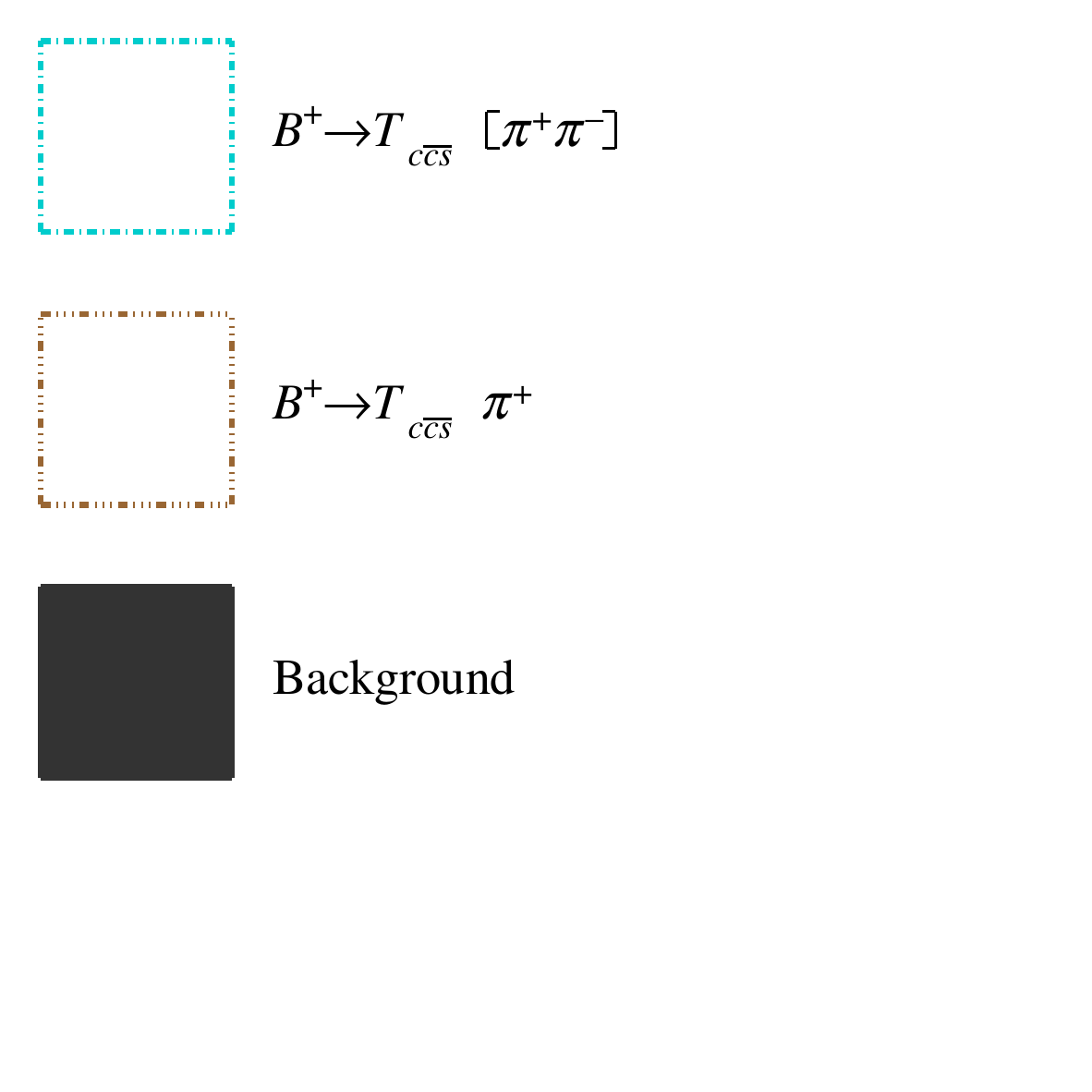}
          
	\caption{Phase-space projections of $\signal$ candidates in the high $m(\Kp\pip\pim)$ region  (points with error bars) and fit projections (solid, blue line) for the \textit{baseline} model.  The displayed $\chi_{\rm 1D}^2/\nu_{\rm 1D}$ value on each projection gives the sum of squared normalised residuals divided by the number of bins minus one. The multi-dimensional $\chi^2$ value is $\chi^2/\nu= 1.00$  with $\nu=426$.}

         \label{fig:fitBest5}

\end{figure}

Table~\ref{tab:resultBaseline} list the moduli and phases of the complex amplitude coefficients obtained by fitting the \textit{baseline} model to the $\signal$ candidates.
The systematic uncertainties are summarised in Tables~\ref{tab:sysFit} and~\ref{tab:sysFit2} for all fit parameters and in Tables~\ref{tab:sysFrac} and~\ref{tab:sysFrac2} for the fit fractions.
Table~\ref{tab:fullFractionsInt} lists the interference fractions ordered by magnitude for the \textit{baseline} model.
Figures~\ref{fig:fitBest2} to~\ref{fig:fitBest5} show additional fit projections of the \textit{baseline} model in slices of the phase space.

Tables~\ref{tab:altModels1_1} to~\ref{tab:altModels2_3} summarise the fit results of the alternative models used for systematic studies.
The code implementation of the \textit{baseline} model and all alternative models, as well as instructions to generate pseudodata from them, can be found in Ref.~\cite{AmpGen-B2VPPP}.

\begin{table}[h]
\centering
\caption{Fit result of the amplitude couplings for the \textit{baseline model}.
Uncertainties are statistical and systematic.
}
 \resizebox{0.8\linewidth}{!}{
	\renewcommand{\arraystretch}{1.05}
\begin{tabular}{l r@{}c@{}r@{}c@{}r r@{}c@{}r@{}c@{}r} 
\hline
\hline
Amplitude coupling & \multicolumn{5}{c}{Magnitude} & \multicolumn{5}{c}{Phase [${}^\circ$]}   \\ 
\hline
$B^{+}[D]\rightarrow \ZAone^{+} K^{*}(892)^{0}\,\text{}$ & $0.85 $&$\pm$&$0.07 $&$\pm$&$0.14 $ & $ -52.6 $&$\pm$&$5.4 $&$\pm$&$4.2$ \\ 
$B^{+}[D]\rightarrow \psi(\text{2S})\,K_{1}(1270)^{+}\,\text{}$ & $0.95 $&$\pm$&$0.07 $&$\pm$&$0.11 $ & $ -210.8 $&$\pm$&$3.9 $&$\pm$&$3.7$ \\ 
$B^{+}[D]\rightarrow \rho(770)^{0}\ZsAone \,\text{}$ & $0.73 $&$\pm$&$0.06 $&$\pm$&$0.16 $ & $ 5.1 $&$\pm$&$5.1 $&$\pm$&$5.6$ \\ 
$B^{+}[P]\rightarrow \ZAone^{+} K^{*}(892)^{0}\,\text{}$ & $1.09 $&$\pm$&$0.09 $&$\pm$&$0.26 $ & $ -124.6 $&$\pm$&$6.0 $&$\pm$&$5.8$ \\ 
$B^{+}[P]\rightarrow \psi(\text{2S})\,K_{1}(1270)^{+}\,\text{}$ & $0.99 $&$\pm$&$0.07 $&$\pm$&$0.11 $ & $ -120.5 $&$\pm$&$3.8 $&$\pm$&$3.5$ \\ 
$B^{+}[P]\rightarrow \psi(\text{2S})\,K_{1}(1400)^{+}\,\text{}$ & $0.33 $&$\pm$&$0.06 $&$\pm$&$0.11 $ & $ 110.6 $&$\pm$&$10.2 $&$\pm$&$7.0$ \\ 
$B^{+}\rightarrow K_{2}^{*}(1430)^{+}\psi(\text{2S})\,\,\text{}$ & $0.93 $&$\pm$&$0.07 $&$\pm$&$0.08 $ & $ 125.9 $&$\pm$&$4.5 $&$\pm$&$5.0$ \\
\hline
\hline
\end{tabular}
  }
\label{tab:resultBaseline}
\end{table}

\begin{table}[h]
\caption{Systematic uncertainties on the fit parameters in units of statistical standard deviations, part 1. 
The different contributions are: 1) fit bias, 2) MC integration precision, 3) acceptance, 4) background, 5) masses and widths of resonances, 6) form factors, 7) $\rho-\omega$ mixing, 8) S-wave model, 9) lineshapes, 10) alternative amplitude models. 
}
\resizebox{0.95\linewidth}{!}{
	\renewcommand{\arraystretch}{1.1}
	\begin{tabular}{l  c  c  c  c  c  c  c  c  c  c  | c }
\hline
\hline
Fit Parameter & 1 & 2 & 3 & 4 & 5 & 6 & 7 & 8 & 9 & 10 &  Total  \\ 
\hline
$B^{+}[D]\rightarrow \ZAone^{+}\, K^{*}(892)^{0}\,\text{Amp}$ & 0.06 & 0.26 & 0.30 & 0.03 & 0.49 & 0.49 & 0.15 & 0.16 & 0.87 & 1.53 & 1.94 \\ 
$B^{+}[D]\rightarrow \ZAone^{+}\, K^{*}(892)^{0}\,\text{Phase}$ & 0.13 & 0.25 & 0.26 & 0.04 & 0.43 & 0.23 & 0.15 & 0.16 & 0.19 & 0.35 & 0.77 \\ 
$B^{+}[D]\rightarrow \psi(\text{2S})\,K_{1}(1270)^{+}\,\text{Amp}$ & 0.29 & 0.34 & 0.29 & 0.04 & 0.36 & 0.38 & 0.21 & 0.30 & 0.62 & 1.22 & 1.60 \\ 
$B^{+}[D]\rightarrow \psi(\text{2S})\,K_{1}(1270)^{+}\,\text{Phase}$ & 0.17 & 0.35 & 0.26 & 0.17 & 0.47 & 0.26 & 0.01 & 0.34 & 0.30 & 0.42 & 0.95 \\ 
$B^{+}[D]\rightarrow \rho(770)^{0}\ZsAone\, \,\text{Amp}$ & 0.30 & 0.32 & 0.27 & 0.06 & 0.33 & 0.50 & 0.03 & 0.17 & 0.53 & 2.53 & 2.71 \\ 
$B^{+}[D]\rightarrow \rho(770)^{0}\ZsAone\, \,\text{Phase}$ & 0.73 & 0.38 & 0.17 & 0.16 & 0.37 & 0.20 & 0.27 & 0.24 & 0.15 & 0.35 & 1.09 \\ 
$B^{+}[P]\rightarrow \ZAone^{+}\, K^{*}(892)^{0}\,\text{Amp}$ & 0.03 & 0.31 & 0.18 & 0.09 & 0.56 & 0.43 & 0.17 & 0.23 & 0.69 & 2.58 & 2.81 \\
\hline
\hline
\end{tabular}
 }
\label{tab:sysFit}
\end{table}

\begin{table}[h]
\caption{Systematic uncertainties on the fit parameters in units of statistical standard deviations, part 2.
The different contributions are: 1) fit bias, 2) MC integration precision, 3) acceptance, 4) background, 5) masses and widths of resonances, 6) form factors, 7) $\rho-\omega$ mixing, 8) S-wave model, 9) lineshapes, 10) alternative amplitude models. }
\resizebox{\linewidth}{!}{
	\renewcommand{\arraystretch}{1.1}
	\begin{tabular}{l  c  c  c  c  c  c  c  c  c  c  | c }
\hline
\hline
Fit Parameter & 1 & 2 & 3 & 4 & 5 & 6 & 7 & 8 & 9 & 10 &  Total  \\ 
\hline
$K_{1}(1270)^{+}[D]\rightarrow K^{*}(892)^{0}\pi^{+}\,\text{Amp}$ & 0.17 & 0.38 & 0.21 & 0.11 & 0.24 & 0.23 & 0.04 & 0.14 & 0.35 & 1.60 & 1.75 \\ 
$K_{1}(1270)^{+}[D]\rightarrow K^{*}(892)^{0}\pi^{+}\,\text{Phase}$ & 0.61 & 0.39 & 0.27 & 0.02 & 0.27 & 0.12 & 0.13 & 0.14 & 0.24 & 0.57 & 1.05 \\ 
$K_{1}(1270)^{+}\rightarrow K^{*}(892)^{0}\pi^{+}\,\text{Amp}$ & 0.09 & 0.38 & 0.28 & 0.07 & 0.40 & 0.44 & 0.02 & 0.16 & 0.36 & 0.94 & 1.28 \\ 
$K_{1}(1270)^{+}\rightarrow K^{*}(892)^{0}\pi^{+}\,\text{Phase}$ & 0.20 & 0.48 & 0.20 & 0.10 & 0.65 & 0.23 & 0.16 & 0.21 & 0.30 & 0.44 & 1.07 \\ 
$K_{1}(1270)^{+}\rightarrow \left[K^{+}\pi^{-}\right]_{\text{S}}\pi^{+}\,\text{Amp}$ & 0.40 & 0.34 & 0.20 & 0.30 & 0.41 & 0.41 & 0.03 & 1.68 & 0.51 & 0.57 & 2.04 \\ 
$K_{1}(1270)^{+}\rightarrow \left[K^{+}\pi^{-}\right]_{\text{S}}\pi^{+}\,\text{Phase}$ & 0.35 & 0.29 & 0.43 & 0.35 & 0.56 & 0.33 & 0.14 & 0.38 & 0.27 & 0.47 & 1.18 \\ 
$K_{1}(1400)^{+}\rightarrow \rho(770)^{0}K^{+}\,\text{Amp}$ & 0.32 & 0.46 & 0.18 & 0.07 & 0.28 & 0.13 & 0.19 & 0.11 & 0.53 & 1.11 & 1.42 \\
\hline
\hline
\end{tabular}
 }
\label{tab:sysFit2}
\end{table}

\begin{table}[h]
\caption{Systematic uncertainties on the fit fractions in units of statistical standard deviations, part 1.
The different contributions are: 1) fit bias, 2) MC integration precision, 3) acceptance, 4) background, 5) masses and widths of resonances, 6) form factors, 7) $\rho-\omega$ mixing, 8) S-wave model, 9) lineshapes, 10) alternative amplitude models. }
\resizebox{\linewidth}{!}{
	\renewcommand{\arraystretch}{1.1}
	\begin{tabular}{l  c  c  c  c  c  c  c  c  c  c  | c }
\hline
\hline
Fit fraction & 1 & 2 & 3 & 4 & 5 & 6 & 7 & 8 & 9 & 10 &  Total  \\ 
\hline
$B^{+}\rightarrow \XSone K^{+}$ & 0.21 & 0.36 & 0.26 & 0.29 & 0.60 & 0.50 & 0.08 & 0.39 & 0.58 & 1.88 & 2.22 \\ 
$B^{+}\rightarrow \psi(\text{2S})\,K^{*}(1680)^{+}$ & 0.02 & 0.34 & 0.30 & 0.15 & 1.42 & 0.44 & 0.03 & 0.15 & 0.29 & 2.15 & 2.68 \\ 
$B^{+}\rightarrow \psi(\text{2S})\,K_{1}(1270)^{+}$ & 0.16 & 0.41 & 0.30 & 0.06 & 0.45 & 0.42 & 0.19 & 0.32 & 0.75 & 1.07 & 1.59 \\ 
$B^{+}[P]\rightarrow \psi(\text{2S})\,K_{1}(1270)^{+}$ & 0.39 & 0.27 & 0.26 & 0.13 & 0.42 & 0.30 & 0.01 & 0.25 & 0.61 & 1.50 & 1.81 \\ 
$B^{+}[D]\rightarrow \psi(\text{2S})\,K_{1}(1270)^{+}$ & 0.26 & 0.30 & 0.25 & 0.11 & 0.07 & 0.06 & 0.14 & 0.13 & 0.17 & 2.59 & 2.64 \\ 
$B^{+}\rightarrow \psi(\text{2S})\,K_{1}(1400)^{+}$ & 0.03 & 0.36 & 0.21 & 0.11 & 0.72 & 0.19 & 0.10 & 0.17 & 0.78 & 0.92 & 1.49 \\ 
$B^{+}\rightarrow \psi(\text{2S})\,K(1460)^{+}$ & 0.48 & 0.30 & 0.42 & 0.10 & 0.46 & 0.78 & 0.08 & 0.48 & 0.27 & 1.31 & 1.83 \\
\hline
\hline
\end{tabular}
 }
\label{tab:sysFrac}
\end{table}

\begin{table}[h]
\caption{Systematic uncertainties on the fit fractions in units of statistical standard deviations, part 2.
The different contributions are: 1) fit bias, 2) MC integration precision, 3) acceptance, 4) background, 5) masses and widths of resonances, 6) form factors, 7) $\rho-\omega$ mixing, 8) S-wave model, 9) lineshapes, 10) alternative amplitude models. }
\resizebox{\linewidth}{!}{
	\renewcommand{\arraystretch}{1.1}
	\begin{tabular}{l  c  c  c  c  c  c  c  c  c  c  | c }
\hline
\hline
Fit fraction & 1 & 2 & 3 & 4 & 5 & 6 & 7 & 8 & 9 & 10 &  Total  \\ 
\hline
$K_{1}(1270)^{+}\rightarrow \rho(770)^{0}K^{+}$ & 0.37 & 0.30 & 0.29 & 0.05 & 0.35 & 0.50 & 0.10 & 0.21 & 0.22 & 1.16 & 1.46 \\ 
$K_{1}(1270)^{+}\rightarrow K^{*}(892)^{0}\pi^{+}$ & 0.06 & 0.37 & 0.21 & 0.06 & 0.43 & 0.33 & 0.09 & 0.14 & 0.46 & 1.14 & 1.42 \\ 
$K_{1}(1270)^{+}\rightarrow \left[K^{+}\pi^{-}\right]_{\text{S}}\pi^{+}$ & 0.35 & 0.40 & 0.27 & 0.36 & 0.43 & 0.35 & 0.00 & 0.65 & 0.57 & 0.76 & 1.46 \\ 
$K_{1}(1270)^{+}[D]\rightarrow K^{*}(892)^{0}\pi^{+}$ & 0.07 & 0.39 & 0.17 & 0.12 & 0.19 & 0.15 & 0.01 & 0.19 & 0.38 & 1.69 & 1.82 \\ 
$\text{Sum } K_{1}(1270)^{+}$ & 0.21 & 0.32 & 0.35 & 0.28 & 0.56 & 0.38 & 0.19 & 0.39 & 0.50 & 1.67 & 2.01 \\ \hline
$K_{1}(1400)^{+}\rightarrow K^{*}(892)^{0}\pi^{+}$ & 0.25 & 0.42 & 0.22 & 0.09 & 0.20 & 0.14 & 0.22 & 0.12 & 0.56 & 1.13 & 1.41 \\ 
$K_{1}(1400)^{+}\rightarrow \rho(770)^{0}K^{+}$ & 0.21 & 0.45 & 0.18 & 0.07 & 0.29 & 0.13 & 0.19 & 0.11 & 0.53 & 1.14 & 1.42 \\
\hline
\hline
\end{tabular}
 }
\label{tab:sysFrac2}
\end{table}

\begin{table}[h]
\centering
\caption{
Interference fractions with $\vert IF_{ij} \vert \ge 0.30\%$ for the \textit{baseline model}.
The uncertainties are statistical only.
}
\resizebox{0.61\linewidth}{!}{
	\renewcommand{\arraystretch}{1.0}
	\begin{tabular}{l l r} 
\hline
\hline
Decay channel $i$ & Decay channel $j$ &  $IF_{ij} [ \% ] $  \\ 
\hline
$B^{+}\rightarrow \XSone K^{+} $ & $ B^{+}\rightarrow \psi(\text{2S})\,K_{1}(1270)^{+}$ & $7.87 \pm 0.61$ \\ 
$B^{+}\rightarrow \ZAone^{+}\, K^{*}(892)^{0} $ & $ B^{+}\rightarrow \psi(\text{2S})\,K_{1}(1400)^{+}$ & $-4.88 \pm 0.68$ \\ 
$B^{+}\rightarrow \XSone K^{+} $ & $ B^{+}\rightarrow \XStwo K^{+}$ & $-4.63 \pm 0.48$ \\ 
$B^{+}\rightarrow \XSone K^{+} $ & $ B^{+}\rightarrow \psi(\text{2S})\,K_{1}(1400)^{+}$ & $4.62 \pm 0.59$ \\ 
$B^{+}\rightarrow \XsAone\pi^{+} $ & $ B^{+}\rightarrow \psi(\text{2S})\,K^{*}(1680)^{+}$ & $-4.62 \pm 1.52$ \\ 
$B^{+}\rightarrow \psi(\text{2S})\,K^{*}(1410)^{+} $ & $ B^{+}\rightarrow \psi(\text{2S})\,K^{*}(1680)^{+}$ & $3.74 \pm 0.68$ \\ 
$B^{+}\rightarrow \XAone K^{+} $ & $ B^{+}\rightarrow \psi(\text{2S})\,K^{*}(1680)^{+}$ & $3.27 \pm 0.56$ \\
\hline
\hline
\end{tabular}
 }
\label{tab:fullFractionsInt}
\end{table}

\clearpage

\clearpage

\begin{table}[h]
\centering
\caption{Fit results of the \textit{baseline} model and of alternative amplitude models 1-5, part 1. Uncertainties are statistical only. }
\resizebox{\linewidth}{!}{
	\renewcommand{\arraystretch}{1.1}
\begin{tabular}{l r r r r r r  } 
\hline
\hline
Decay channel $F_{i} [ \% ] $  & \textit{Baseline} & Model 1 & Model 2 & Model 3 & Model 4 & Model 5  \\
\hline
\hline

$B^{+}\rightarrow \XSone K^{+}$  & $ 18.45 \pm 1.31 $  & $ 18.09 \pm 1.21 $  & $ 16.42 \pm 1.79 $  & $ 17.21 \pm 1.20 $  & $ 17.66 \pm 1.13 $  & $ 18.52 \pm 1.20 $   \\
$B^{+}\rightarrow \psi(\text{2S})\,K^{*}(1680)^{+}$  & $ 8.15 \pm 1.31 $  & $ 10.22 \pm 1.64 $  & $ 7.04 \pm 1.26 $  & $ 12.10 \pm 1.79 $  & $ 13.76 \pm 1.69 $  & $ 7.97 \pm 1.29 $   \\
$B^{+}\rightarrow \psi(\text{2S})\,K_{1}(1270)^{+}$  & $ 7.60 \pm 0.85 $  & $ 7.48 \pm 0.92 $  & $ 8.18 \pm 0.96 $  & $ 7.76 \pm 0.76 $  & $ 8.30 \pm 0.81 $  & $ 7.58 \pm 0.84 $   \\
$B^{+}[P]\rightarrow \psi(\text{2S})\,K_{1}(1270)^{+}$  & $ 7.52 \pm 0.60 $  & $ 6.75 \pm 0.63 $  & $ 7.11 \pm 0.61 $  & $ 6.58 \pm 0.63 $  & $ 5.40 \pm 0.50 $  & $ 7.70 \pm 0.61 $   \\
$B^{+}[D]\rightarrow \psi(\text{2S})\,K_{1}(1270)^{+}$  & $ 6.81 \pm 0.45 $  & $ 6.41 \pm 0.44 $  & $ 6.80 \pm 0.47 $  & $ 6.58 \pm 0.46 $  & $ 6.21 \pm 0.42 $  & $ 6.78 \pm 0.44 $   \\
$B^{+}\rightarrow \psi(\text{2S})\,K_{1}(1400)^{+}$  & $ 5.78 \pm 0.62 $  & $ 5.94 \pm 0.64 $  & $ 4.58 \pm 0.71 $  & $ 5.68 \pm 0.65 $  & $ 4.90 \pm 0.57 $  & $ 5.47 \pm 0.62 $   \\
$B^{+}\rightarrow \psi(\text{2S})\,K(1460)^{+}$  & $ 5.26 \pm 0.48 $  & $ 5.31 \pm 0.43 $  & $ 4.90 \pm 0.43 $  & $ 5.36 \pm 0.44 $  & $ 4.86 \pm 0.43 $  & $ 5.15 \pm 0.42 $   \\
\hline
\hline
\end{tabular}
 }
\label{tab:altModels1_1}
\end{table}

\begin{table}[h]
\centering
\caption{Fit results of the \textit{baseline} model and of alternative amplitude models 1-5, part 2. Uncertainties are statistical only.}
\resizebox{0.88\linewidth}{!}{
	\renewcommand{\arraystretch}{1.}
	\begin{tabular}{l r@{}c@{}l r@{}c@{}l r@{}c@{}l r@{}c@{}l r@{}c@{}l r@{}c@{}l} 
\hline
\hline
Decay channel $F_{i} [ \% ] $  & \multicolumn{3}{c}{\textit{Baseline}} & \multicolumn{3}{c}{Model 1} & \multicolumn{3}{c}{Model 2} & \multicolumn{3}{c}{Model 3} & \multicolumn{3}{c}{Model 4} & \multicolumn{3}{c}{Model 5}  \\
\hline
\hline

$K_{1}(1270)^{+}\rightarrow \rho(770)^{0}K^{+}$  & $ 50.71 $&$\pm$&$2.18 $  & $ 54.95 $&$\pm$&$3.36 $  & $ 49.38 $&$\pm$&$2.22 $  & $ 47.56 $&$\pm$&$2.16 $  & $ 48.89 $&$\pm$&$2.28 $  & $ 51.64 $&$\pm$&$2.18 $   \\
$K_{1}(1270)^{+}\rightarrow K^{*}(892)^{0}\pi^{+}$  & $ 19.86 $&$\pm$&$1.44 $  & $ 20.13 $&$\pm$&$1.71 $  & $ 20.33 $&$\pm$&$1.51 $  & $ 21.98 $&$\pm$&$1.56 $  & $ 22.07 $&$\pm$&$1.57 $  & $ 20.17 $&$\pm$&$1.45 $   \\
$K_{1}(1270)^{+}\rightarrow \left[K^{+}\pi^{-}\right]_{\text{S}}\pi^{+}$  & $ 11.35 $&$\pm$&$1.45 $  & $ 13.76 $&$\pm$&$3.25 $  & $ 11.44 $&$\pm$&$1.29 $  & $ 11.01 $&$\pm$&$1.35 $  & $ 10.05 $&$\pm$&$1.30 $  & $ 10.26 $&$\pm$&$1.25 $   \\
$K_{1}(1270)^{+}[D]\rightarrow K^{*}(892)^{0}\pi^{+}$  & $ 8.32 $&$\pm$&$0.85 $  & $ 8.13 $&$\pm$&$0.89 $  & $ 8.04 $&$\pm$&$0.88 $  & $ 7.96 $&$\pm$&$0.86 $  & $ 7.10 $&$\pm$&$0.83 $  & $ 8.26 $&$\pm$&$0.84 $   \\
$K_{1}(1270)^{+}\rightarrow \rho(1450)^{0}K^{+}$  & \multicolumn{3}{c}{$-$}  & $ 0.69 $&$\pm$&$1.03 $  & \multicolumn{3}{c}{$-$}  & \multicolumn{3}{c}{$-$}  & \multicolumn{3}{c}{$-$}  & \multicolumn{3}{c}{$-$}   \\
$K_{1}(1270)^{+}\rightarrow K^{*}(1410)^{0}\pi^{+}$  & \multicolumn{3}{c}{$-$}  & $ 0.00 $&$\pm$&$0.02 $  & \multicolumn{3}{c}{$-$}  & \multicolumn{3}{c}{$-$}  & \multicolumn{3}{c}{$-$}  & \multicolumn{3}{c}{$-$}   \\
$\text{Sum } K_{1}(1270)^{+}$  & $ 90.24 $&$\pm$&$1.83 $  & $ 97.66 $&$\pm$&$6.26 $  & $ 89.18 $&$\pm$&$1.70 $  & $ 88.50 $&$\pm$&$1.69 $  & $ 88.10 $&$\pm$&$1.65 $  & $ 90.34 $&$\pm$&$1.68 $   \\
\hline

\hline
\hline
\end{tabular}
 }
\label{tab:altModels1_2}
\end{table}

\begin{table}[h]
\centering
\caption{Fit results of the \textit{baseline} model and of alternative amplitude models 1-5, part 3. Uncertainties are statistical only.}
\resizebox{\linewidth}{!}{
	\renewcommand{\arraystretch}{1.1}
	\begin{tabular}{l l@{}r@{}r l@{}r@{}r l@{}r@{}r l@{}r@{}r l@{}r@{}r l@{}r@{}r} 
\hline
\hline
Fit parameter   & \multicolumn{3}{c}{\textit{Baseline}} & \multicolumn{3}{c}{Model 1} & \multicolumn{3}{c}{Model 2} & \multicolumn{3}{c}{Model 3} & \multicolumn{3}{c}{Model 4}  & \multicolumn{3}{c}{Model 5}  \\
\hline
\hline
$\XSone  \text{ mass [MeV]}$  & $ 4474.5 $&$\pm$&$6.9 $  & $ 4470.5 $&$\pm$&$6.0 $  & $ 4473.1 $&$\pm$&$7.0 $  & $ 4471.0 $&$\pm$&$5.2 $  & $ 4471.2 $&$\pm$&$5.8 $  & $ 4474.3 $&$\pm$&$6.1 $   \\
$\XSone  \text{ width [MeV]}$  & $ 230.8 $&$\pm$&$19.5 $  & $ 228.7 $&$\pm$&$17.1 $  & $ 210.5 $&$\pm$&$18.2 $  & $ 218.6 $&$\pm$&$15.2 $  & $ 218.4 $&$\pm$&$15.7 $  & $ 233.2 $&$\pm$&$17.7 $   \\
$\XAone \text{ mass [MeV]}$  & $ 4652.5 $&$\pm$&$14.4 $  & $ 4634.3 $&$\pm$&$21.1 $  & $ 4644.3 $&$\pm$&$15.2 $  & $ 4687.9 $&$\pm$&$12.7 $  & $ 4704.9 $&$\pm$&$14.9 $  & $ 4655.0 $&$\pm$&$14.0 $   \\
$\XAone \text{ width [MeV]}$  & $ 227.0 $&$\pm$&$26.3 $  & $ 246.7 $&$\pm$&$154.0 $  & $ 234.1 $&$\pm$&$28.0 $  & $ 249.4 $&$\pm$&$162.7 $  & $ 238.0 $&$\pm$&$32.5 $  & $ 223.2 $&$\pm$&$26.2 $   \\
$\XStwo  \text{ mass [MeV]}$  & $ 4710.2 $&$\pm$&$4.4 $  & $ 4710.2 $&$\pm$&$4.2 $  & $ 4710.5 $&$\pm$&$4.4 $  & $ 4707.0 $&$\pm$&$4.4 $  & $ 4707.4 $&$\pm$&$5.0 $  & $ 4713.1 $&$\pm$&$3.8 $   \\
$\XStwo  \text{ width [MeV]}$  & $ 63.6 $&$\pm$&$8.7 $  & $ 53.9 $&$\pm$&$8.0 $  & $ 61.2 $&$\pm$&$8.7 $  & $ 73.5 $&$\pm$&$9.3 $  & $ 66.6 $&$\pm$&$9.1 $  & $ 56.4 $&$\pm$&$7.4 $   \\
$\XVone \text{ mass [MeV]}$  & $ 4784.6 $&$\pm$&$37.4 $  & $ 4751.3 $&$\pm$&$29.3 $  & $ 4765.6 $&$\pm$&$33.2 $  & $ 4781.0 $&$\pm$&$41.8 $  & $ 4784.0 $&$\pm$&$36.5 $  & $ 4706.6 $&$\pm$&$29.9 $   \\
\hline
\hline
\end{tabular}
 }
\label{tab:altModels1_3}
\end{table}

\begin{table}[h]
\centering
\caption{Fit results of alternative amplitude models 6-11, part 1. Uncertainties are statistical only.}
\resizebox{\linewidth}{!}{
	\renewcommand{\arraystretch}{1.1}
	\begin{tabular}{l r r r r r r  } 
\hline
\hline 
Decay channel $F_{i} [ \% ] $ & Model 6 & Model 7 & Model 8 & Model 9 & Model 10 & Model 11 \\
\hline 
$B^{+}\rightarrow \XSone K^{+}$ & $ 13.29 \pm 1.61 $  & $ 20.50 \pm 1.35 $  & $ 22.65 \pm 1.57 $  & $ 18.12 \pm 1.19 $  & $ 15.38 \pm 1.11 $  & $ 16.86 \pm 1.81 $  \\
$B^{+}\rightarrow \psi(\text{2S})\,K^{*}(1680)^{+}$ & $ 9.07 \pm 1.46 $  & $ 6.89 \pm 1.19 $  & $ 8.09 \pm 1.36 $  & $ 7.40 \pm 1.32 $  & $ 15.22 \pm 2.20 $  & $ 10.53 \pm 2.01 $  \\
$B^{+}\rightarrow \psi(\text{2S})\,K_{1}(1270)^{+}$ & $ 10.08 \pm 0.97 $  & $ 6.95 \pm 0.78 $  & $ 8.06 \pm 0.88 $  & $ 9.44 \pm 0.84 $  & $ 8.90 \pm 1.05 $  & $ 8.64 \pm 1.52 $  \\
$B^{+}[P]\rightarrow \psi(\text{2S})\,K_{1}(1270)^{+}$ & $ 6.31 \pm 0.55 $  & $ 8.67 \pm 0.55 $  & $ 7.55 \pm 0.55 $  & $ 6.30 \pm 0.56 $  & $ 6.67 \pm 0.62 $  & $ 7.76 \pm 0.69 $  \\
$B^{+}[D]\rightarrow \psi(\text{2S})\,K_{1}(1270)^{+}$ & $ 5.68 \pm 0.40 $  & $ 6.84 \pm 0.44 $  & $ 6.45 \pm 0.43 $  & $ 6.49 \pm 0.43 $  & $ 8.85 \pm 0.47 $  & $ 9.51 \pm 0.51 $  \\
$B^{+}\rightarrow \psi(\text{2S})\,K_{1}(1400)^{+}$ & $ 4.52 \pm 0.62 $  & $ 5.29 \pm 0.70 $  & $ 5.77 \pm 0.62 $  & $ 4.77 \pm 0.61 $  & $ 5.76 \pm 0.66 $  & $ 6.11 \pm 0.86 $  \\
$B^{+}\rightarrow \psi(\text{2S})\,K(1460)^{+}$ & $ 3.45 \pm 0.39 $  & $ 4.80 \pm 0.42 $  & $ 3.71 \pm 0.29 $  & $ 5.17 \pm 0.41 $  & $ 4.94 \pm 0.42 $  & $ 4.57 \pm 0.43 $  \\
\hline
\hline
\end{tabular}
 }
\label{tab:altModels2_1}
\end{table}

\begin{table}[h]
\centering
\caption{Fit results of alternative amplitude models 6-11, part 2. Uncertainties are statistical only.}
\resizebox{\linewidth}{!}{
	\renewcommand{\arraystretch}{1.}
	\begin{tabular}{l r@{}c@{}l r@{}c@{}l r@{}c@{}l r@{}c@{}l r@{}c@{}l r@{}c@{}l} 
\hline
\hline
Decay channel $F_{i} [ \% ] $  & \multicolumn{3}{c}{Model 6} & \multicolumn{3}{c}{Model 7} & \multicolumn{3}{c}{Model 8} & \multicolumn{3}{c}{Model 9} & \multicolumn{3}{c}{Model 10}  & \multicolumn{3}{c}{Model 11}  \\
\hline 
$K_{1}(1270)^{+}\rightarrow \rho(770)^{0}K^{+}$ & $ 49.34 $&$\pm$&$2.18 $  & $ 49.37 $&$\pm$&$2.11 $  & $ 50.84 $&$\pm$&$2.15 $  & $ 44.94 $&$\pm$&$2.10 $  & $ 48.21 $&$\pm$&$2.34 $  & $ 48.42 $&$\pm$&$2.53 $  \\
$K_{1}(1270)^{+}\rightarrow K^{*}(892)^{0}\pi^{+}$ & $ 18.91 $&$\pm$&$1.56 $  & $ 21.47 $&$\pm$&$1.44 $  & $ 24.09 $&$\pm$&$1.48 $  & $ 19.63 $&$\pm$&$1.47 $  & $ 18.26 $&$\pm$&$1.34 $  & $ 20.06 $&$\pm$&$1.54 $  \\
$K_{1}(1270)^{+}\rightarrow \left[K^{+}\pi^{-}\right]_{\text{S}}\pi^{+}$ & $ 10.10 $&$\pm$&$1.22 $  & $ 11.22 $&$\pm$&$1.22 $  & $ 10.54 $&$\pm$&$1.18 $  & $ 10.90 $&$\pm$&$1.19 $  & $ 12.37 $&$\pm$&$1.29 $  & $ 10.88 $&$\pm$&$1.22 $  \\
$K_{1}(1270)^{+}[D]\rightarrow K^{*}(892)^{0}\pi^{+}$ & $ 9.59 $&$\pm$&$0.92 $  & $ 8.01 $&$\pm$&$0.83 $  & $ 7.68 $&$\pm$&$0.81 $  & $ 12.48 $&$\pm$&$1.01 $  & $ 8.70 $&$\pm$&$0.90 $  & $ 8.81 $&$\pm$&$1.01 $  \\
$\text{Sum } K_{1}(1270)^{+}$ & $ 87.94 $&$\pm$&$1.56 $  & $ 90.08 $&$\pm$&$1.67 $  & $ 93.15 $&$\pm$&$1.65 $  & $ 87.94 $&$\pm$&$1.59 $  & $ 87.54 $&$\pm$&$1.56 $  & $ 88.17 $&$\pm$&$1.54 $  \\ \hline
$K_{1}(1400)^{+}\rightarrow K^{*}(892)^{0}\pi^{+}$ & $ 90.40 $&$\pm$&$3.93 $  & $ 76.80 $&$\pm$&$4.75 $  & $ 88.49 $&$\pm$&$3.97 $  & $ 84.07 $&$\pm$&$4.22 $  & $ 80.86 $&$\pm$&$4.56 $  & $ 80.71 $&$\pm$&$7.56 $  \\
$K_{1}(1400)^{+}\rightarrow \rho(770)^{0}K^{+}$ & $ 18.42 $&$\pm$&$4.92 $  & $ 30.89 $&$\pm$&$5.05 $  & $ 20.08 $&$\pm$&$4.70 $  & $ 25.40 $&$\pm$&$4.88 $  & $ 31.86 $&$\pm$&$5.12 $  & $ 31.58 $&$\pm$&$8.05 $  \\
\hline
\hline
\end{tabular}
 }
\label{tab:altModels2_2}
\end{table}

\begin{table}[h]
\centering
\caption{Fit results of alternative amplitude models 6-11, part 3. Uncertainties are statistical only.}
\resizebox{\linewidth}{!}{
	\renewcommand{\arraystretch}{1.1}
	\begin{tabular}{l l@{}r@{}r l@{}r@{}r l@{}r@{}r l@{}r@{}r l@{}r@{}r l@{}r@{}r} 
\hline
\hline
Fit parameter  & \multicolumn{3}{c}{Model 6} & \multicolumn{3}{c}{Model 7} & \multicolumn{3}{c}{Model 8} & \multicolumn{3}{c}{Model 9} & \multicolumn{3}{c}{Model 10}  & \multicolumn{3}{c}{Model 11}  \\
\hline 
$\XSone  \text{ mass [MeV]}$ & $ 4461.1 $&$\pm$&$ 6.7 $  & $ 4474.2 $&$\pm$&$ 5.4 $  & $ 4483.1 $&$\pm$&$ 5.9 $  & $ 4478.2 $&$\pm$&$ 6.1 $  & $ 4461.4 $&$\pm$&$ 5.6 $  & $ 4468.4 $&$\pm$&$ 7.5 $  \\
$\XSone  \text{ width [MeV]}$ & $ 193.7 $&$\pm$&$ 17.7 $  & $ 227.8 $&$\pm$&$ 16.3 $  & $ 248.4 $&$\pm$&$ 17.4 $  & $ 226.2 $&$\pm$&$ 16.3 $  & $ 213.2 $&$\pm$&$ 15.5 $  & $ 232.3 $&$\pm$&$ 22.4 $  \\
$\XAone \text{ mass [MeV]}$ & $ 4691.7 $&$\pm$&$ 14.4 $  & $ 4662.0 $&$\pm$&$ 13.8 $  & $ 4673.1 $&$\pm$&$ 12.9 $  & $ 4680.2 $&$\pm$&$ 12.0 $  & $ 4628.5 $&$\pm$&$ 18.4 $  & $ 4651.0 $&$\pm$&$ 15.9 $  \\
$\XAone \text{ width [MeV]}$ & $ 250.0 $&$\pm$&$ 151.6 $  & $ 243.2 $&$\pm$&$ 26.6 $  & $ 219.9 $&$\pm$&$ 26.8 $  & $ 202.8 $&$\pm$&$ 24.5 $  & $ 250.0 $&$\pm$&$ 176.8 $  & $ 239.3 $&$\pm$&$ 30.1 $  \\
$\XStwo  \text{ mass [MeV]}$ & $ 4702.7 $&$\pm$&$ 5.9 $  & $ 4710.8 $&$\pm$&$ 3.9 $  & $ 4705.6 $&$\pm$&$ 4.4 $  & $ 4708.6 $&$\pm$&$ 4.1 $  & $ 4702.8 $&$\pm$&$ 4.8 $  & $ 4706.1 $&$\pm$&$ 6.7 $  \\
$\XStwo  \text{ width [MeV]}$ & $ 72.0 $&$\pm$&$ 10.0 $  & $ 68.3 $&$\pm$&$ 8.9 $  & $ 71.0 $&$\pm$&$ 9.3 $  & $ 58.0 $&$\pm$&$ 7.2 $  & $ 75.0 $&$\pm$&$ 10.3 $  & $ 66.7 $&$\pm$&$ 11.5 $  \\
$\XVone \text{ mass [MeV]}$ & $ 4727.4 $&$\pm$&$ 23.6 $  & $ 5100.0 $&$\pm$&$ 30.8 $  & $ 4767.1 $&$\pm$&$ 31.7 $  & $ 4904.1 $&$\pm$&$ 38.1 $  & $ 4779.9 $&$\pm$&$ 33.9 $  & $ 4798.2 $&$\pm$&$ 44.2 $  \\
\hline
\hline
\end{tabular}
 }
\label{tab:altModels2_3}
\end{table}

\clearpage
\section{Resonant substructure of the $K_1(1270)^+$}
\label{a:k1Results}

\renewcommand{\thefigure}{E.\arabic{figure}}
\renewcommand{\thetable}{E.\arabic{table}}
\setcounter{table}{0}
\setcounter{figure}{0}

The $K_1(1270)^+$ resonance is the prominent contribution to $\signal$ decays and thus serves as a reference channel for the exotic contributions.
Table~\ref{tab:k1Bfs} compares the branching fractions of the $K_1(1270)$ resonances obtained from the baseline model to those from a $\norm$ amplitude analysis by Belle~\cite{BELLE}
and to those from scattering experiments that constitute the basis for the PDG values~\cite{PDG2022}.
The largest difference between the PDG and Belle branching fractions is observed for the $K\pi$ S-wave, which was considered to be a $K_0^*(1430)^0$ resonance in those measurements.
We use a more sophisticated K-matrix approach instead
and obtain a reasonable result between the PDG and Belle results.

To translate the fit fractions from Table~\ref{tab:fitBest} to branching fractions, we include isospin factors~\cite{BELLE} and
combine the $K_1(1270)^+[S] \to K^*(892)^0\pip$ and $K_1(1270)^+[D] \to K^*(892)^0\pip$ decay fractions taking their interference into account.
We describe $\rho(770)^0 - \omega$ mixing with a combined lineshape~\ref{eq:rhoOmegaT}.
For the comparison of the branching fractions, it is instructive to separate the $\rho(770)^0$ and $\omega$ contributions.
We adapt the procedure from the \mbox{$\chi_{c1}(3872) \to J/\psi \rho(770)^0/\omega$} analysis~\cite{LHCb-PAPER-2021-045}
to quantify the relative rate of the $\rho(770)^0$ and $\omega$ contributions by computing the integrals
\begin{align}
I(s_{123} \vert a, \delta) &= \int_{(2m_\pi)^2}^{(\sqrt{s_{123}}-m_K)^2} \, \vert A(s_{123}, s_{23} \vert a, \delta) \vert^2 \, ds_{23},
\end{align}
where $s_{123}$ and $s_{23}$ refer to the $\Kp\pip\pim$ and $\pip\pim$ invariant-mass squared.
The amplitude is given by $A(s_{123}, s_{23} \vert a, \delta) = q(s_{123}, s_{23}) \, F_1(\Tilde{q}(s_{23})) \, \Tilde{q}(s_{23})^3 \, T_{GS}(s_{23}) \left( a + \delta \frac{s_{23}}{m_\omega^2} T_\omega(s_{23}) \right)$, 
where $q$ and $\Tilde{q}$ refer to the break-up momentum of the $K_1(1270)^+$ and $\rho(770)^0/\omega$ decays.
In contrast to Ref.~\cite{LHCb-PAPER-2021-045} which evaluates the integral at the resonance mass, \ie \mbox{$I(s_{123}=m_{K_1(1270)^+}^2 \vert a, \delta)$}, we take the broad lineshape of the $K_1(1270)^+$ into account by
weighting with the $K_1(1270)^+$ Breit--Wigner, $T_{K_1(1270)^+}(s_{123})$, as follows:
\begin{align}
\langle I(a, \delta) \rangle &= \int_{(m_K+2m_\pi)^2}^{(m_B-m_\psitwos)^2} \, I(s_{123}\vert a, \delta) \, \vert T_{K_1(1270)^+}(s_{123})\vert^2\, ds_{123}.
\end{align}
A measure of the $\omega$ ($\rho(770)^0$) contribution is then obtained as 
$R_{\omega} = \frac{\langle I(a=0, \delta) \rangle}{\langle I(a=1, \delta) \rangle}$
($R_{\rho} = 1-R_{\omega})$.
We obtain \mbox{$R_{\omega} = (0.49 \pm 0.24) \%$}, which is then corrected with \mbox{$\mathcal{B}(\omega\to\pi\pi)=(1.53\pm0.12)\%$}~\cite{PDG2022} to determine the $K_1(1270) \to K \omega$ branching fraction.
\begin{table}[b]
\centering
\caption{Comparison of branching fractions for \( K_1(1270) \) decays according to the PDG~\cite{PDG2022} and to the Belle amplitude analysis~\cite{BELLE} with fixed (Fit 1) and free (Fit 2) mass and width of the \( K_1(1270) \). }
\label{tab:k1Bfs}
\small
\begin{tabular}{l r@{}c@{}l r@{}c@{}l r@{}c@{}l r@{}c@{}l}
\hline
\hline
Decay mode & \multicolumn{3}{l}{PDG} (\%) & \multicolumn{3}{l}{Belle Fit 1 (\%)} & \multicolumn{3}{l}{Belle Fit 2 (\%)} & \multicolumn{3}{l}{\textit{Baseline} (\%)} \\
\hline
\( K \rho(770) \)            & $ 42 $&$\pm$&$ 6 $    & $57.3 $&$\pm$&$ 3.5 $ & $ 58.4 $&$\pm$&$ 4.3 $ &  $ 58.8 $&$\pm$&$ 4.5 $ \\
\( K \omega \)          & $ 11 $&$\pm$&$ 2 $    & $ 14.8 $&$\pm$&$ 4.7 $  & $ 22.5 $&$\pm$&$ 5.2 $ &   $ 6.3 $&$\pm$&$ 3.1 $ \\
\( K^*(892)\pi \)       & $ 16 $&$\pm$&$ 5 $    & $ 26.0 $&$\pm$&$ 2.1 $  & $ 17.1 $&$\pm$&$ 2.3 $ & $ 25.0 $&$\pm$&$ 6.3 $ \\
\( [K\pi]_S \pi \)      & $ 28 $&$\pm$&$ 4 $    & $ 1.90 $&$\pm$&$ 0.66 $ & $ 2.01 $&$\pm$&$ 0.64 $ &   $ 9.9 $&$\pm$&$ 2.2 $ \\
\( K f_0(1370) \)       & $ 3 $&$\pm$&$ 2 $     &  \multicolumn{3}{c}{$-$}   & \multicolumn{3}{c}{$-$} & \multicolumn{3}{c}{$-$} \\
\hline
\hline
\end{tabular}
\end{table}

The LHCb collaboration has previously studied the $K_1(1270)^+$ substructure in 
\mbox{$\Dz \to \Kp \Km \pip \pim$}~\cite{LHCB-PAPER-2018-041},
\mbox{$\Dzb \to \Kp \pim \pip \pim$}~\cite{LHCb-PAPER-2017-040}
and \mbox{$B_s^0 \to D_s^- \Kp \pip\pim$}~\cite{LHCB-PAPER-2020-030} decays.
There are several notable differences in the parameterisations used.
The charm amplitude analyses model the $\pi\pi$ P-wave as the sum of
$\rho(770)^0$, $\omega$ and $\rho(1450)^0$ Breit--Wigner functions.
In contrast, a special $\rho(770)^0-\omega$ mixing lineshape is used for 
$B_s^0 \to D_s^- \Kp \pip\pim$ and $\signal$ decays, without a $\rho(1450)^0$ contribution.
The LASS parameterisation~\cite{Aston:1987ir,Aubert:2005ce} is used for the $K\pi$ S-wave 
in $B_s^0 \to D_s^- \Kp \pip\pim$ decays,
while the others use a K-matrix approach.

Table~\ref{tab:k1FFs} compares the $K_1(1270)^+$ fit fractions from those measurements to 
the results of the $\signal$ \textit{baseline} model.
At first glance, the results appear rather inconsistent.
There is a large $\rho(1450)^0$ contribution in $\Dzb \to \Kp \pim \pip \pim$ decays, in contrast to the other measurements.
The $\rho(1450)^0$ resonance has a huge negative interference with the $\rho(770)^0$ resonance that may hint at overfitting.
In $\Dz \to \Kp \Km \pip \pim$ decays, $F_i(\Kp \rho(770)^0) \lessapprox F_i(K^*(892)^0 \pip)$ is observed.
The other decays have $F_i(\Kp \rho(770)^0) \gg F_i(K^*(892)^0 \pip)$.
As the fit fractions are obtained by integrating over the respective decay phase space,
this can be understood as a consequence of different phase-space limits.
The kinematically allowed $m(\Kp\pip\pim)$ range extends up to 
1367\,MeV, 1725\,MeV, 3398\,MeV and 1593\,MeV
for $\Dz \to \Kp \Km \pip \pim$, $\Dzb \to \Kp \pim \pip \pim$, 
$B_s^0 \to D_s^- \Kp \pip\pim$
and $\signal$ decays, respectively.\footnote{The considered
$B_s^0 \to D_s^- \Kp \pip\pim$ phase-space region is limited to $m(K\pi\pi)<1950 \mev$ in Ref.~\cite{LHCB-PAPER-2020-030}.}
The $\Kp\rho(770)^0$ threshold is close to the $K_1(1270)^+$ mass,
which leads to a highly asymmetric contribution of the 
$\Kp\rho(770)^0$ channel to the $K_1(1270)^+$ lineshape. 
Therefore a kinematic cut-off in the $m(\Kp\pip\pim)$ range removes more of the $\Kp\rho(770)^0$
than of the $K^*(892)^0 \pip$ contribution.
This effect becomes negligible when the cut-off is sufficiently far away from the 
$K_1(1270)^+$ mass.

\begin{figure}[h]
       	 \includegraphics[width=0.329\textwidth,height=!]{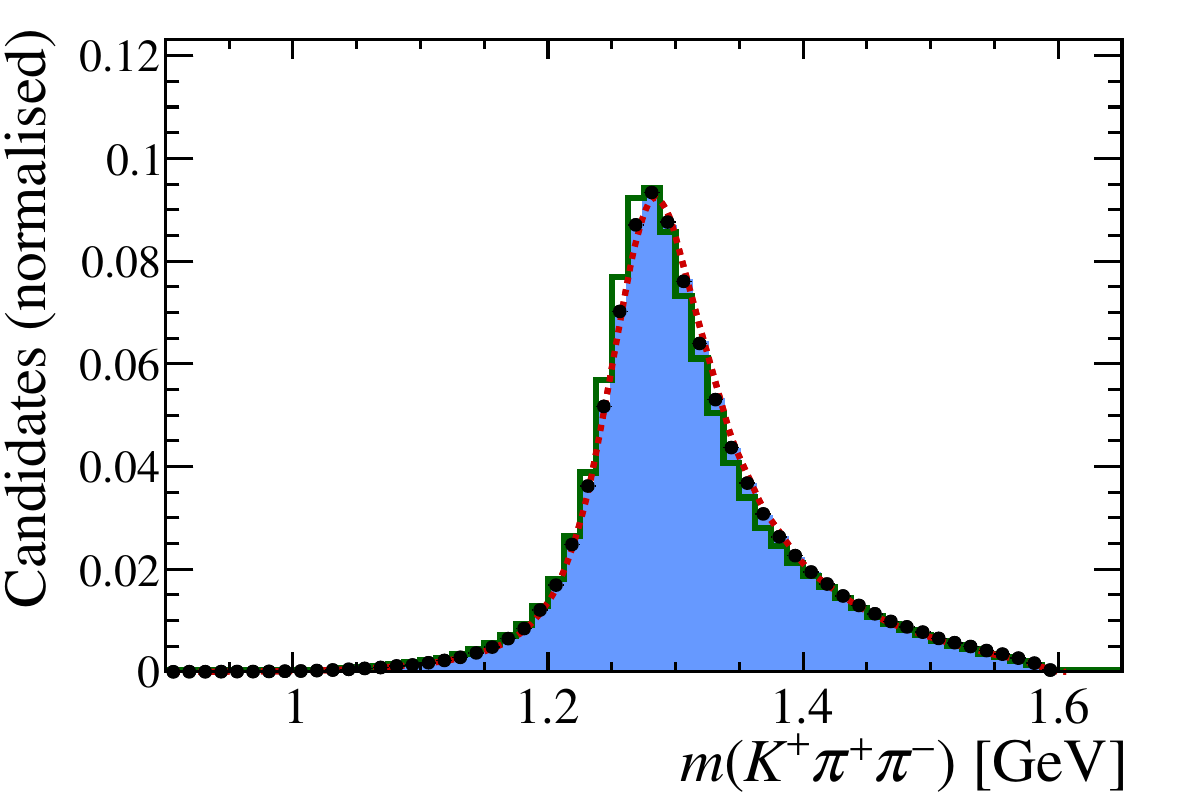}
       	 \includegraphics[width=0.329\textwidth,height=!]{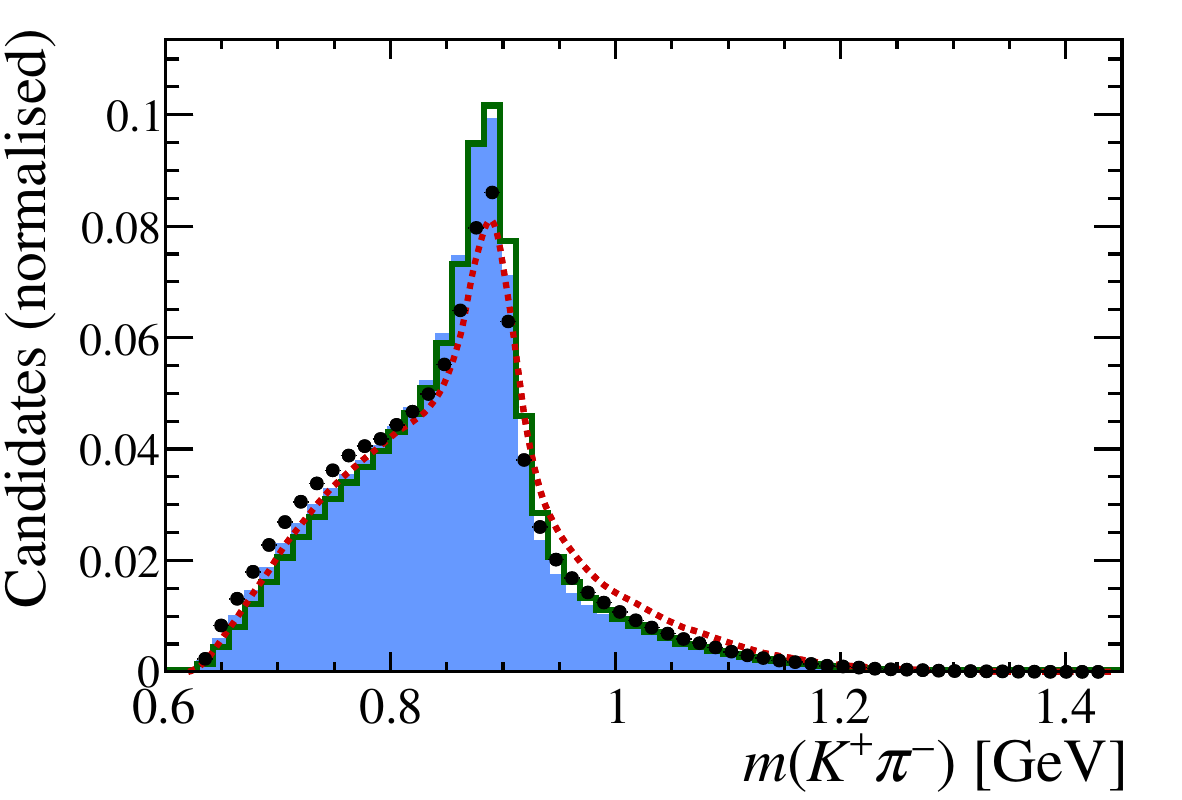}
       	 \includegraphics[width=0.329\textwidth,height=!]{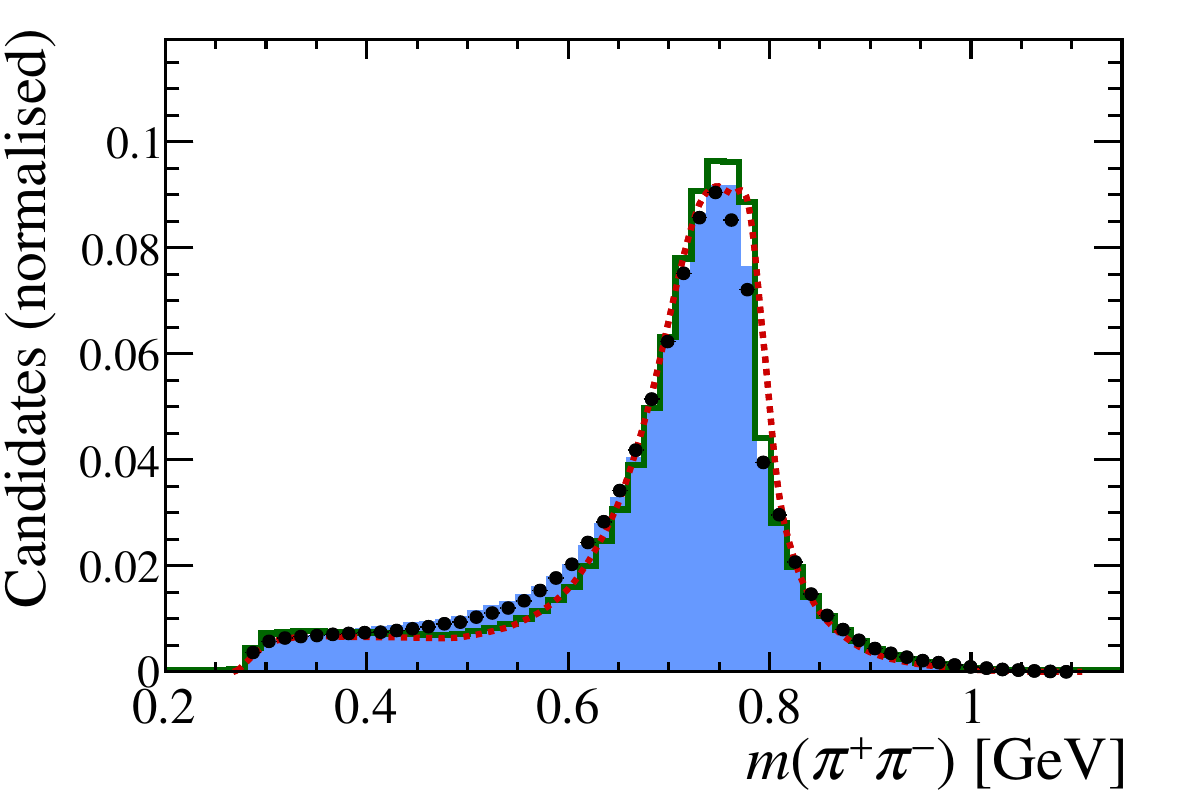}

       	 \includegraphics[width=0.329\textwidth,height=!]{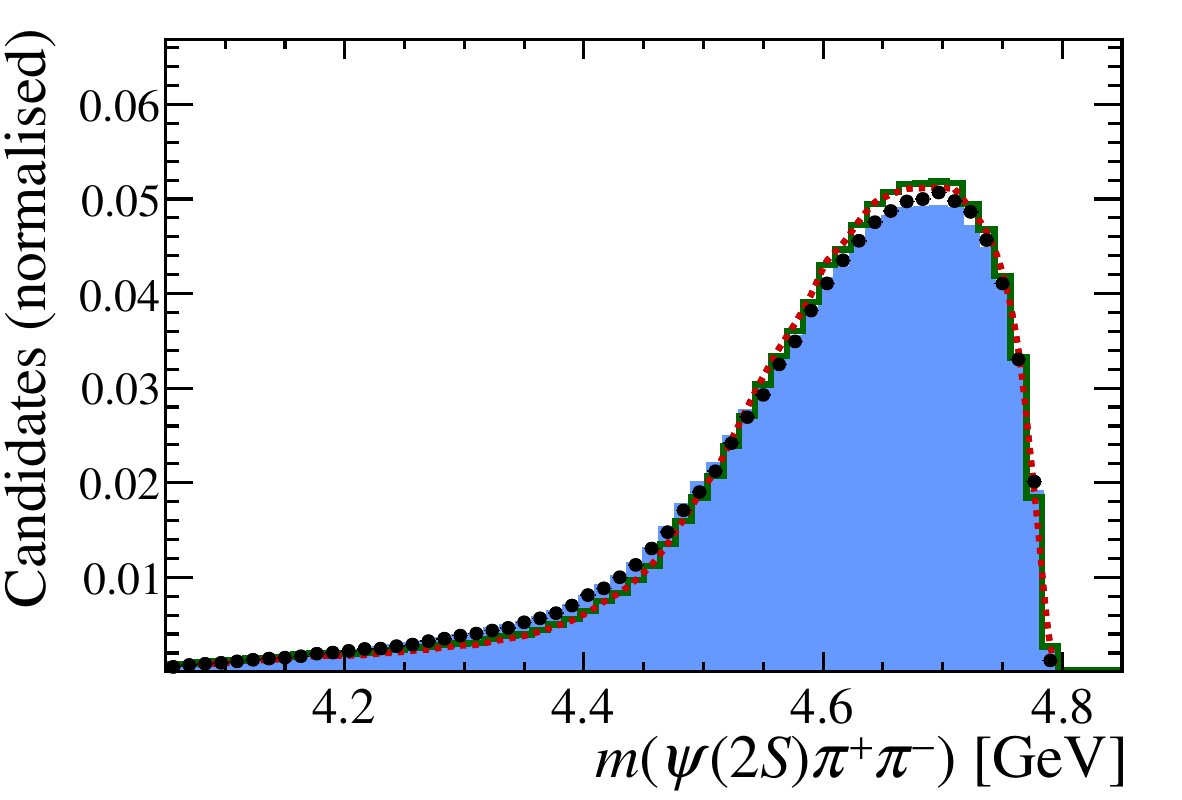}
       	 \includegraphics[width=0.329\textwidth,height=!]{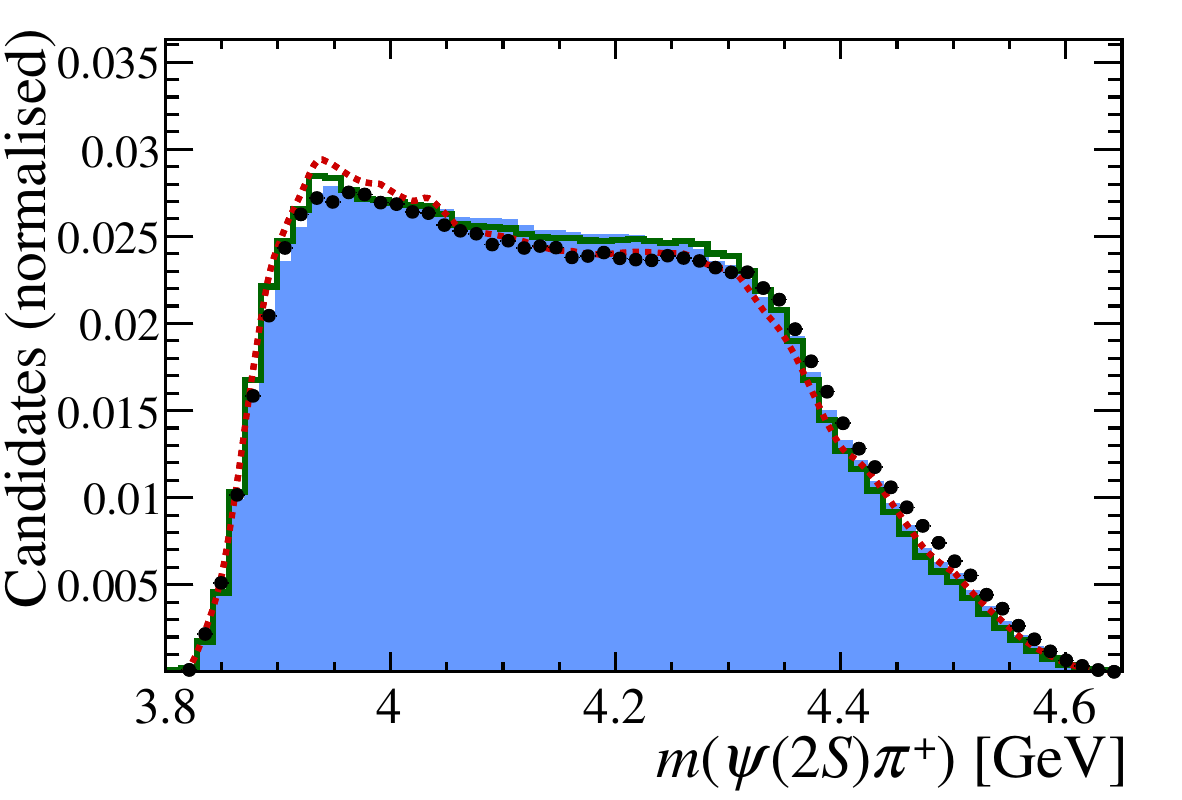}
       	 \includegraphics[width=0.329\textwidth,height=!]{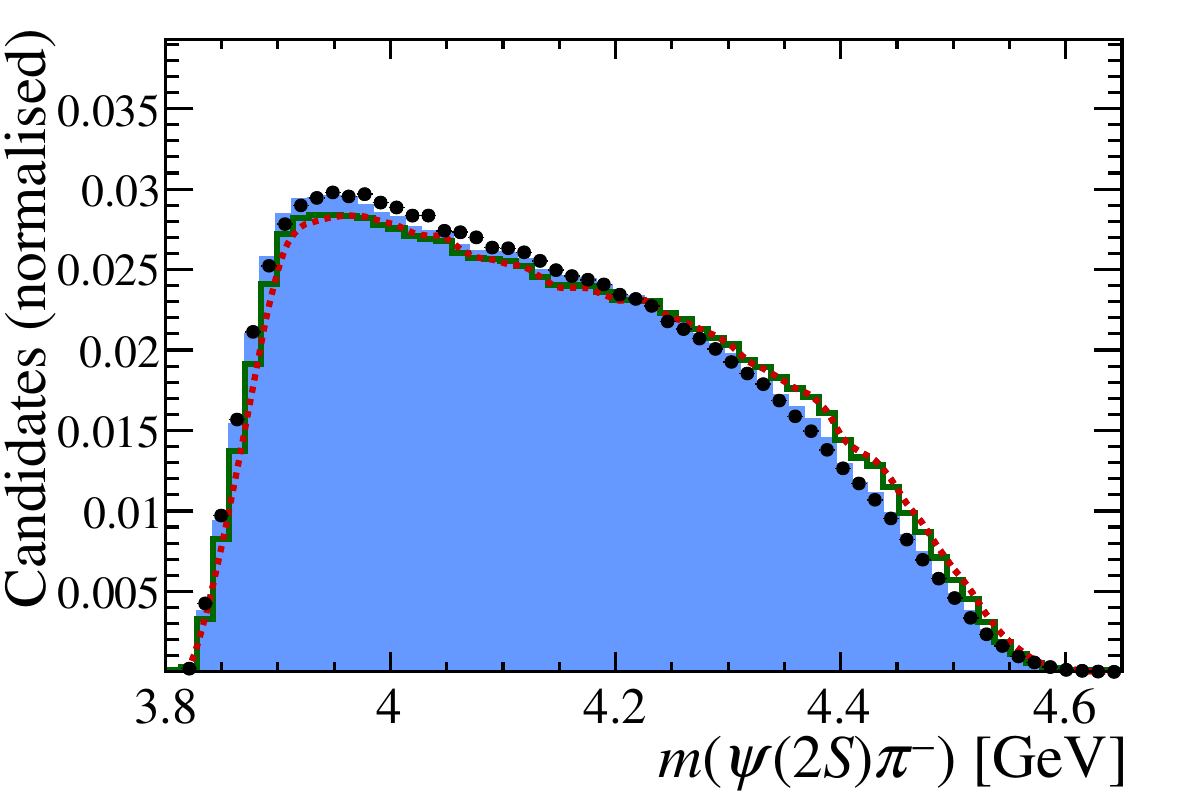}

       	 \includegraphics[width=0.329\textwidth,height=!]{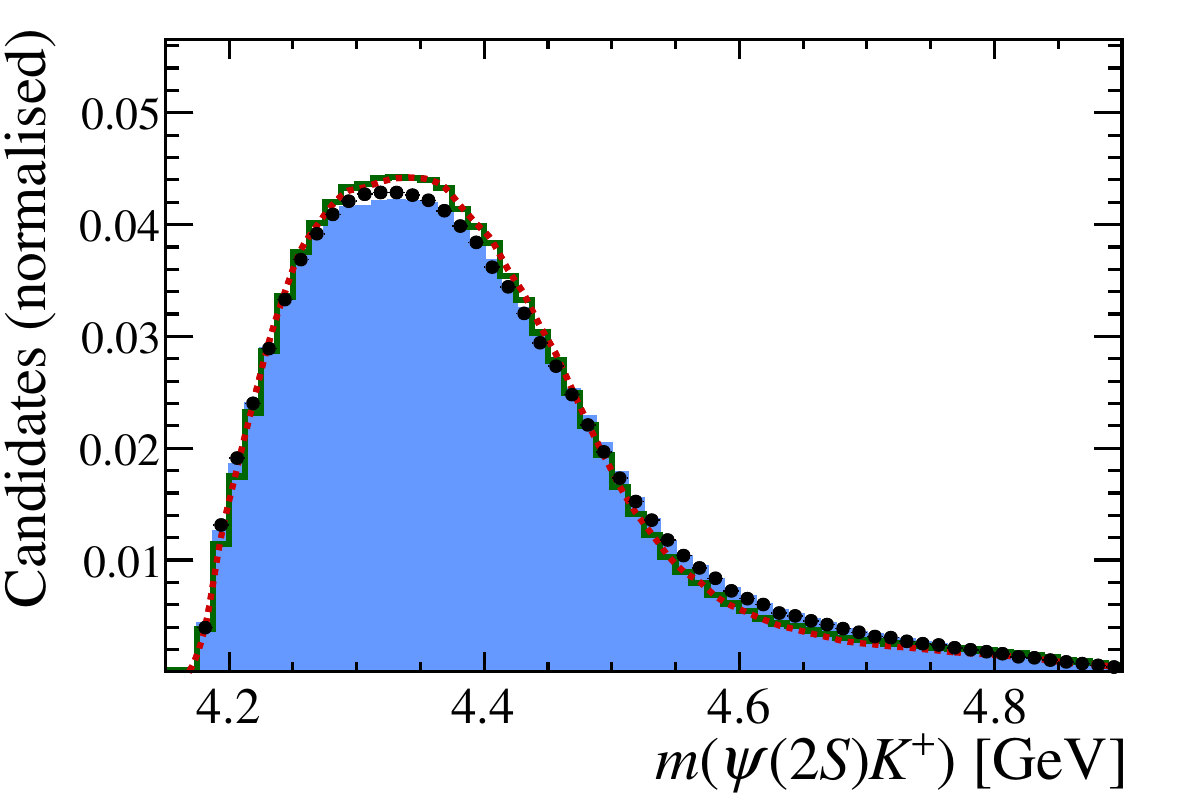}
       	 \includegraphics[width=0.329\textwidth,height=!]{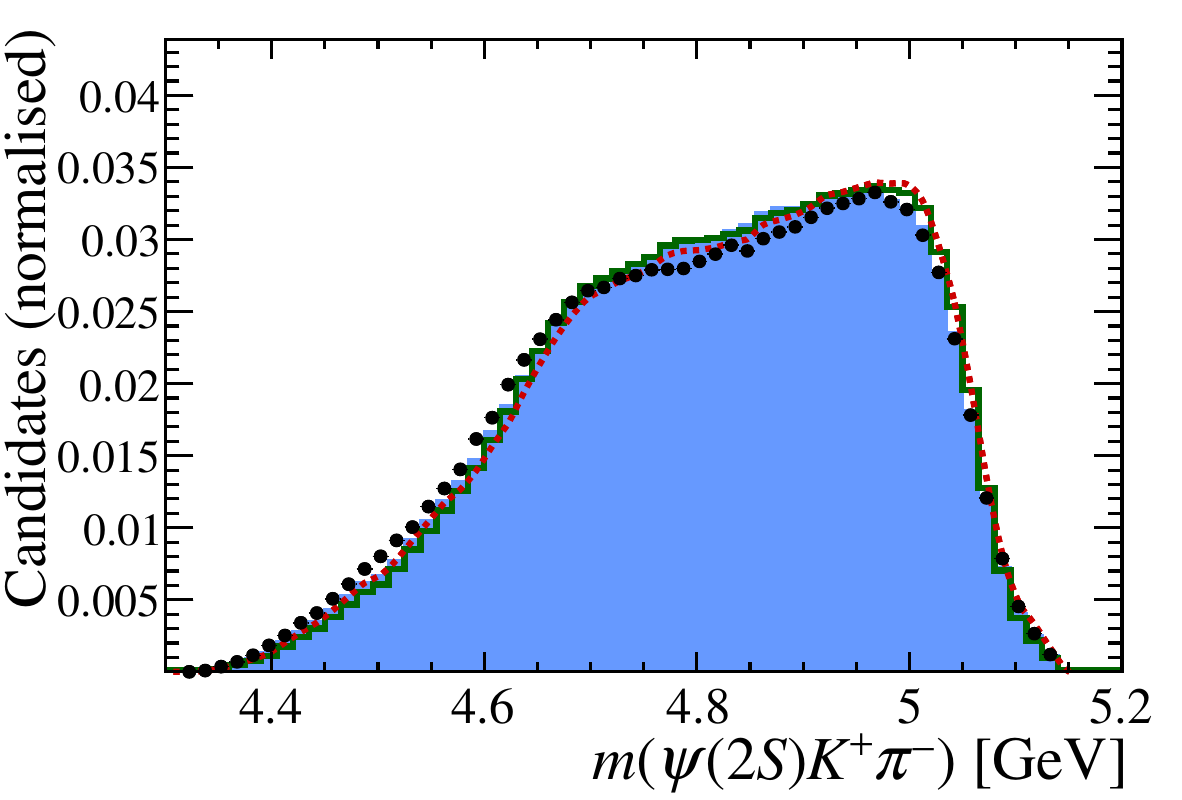}
       	 \includegraphics[width=0.329\textwidth,height=!]{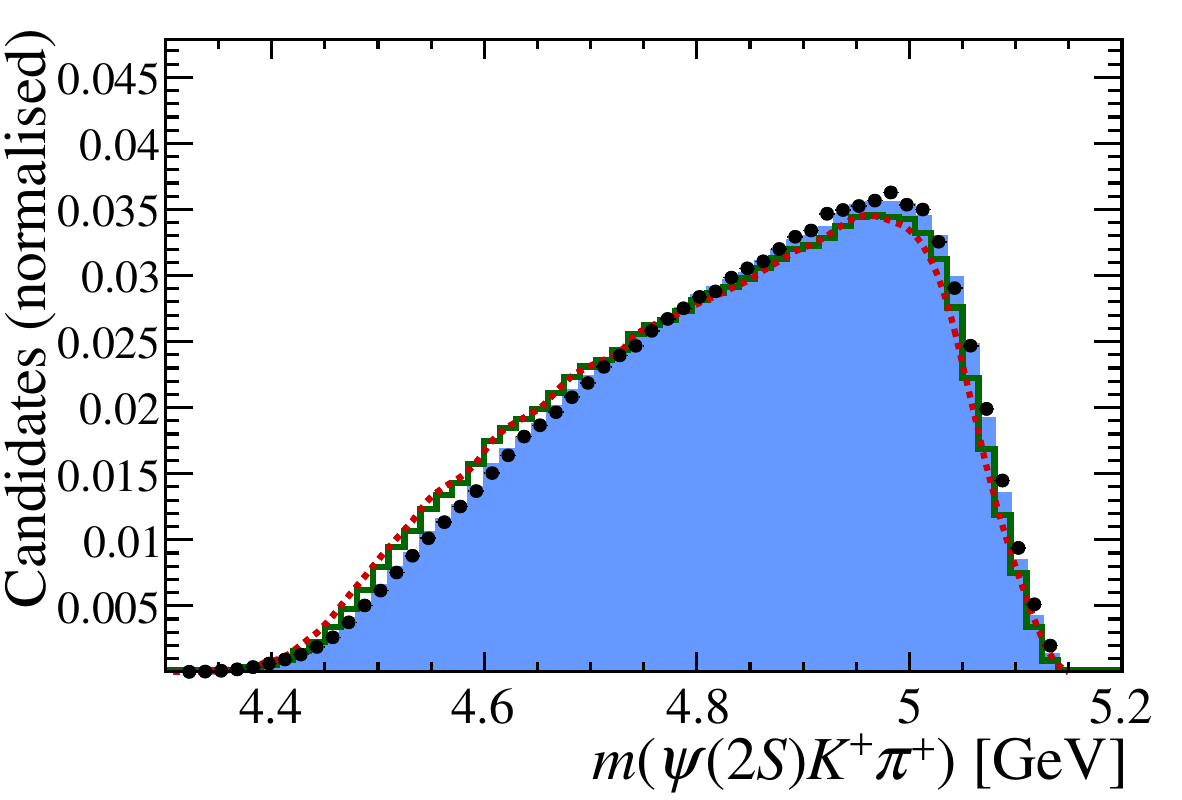}
              
       	 \includegraphics[width=0.329\textwidth,height=!]{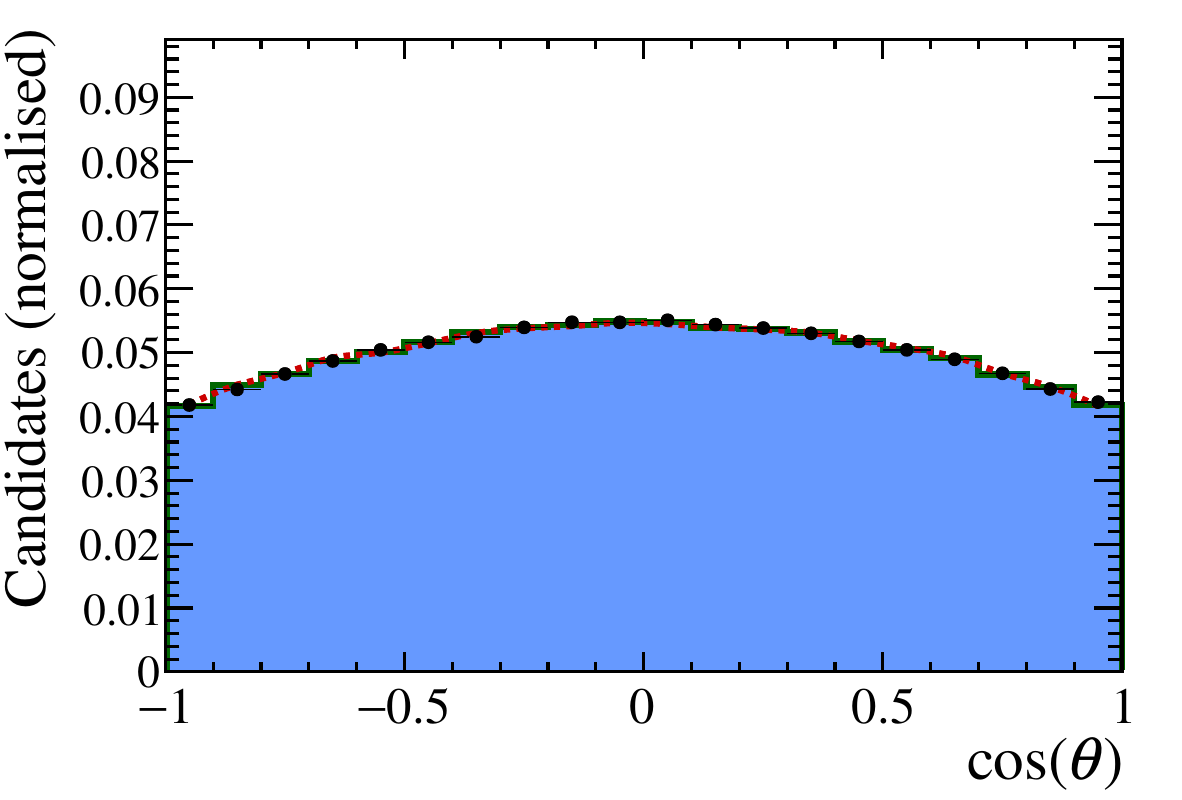}
       	 \includegraphics[width=0.329\textwidth,height=!]{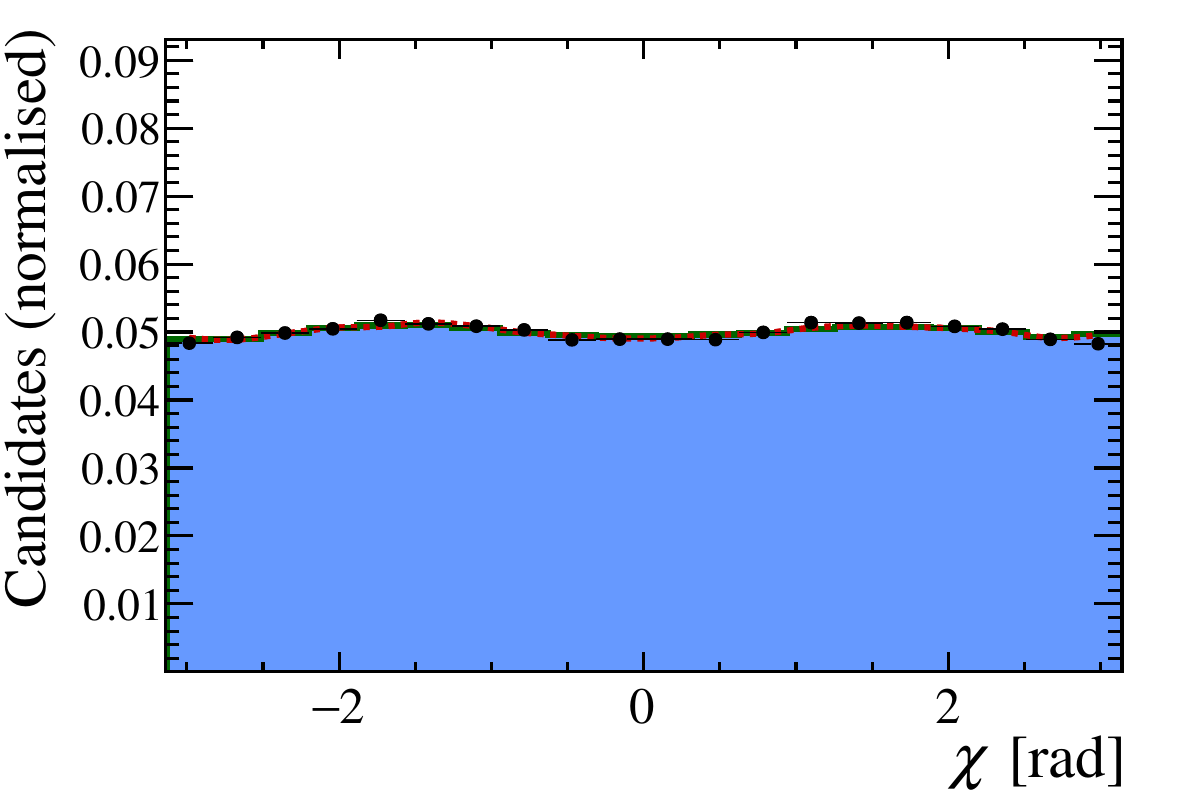}
       	 \includegraphics[width=0.329\textwidth,height=!]{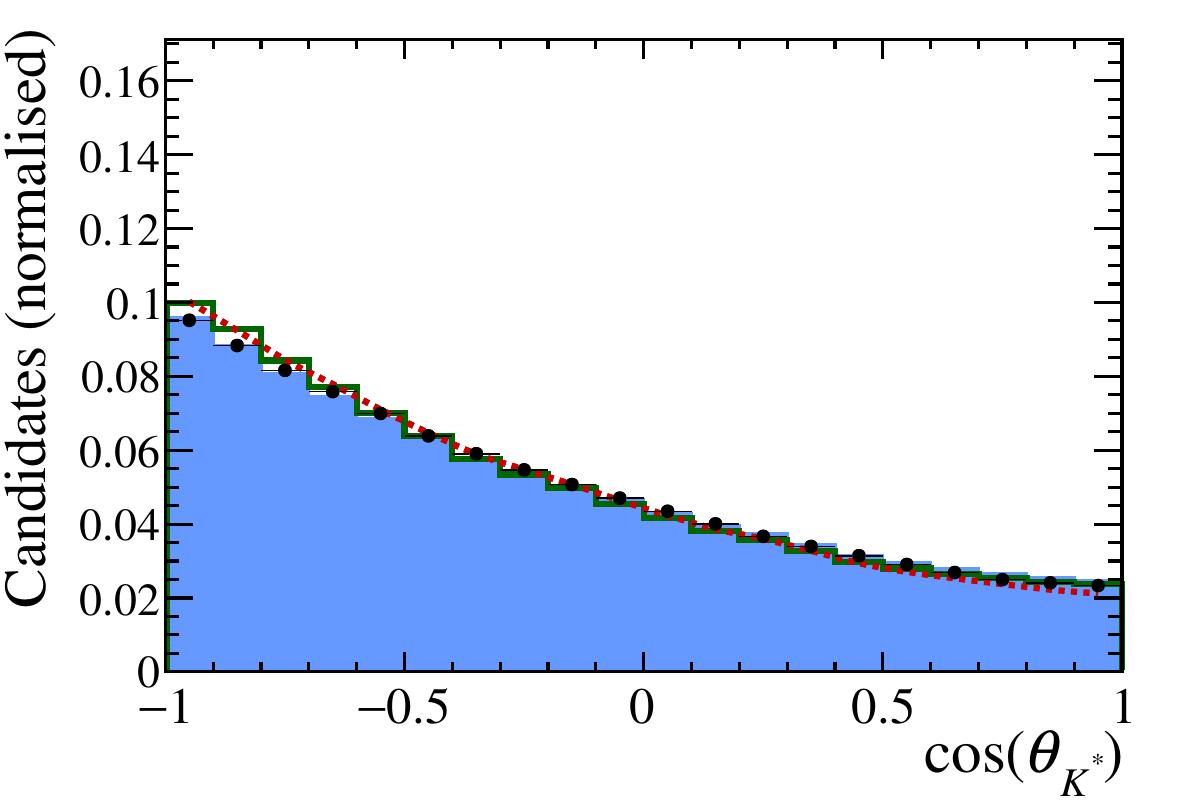}
         
	\caption{
 Phase-space distribution of pseudo $B^+\to K_1(1270)^+ \psitwos$ events
generated using the $\signal$
\textit{baseline} model (black data points)
or the amplitude models from the 
$\Dz \to \Kp \Km \pip \pim$~\cite{LHCB-PAPER-2018-041} (green histograms),
$\Dzb \to \Kp \pim \pip \pim$~\cite{LHCb-PAPER-2017-040} (dashed, red lines)
and $B_s^0 \to D_s^- \Kp \pip\pim$~\cite{LHCB-PAPER-2020-030} (filled, blue histograms) analyses.
    }
         \label{fig:k1Toys}

\end{figure}

For a better comparison of the results, 
we thus recompute the fit fractions in a consistent phase-space region.
First, we take the $K_1(1270)^+$ amplitude models from the $\Dz \to \Kp \Km \pip \pim$,
$\Dzb \to \Kp \pim \pip \pim$
and $B_s^0 \to D_s^- \Kp \pip\pim$ measurements
and generate pseudoexperiments for $B^+\to K_1(1270)^+ \psitwos$ decays.
Figure~\ref{fig:k1Toys} compares the resulting phase-space distributions.
Given that these models were obtained from entirely different decays 
that have a multitude of non-$K_1(1270)^+$ components
and use different parameterisations, 
the distributions agree reasonably well.
The corresponding fit fractions, now evaluated in the $\signal$ phase space,
are given in Table~\ref{tab:k1FFs2}.
Here, we compute a combined $\pip\pim$ P-wave fit fraction and 
a combined $[K^*(892)^0 \pi]_S + [K^*(892)^0 \pi]_D$ fit fraction
for a better comparability.
The obtained fit fractions are fairly consistent.
One outlier is a large $[K\pi]_S \pi$ fit fraction from the $\Dzb \to \Kp \pim \pip \pim$ model.

\begin{table}[b]
\centering
\caption{Fit fractions of the $K_1(1270)^+$ substructure in 
\mbox{$\Dz \to \Kp \Km \pip \pim$}~\cite{LHCB-PAPER-2018-041},
\mbox{$\Dzb \to \Kp \pim \pip \pim$}~\cite{LHCb-PAPER-2017-040}
and \mbox{$B_s^0 \to D_s^- \Kp \pip\pim$}~\cite{LHCB-PAPER-2020-030} decays
compared to the results of the $\signal$ \textit{baseline} model.
}
\label{tab:k1FFs}

\resizebox{1.\linewidth}{!}{
		 \footnotesize
		\centering
		\renewcommand{\arraystretch}{1.25}
\begin{tabular}{l| r@{}c@{}c@{}c@{}l r@{}c@{}c@{}c@{}l r@{}c@{}c@{}c@{}c@{}c@{}l  r@{}c@{}c@{}c@{}l }
\hline
\hline
Fit fraction ($\%$) & \multicolumn{5}{c}{$\Dz \to \Kp \Km \pip \pim$} & \multicolumn{5}{c}{$\Dzb \to \Kp \pim \pip \pim$} & \multicolumn{7}{c}{$B_s^0 \to D_s^- \Kp \pip\pim$} & \multicolumn{5}{c}{$\signal$} \\
\hline

\( \Kp [\rho(770)^0/\omega] \) &  $49.58 $&$\pm$&$ 1.99 $&$\pm$&$ 4.35$ &  \multicolumn{5}{c}{$-$} &  $52.5 $&$\pm$&$  4.6 $&$\pm$&$ 5.9  $&$\pm$&$ 6.9$ & $50.71 $&$\pm$&$ 2.18 $&$\pm$&$ 3.19$  \\

\( \Kp \rho(770)^0 \) &  \multicolumn{5}{c}{$-$} &  $96.30 $&$\pm$&$ 1.64 $&$\pm$&$ 6.61$ &  \multicolumn{7}{c}{$-$} &  \multicolumn{5}{c}{$-$}  \\

\( \Kp \omega \) &  \multicolumn{5}{c}{$-$} &  $1.65 $&$\pm$&$ 0.11 $&$\pm$&$ 0.16$ & \multicolumn{7}{c}{$-$} &   \multicolumn{5}{c}{$-$}  \\

\( \Kp \rho(1450)^0 \) &  $1.50 $&$\pm$&$ 0.47 $&$\pm$&$ 1.04$ &  $49.09 $&$\pm$&$ 1.58 $&$\pm$&$ 11.54$ &  \multicolumn{7}{c}{$-$}  &   \multicolumn{5}{c}{$-$}  \\

\( [K^*(892)^0 \pip]_S \) &  $51.22 $&$\pm$&$ 1.06 $&$\pm$&$ 3.21$ &  $27.08 $&$\pm$&$ 0.64 $&$\pm$&$ 2.82$ & $42.7  $&$\pm$&$ 7.9 $&$\pm$&$ 8.9 $&$\pm$&$ 11.2$ &  $19.86 $&$\pm$&$ 1.44 $&$\pm$&$ 2.05$  \\

\( [K^*(892)^0 \pip]_D \) &  $2.03 $&$\pm$&$ 0.17 $&$\pm$&$ 0.20$ &  $3.47 $&$\pm$&$ 0.17 $&$\pm$&$ 0.31$ & \multicolumn{7}{c}{$-$}  &   $8.32 $&$\pm$&$ 0.85 $&$\pm$&$ 1.54$  \\

\( [K\pi]_S \pip \) &  $6.27 $&$\pm$&$ 0.48 $&$\pm$&$ 1.66$ &  $22.90 $&$\pm$&$ 0.72 $&$\pm$&$ 1.89$ &  $1.2 $&$\pm$&$  1.6 $&$\pm$&$ 3.3  $&$\pm$&$ 1.3$ &  $11.35 $&$\pm$&$ 1.45 $&$\pm$&$ 2.11$ \\ \hline

Sum&  $110.60 $&$\pm$&$ 2.20 $&$\pm$&$ 5.76$ &  \multicolumn{5}{c}{$200.49$} &  \multicolumn{7}{c}{$106.4$} &  $90.24 $&$\pm$&$ 1.83 $&$\pm$&$ 3.67$  \\

\hline
\hline

\end{tabular}
}
\end{table}

\begin{table}[h]
\centering
\caption{Fit fractions of the $K_1(1270)^+$ substructure in $B^+\to K_1(1270)^+ \psitwos$ decays using the amplitude models from 
\mbox{$\Dz \to \Kp \Km \pip \pim$}~\cite{LHCB-PAPER-2018-041},
\mbox{$\Dzb \to \Kp \pim \pip \pim$}~\cite{LHCb-PAPER-2017-040}
and \mbox{$B_s^0 \to D_s^- \Kp \pip\pim$}~\cite{LHCB-PAPER-2020-030} decays.
The results of the $\signal$ \textit{baseline} model are shown as comparison.}
\label{tab:k1FFs2}

 \resizebox{1.\linewidth}{!}{
		 \footnotesize
		\centering
		\renewcommand{\arraystretch}{1.25}
\begin{tabular}{l| r@{}c@{}l r@{}c@{}l r@{}c@{}l r@{}c@{}l}
\hline
\hline
Fit fraction ($\%$) & \multicolumn{3}{l}{$\Dz \to \Kp \Km \pip \pim$} & \multicolumn{3}{l}{$\Dzb \to \Kp \pim \pip \pim$} & \multicolumn{3}{l}{$B_s^0 \to D_s^- \Kp \pip\pim$} & \multicolumn{3}{l}{$\signal$} \\
\hline

\( \Kp [\rho(770)^0/\omega/\rho(1450)^0] \) &  $53.6 $&$\pm$&$ 4.9$ &  $58.01 $&$\pm$&$ 6.7$ &  $63.8 $&$\pm$&$ 10.2$ &   $50.7 $&$\pm$&$ 3.9$  \\

\( K^*(892)^0 \pip \) &  $30.3 $&$\pm$&$ 3.4$ &  $22.1 $&$\pm$&$ 2.9$ &  $33.5 $&$\pm$&$ 16.3$ &   $28.4 $&$\pm$&$ 3.1$  \\

\( [K\pi]_S \pip \) &  $7.0 $&$\pm$&$ 1.7$ &  $24.85 $&$\pm$&$ 2.0$ &  $9.0 $&$\pm$&$ 3.9$ &    $11.4 $&$\pm$&$ 2.6$ \\ \hline

Sum&  $89.2 $&$\pm$&$ 6.1$ &  \multicolumn{3}{l}{$103.5$} &  \multicolumn{3}{l}{$106.3$} &    $90.6 $&$\pm$&$ 4.1$  \\

\hline
\hline

\end{tabular}
 }
\end{table}

\clearpage
\section{Quasi-model-independent lineshapes}
\label{a:MIPW}

\renewcommand{\thefigure}{F.\arabic{figure}}
\renewcommand{\thetable}{F.\arabic{table}}
\setcounter{table}{0}
\setcounter{figure}{0}

\begin{figure}[h]
\centering
\begin{subfigure}[t]{\textwidth}
       	 \includegraphics[width=0.31\textwidth,height=!]{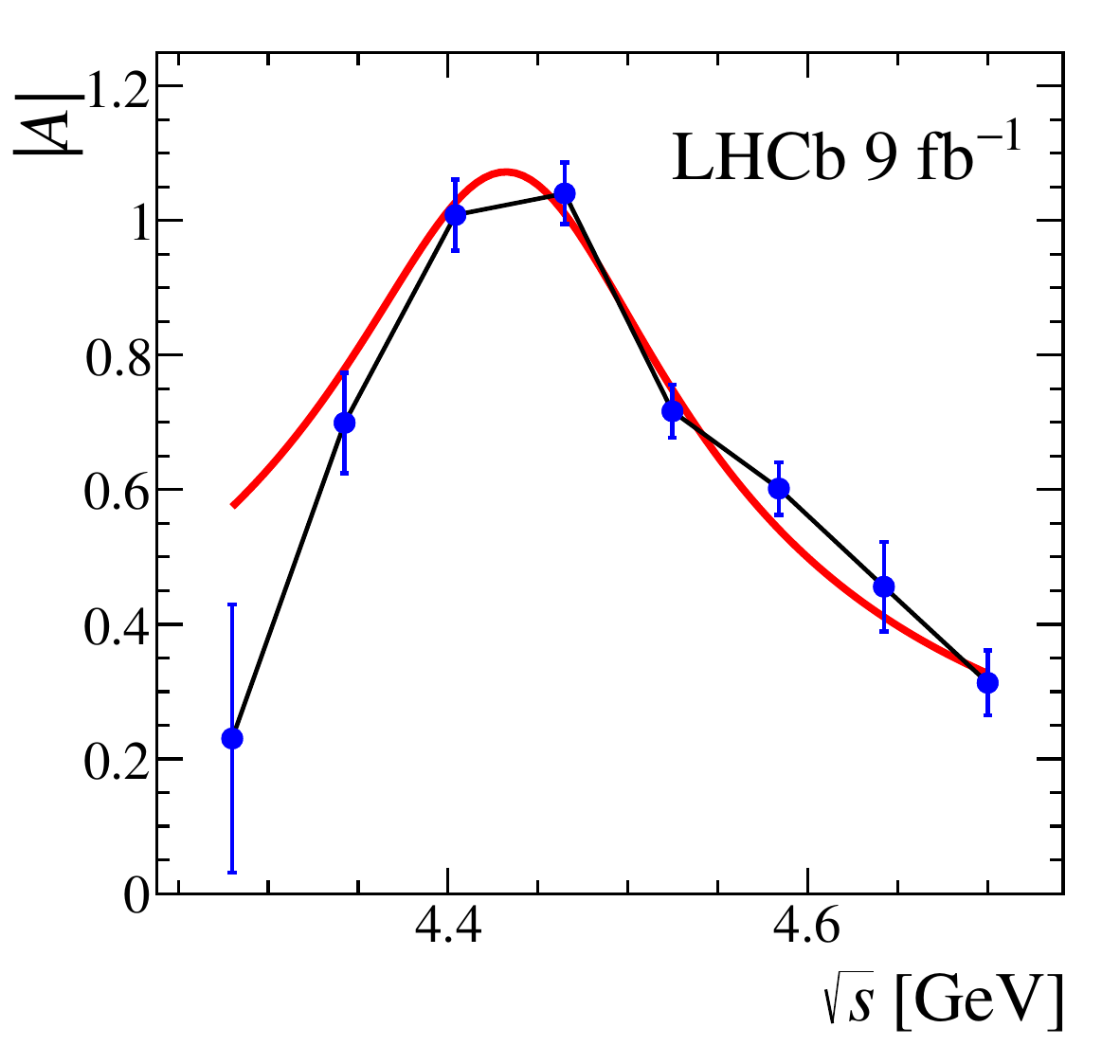}
       	 \includegraphics[width=0.31\textwidth,height=!]{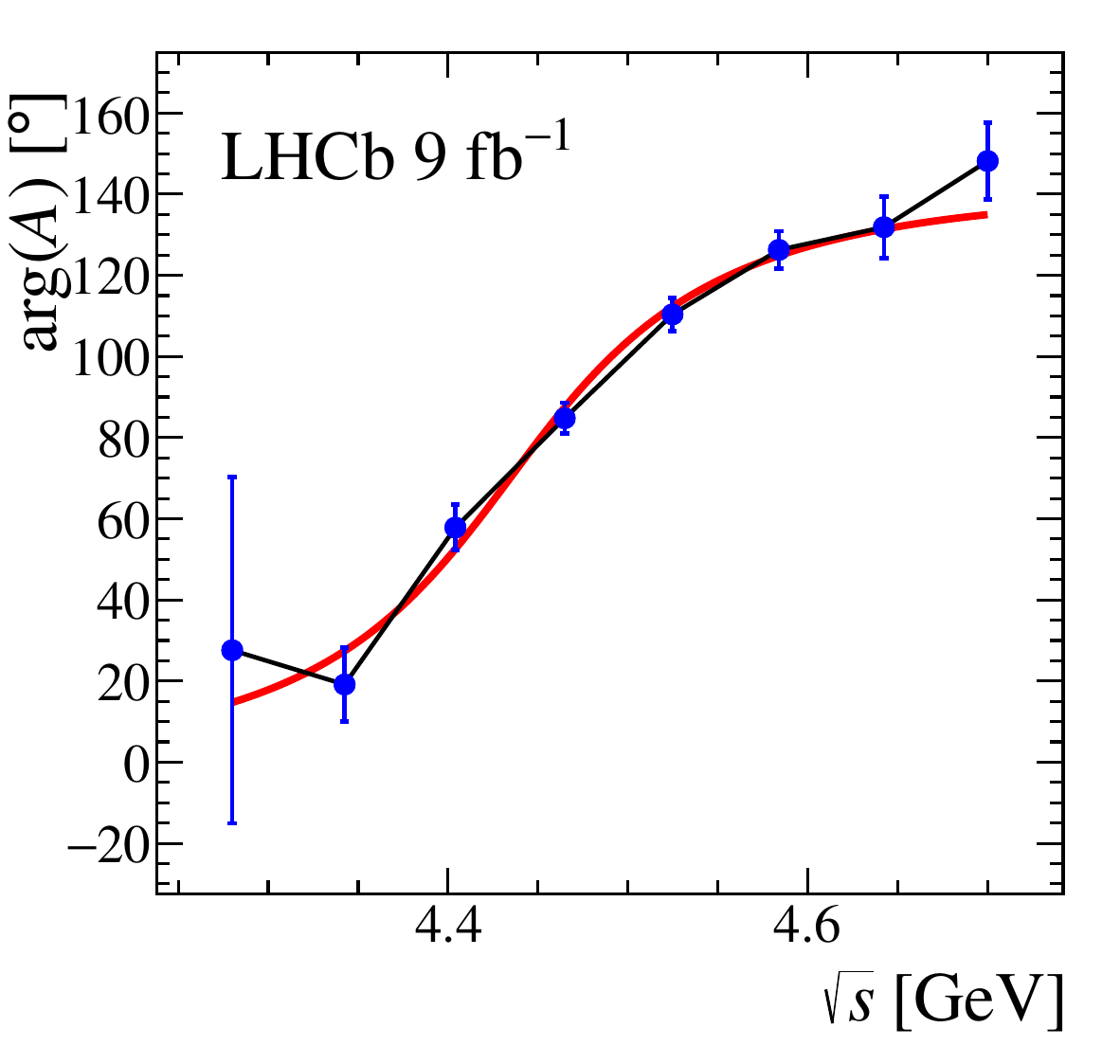}
       	 \includegraphics[width=0.31\textwidth,height=!]{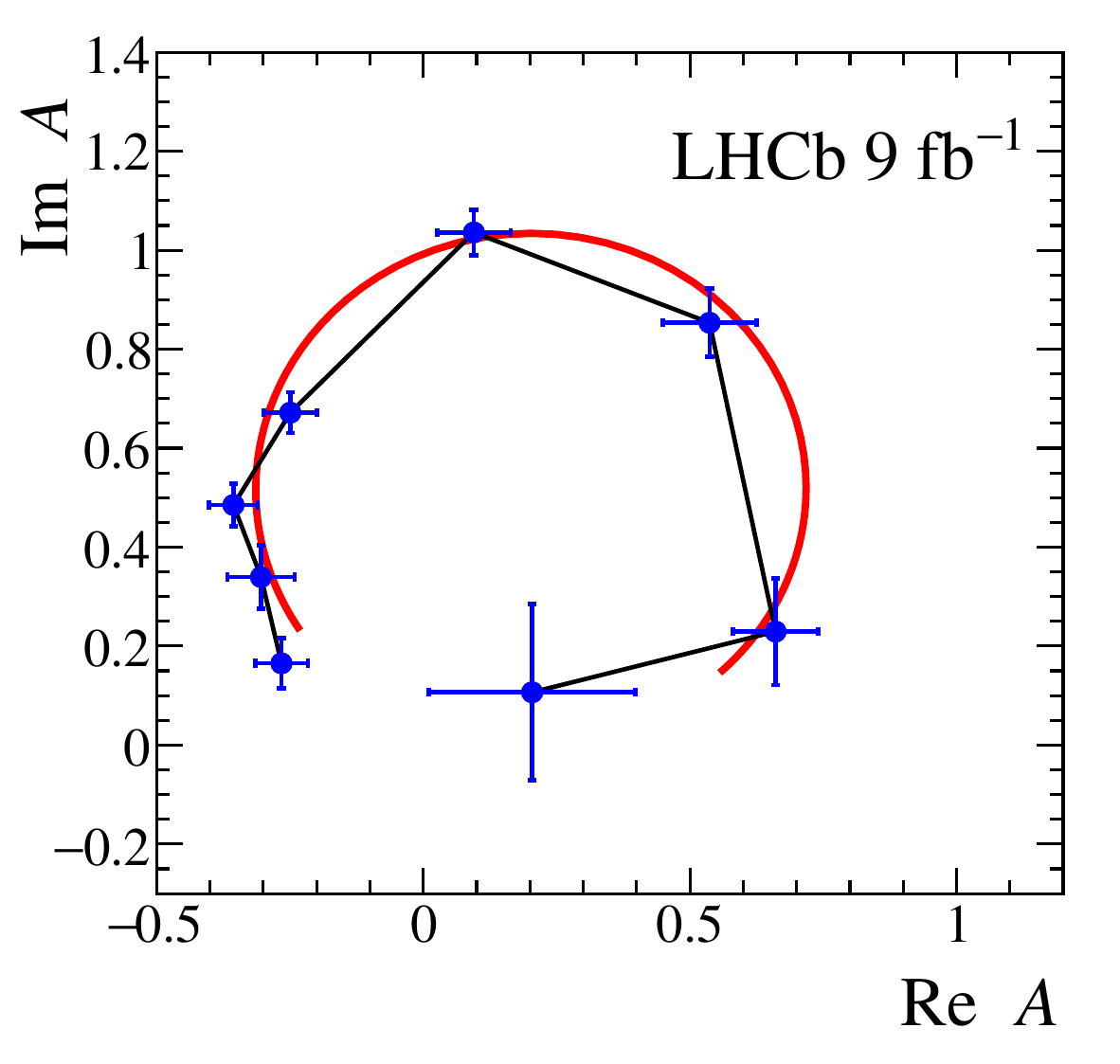}
	\caption{Lineshape of the $\XSone$ resonance.}
\end{subfigure}

\begin{subfigure}[t]{\textwidth}
       	 \includegraphics[width=0.31\textwidth,height=!]{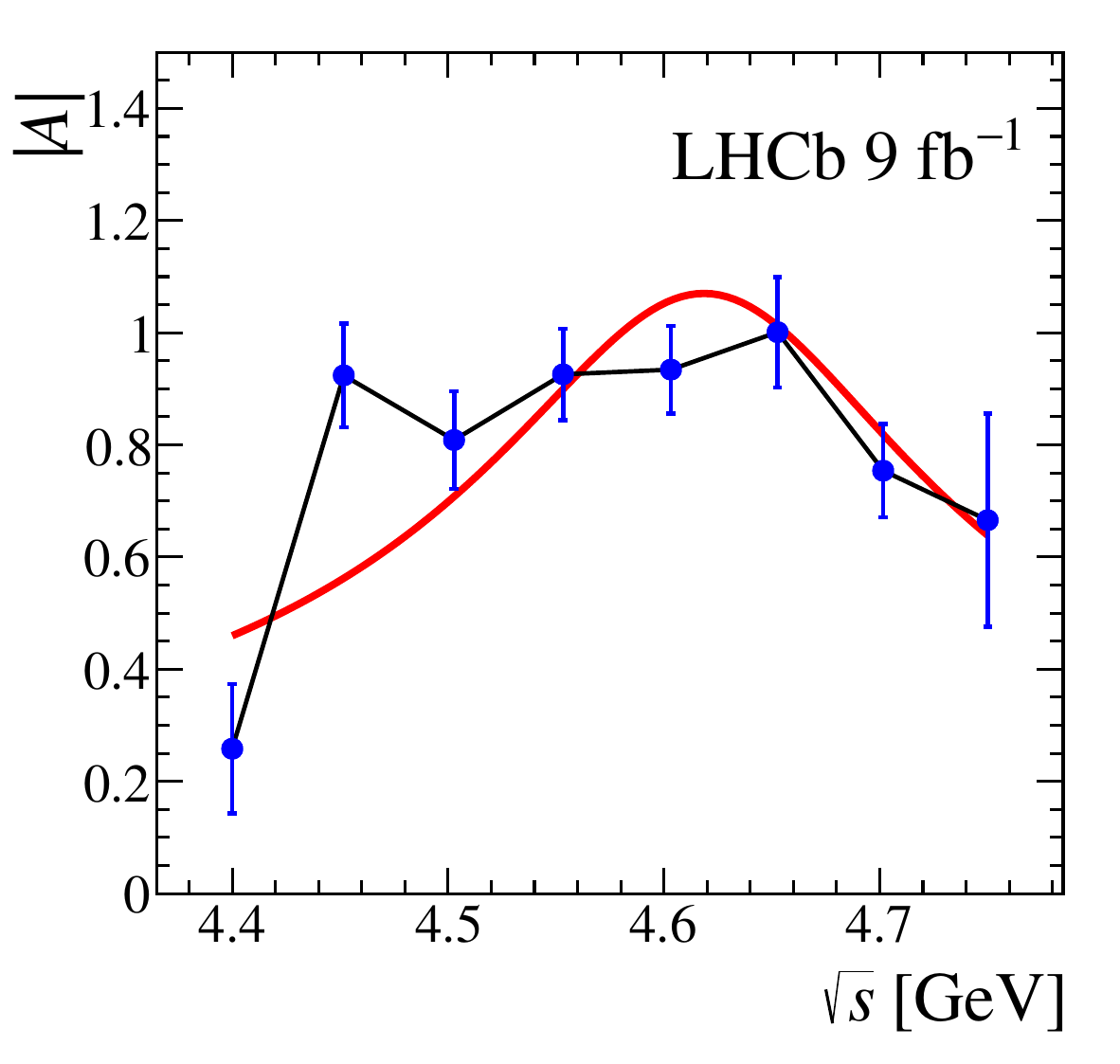}
       	 \includegraphics[width=0.31\textwidth,height=!]{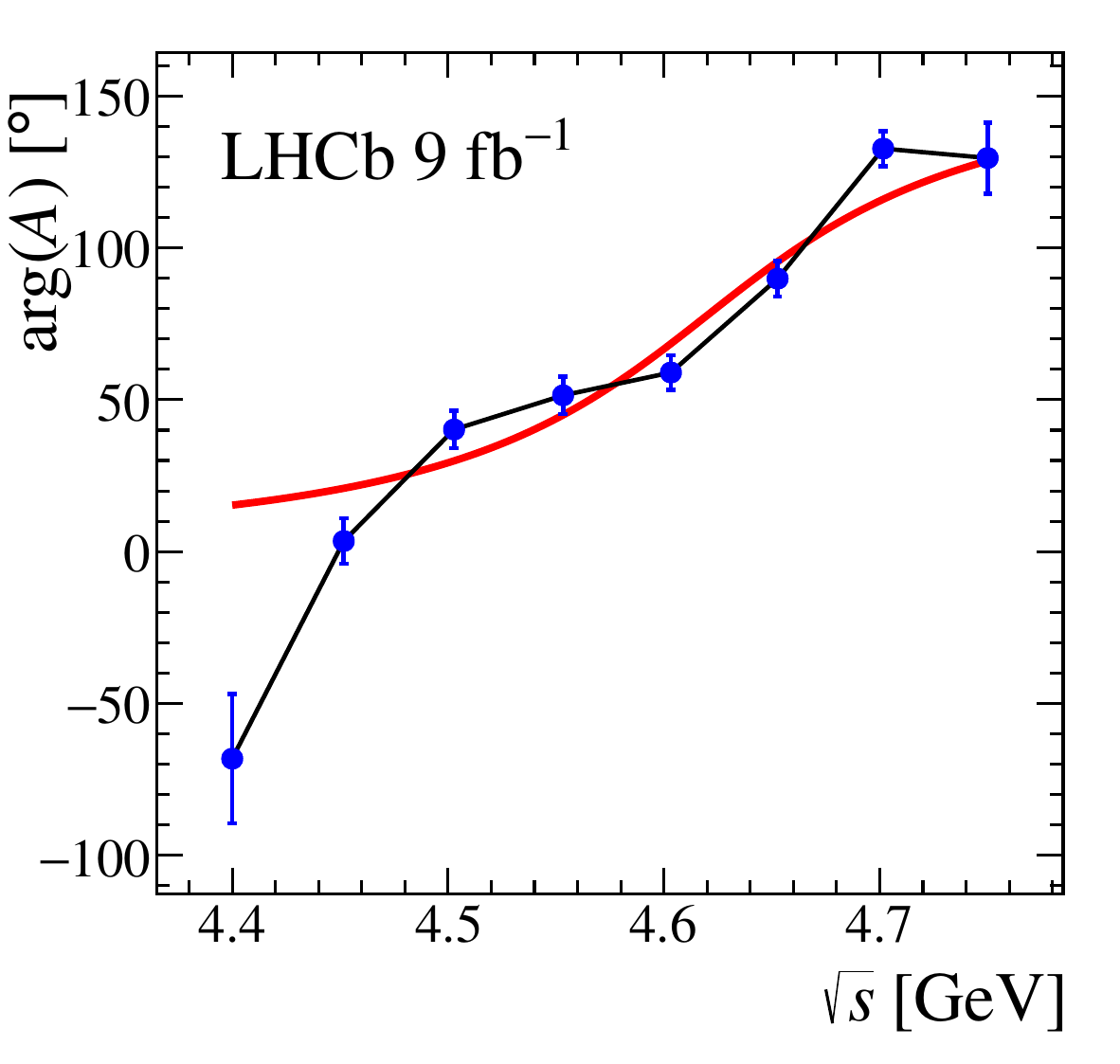}
       	 \includegraphics[width=0.31\textwidth,height=!]{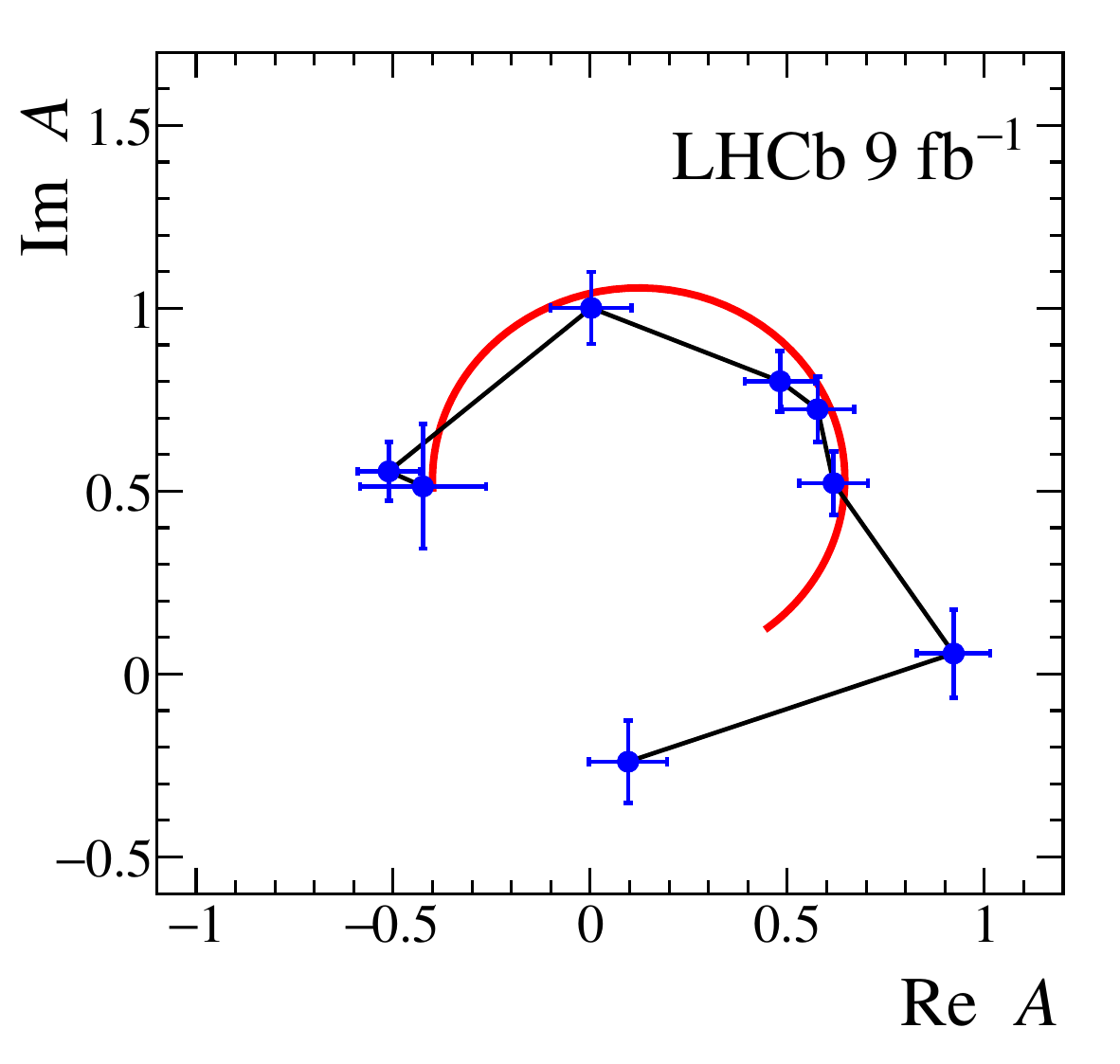}
	\caption{Lineshape of the $\XAone$ resonance.}
\end{subfigure}

\begin{subfigure}[t]{\textwidth}
       	 \includegraphics[width=0.31\textwidth,height=!]{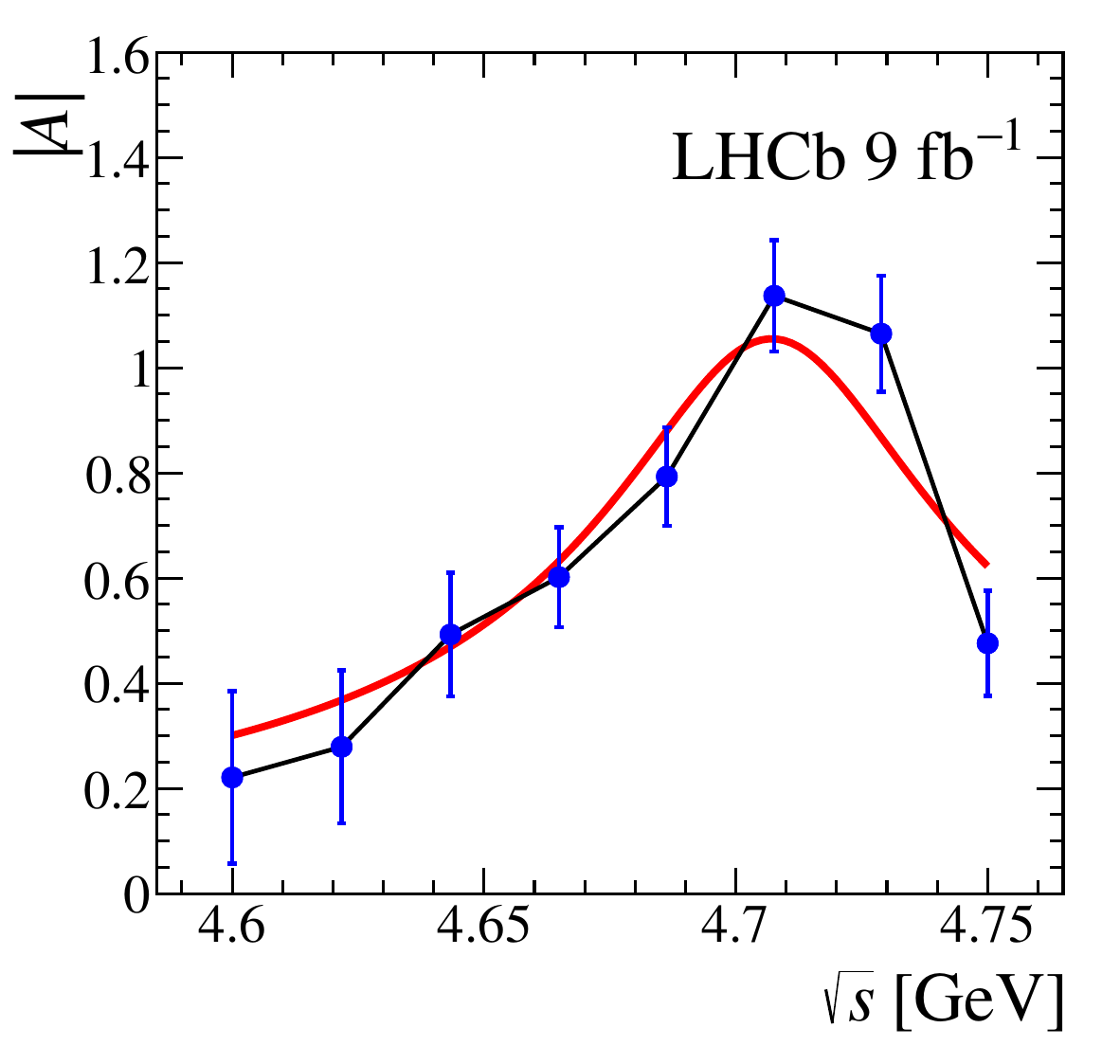}
       	 \includegraphics[width=0.31\textwidth,height=!]{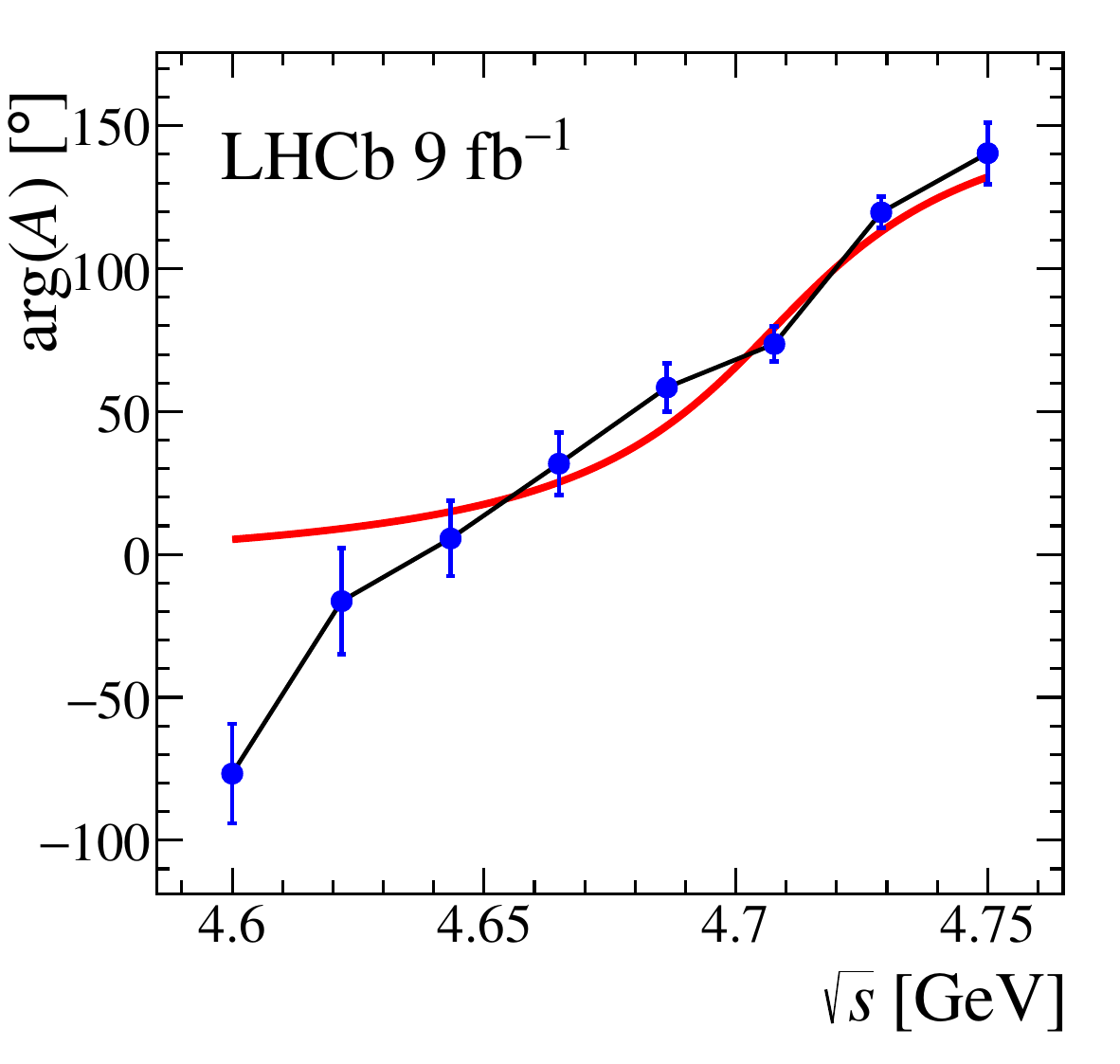}
       	 \includegraphics[width=0.31\textwidth,height=!]{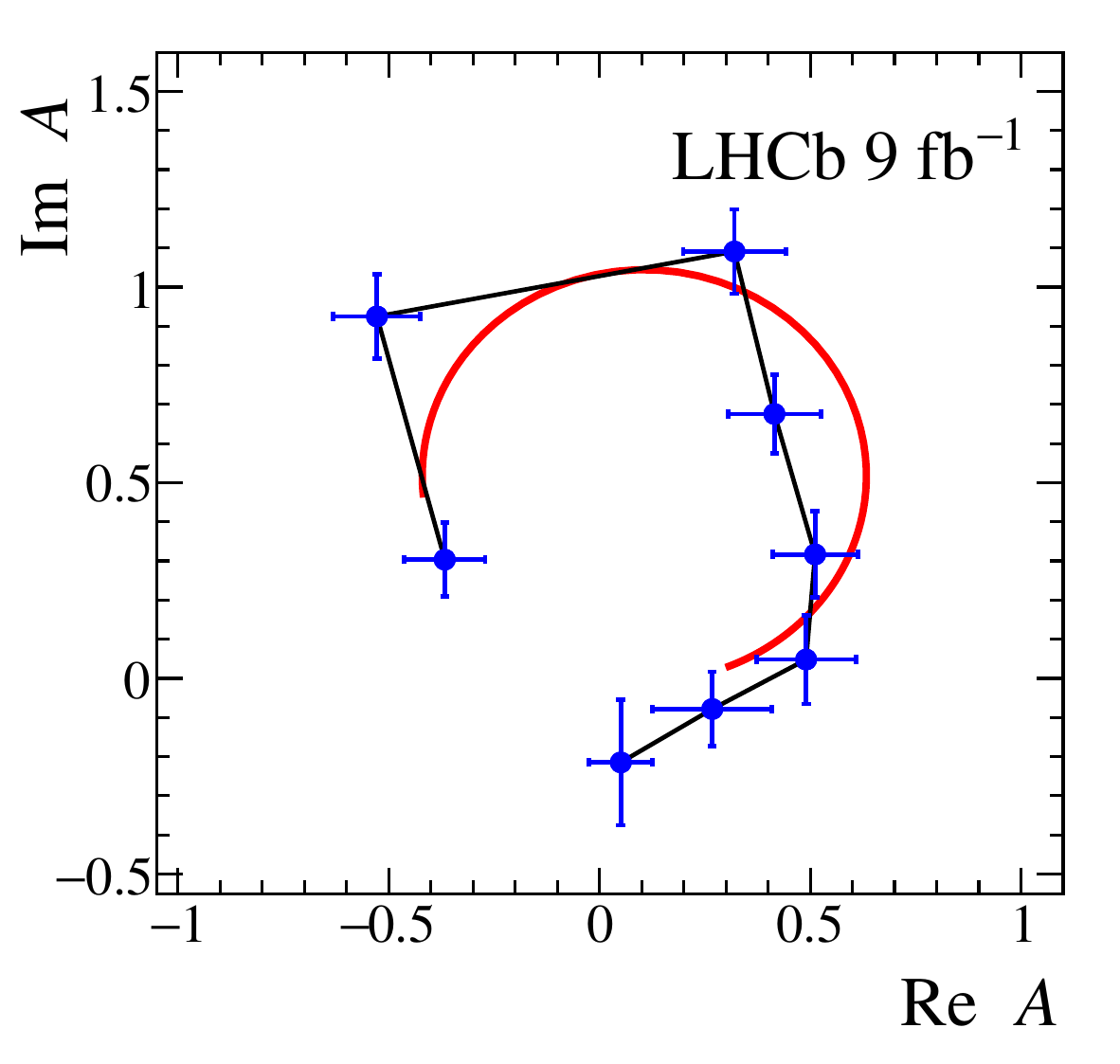}
	\caption{Lineshape of the $\XStwo$ resonance.}
\end{subfigure}

\begin{subfigure}[t]{\textwidth}
       	 \includegraphics[width=0.31\textwidth,height=!]{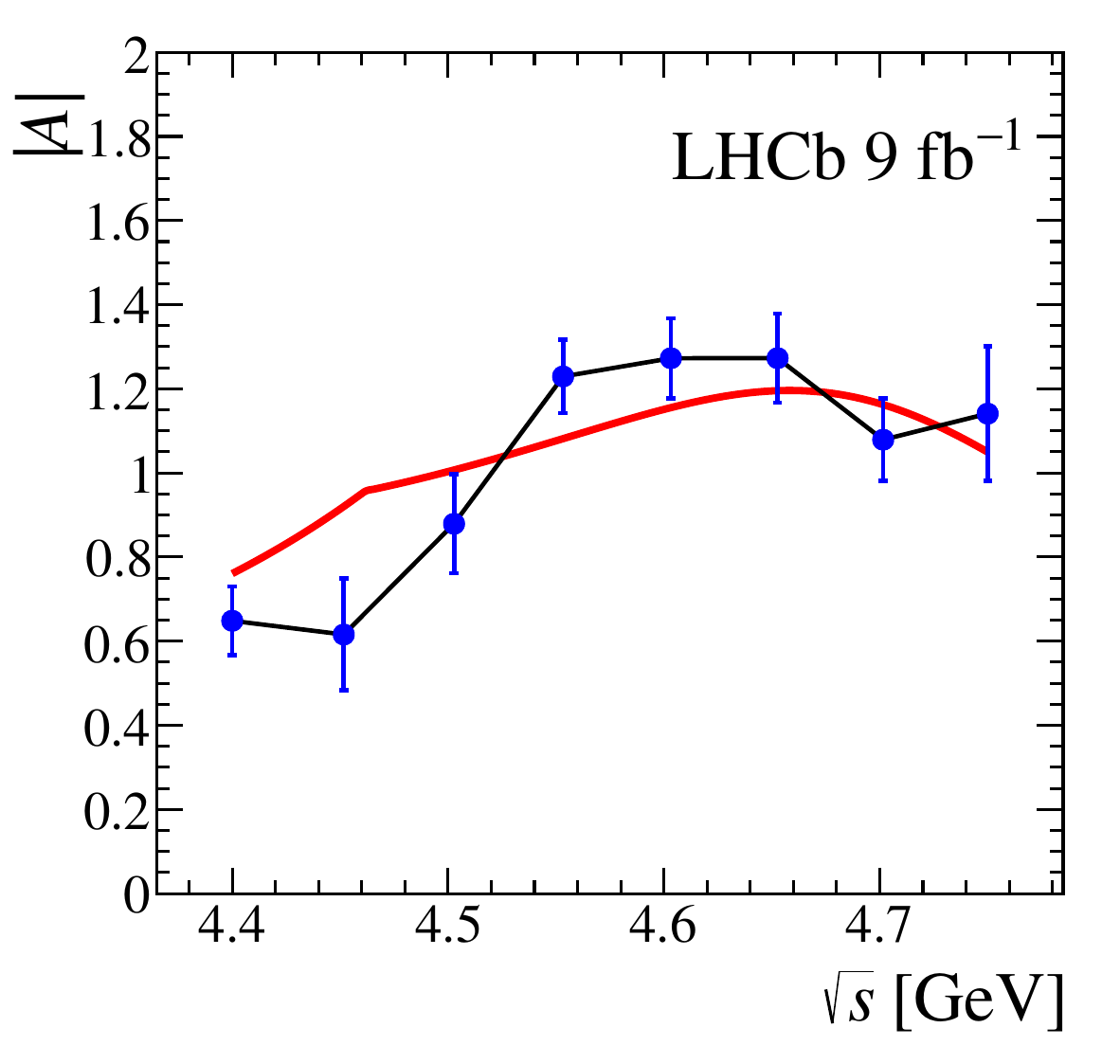}
       	 \includegraphics[width=0.31\textwidth,height=!]{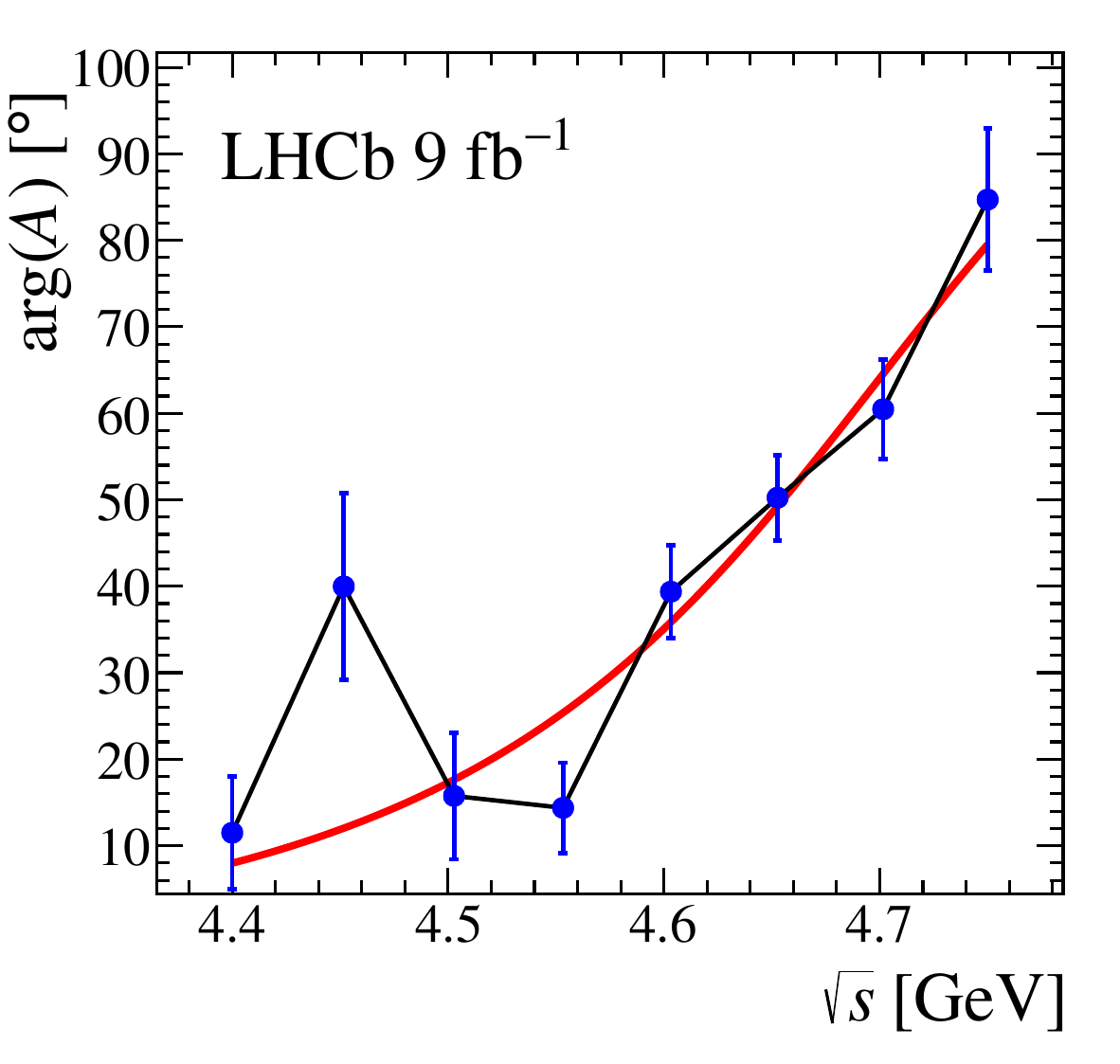}
       	 \includegraphics[width=0.31\textwidth,height=!]{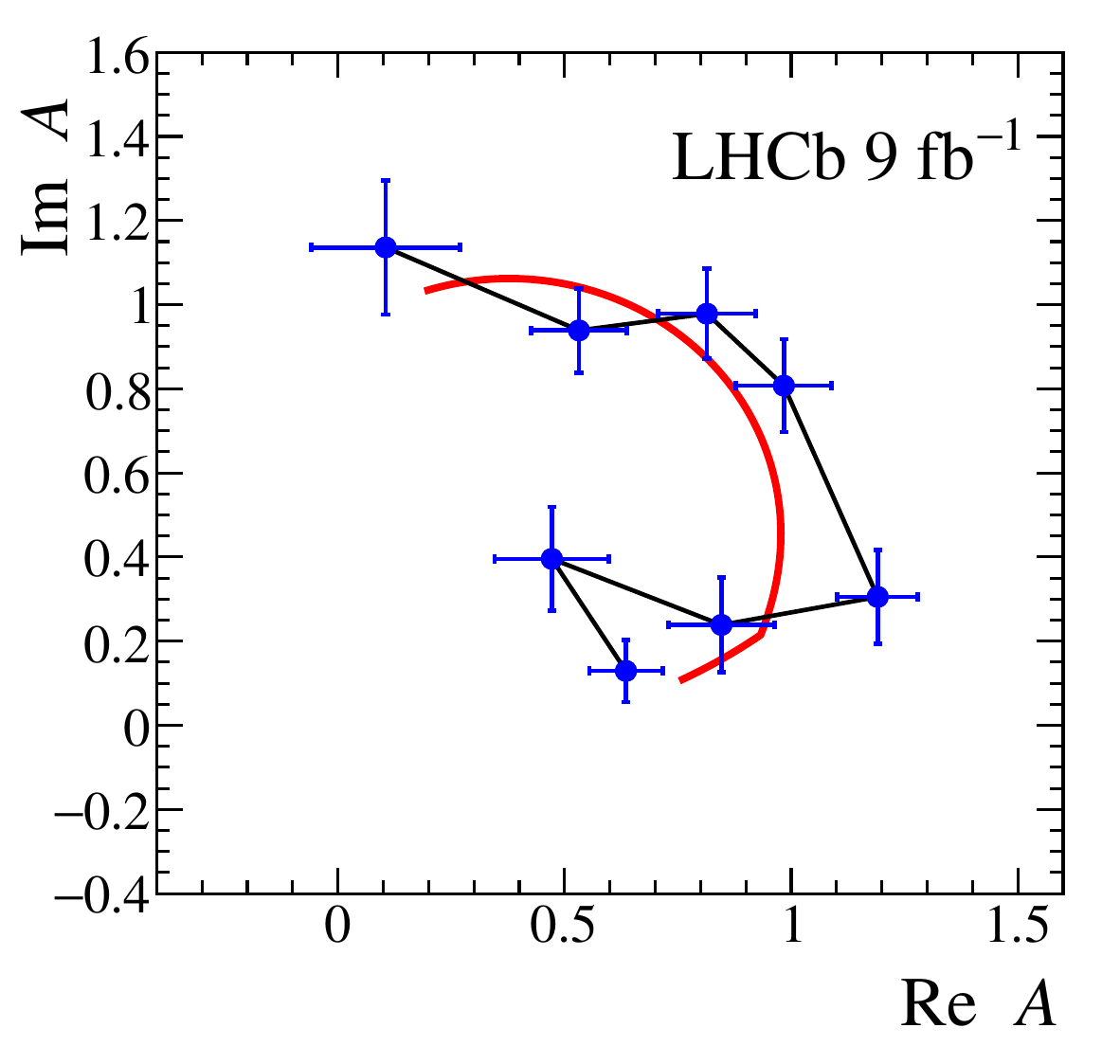}
	\caption{Lineshape of the $\XVone$ resonance.}
\end{subfigure}

	\caption{(Left) Magnitude, (middle) phase and (right) Argand diagram of the quasi-model-independent lineshape for $\Xz \to \psitwos \pip \pim$ states. 
	The fitted knots are displayed as connected points with error bars.
	The Breit--Wigner lineshape with the mass and width from the nominal fit is superimposed (red line). }
	\label{fig:argand}
\end{figure}

Figure~\ref{fig:argand} shows the determined quasi-model-independent lineshapes
for the \mbox{$\Xz \to \psitwos \pip \pim$} resonances,
where the expectations from a Breit--Wigner propagator 
with the mass and width from the nominal fit are superimposed. 
The $\XSone$ state has the highest fit fraction of any of the exotic states.
Its Argand diagram in Fig.~\ref{fig:argand}(a) shows a clear circular, counter-clockwise trajectory; which is the expected behaviour of a resonant state.
The quasi-model-independent lineshapes of the $\XAone$ and $\XStwo$
are qualitatively consistent with the expectations as well.
The phase motion of the 
$\XAone$ at low mass hints at a possible second $1^+$ resonance 
that is also part of alternative model 4, see Tables~\ref{tab:altModels1_1} to~\ref{tab:altModels1_3}.
Note that the high-mass tail of the broad $\XVone$ state is outside of the phase-space boundary 
such that the full phase motion cannot be investigated. 

\begin{figure}[h]
	\centering

\begin{subfigure}[t]{\textwidth}
       	 \includegraphics[width=0.31\textwidth,height=!]{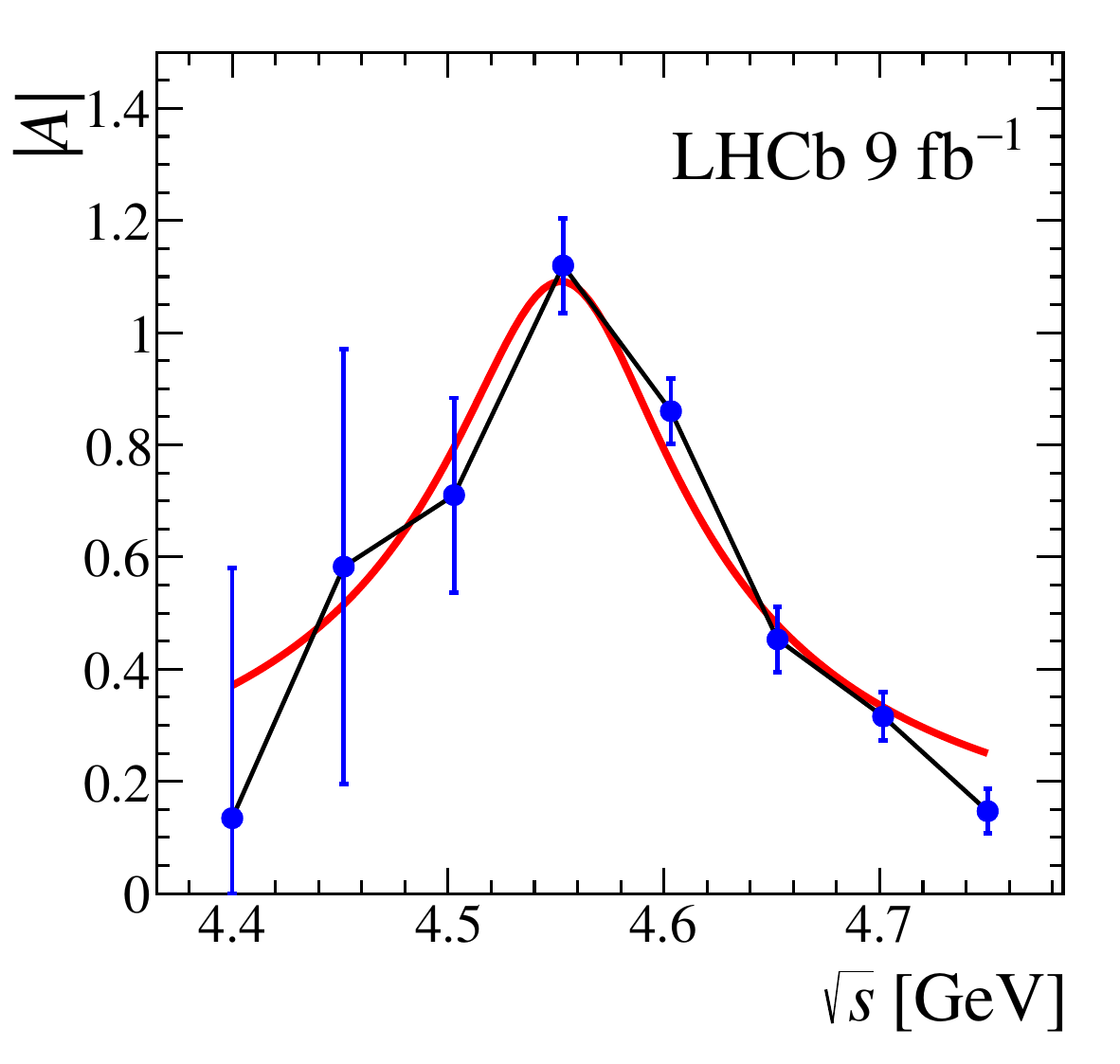}
       	 \includegraphics[width=0.31\textwidth,height=!]{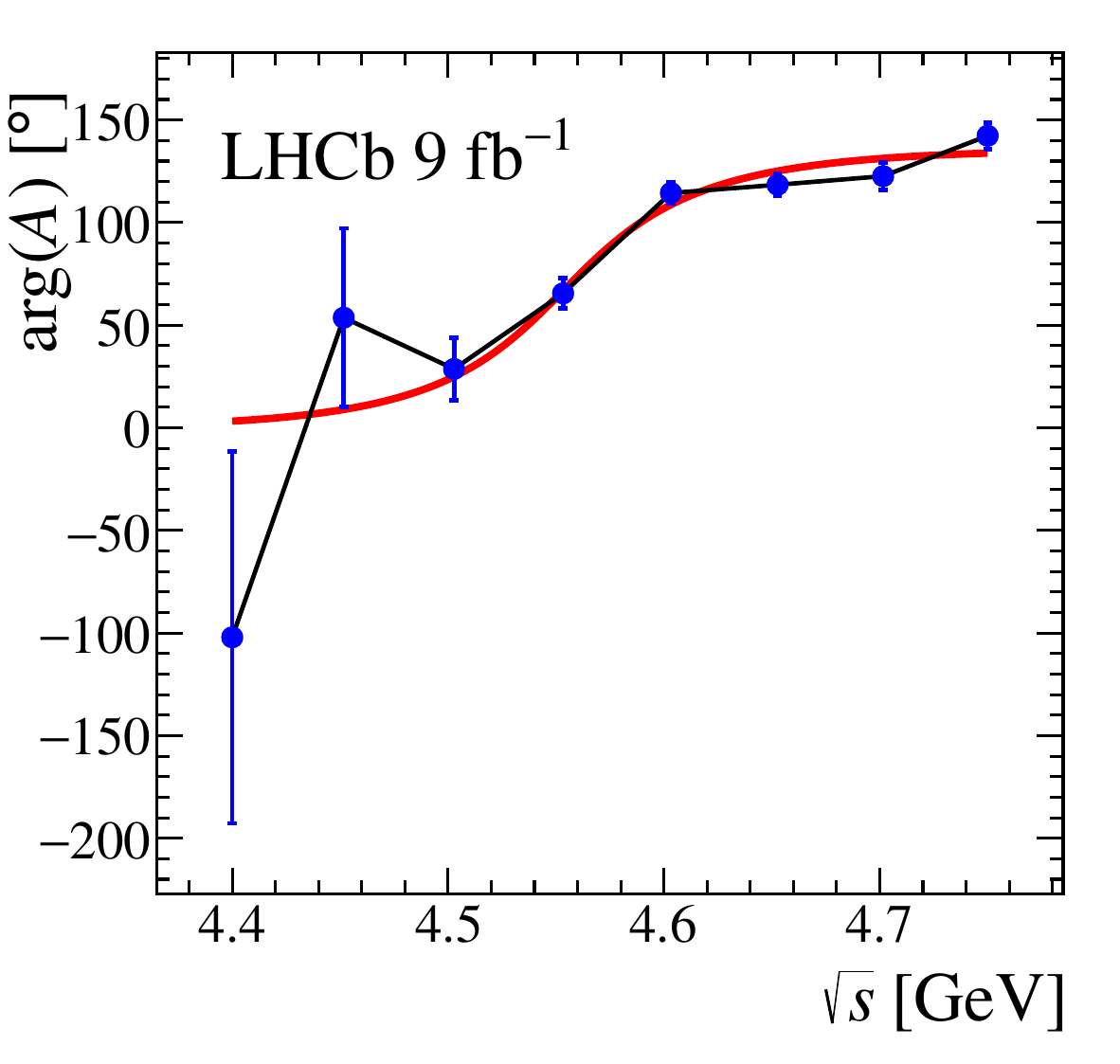}
       	 \includegraphics[width=0.31\textwidth,height=!]{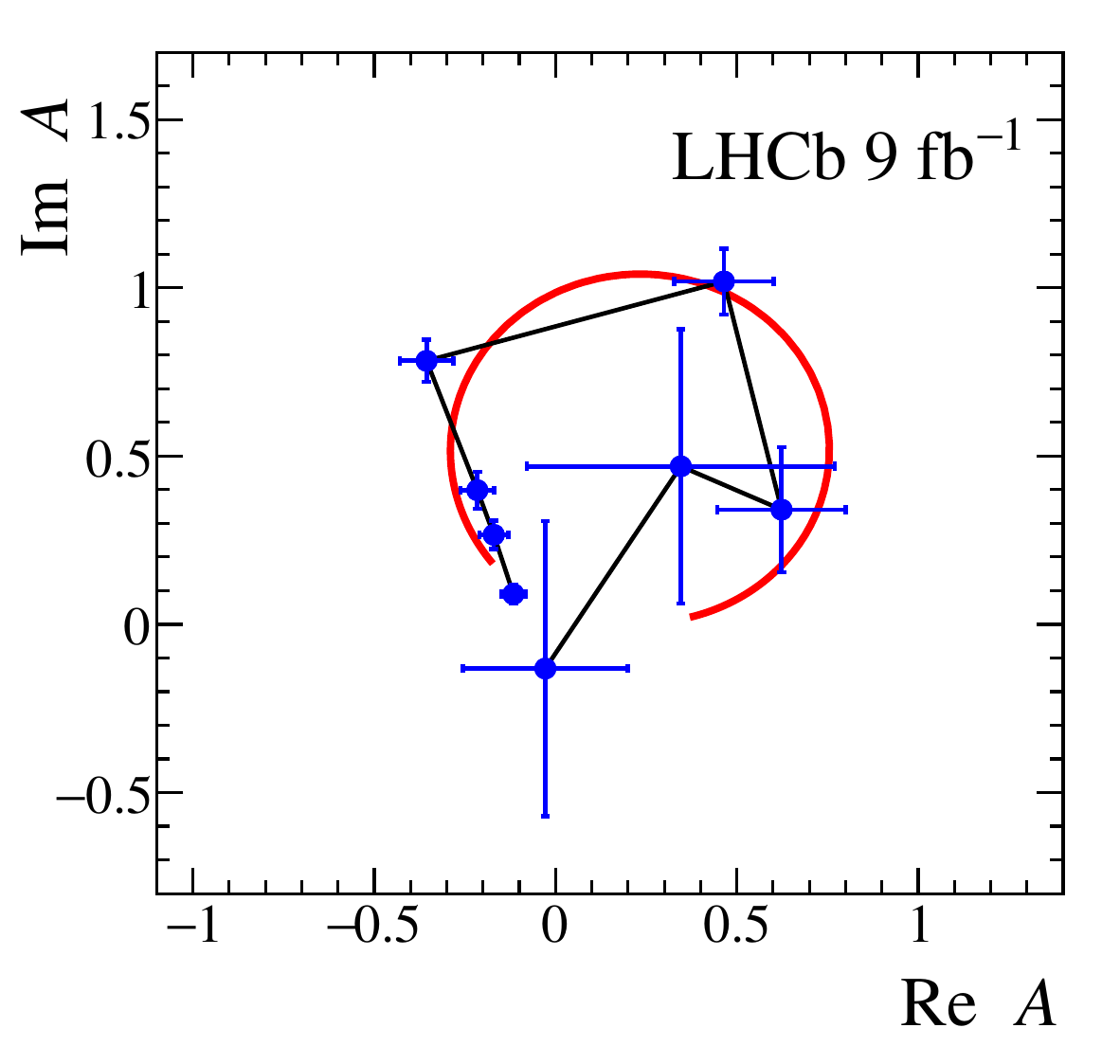}
	\caption{Lineshape of the $\XsAone$  resonance.}
\end{subfigure}

\begin{subfigure}[t]{\textwidth}
       	 \includegraphics[width=0.31\textwidth,height=!]{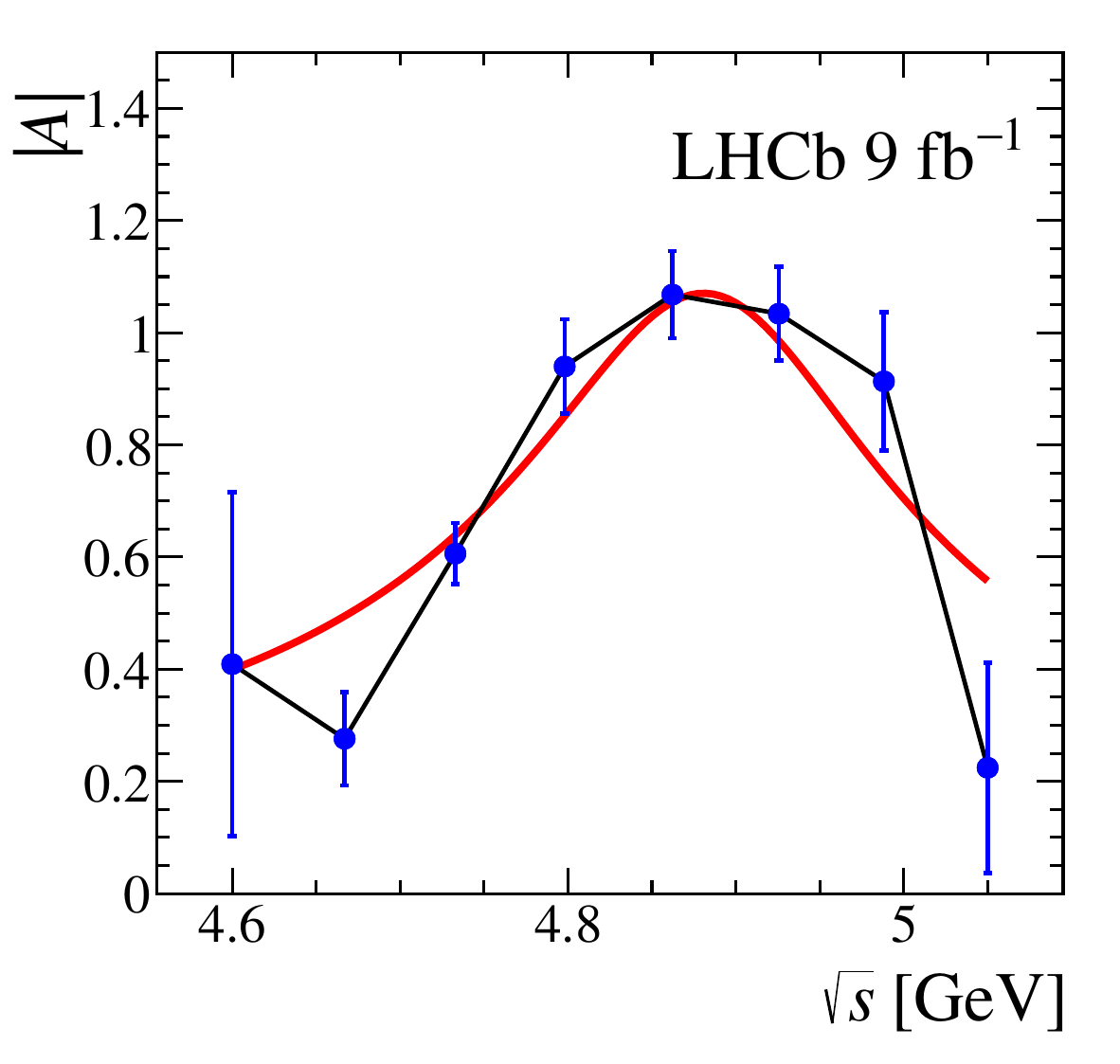}
       	 \includegraphics[width=0.31\textwidth,height=!]{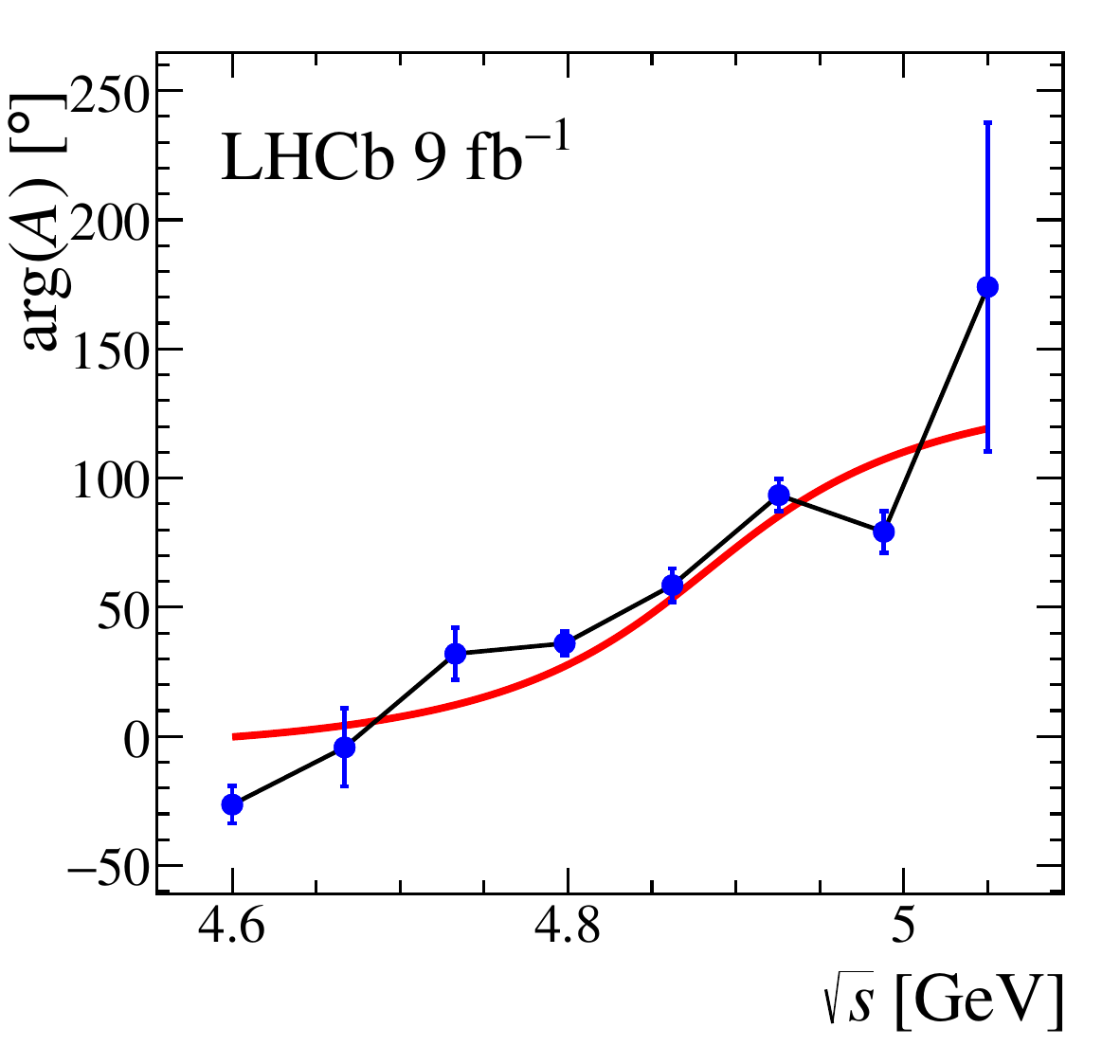}
       	 \includegraphics[width=0.31\textwidth,height=!]{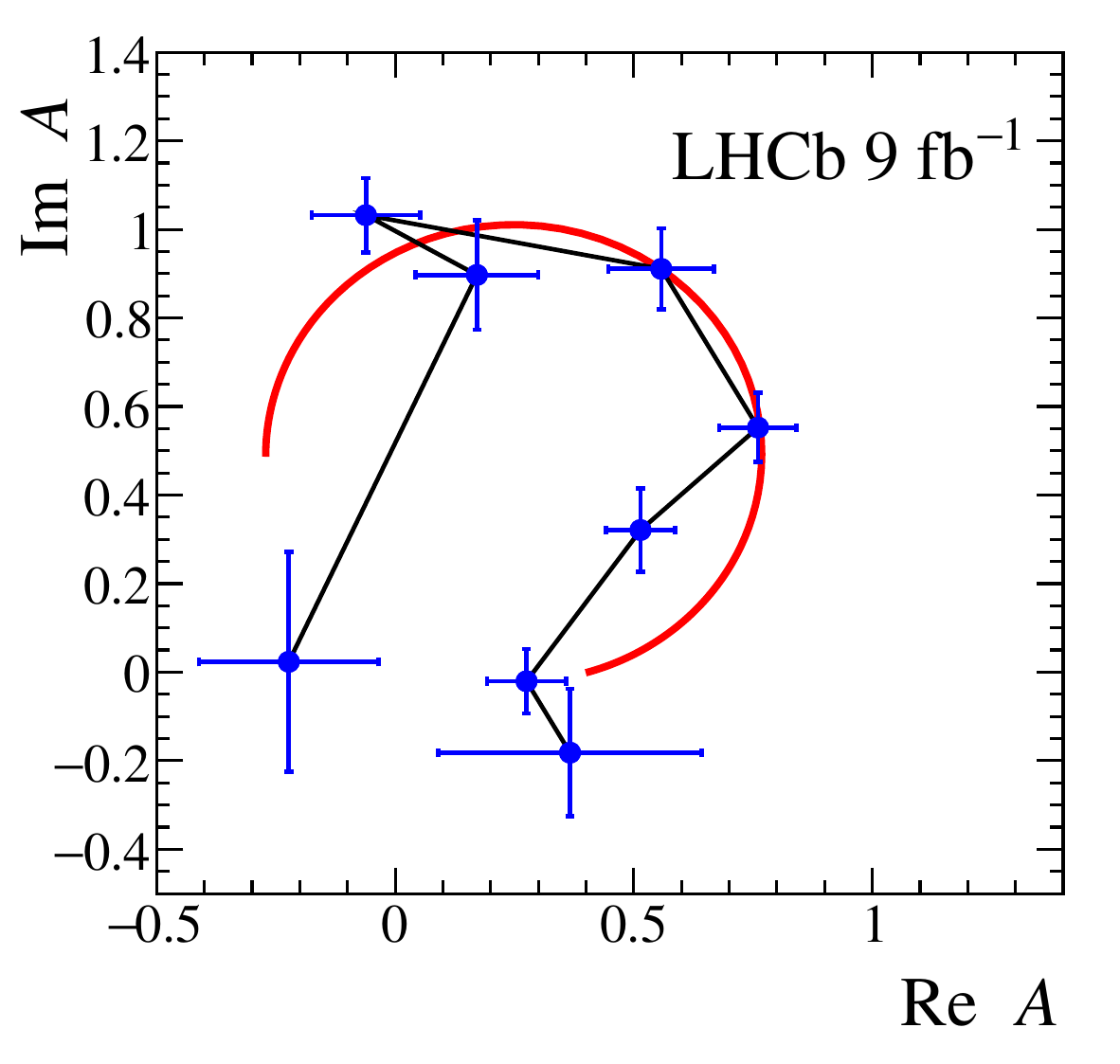}
	\caption{Lineshape of the $\XsAtwo$ resonance.}
\end{subfigure}

\begin{subfigure}[t]{\textwidth}
       	 \includegraphics[width=0.31\textwidth,height=!]{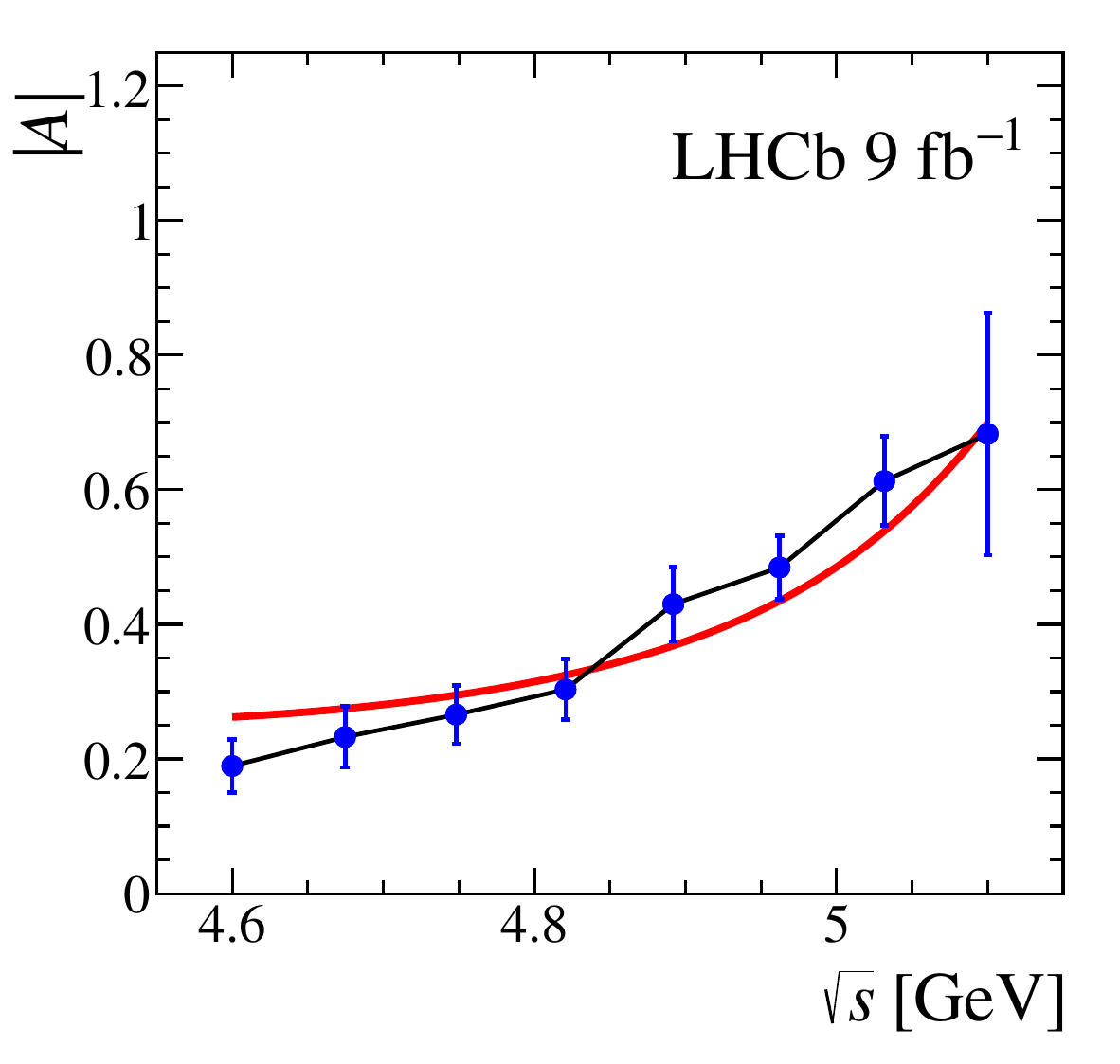}
       	 \includegraphics[width=0.31\textwidth,height=!]{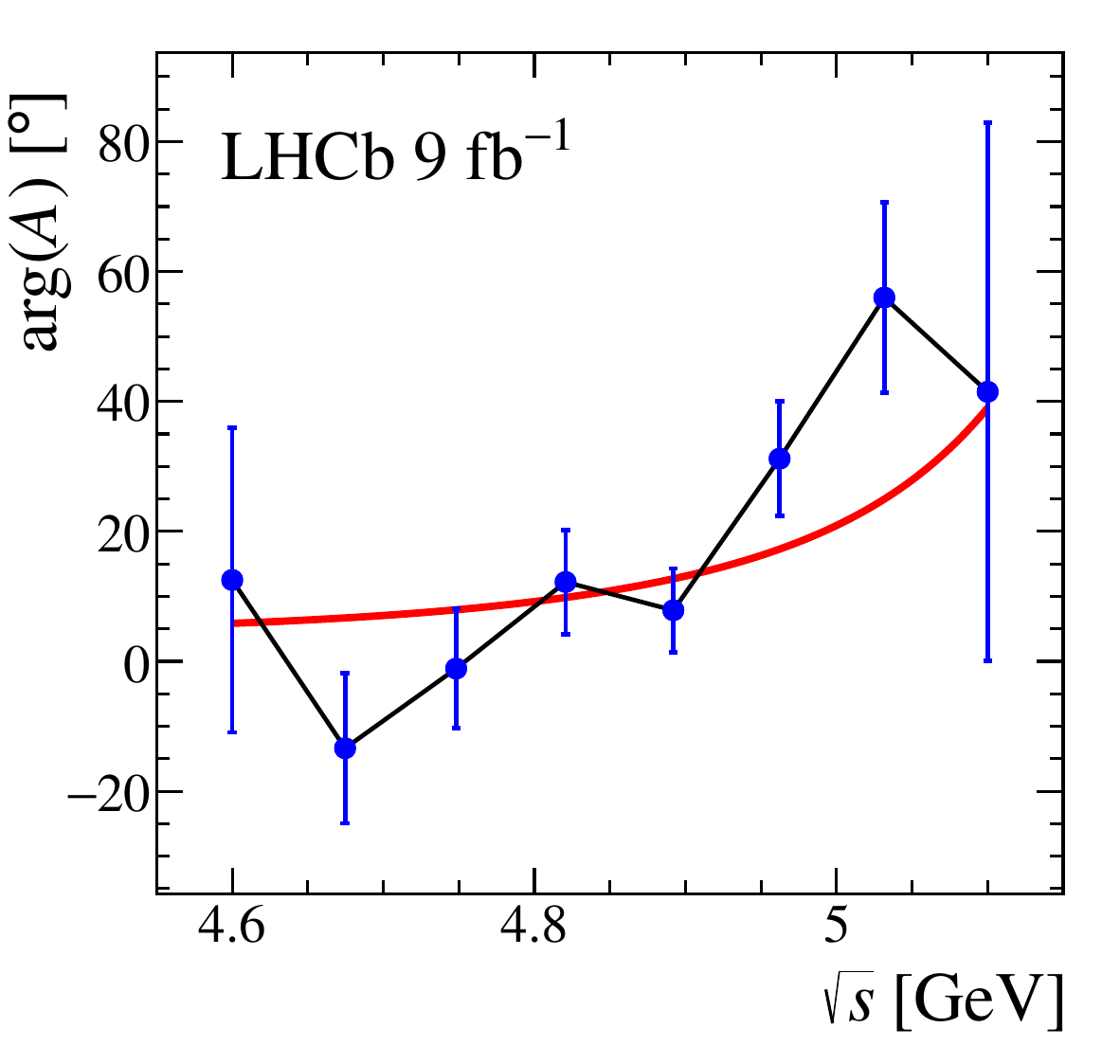}
       	 \includegraphics[width=0.31\textwidth,height=!]{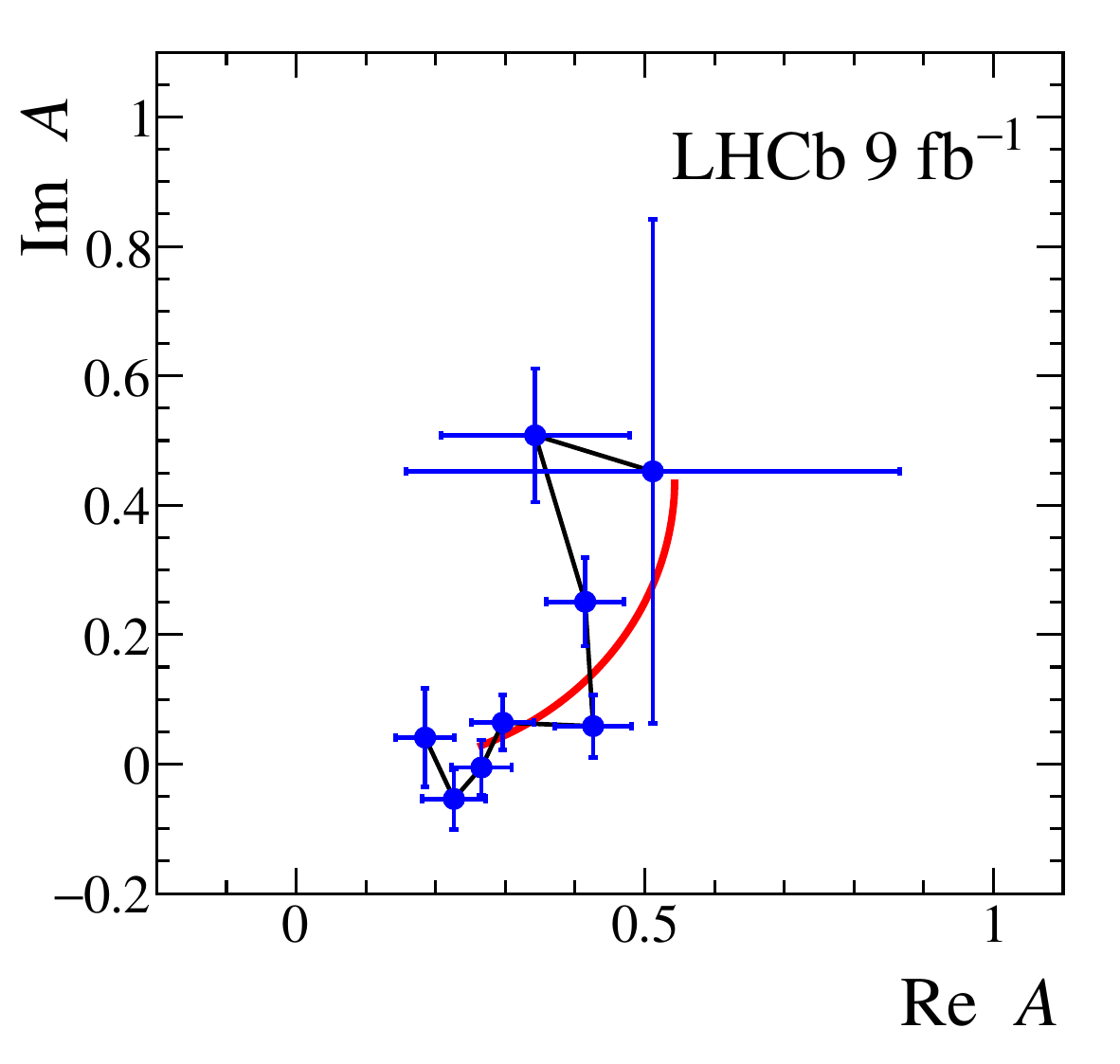}
	\caption{Lineshape of the $\XsVone$ resonance.}
\end{subfigure}
         
	\caption{(Left) Magnitude, (middle) phase and (right) Argand diagram of the quasi-model-independent lineshape for $\Xsz \to \psitwos \Kp \pim$ states. 
	The fitted knots are displayed as connected points with error bars.
	The Breit--Wigner lineshape with the mass and width from the nominal fit is superimposed (red line). }
	\label{fig:argand2}
\end{figure}

The quasi-model-independent lineshapes for the $\Xsz \to \psitwos \Kp \pim$ resonances are shown in Fig.~\ref{fig:argand2}. 
The $\XsAone$ has the highest fit fraction of these states and is shown in Fig.~\ref{fig:argand2}(a). 
A clear, rapid phase-motion around the resonance pole is observed.
The $\XsAtwo$ shows an indicative resonant phase-motion as well, see Fig.~\ref{fig:argand2}(b).
As the $\XsVone$ state, shown in Fig.~\ref{fig:argand2}(c), has a low fit fraction and is close to the phase-space boundary, 
no conclusive statements can be made. 

\begin{figure}[h]
	\centering
\begin{subfigure}[t]{\textwidth}
       	 \includegraphics[width=0.31\textwidth,height=!]{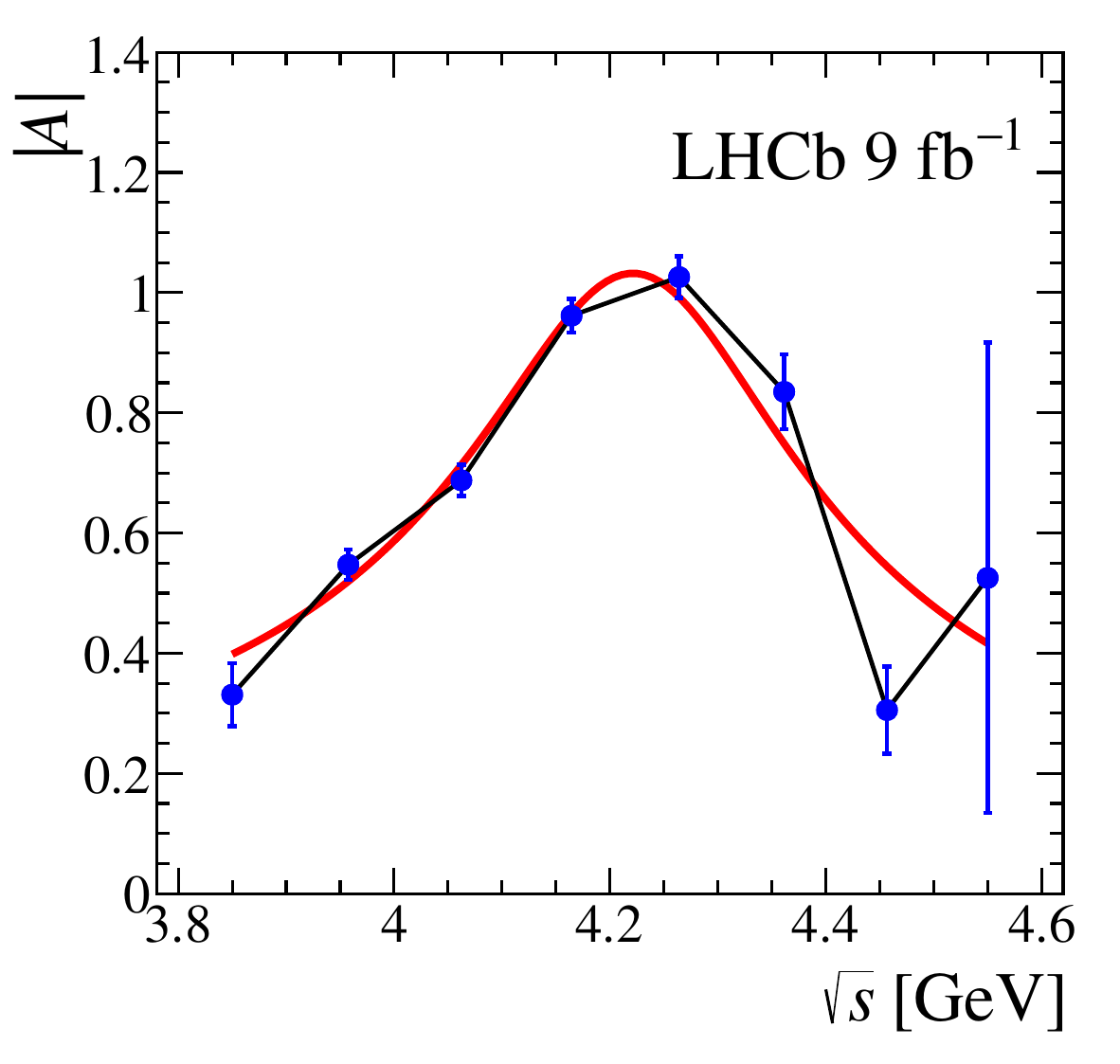}
       	 \includegraphics[width=0.31\textwidth,height=!]{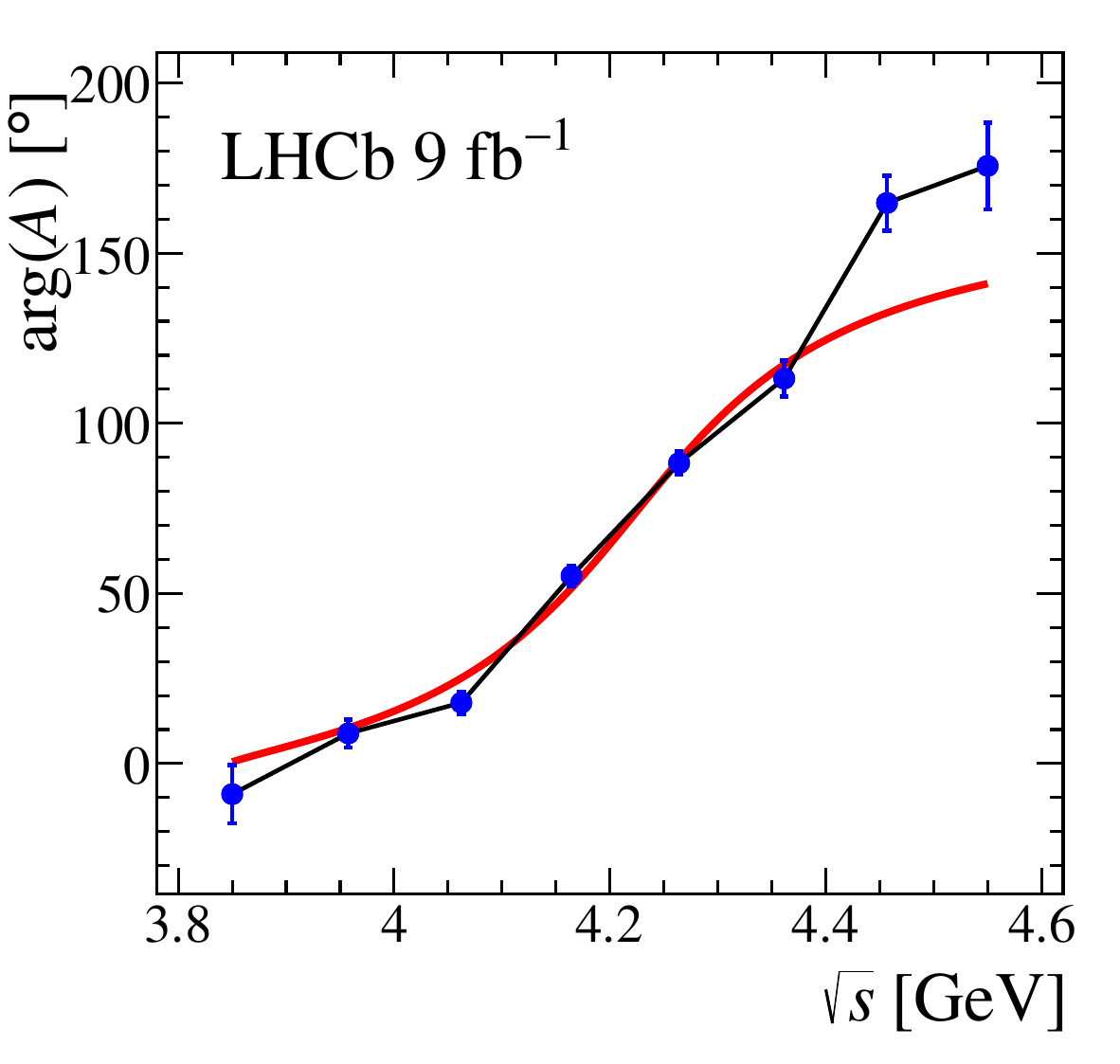}
       	 \includegraphics[width=0.31\textwidth,height=!]{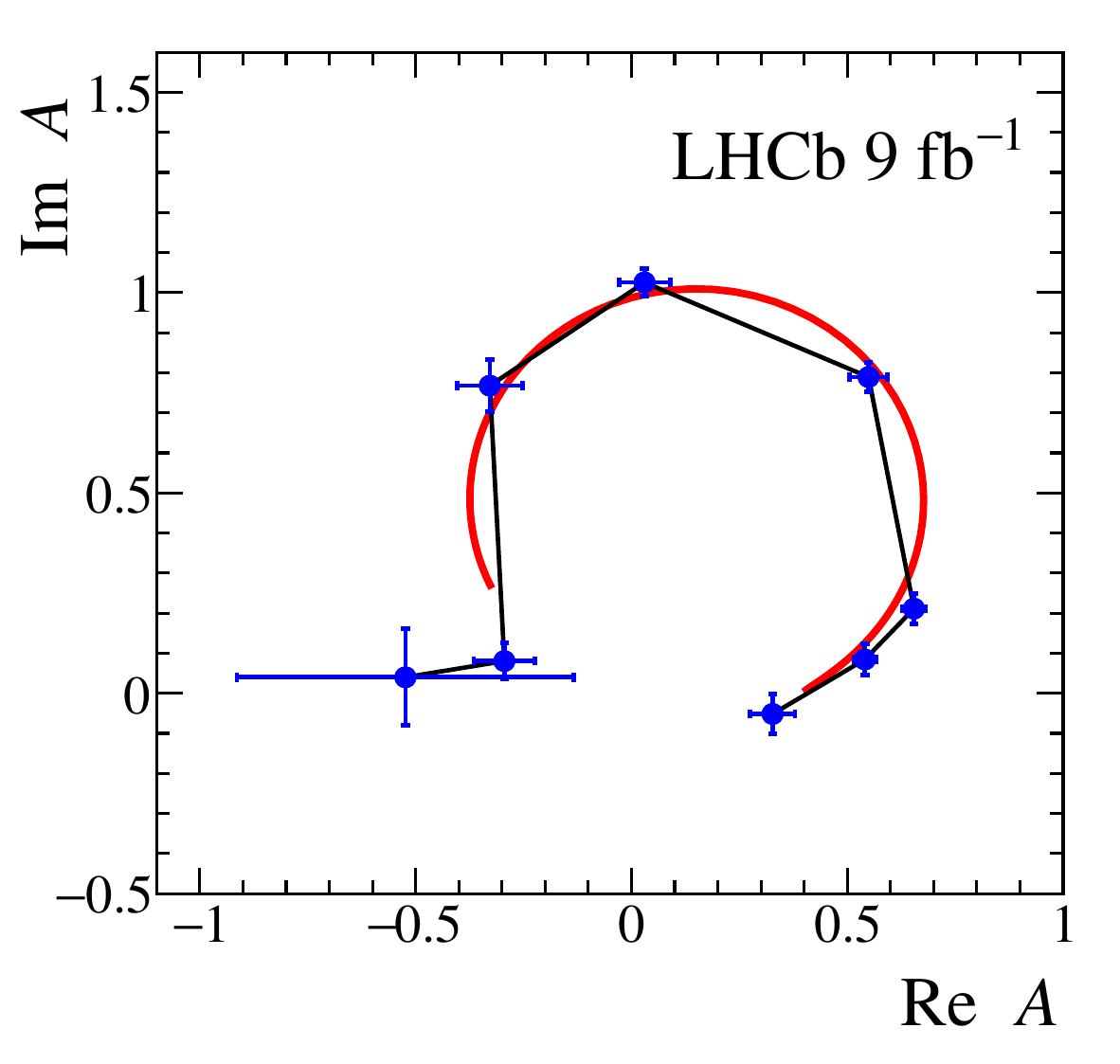}
	\caption{Lineshape of the $\ZAone^\pm$  resonance.}
\end{subfigure}

\begin{subfigure}[t]{\textwidth}
       	 \includegraphics[width=0.31\textwidth,height=!]{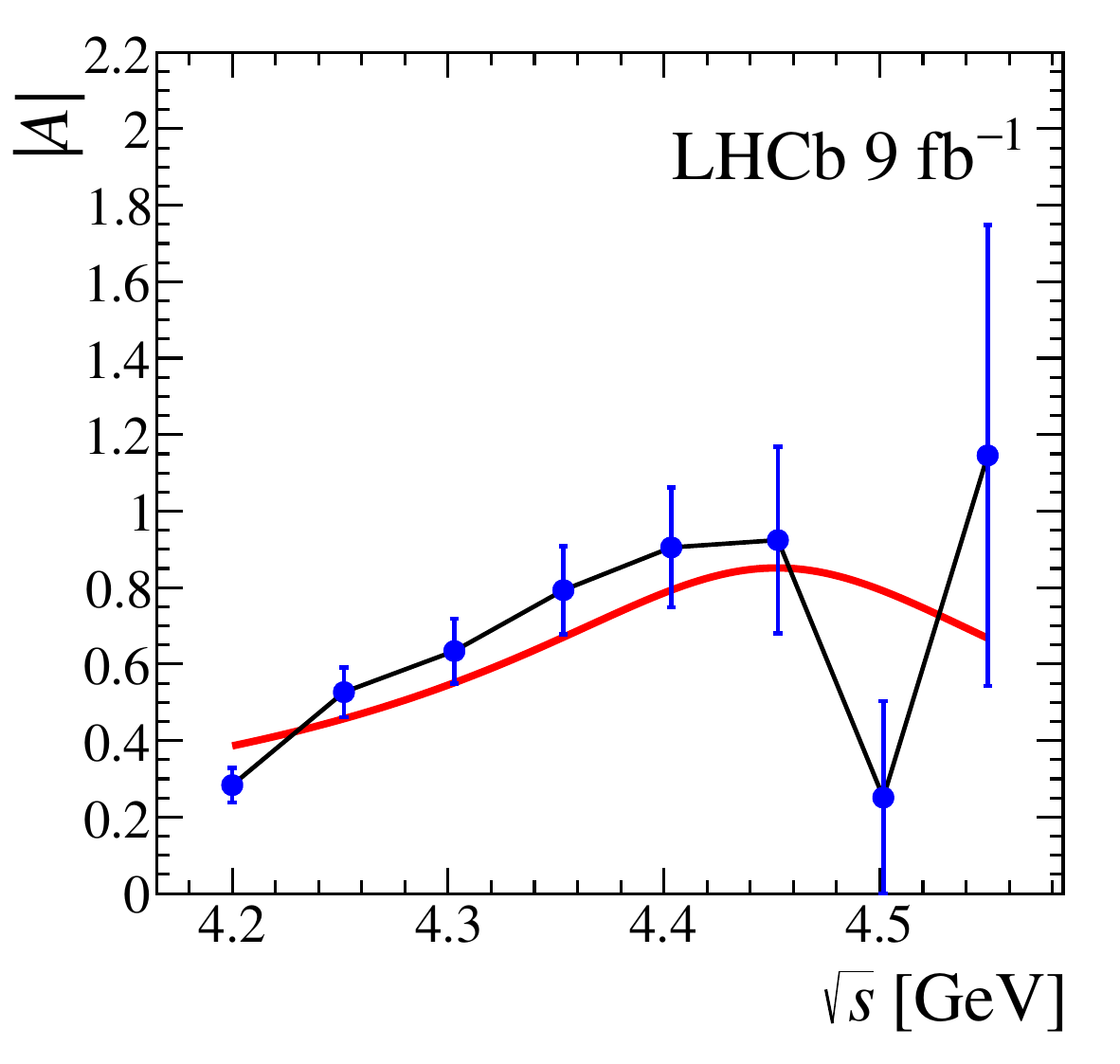}
       	 \includegraphics[width=0.31\textwidth,height=!]{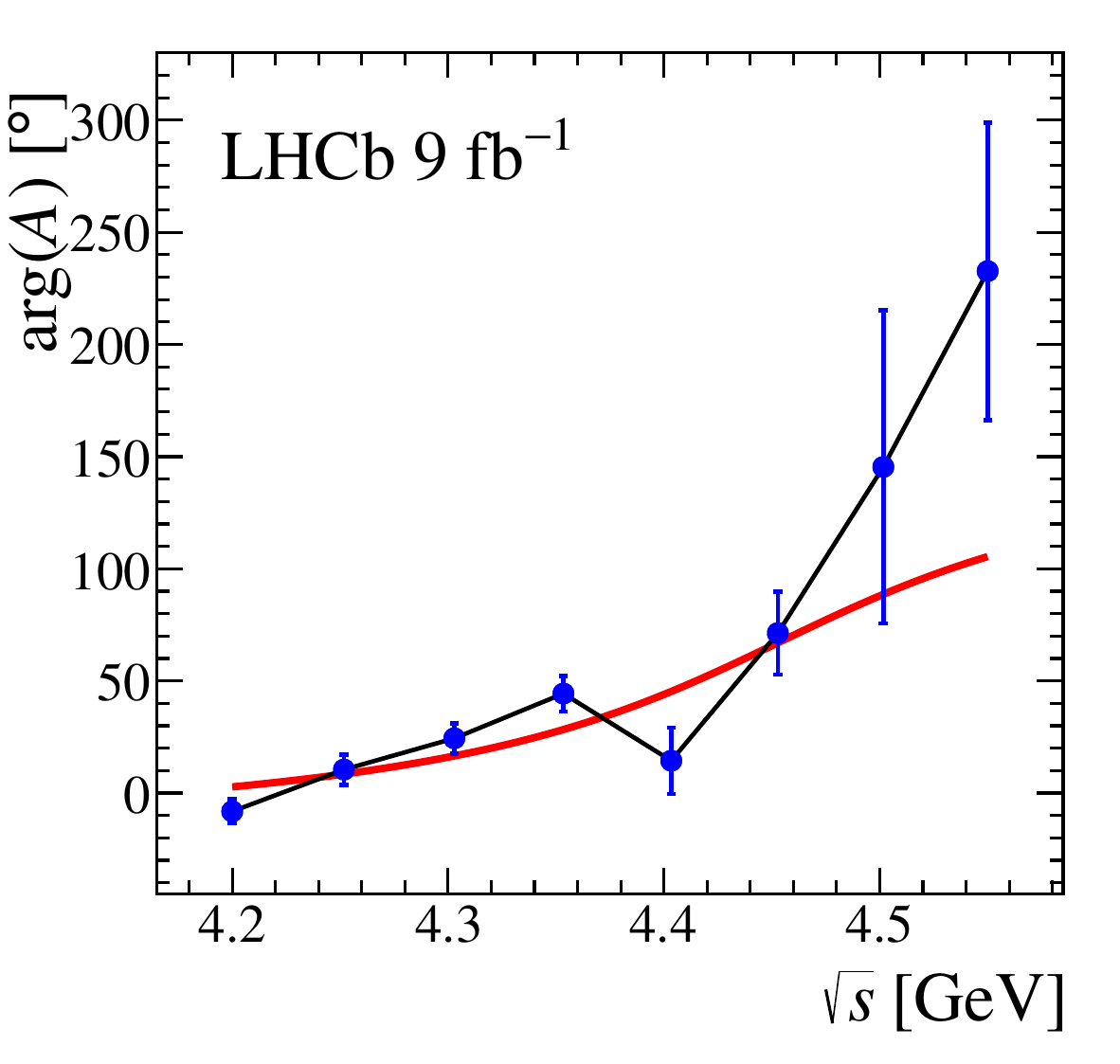}
       	 \includegraphics[width=0.31\textwidth,height=!]{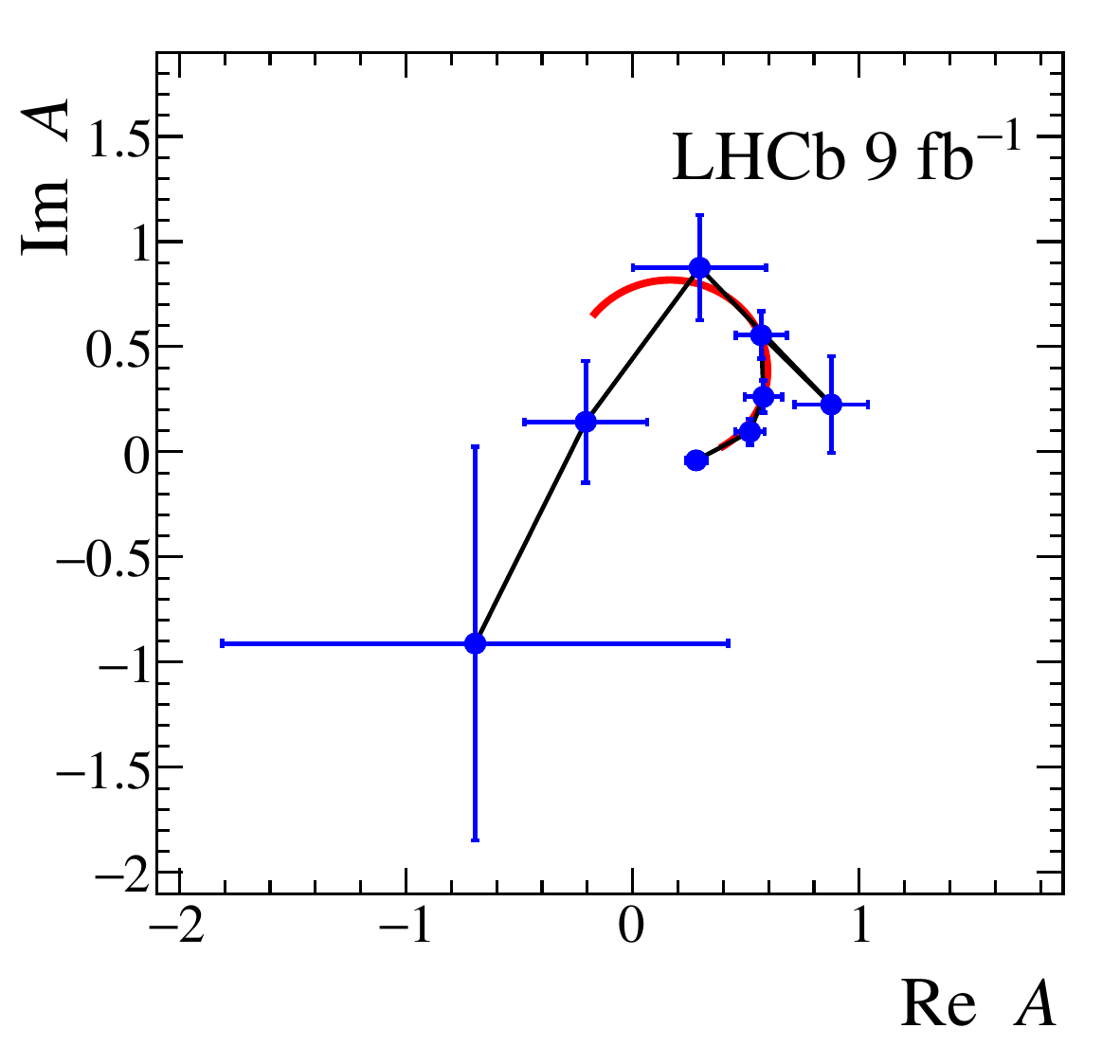}
	\caption{Lineshape of the $\ZAtwo^\pm$  resonance.}
\end{subfigure}

\begin{subfigure}[t]{\textwidth}
       	 \includegraphics[width=0.31\textwidth,height=!]{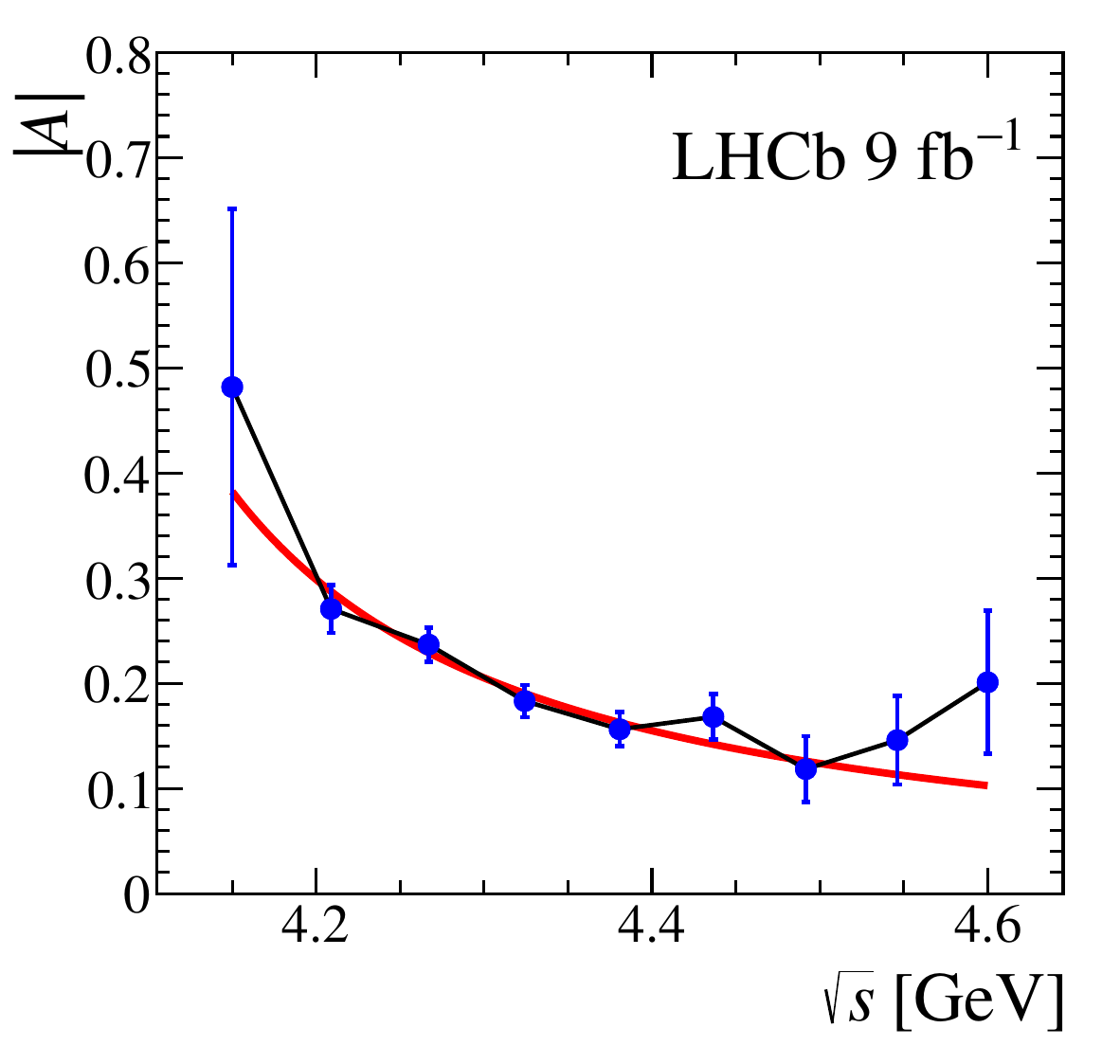}
       	 \includegraphics[width=0.31\textwidth,height=!]{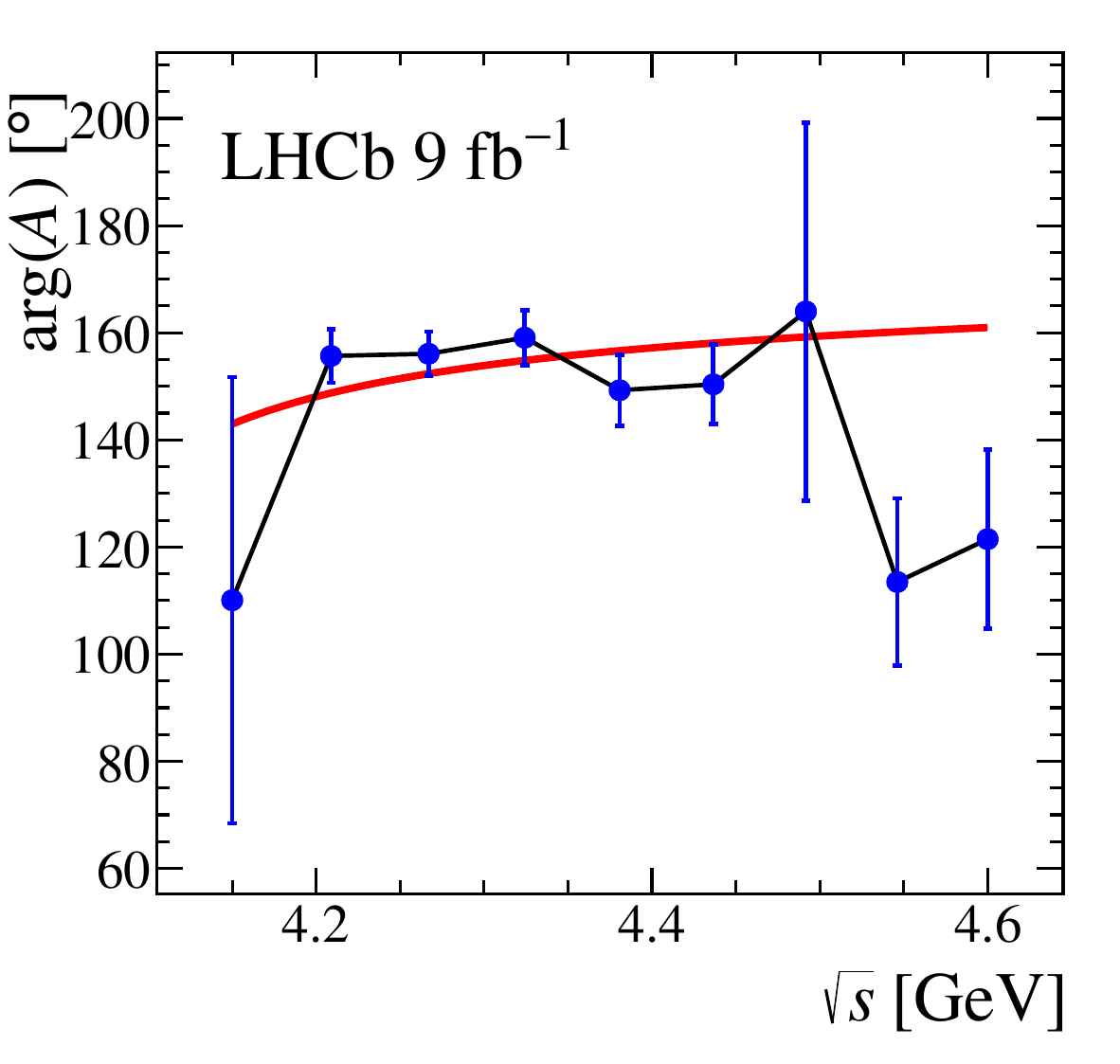}
       	 \includegraphics[width=0.31\textwidth,height=!]{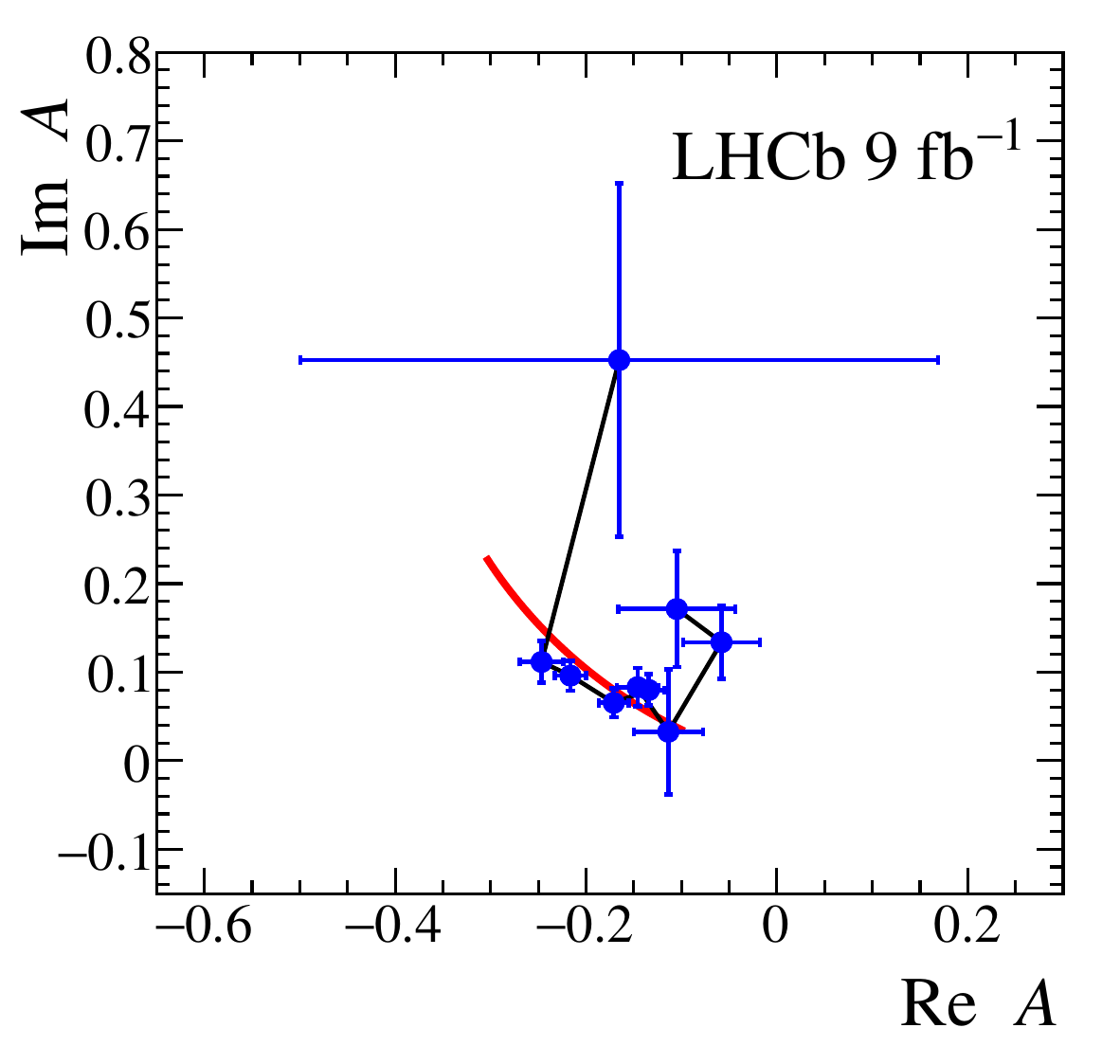}
	 	\caption{Lineshape of the   $\ZsAone$  resonance.}
\end{subfigure}

	\caption{(Left) Magnitude, (middle) phase and (right) Argand diagram of the quasi-model-independent lineshape for $\Zpm \to \psitwos \pi^\pm $ and $\Zsp\to \psitwos \Kp $ states. 
	The fitted knots are displayed as connected points with error bars.
	The Breit--Wigner lineshape with the mass and width from the nominal fit is superimposed (red line). 
 }
	\label{fig:argand3}
\end{figure}

The $\ZAone^+$ state has the second highest fit fraction among the exotic states.
The Argand diagram of the $\ZAone^+$, shown in Fig.~\ref{fig:argand3}(a), is consistent with a circular trajectory.
The quasi-model-independent lineshape of the $\ZAtwo^+$ state in Fig.~\ref{fig:argand3}(b)
is more difficult to interpret, as the mass of the $\ZAtwo^+$ state is close to the phase-space boundary and because of the large correlation with the $\ZAone^+$ state.
The pole mass of the $\ZsAone$ state is outside of the phase space such that only its tail contributes.
The corresponding quasi-model-independent lineshape is consistent with the tail of a Breit--Wigner function, see Fig.~\ref{fig:argand3}(c). 

As the investigated resonances have large decay widths and the interpolated spline function requires an extensive amount of free fit parameters (12 more than the Breit--Wigner function), 
the quasi-model-independent approach is fairly sensitive to statistical fluctuations in the data,
especially near the phase-space boundaries. 
With this in mind, the agreement with the Breit--Wigner expectation
can be considered as reasonable in all cases.

\clearpage
\addcontentsline{toc}{section}{References}
\bibliographystyle{LHCb}
\ifx\mcitethebibliography\mciteundefinedmacro
\PackageError{LHCb.bst}{mciteplus.sty has not been loaded}
{This bibstyle requires the use of the mciteplus package.}\fi
\providecommand{\href}[2]{#2}

\newpage
\centerline
{\large\bf LHCb collaboration}
\begin
{flushleft}
\small
R.~Aaij$^{36}$\lhcborcid{0000-0003-0533-1952},
A.S.W.~Abdelmotteleb$^{55}$\lhcborcid{0000-0001-7905-0542},
C.~Abellan~Beteta$^{49}$,
F.~Abudin{\'e}n$^{55}$\lhcborcid{0000-0002-6737-3528},
T.~Ackernley$^{59}$\lhcborcid{0000-0002-5951-3498},
A. A. ~Adefisoye$^{67}$\lhcborcid{0000-0003-2448-1550},
B.~Adeva$^{45}$\lhcborcid{0000-0001-9756-3712},
M.~Adinolfi$^{53}$\lhcborcid{0000-0002-1326-1264},
P.~Adlarson$^{79}$\lhcborcid{0000-0001-6280-3851},
C.~Agapopoulou$^{13}$\lhcborcid{0000-0002-2368-0147}
% ...

{\footnotesize \it

% ...
$^{13}$Universit{\'e} Paris-Saclay, CNRS/IN2P3, IJCLab, Orsay, France\\
% ...
$^{36}$Nikhef National Institute for Subatomic Physics, Amsterdam, Netherlands\\
% ...
$^{45}$Instituto Galego de F{\'\i}sica de Altas Enerx{\'\i}as (IGFAE), Universidade de Santiago de Compostela, Santiago de Compostela, Spain\\
% ...
$^{49}$Physik-Institut, Universit{\"a}t Z{\"u}rich, Z{\"u}rich, Switzerland\\
% ...
$^{53}$H.H. Wills Physics Laboratory, University of Bristol, Bristol, United Kingdom\\
% ...
$^{55}$Department of Physics, University of Warwick, Coventry, United Kingdom\\
% ...
$^{58}$School of Physics and Astronomy, University of Glasgow, Glasgow, United Kingdom\\
$^{59}$Oliver Lodge Laboratory, University of Liverpool, Liverpool, United Kingdom\\
% ...
$^{67}$Syracuse University, Syracuse, NY, United States\\
% ...
$^{79}$Department of Physics and Astronomy, Uppsala University, Uppsala, Sweden, associated to $^{58}$\\
% ...
}
\end{flushleft}

\end{document}